\pdfoutput=1
\documentclass[12pt, a4paper, twoside, openright]{report}
\newcommand{\thesisType}{M} % M - master, B - bachelor
\newcommand{\thesisAuthor}{Joel Karlsson}
\newcommand{\thesisMonth}{June}
\newcommand{\thesisYear}{2021}

\newcommand{\thesisStatus}{f} % d for draft, f for final
\newcommand{\thesisArXiv}{y} % y for arXiv version, n otherwise
\newcommand{\thesisPrint}{n} % y for print version, n otherwise

\if\thesisStatus f
    \newcommand{\thesisLayout}{2}
\else
    \newcommand{\thesisLayout}{1}
\fi

\newcommand{\thesisTitle}{Compactifications of String/M-Theory and\\[0.5ex] the Swampland}
\newcommand{\thesisImprintTitle}{Compactifications of String/M-Theory and the Swampland}

\newcommand{\thesisSubtitle}{A Study of the AdS\textsubscript{4} Mass Spectrum of Eleven-Dimensional\\[0.25ex] Supergravity on the Squashed Seven-Sphere}
\newcommand{\thesisImprintSubtitle}{A Study of the AdS\textsubscript{4} Mass Spectrum of Eleven-Dimensional\\ Supergravity on the Squashed Seven-Sphere}

\newcommand{\thesisDepartment}{Department of Physics}
\newcommand{\thesisDivision}{Division of Subatomic, High Energy and Plasma Physics}
\newcommand{\thesisGroup}{Group of Mathematical Physics}

\newcommand{\thesisUniversity}{Chalmers University of Technology}

\newcommand{\thesisCity}{Gothenburg}
\newcommand{\thesisCountry}{Sweden}
\newcommand{\thesisLocation}{\thesisCity, \thesisCountry}

\newcommand{\thesisSupervisor}{Bengt E. W. Nilsson, Department of Physics}
\newcommand{\thesisPrintedBy}{Chalmers Digital Printing} % remove this line to remove it on the imprint page

\newcommand{\thesisImprintLocation}{SE-412 96 Gothenburg}
\newcommand{\thesisUniversityTel}{+46 31 772 1000}

\newcommand{\thesisKeywords}{squashed seven-sphere, mass spectrum, flux compactification, M-theory, string theory}

\usepackage[english]{babel}         % Language settings
\usepackage[utf8]{inputenc}         % Input settings
\usepackage[T1]{fontenc}            % Output settings
\usepackage{lmodern}                % Latin modern font
\usepackage{textcomp}               % Fonts, symbols etc.
\usepackage{microtype}              % Improved micro-typography
\usepackage[top=3cm, bottom=3cm, inner=3cm, outer=3cm]{geometry}

\DeclareMicrotypeSetDefault[expansion]{alltext-nott}

\usepackage{amsmath}                % Mathematical expressions
\usepackage{amssymb}                % Mathematical symbols
\usepackage{mathtools}              % (TODO)
\usepackage{commath}                % E.g. \abs{} och \norm{}
\usepackage{dsfont}                 % (TODO)
\usepackage{bm}                     % Bold math symbols
\usepackage{gensymb}                % (TODO)
\usepackage{mathrsfs}               % (TODO)
\usepackage{tensor}                 % Index notation
\usepackage{accents}                % Math accents
\usepackage{slashed}                % Feynman slash notation
\usepackage{braket}                 % Dirac bra-ket notation
\usepackage{ytableau}               % Young tableaux

\usepackage{graphicx}               % Figures
\usepackage{float}                  % Position enforcement using [H]
\usepackage{pdfpages}               % Include PDF-pages

\usepackage{array}                  % (TODO)
\usepackage{tabularx}               % (TODO)
\usepackage{diagbox}                % Slash in tables
\usepackage{booktabs}               % Improved rules in tables

\usepackage[
    labelfont       = bf,
    font            = normalsize,
    width           = 0.94\textwidth,
    justification   = justified,
    singlelinecheck = true
]{caption}
\usepackage{silence}  % Suppress warnings (manually)
\newcommand{\thesisUseDoi}{true}
\if\thesisArXiv y
    \renewcommand{\thesisUseDoi}{false}
\fi

\usepackage[
    backend     = biber,
    style       = ieee,
    dashed      = false,
    maxnames    = 6,
    natbib      = true,
    urldate     = iso,
    seconds     = true,
    isbn        = false,
    doi         = \thesisUseDoi
]{biblatex}

\DefineBibliographyStrings{english}{%
  mathesis = {MSc thesis},
}

\usepackage{siunitx}

\DeclareSIUnit\clight{\text{\ensuremath{c}}}
\DeclareSIUnit\eVperc{\eV\per\clight}
\DeclareSIUnit\au{\text{a.u.}}

\usepackage{fancyhdr}
\usepackage{chappg}

\if\thesisLayout 1 % One-sided
\fi

\if\thesisLayout 2 % Two-sided
	\fancypagestyle{plain}{% Redefine the plain page style
	\fancyhf{}
	
	\fancyfoot[LE,RO]{\thepage}}
\fi

\usepackage[bottom, hang]{footmisc}
\newcommand{\footnotemarksep}{\textsuperscript{,}}
\makeatletter
\renewcommand*{\footnoterule}{\kern-3\p@ \hrule \kern2.6\p@}
\makeatother

\usepackage[raggedright]{titlesec}
\usepackage{parskip}                % Enables vertical spaces correctly

\usepackage{titletoc}
\usepackage[title]{appendix}

\newcommand*{\thesisChapterStyle}{1} % 0 for default

\ifnum\thesisChapterStyle=1
    \titleformat{\chapter}[hang]{\fontsize{30}{10}\selectfont}
    {{\fontsize{30pt}{1em}\vspace{-5.2ex}\selectfont \textnormal{\thechapter. \hspace{1pt}}}}
    {.5ex}{\raggedright}[\rule{\textwidth}{0.3pt}]
    \titlespacing{\chapter}{0pt}{0pt}{\parskip}
\fi

\ifnum\thesisChapterStyle=2
    \titleformat{\chapter}[display]
    {\Huge\bfseries\filcenter}
    {{\fontsize{50pt}{1em}\vspace{-4.2ex}\selectfont \textnormal{\thechapter}}}{1ex}{}[]
\fi

\if\thesisLayout 1
    \renewcommand{\cleardoublepage}{\clearpage}
\fi

\addto{\captionsenglish}{\renewcommand{\contentsname}{Table of contents}}
\addto{\captionsenglish}{}
\addto{\captionsenglish}{\renewcommand{\listtablename}{List of tables}}
\newcommand{\thesisBibName}{References}
\addto{\captionsenglish}{}

\usepackage{eso-pic}

\definecolor{headerBrown}{RGB}{144,102,78}

\if\thesisType B
    \definecolor{thesisHeaderColor}{RGB}{126,180,56} % Green
\fi

\if\thesisType M
    \definecolor{thesisHeaderColor}{cmyk}{0.14,0,0,0.65} % Gray
\fi

\makeatletter
\patchcmd\@include\@input@\input{}{}
\makeatother

\usepackage{csquotes}               % Quotations, using the command
\usepackage{lipsum}                 % Generating Lorem Ipsum
\usepackage{enumitem}               % Provides control over lists
\usepackage{xcolor}                 % Coloured text
\usepackage{hyphenat}               % Hyphenation commands

\usepackage[yyyymmdd]{datetime}     % Dates
\if\thesisStatus f
    \usepackage[textsize=tiny,disable]{todonotes}
\else
    \if\thesisStatus d
        \usepackage[textsize=tiny]{todonotes}
    \fi
\fi
\if\thesisStatus f
    \newcommand{\thesisColorlinks}{false}
\else
    \newcommand{\thesisColorlinks}{true}
\fi

\usepackage[
    hidelinks,
    linktoc     = all,
    colorlinks  = \thesisColorlinks,
    filecolor   = blue,
    linkcolor   = blue,
    urlcolor    = blue,
    citecolor   = blue,
    anchorcolor = blue,
    bookmarksopen = false,
]{hyperref}
\hypersetup{pdfinfo = {
    Author   = {\thesisAuthor},
    Title    = {\thesisImprintTitle},
    Keywords = {\thesisKeywords}
}}

\usepackage[numbered=false]{bookmark}                   % Improve bookmarks
\usepackage{url}                                        % Clickable url links
\usepackage[noabbrev]{cleveref}                         % Clever references
\crefname{equation}{}{}  % "(1.1)" instead of "equation (1.1)"

\numberwithin{equation}{chapter}    % Number equations within chapter
\numberwithin{figure}{chapter}      % Number figures within chapter
\numberwithin{table}{chapter}       % Number tables within chapter
\counterwithout{footnote}{chapter}  % Number footnotes without chapter

\renewcommand*{\i}{\mathrm{i}}
\renewcommand*{\j}{\mathrm{j}}
\renewcommand*{\k}{\mathrm{k}}

\renewcommand{\Re}{\operatorname{Re}}
\renewcommand{\Im}{\operatorname{Im}}

\newcommand{\sign}{\operatorname{sign}}
\newcommand{\trace}{\operatorname{tr}}
\newcommand{\transpose}{\mathrm{T}}
\newcommand{\Ordo}{\mathcal{O}}
\newcommand{\im}{\operatorname{im}}     % image
\newcommand{\id}{\operatorname{id}}

\newcommand*{\e}{\mathrm{e}}

\newcommand*{\1}{\mathds{1}}    % unit matrix/id operator
\renewcommand*{\vec}[1]{\boldsymbol{#1}}
\newcommand*{\diag}{\operatorname{diag}}

\renewcommand*{\d}{\mathrm{d}}        % Differential d
\newcommand*{\dd}{\delta}             % Codifferential
\newcommand*{\D}{\mathcal{D}}         % Covariant exterior derivative
\declareslashed{}{/}{.05}{0}{\D}      % Fine-tune slashed
\newcommand{\Dg}{\check{\D}}               % G2 cov. der.
\newcommand{\DDg}{\check{\square}}         % G2 cov. box
\newcommand{\Dop}{\mathfrak{D}}            % Various diff. ops.

\newcommand{\M}{\mathcal{M}}      % Manifold
\newcommand{\m}{\mathfrak{m}}     % Tangent space in coset construction
\newcommand{\R}{R}                % Curvature 2-form
\newcommand{\affine}{\varGamma}   % Christoffel symbols of affine connection

\newcommand{\N}{\mathcal{N}}

\newcommand*{\hati}{\hat{\imath}}
\newcommand*{\hatj}{\hat{\jmath}}

\newcommand{\vol}{\mathrm{vol}}
\newcommand{\uperm}{\bar{\varepsilon}}                  % Contravariant permutation symbol
\newcommand{\dperm}{\underaccent{\bar}{\varepsilon}}    % Covariant permutation symbol

\newcommand{\from}{\colon}              % Function *from*
\newcommand{\hodge}{\mathop{}\!\mathord{\star}}    % Hodge star operator

\newcommand{\SU}{{\mathrm{SU}}}
\renewcommand{\O}{{\mathrm{O}}}
\newcommand{\SO}{{\mathrm{SO}}}
\newcommand{\Spin}{{\mathrm{Spin}}}
\newcommand{\Sp}{{\mathrm{Sp}}}
\newcommand{\U}{{\mathrm{U}}}
\newcommand{\GL}{{\mathrm{GL}}}
\newcommand{\SL}{{\mathrm{SL}}}

\newcommand{\Hol}{\operatorname{Hol}}

\newcommand{\su}{{\mathfrak{su}}}
\newcommand{\so}{{\mathfrak{so}}}
\renewcommand{\sp}{{\mathfrak{sp}}}
\newcommand{\gl}{{\mathfrak{gl}}}

\newcommand{\osp}{{\mathfrak{osp}}}

\newcommand{\der}{\mathop{\mathfrak{der}}}
\newcommand{\g}{\mathfrak{g}}
\newcommand{\h}{\mathfrak{h}}
\newcommand{\hol}{\mathfrak{hol}}

\newcommand{\C}{\mathcal{C}}    % Casimir
\newcommand{\UU}{\mathcal{U}}   % Universal enveloping algebra
\newcommand{\ad}{\operatorname{ad}}     % adjoint representation

\newcommand{\ZZ}{\mathbb{Z}}

\newcommand{\RR}{\mathbb{R}}
\newcommand{\CC}{\mathbb{C}}
\newcommand{\HH}{\mathbb{H}}
\newcommand{\OO}{\mathbb{O}}

\makeatletter
\DeclareRobustCommand*\uell{\mathpalette\@uell\relax}
\newcommand*\@uell[2]{
  \setbox0=\hbox{$#1\ell$}
  \setbox1=\hbox{\rotatebox{10}{$#1\ell$}}
  \dimen0=\wd0 \advance\dimen0 by -\wd1 \divide\dimen0 by 2
  \mathord{\lower 0.1ex \hbox{\kern\dimen0\unhbox1\kern\dimen0}}
}
\makeatother

\newcommand{\cliff}{\mathrm{C}\uell}

\makeatletter
\newlength{\negph@wd}
\DeclareRobustCommand{\negphantom}[1]{%
  \ifmmode
    \mathpalette\negph@math{#1}%
  \else
    \negph@do{#1}%
  \fi
}
\newcommand{\negph@math}[2]{\negph@do{$\m@th#1#2$}}
\newcommand{\negph@do}[1]{%
  \settowidth{\negph@wd}{#1}%
  \hspace*{-\negph@wd}%
}
\makeatother

\newcommand{\eqspace}{\hphantom{{}={}}\mskip-2\thinmuskip\mathop{}}

\newcommand{\varpm}{\mathchoice%
  {\mathbin{\ooalign{\hfil$\displaystyle\pm$\hfil\cr\raise-0.20ex\hbox{\scalebox{0.4}{$\displaystyle(\mkern-1mu $}$\displaystyle\hphantom{{\pm}}$\scalebox{0.4}{$\displaystyle\mkern-1mu )$}}\cr}}}
  {\mathbin{\ooalign{\hfil$\textstyle\pm$\hfil\cr\raise-0.22ex\hbox{\scalebox{0.4}{$\textstyle(\mkern-1mu $}$\textstyle\hphantom{{\pm}}$\scalebox{0.4}{$\textstyle\mkern-1mu )$}}\cr}}}
  {\mathbin{\ooalign{\hfil$\scriptstyle\pm$\hfil\cr\raise-0.18ex\hbox{\scalebox{0.4}{$\scriptstyle(\mkern-1mu $}$\scriptstyle\hphantom{{\pm}}$\scalebox{0.4}{$\scriptstyle\mkern-1mu )$}}\cr}}}
  {\mathbin{\ooalign{\hfil$\scriptscriptstyle\pm$\hfil\cr\raise-0.145ex\hbox{\scalebox{0.4}{$\scriptscriptstyle(\mkern-1mu $}$\scriptscriptstyle\hphantom{{\pm}}$\scalebox{0.4}{$\scriptscriptstyle\mkern-1mu )$}}\cr}}}
}

\newcommand{\varmp}{\mathchoice%
  {\mathbin{\ooalign{\hfil$\displaystyle\mp$\hfil\cr\raise+0.00ex\hbox{\scalebox{0.5}{$\displaystyle(\mkern-1mu $}$\displaystyle\hphantom{{\mp}}$\scalebox{0.5}{$\displaystyle\mkern-1mu )$}}\cr}}}
  {\mathbin{\ooalign{\hfil$\textstyle\mp$\hfil\cr\raise+0.00ex\hbox{\scalebox{0.5}{$\textstyle(\mkern-1mu $}$\textstyle\hphantom{{\mp}}$\scalebox{0.5}{$\textstyle\mkern-1mu )$}}\cr}}}
  {\mathbin{\ooalign{\hfil$\scriptstyle\mp$\hfil\cr\raise-0.05ex\hbox{\scalebox{0.5}{$\scriptstyle(\mkern-1mu $}$\scriptstyle\hphantom{{\mp}}$\scalebox{0.5}{$\scriptstyle\mkern-1mu )$}}\cr}}}
  {\mathbin{\ooalign{\hfil$\scriptscriptstyle\mp$\hfil\cr\raise-0.10ex\hbox{\scalebox{0.6}{$\scriptscriptstyle(\mkern-1mu $}$\scriptscriptstyle\hphantom{{\mp}}$\scalebox{0.6}{$\scriptscriptstyle\mkern-1mu )$}}\cr}}}
}

\if\thesisStatus d % only effective in draft mode
\fi

\begin{document}

% COVER PAGE, TITLE PAGE AND IMPRINT PAGE
\if\thesisArXiv y
\else
\if\thesisPrint y
\else
\if\thesisStatus f
    \include{template/pages/cover}
\fi
\fi
\fi
\pagenumbering{roman}
% TITLE PAGE
\begingroup % make parskip changes local
\newpage
\thispagestyle{empty}
\begin{center}
    \if\thesisStatus f
        \textsc{\large
        \if\thesisType M
            Master's thesis
        \else
            Kandidatarbete
        \fi
        \thesisYear
        }
    \else
        \mbox{}
    \fi
    \\[4cm]
    \textbf{\Large \thesisTitle} \\[1cm]
    {\large \thesisSubtitle}\\[1cm]
    {\large \thesisAuthor}

    \if\thesisStatus f
        \vfill
        % Logotype on titlepage
        \begin{figure}[H]
        \centering
        % Remove this figure to remove the titlepage logotype
        \if\thesisType M
        \includegraphics[width=0.2\pdfpagewidth]{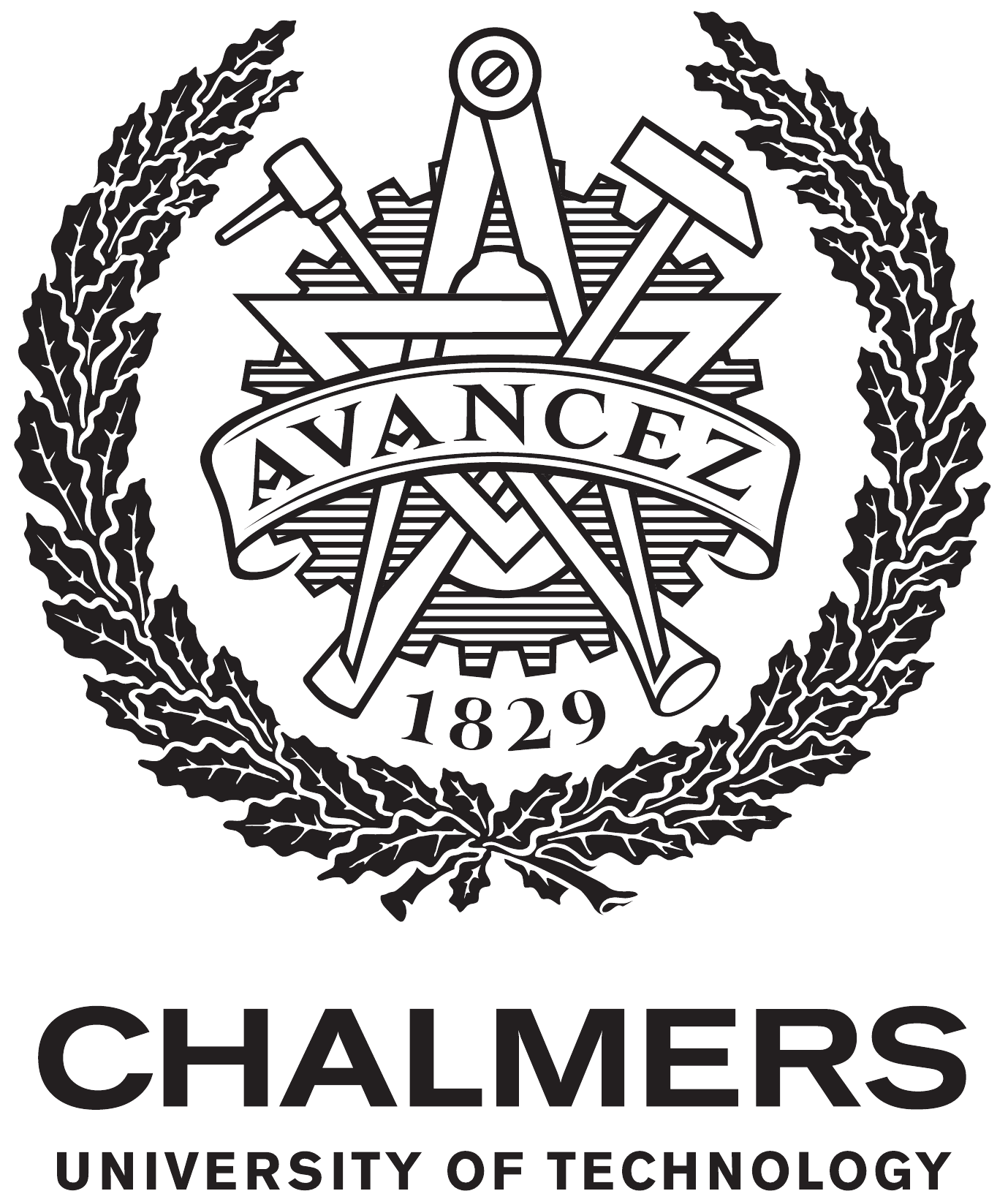} \\
        \fi
        \if\thesisType B
        \includegraphics[width=0.2\pdfpagewidth]{template/figures/AvancezChalmers_black_centered.eps} \\
        \fi
        \end{figure}
        \vspace{5mm}

        \thesisDepartment\\
        \emph{\thesisDivision}\\
        \ifx\thesisGroup\undefined
        \else
            \thesisGroup\\
        \fi
        \textsc{\thesisUniversity}\\
        \thesisLocation\ \thesisYear\\
    \else
        \vspace{5cm}
        \textbf{\Huge [DRAFT]}
    \fi
\end{center}
\endgroup

\if\thesisStatus f
    % IMPRINT PAGE (BACK OF TITLE PAGE)
\begingroup % make parskip changes local
\newpage
\thispagestyle{empty}
\vspace*{4.5cm}
\thesisImprintTitle:\\
\thesisImprintSubtitle\\[1ex]
\thesisAuthor \setlength{\parskip}{2\baselineskip}

\copyright~\thesisAuthor, \thesisYear.

\if\thesisType M
    Supervisor:
\else
    Handledare:
\fi
\thesisSupervisor\\

\if\thesisType M
    Master's Thesis
\else
    Kandidatarbete
\fi
\thesisYear\\
\thesisDepartment\\
\thesisDivision\\
\ifx\thesisGroup\undefined
\else
\thesisGroup\\
\fi
\thesisUniversity\\
\thesisImprintLocation\\
\if\thesisType M
    Telephone:
\else
    Telefon:
\fi
\thesisUniversityTel

\vfill
\ifx\thesisCoverFigure\undefined
\else
    \if\thesisType M
        Cover:
    \else
        Omslag:
    \fi
    \thesisCoverFigureCaption
    \setlength{\parskip}{\baselineskip}
\fi

\if\thesisType M
    Typeset in
\else
    Typsatt i
\fi
\LaTeX\\
\ifx\thesisPrintedBy\undefined
\else
    \if\thesisType M
        Printed by
    \else
        Tryckt av
    \fi
    \thesisPrintedBy\\
\fi
\thesisLocation\ \thesisYear
\endgroup

\fi

% ABSTRACT & ACKNOWLEDGEMENTS
\thesisImprintTitle:\\
\thesisImprintSubtitle\\[1ex]
\thesisAuthor\\
\thesisDepartment\\
\thesisUniversity

\thispagestyle{plain}           % Suppress header
\section*{\abstractname}
The landscape of possible four-dimensional low-energy effective theories arising from compactifications of string/M-theory seems vast.
This might lead one to believe that any consistent-looking effective field theory coupled to gravity can be obtained as a low-energy limit of string theory.
However, a set of ``swampland'' conjectures suggests that this is not true and that, in fact, there is an even larger set of effective field theories that cannot be obtained in this way.
In particular, the AdS instability swampland conjecture asserts that nonsupersymmetric anti-de Sitter vacua are unstable.
These swampland criteria can have implications for, for instance, low-energy physics and cosmology.

M-theory is a nonperturbative unification of all superstring theories.
Its low-energy limit, eleven-dimensional supergravity, admits two compactifications on the squashed seven-sphere.
One of the solutions has one unbroken supersymmetry ($\N = 1$) while the other has none ($\N = 0$).
Due to the AdS instability swampland conjecture, the latter should be unstable.
However, this has not been demonstrated explicitly.
To study the stability of the $\N = 0$ vacuum, we investigate the mass spectrum of the theory.
The main advancement compared to previous attempts is the realisation that all mass operators in the Freund--Rubin compactification are related to a universal Laplacian, allowing us to relate Weyl tensor terms to group invariants.
One limitation of the group-theoretical method we employ is that it can lead to false roots.
This is remedied, at least in part, by demanding that the fields form supermultiplets in the $\N = 1$ case.
Although we arrive at an eigenvalue spectrum for all operators of interest, there is a hint that the results may be incomplete.
Thus, we do not reach a decisive conclusion regarding the investigated type of instability.

\if\thesisType M
    \textbf{Keywords:}
\else
    \textbf{Nyckelord:}
\fi
\thesisKeywords.

% NOTE: this needs modification if the abstract is longer than one page
% (which it shouldn't be)
\if\thesisLayout 2
\newpage                % Create blank page
\thispagestyle{empty}
\mbox{}
\fi

\thispagestyle{plain}           % Suppress header
\section*{Acknowledgements}
I would like to thank my supervisor, Bengt E. W. Nilsson, for his support and for suggesting the project.
Bengt has always taken the time to answer and discuss the questions I have had.
I would also like to thank my fellow students for interesting discussions on mathematical and physical subjects, in particular Rolf Andréasson, Markus Klyver, Ludvig Svensson and Eric Nilsson.
An additional thank you to Ludvig for providing useful comments on my text.
Lastly, a big thank you to all of my friends and family for their support.

%\vspace{0.25cm}
\hfill
\thesisAuthor, \thesisCity, \thesisMonth\ \thesisYear

% NOTE: this needs modification if there is more than one page of
% acknowledgements (which is probably too much)
\if\thesisLayout 2
\newpage                % Create blank page
\thispagestyle{empty}
\mbox{}
\fi

% CONTENTS LIST (TABLE OF CONTENTS, FIGURES, TABLES)
% Use a separate file to be able to exclude with \includeonly.
\cleardoublepage
\pdfbookmark[chapter]{\contentsname}{toc}
\begingroup
\titlespacing*{\chapter}{0pt}{0pt}{2.5pt}
\tableofcontents
\endgroup

% List of figures (add to table of contents)
% \cleardoublepage
% \phantomsection
% \addcontentsline{toc}{chapter}{\listfigurename}
% \listoffigures

% List of tables (add to table of contents)
\cleardoublepage
\phantomsection
\addcontentsline{toc}{chapter}{\listtablename}
\listoftables

% The page numbering is changed after this, whence we need a \cleardoublepage.
% We put it here to be able to exclude it with \includeonly.
\cleardoublepage

% BEGINNING OF MAIN DOCUMENT
\pagenumbering{arabic}  % Arabic numbering starting from 1 (one)

% CHAPTERS
\chapter{Introduction}
The perhaps greatest challenge in fundamental physics is quantum gravity.
During the twentieth century, two theories transformed the scientific worldview radically.
The first, Einstein's theory of relativity, tells us that space and time are not absolute but relative to the observer and that only spacetime, not space and time separately, is physically meaningful (special relativity).
Furthermore, spacetime is dynamical: it curves as a result of the presence of matter and energy and gravity is a manifestation of this curvature (general relativity).
The second is quantum theory%
\footnote{Quantum theory is not a physical theory but a theoretical framework. Similarly, there is a theoretical framework subsuming general relativity.},
describing matter by wavefunctions with nonclassical properties such as entanglement and only giving probabilistic predictions for observable measurement outcomes.

General relativity and the Standard Model of particle physics, a (special-)relativistic quantum field theory, are often considered the two most successful theories in all of physics.
Yet, they seem incompatible \cite{ref:Becker--Becker--Schwarz}.
Due to the nonclassical properties of reality, demonstrated for instance by experiments that violate the Bell inequality \cite{ref:Bell}, the consensus is that gravity has to be quantum or emergent from more fundamental quantum degrees of freedom.
Still, the quantum nature of gravity remains unclear.
There are, however, several approaches to quantum gravity, including string/M-theory \cite{ref:Becker--Becker--Schwarz} and loop quantum gravity \cite{ref:Rovelli}.
This thesis is concerned with the former.

String theory is a framework that generalises quantum mechanics to extended one-dimensional objects, strings.%
\footnote{String theory also contains higher-dimensional extended objects known as branes.}
There are five (critical) string theories with fermions, known as Type I, Type IIA, Type IIB, heterotic $\SO(32)$ and heterotic $E_8 \times E_8$, which all live in ten dimensions \cite{ref:Becker--Becker--Schwarz}.
These are related by various dualities, for instance T-duality (inverting the radius of a compactified dimension) and S-duality (inverting the string coupling constant) \cite{ref:Witten:revolution}.
Thus, they may be viewed as perturbative regimes of an underlying nonperturbative theory known as M-theory \cite{ref:Duff:M-theory}.
Remarkably, the low-energy limit of M-theory is eleven-dimensional supergravity, which, as the name suggests, is formulated in eleven dimensions rather than the ten of the superstring theories \cite{ref:Witten:revolution}.
Eleven is special in the sense that it is the maximum number of dimensions in which one can have supersymmetry without particles with spins greater than two and the highest dimension that admits super $p$-branes~\cite{ref:Duff:M-theory}.%
\footnote{This depends crucially on the assumption of Minkowski signature and can be avoided in twelve dimensions with two timelike directions, a fact that is used in F-theory \cite{ref:Duff:M-theory,ref:Vafa--Brennan--Carta}.}

One may ask how string theory in ten dimensions and supergravity in eleven can have any prospect of describing the four-dimensional universe we perceive, with three dimensions of space and one of time.
One possible answer is that some of the dimensions are small and compact and, therefore, not observed in experiments.
This is known as Kaluza--Klein compactification \cite{ref:Becker--Becker--Schwarz}.
A useful analogy is the surface of a rope, which looks one-dimensional from afar but is really two-dimensional.
Another, not mutually exclusive, possibility is the brane-world scenario, in which our four-dimensional spacetime corresponds to a defect created by branes \cite{ref:Becker--Becker--Schwarz}.%
\footnote{In such scenarios, there may be noncompact extra dimensions. If the space is a warped product, one can still get four-dimensional gravity with an inverse-square law \cite{ref:Becker--Becker--Schwarz}.}

Although compact dimensions can solve the problem of why we do not observe ten or eleven dimensions in experiments, it also gives rise to another problem, namely, the problem of choosing a compact manifold.
When compactifying M-theory to four dimensions, one obtains a low-energy effective theory which depends on the specifics of the compactification.
One estimate suggests that there is at least \num{e272000} such vacua \cite{ref:Taylor:10^272000_F-vacua}, although it is unclear whether they are all distinct or if some are related by dualities \cite{ref:Vafa--Brennan--Carta}.
It is then reasonable to ask whether any consistent-looking effective field theory coupled to gravity can be obtained as a low-energy limit of an M-theory vacuum.
A set of ``swampland'' conjectures claims, to the contrary, that the space of effective field theories that cannot be obtained in this way is even larger \cite{ref:Vafa:swampland_original,ref:Vafa--Brennan--Carta}.
In contrast to the string landscape, these are said to belong to the swampland.

When compactifying eleven-dimensional supergravity to four dimensions, the internal space, that is, the compact dimensions, is of course seven-dimensional.
Thus, the seven-sphere, $S^7$, may be used as internal space.
In the Freund--Rubin ansatz, a special form of flux compactification, the compact manifold has to be an Einstein space, $R_{mn} \propto g_{mn}$, to satisfy the supergravity field equations.
The seven-sphere admits two Einstein metrics, the usual maximally symmetric \emph{round} one and a \emph{squashed} metric with fewer isometries.
There are two vacua with the squashed seven-sphere as internal space, related by ``skew-whiffing'', that is, by reversing the direction of the flux.
One of these has one unbroken supersymmetry while the other has none \cite{ref:Duff--Nilsson--Pope:squashed_S7}.
We will refer to the $\N=1$ solution as the left-squashed vacuum and the $\N=0$ solution as right-squashed.
Since the metric on the internal space is a spacetime scalar, the four-dimensional theories obtained from the squashed sphere can be viewed as spontaneously broken phases of the $\N=8$ maximally supersymmetric theory obtained from the round $S^7$ \cite{ref:Duff--Nilsson--Pope:squashed_S7,ref:Nilsson--Padellaro--Pope}.%
\todo[disable]{}

One swampland conjecture, which we will refer to as the AdS instability swampland conjecture, asserts that nonsupersymmetric anti-de Sitter (AdS) vacua are unstable and, hence, belong to the swampland \cite{ref:Ooguri--Vafa:AdS}.
Accordingly, the right-squashed vacuum described above, which is AdS, is expected to be unstable.
However, no instability has been explicitly demonstrated for this vacuum.
For instance, due to the relation with the left-squashed $\N=1$ vacuum, the Breitenlohner-Freedman bound \cite{ref:Breitenlohner--Freedman:short,ref:Breitenlohner--Freedman:long} is not violated \cite{ref:Duff--Nilsson--Pope:vacuum_stability}.
If present, the instability must, therefore, arise in some other way.
One possibility is that the vacuum is shifted significantly by an instability indicated by a tadpole.
For the shift to be significant, the field with the tadpole has to correspond to a global singlet marginal operator (GSMO) in the conformal field theory (CFT) dual to the supergravity theory \cite{ref:Berkooz--Rey}.
Other possibilities for instabilities include the formation of a ``bubble of nothing'' \cite{ref:Witten:bubble_of_nothing} and brane-jet instabilities~\cite{ref:Bena--Pilch--Warner}.

In this thesis, we focus on tadpole instabilities related to GSMOs.
As explained in \cite{ref:Murugan}, such instabilities can occur not only for the elementary scalar fields in the theory but also for composite fields.
Hence, one needs detailed information of considerable parts of the Kaluza--Klein mass spectrum to investigate whether any GSMO-related instability occurs.
Since the masses are determined by the eigenvalue spectra of certain differential operators on the internal space \cite{ref:Duff--Nilsson--Pope:KK}, we aim to derive said spectra.

\section{Thesis outline} \label{sec:outline}%
The thesis is structured as follows.
In the remainder of this chapter, we provide some context for the problem and elaborate on the kind of instability we are going to investigate.
The reader is assumed to be familiar with quantum field theory and general relativity.
Also, familiarity with some aspects of string theory is assumed, at least on a conceptual level.
Detailed knowledge of supersymmetry and supergravity is not a prerequisite and we, therefore, present some relevant background, including the superspace construction of eleven-dimensional supergravity, in \cref{chap:susy_sugra}.

In \cref{chap:sugra_comp}, we discuss compactifications of eleven-dimensional supergravity, in particular the Freund--Rubin ansatz.
We also discuss the concept of mass in AdS and present the expressions for the mass operators, in terms of differential operators on the internal space, and the $\N=1$ supermultiplets in $\mathrm{AdS}_4$.
In the last section of the chapter, \cref{sec:ads_susy:Laplacian}, we relate the mass operators to a universal Laplacian that later proves to be of great use.
This realisation is not present in the literature as far as we know.

In \cref{chap:coset}, we review aspects of the geometry of and harmonic analysis on homogeneous spaces $G/H$.
Some of the material is based on \cite{ref:Bais--Nicolai--van_Nieuwenhuizen} but we generalise the discussion to an arbitrary $G$-invariant metric.
We also derive the equations, in \cref{sec:coset:master_eq}, that are later used to find the eigenvalue spectra of the squashed seven-sphere.

Two constructions of the squashed-seven sphere are presented in \cref{chap:squashed_geometry}.
We use the above to realise the squashed $S^7$ with arbitrary squashing parameter as a coset space $G/H$ with $G=\Sp(2)\times\Sp(1)$ and $H\simeq\Sp(1)\times\Sp(1)$.
The Einstein-squashed seven-sphere, on which we compactify eleven-dimensional supergravity, is of course of particular interest.
However, we also see that the round $S^7$ comes from a metric of indefinite signature on $G$.

The eigenvalue spectra of all operators of interest on the (Einstein-)squashed seven-sphere are derived in \cref{chap:squashed_spectrum}, using the coset construction of the previous chapter and the equations derived in \cref{sec:coset:master_eq}.
We do not solve any differential equations explicitly but rather use group theoretical techniques.
A limitation of the method is that it can lead to false roots.

Recently, large parts of the eigenvalue spectra of the squashed seven-sphere were uploaded to arXiv \cite{ref:Ekhammar--Nilsson}, although substantial parts were known before that as well \cite{ref:Nilsson--Pope,ref:Yamagishi,ref:Duff--Nilsson--Pope:KK}.
Our calculations, which are completely independent, agree with all previous results and extend them by providing eigenvalues for $\i\slashed\D_{3/2}$ independent of supersymmetry requirements.

Having investigated the squashed seven-sphere in detail, we return to eleven-dimen\-sion\-al supergravity and derive the mass spectrum and supermultiplet structure upon compactification on the squashed $S^7$ in \cref{chap:masses_and_susy}.
We focus on the left-squashed $\N=1$ vacuum and use the fact that the fields must fall into supermultiplets to eliminate false roots.
In this way, we arrive at eigenvalue spectra that are consistent with supersymmetry.
However, there seem to be degeneracies that we have not been able to explain.
This could be taken as an argument for our results being incomplete.

The analysis regarding whether GSMO-related instabilities can occur in the right-squashed vacuum is not completed.
We end with some concluding remarks and a discussion on what is needed for this in \cref{chap:conclusions}.
Further results will hopefully be presented in a future publication \cite{ref:Karlsson--Nilsson}.

Conventions, notation and some mathematical preliminaries are presented in \cref{app:conventions,app:spinors,app:octonions,app:conventions:diff_forms,app:bundles_gauge_theory_gravity,app:Grassmann}.
In \cref{app:sugra_calc} we solve the Bianchi identities of eleven-dimensional supergravity, a calculation too long to fit in the main text.

We do not attempt to always give complete lists of original references.
In particular, several textbooks are used for well-known results, especially when of mathematical nature.
Also, the review article \cite{ref:Duff--Nilsson--Pope:KK} is used for large parts of the background material regarding compactifications of eleven-dimensional supergravity.

\section{The swampland}
As already mentioned, the swampland program aims to demonstrate that not all consistent-looking effective field theories (EFTs) coupled to gravity can be obtained from M-theory compactifications.
Theories that cannot be obtained in this way are thought to not admit a finite UV-completion and are therefore considered inconsistent \cite{ref:Vafa--Brennan--Carta}.
Some swampland conjectures are concerned with quantum gravity more generally while others are concerned with string/M-theory more specifically.
In this section, we provide some background regarding the swampland program to put the project in a wider context and expand on some details concerning GSMO-related instabilities.

Before turning to specific conjectures and their consequences, we note that many of the swampland conjectures have not been rigorously proven.
Indeed, the lack of a complete, nonperturbative definition of M-theory is a significant obstruction to such proofs.
Instead, the conjectures are motivated by, among other things, realisations and examples from string theory and black hole physics \cite{ref:Vafa--Brennan--Carta}.
To be able to claim predictions from string theory based on swampland conjectures, one must be confident enough of the validity of the relevant conjectures, within the theoretical framework, to believe that the fault lies in string theory itself, and not the conjectures, in the event that repeated experiments would violate the prediction.
Otherwise, the experiments can only be said to test the swampland conjectures and not string theory.
This motivates further theoretical study of the subject to refine and strengthen the conjectures.

\subsection{The weak gravity and AdS instability swampland conjectures}
The weak gravity conjecture states, loosely, that gravity is the weakest force in any consistent theory of quantum gravity \cite{ref:Arkani-Hamed--et_al:WGC}.
More specifically, for a $\U(1)$ gauge field, the conjecture asserts that there must be a state of mass $M$ and charge $Q$ such that
\begin{equation}
\label{eq:swampland:WGC}
    \frac{M}{M_\mathrm{Pl}} \leq |Q|,
\end{equation}
where $M_\mathrm{Pl}$ is the Planck mass \cite{ref:Arkani-Hamed--et_al:WGC}.%
\todo[disable]{}
Several arguments motivate the conjecture \cite{ref:Arkani-Hamed--et_al:WGC,ref:Vafa--Brennan--Carta,ref:Ooguri--Vafa:AdS}.
Firstly, it agrees with what is observed in nature and known string theory compactifications \cite{ref:Arkani-Hamed--et_al:WGC}.
It can also be motivated by black hole physics as follows.
A Reissner--Nordström black hole, that is, a charged black hole in 4 dimensions, is described by the metric
\begin{equation}
    \d s^2
    = - \Bigl(1 - 2\:\! \frac{r_M}{r} + \frac{r_Q^2}{r^2} \Bigr) \d t^2
    + \Bigl(1 - 2\:\! \frac{r_M}{r} + \frac{r_Q^2}{r^2} \Bigr)^{-1} \d r^2
    + r^2 \d \Omega_2,
\end{equation}
where $r_M = \ell_\mathrm{Pl}\, M/M_\mathrm{Pl} = r_\mathrm{S}/2$ is half of the Schwarzschild radius, $r_Q = \ell_\mathrm{Pl} Q$ is the length scale associated with the charge and $\d \Omega_2$ is the usual metric on $S^2$.
We see from the metric that there are horizons at
\begin{equation}
    r = r_M \pm \sqrt{r_M^2 - r_Q^2}.
\end{equation}
If we imagine turning up the charge from $Q=0$, corresponding to a Schwarzschild black hole, the outer horizon shrinks while the inner one grows.
When $r_Q^2 = r_M^2$, corresponding to $M/M_\mathrm{Pl} = |Q|$, the two coincide and the black hole is said to be extremal.
For larger charges, there is no horizon and we get a naked singularity, violating the cosmic censorship hypothesis.
Thus, for extremal black holes to be able to evaporate via Hawking radiation without violating the cosmic censorship hypothesis, there has to be a state satisfying \cref{eq:swampland:WGC}.
For macroscopic black holes with $M \gg M_\mathrm{Pl}$, such evaporation is expected to be possible to avoid large numbers of Planck scale black hole remnants and large numbers of exactly stable objects not protected by symmetry \cite{ref:Arkani-Hamed--et_al:WGC}.
This does not apply to Bogomol'nyi--Prasad--Sommerfield (BPS) states, which saturate \cref{eq:swampland:WGC} and whose stability is protected by supersymmetry \cite{ref:Ooguri--Vafa:AdS,ref:Witten:revolution}.
The weak gravity conjecture applies to charged branes in string theory as well; the gravitational attraction is conjectured to be weaker or equally strong as the electric repulsion \cite{ref:Ooguri--Vafa:AdS}.

There is a sharpened version of this conjecture that states that equality only occurs for BPS states in supersymmetric theories \cite{ref:Ooguri--Vafa:AdS}.
This can be motivated by the argument presented above since the phase space of the emission of particles saturating \cref{eq:swampland:WGC} vanishes \cite{ref:Ooguri--Vafa:AdS}.
Note that not all states must satisfy \cref{eq:swampland:WGC}; it is only required that there exists some state satisfying it.

The sharpened weak gravity conjecture has a consequence related to nonsupersymmetric AdS and holography.
To see this, we consider Maldacena's \cite{ref:Maldacena} original construction of the AdS/CFT correspondence.
The conformal field theory lives on a stack of $N$ coincident branes and decouples from the bulk theory in the low-energy limit.
The geometry is described by a black brane supergravity solution, valid for large $N$, whose near-horizon limit typically reduces to a product of AdS and a sphere.
In the nonsupersymmetric setting, the electric repulsion between the branes is stronger than the gravitational attraction, by the sharpened weak gravity conjecture \cite{ref:Ooguri--Vafa:AdS}.
This renders the system unstable.
Furthermore, the lifetime approaches zero in the near-horizon limit due to the gravitational time dilation \cite{ref:Ooguri--Vafa:AdS}.

Although the above depends somewhat on the specific construction, \cite{ref:Ooguri--Vafa:AdS} conjectures that nonsupersymmetric AdS holography with a low-energy description in terms of finitely many matter fields coupled to Einstein gravity belongs to the swampland.
If true, this means that all nonsupersymmetric AdS vacua of M-theory are unstable.
We refer to this as the AdS instability swampland conjecture%
\footnote{Not to be confused with the (gravitational) AdS instability conjecture, which asserts that AdS is unstable to black hole formation under arbitrary small perturbations \cite{ref:Bizon--Rostworowski,ref:Bizon} and has been proven for some gravity-matter systems \cite{ref:Moschidis:1,ref:Moschidis:2}.}.

\subsection{The swampland, the Standard Model and cosmology}
Some swampland conjectures can be related to the Standard Model of particle physics or cosmology.
Here, we describe a couple of examples of this.
The first example comes from considering compactifications of the Standard Model to three or two dimensions.
At first, this might seem peculiar but if the Standard Model can be obtained from M-theory so can its compactifications.
Thus, if stable AdS vacua can be obtained from compactifications of the Standard Model, the Standard Model itself belongs to the swampland according to the AdS instability swampland conjecture.
Aspects of this are investigated in \cite{ref:Arkani-Hamed--et_al:SM_compactification,ref:Ibanez,ref:Hamada--Shiu}.
In particular, \cite{ref:Arkani-Hamed--et_al:SM_compactification,ref:Ibanez} find that the Standard Model augmented with Majorana neutrino masses gives rise to AdS vacua after compactification with current values of the neutrino masses and the cosmological constant.
However, \cite{ref:Ibanez} also finds that this can be avoided by adding a light beyond-the-Standard-Model particle or if the neutrinos are Dirac fermions and the lightest neutrino is sufficiently light.%
\footnote{Note that the smallest neutrino mass is not bounded from below by current neutrino oscillation experiments.}
As pointed out in \cite{ref:Ibanez}, the above argument can be reversed to provide a lower bound on the cosmological constant based on the neutrino masses.
Note that the above constraints only apply if the AdS vacuum obtained from compactification of the Standard Model is stable, as discussed in~\mbox{\cite{ref:Ibanez,ref:Hamada--Shiu}}.
\clearpage

The implications of swampland conjectures have also been studied in the context of cosmology.
This is based, for instance, on the swampland conjecture known as the refined de Sitter conjecture which states that the effective low-energy potential $V(\phi)$ for scalar fields $\phi_i$ must satisfy
\begin{equation}
\label{eq:swampland:dS}
    \| \nabla V \| \geq \frac{c}{M_\mathrm{Pl}} V
    \qquad\text{or}\qquad
    \min \nabla^2 V \leq - \frac{c'}{M_\mathrm{Pl}^2} V,
\end{equation}
for universal positive constants $c, c' \sim \Ordo(1)$, in any consistent theory of quantum gravity \cite{ref:Garg--Krishnan,ref:Obied--el_al,ref:Ooguri--et_al}.
In these expressions, $\| \nabla V \|^2 = g^{ij}(\phi) \nabla_i V \nabla_j V$ where $g^{ij}(\phi)$ is the field-space metric from the kinetic term in the Lagrangian, that is, $\mathcal{L}_\mathrm{kin.} = - g^{ij}(\phi)/2\, \partial_\mu \phi_i \partial^\mu \phi_j$, and the minimum refers to the minimum eigenvalue of the Hessian $\nabla^2 V$ in an orthonormal field-frame.
The name of the conjecture comes from the fact that it excludes (meta-)stable de Sitter vacua \cite{ref:Garg--Krishnan,ref:Ooguri--et_al}.
If true, this would imply that the state of the universe is unstable.
One possibility is a quintessence model where the cosmological ``constant'' asymptotically goes to zero with cosmic time, which would have consequences for the dark energy equation of state \cite{ref:Vafa--Brennan--Carta}.
Notably, recent studies based on observations suggest an evolving equation of state for dark energy \cite{ref:Demianski}.
However, in \cite{ref:Akrami--et_al}, it is argued that a metastable de Sitter vacuum produced by the KKLT (Kachru--Kallosh--Linde--Trivedi) mechanism \cite{ref:KKLT} avoids the problems motivating the swampland conjecture and KKLT could, therefore, still be a viable mechanism for producing de Sitter vacua in string theory.
They further note that many proposed quintessence models, including those presented in \cite{ref:Obied--el_al}, are excluded at high significance by cosmological data.

In \cite{ref:Kinney--Vagnozzi--Visinelli}, the refined de Sitter conjecture is applied to single-field inflation models.
They consider, in particular, the ratio between scalar and tensor modes in primordial fluctuations, $r$, and the scalar spectral index, $n_\mathrm{s}$, parameterising the scale dependence of the scalar fluctuations.
For consistency between observational data and the single-field slow-roll inflation model, they find that $c \lesssim \Ordo(0.1)$ or $c' \lesssim \Ordo(0.01)$ depending on which inequality in \cref{eq:swampland:dS} applies to the inflaton potential.
Depending on the precision of the statement that $c, c' \sim \Ordo(1)$ in the swampland conjecture, this result is in considerable tension with the conjecture.
As stated above, the refined de Sitter conjecture should apply to this model.
The validity of this application has however been questioned in \cite{ref:Akrami--et_al}.

\subsection{Unstable nonsupersymmetric AdS vacua and GSMOs} \label{sec:GSMO}%
As already mentioned, the AdS instability swampland conjecture asserts that nonsupersymmetric AdS belongs to the swampland.
Here, we elaborate on the kind of instability we aim to investigate for the squashed seven-sphere compactification of eleven-dimensional supergravity, namely, instabilities related to tadpoles and global singlet marginal operators (GSMOs).

Consider an M-theory vacuum dual to a conformal field theory (CFT) in the $N\to\infty$ limit.
At large $N < \infty$, there can be tadpoles in the supergravity theory, corresponding to $1/N$ corrections of the $\beta$-functions of the dual CFT, that signifies a shift of the true vacuum \cite{ref:Murugan}.
This does not happen in supersymmetric cases but can lead to an instability for nonsupersymmetric theories \cite{ref:Murugan}.
The question is then whether there is a $1/N$ perturbed vacuum close by in parameter space or not, the latter signalling instability.
In \cite{ref:Berkooz--Rey}, it is argued that only fields dual to operators that are marginal in the $N \to \infty$ limit can give rise to such instabilities.
This is based on the $1/N$ expansion of the $\beta$-functions, which is of the form
\begin{equation}
\label{eq:GSMO:beta}
    \beta(g) \sim (\Delta - d) (g- g_\ast) + \frac{a}{N} + \hdots,
\end{equation}
for a coupling constant $g\sim g_\ast$ of some operator $O$, where $g=g_\ast$ at the $N\to\infty$ conformal fixed point, $d$ is the spacetime dimension of the CFT and $\Delta$ is the scaling dimension of the operator $O$ \cite{ref:Murugan}.
From this, we see that the correction to the fixed point is of order $1/N$ and, thus, goes to zero in the large $N$ limit, as long as $\Delta - d \neq 0$.
For a marginal operator, $\Delta = d$ and the $1/N$ correction of the $\beta$-function may eliminate the conformal fixed point completely.
This implies that the limit $N\to\infty$ cannot be taken smoothly since the theory would flow to a point far from $g_\ast$ for all finite $N$ \cite{ref:Murugan}.

If the $\beta$-function of a marginal operator has a saddle point at $g_\ast$ or the $1/N$ correction is in the right direction in the case of a local extremum, the fixed point would only receive a small $1/N$ correction \cite{ref:Berkooz--Rey}.
Thus, the presence of a marginal operator does not imply that the vacuum is unstable \cite{ref:Murugan}.

A tadpole instability can only develop for fields that are neutral with respect to the gauge symmetries in the supergravity theory since it would, otherwise, explicitly break gauge invariance \cite{ref:Berkooz--Rey}.
Thus, the corresponding operator need not only be marginal but also invariant under the global symmetries of the CFT, that is, it has to be a GSMO \cite{ref:Berkooz--Rey,ref:Murugan}.
However, as emphasised in \cite{ref:Murugan}, the GSMO can be a multi-trace operator, corresponding to a composite field in the supergravity theory.

To be able to examine the presence of GSMO-related instabilities from the supergravity side, we need the relation between the scaling dimension $\Delta$ and properties of the fields in AdS.
For this, the picture of the AdS/CFT correspondence presented in \cite{ref:Witten:AdS/CFT,ref:Gubser--Klebanov--Polyakov}, where it was realised that the CFT lives on the conformal boundary of AdS, is useful.%
\footnote{Note that the Lie algebras of the isometry group $\SO(d-1, 2)$ of $\mathrm{AdS}_d$ and the conformal group $\mathrm{Conf}(d-2,1)$ are isomorphic.}
The scaling dimension of an operator is then seen to be related to the asymptotic behaviour of the dual field \cite{ref:Witten:AdS/CFT} and coincides with the dimensionless energy $E_0$ of the field \cite{ref:Murugan}.
Thus, in the case of an $\mathrm{AdS}_4$ vacuum, GSMOs correspond to (possibly composite) fields with $E_0 = 3$.
In \cite{ref:Murugan}, an argument that such a GSMO is always present in nonsupersymmetric vacua related to $\N \geq 2$ vacua by skew-whiffing is presented.
However, the situation for the $\N=0$ vacuum of eleven-dimensional supergravity compactified on the squashed seven-sphere, whose skew-whiffed partner has $\N=1$, remains unclear.

\chapter{Supersymmetry and supergravity} \label{chap:susy_sugra}
Supersymmetry is a symmetry that relates bosons and fermions, that is, commuting and anticommuting fields.
It has been proposed as a possible solution to a number of current problems, for instance, dark matter, and is needed in string theories with fermions \cite{ref:Becker--Becker--Schwarz}.
Also, supersymmetry can ensure stability and finiteness of a theory.
For example, $\N=4$ super-Yang--Mills, a supersymmetric quantum field theory, is finite to all orders in perturbation theory \cite{ref:Brink--Lindgren--Nilsson}.
Although some argued that superpartners, particles predicted by supersymmetry, would be experimentally discovered at the Large Hadron Collider \cite{ref:Becker--Becker--Schwarz}, there is, as of today, no experimental evidence for supersymmetry \cite{ref:Baer--et_al,ref:Canepa}.
Still, supersymmetry remains a large area of interest for theories beyond the Standard Model.

Supergravity combines supersymmetry with ideas from general relativity.
These theories can be formulated on supermanifolds, a generalisation of ordinary manifolds that uses both ordinary (commuting) and fermionic (anticommuting) coordinates, or as field theories with gauged supersymmetry on an ordinary spacetime manifold.
Although there are renormalisable and, as noted above, even finite supersymmetric quantum field theories, supergravity theories are in general nonrenormalisable and, therefore, considered as effective field theories \cite{ref:Vafa--Brennan--Carta}.
Some of these, in the landscape, arise as effective low-energy descriptions of (possibly compactified) M-theory while others, in the swampland, are thought to not admit a finite UV-completion and are, thus, deemed inconsistent as quantum theories \cite{ref:Vafa--Brennan--Carta}.
Specifically, five supergravity theories in $D=10$ are the massless tree-level approximations of the five consistent superstring theories \cite{ref:Becker--Becker--Schwarz}.
Apart from these ten-dimensional supergravity theories, there is an eleven-dimensional supergravity theory which is an effective low-energy limit of M-theory.
M-theory is a quantum theory, first conjectured by Witten, that unifies the string theories and whose quantum structure is inherently nonperturbative and remains largely unknown \cite{ref:Duff:M-theory}.
The eleven-dimensional supergravity theory is related to the ten-dimensional supergravity theories via duality transformations.
The simplest such relation is that type IIA supergravity can be obtained by dimensional reduction of $D=11$ supergravity \cite{ref:Becker--Becker--Schwarz} and the relations to other supergravity theories can be understood via string dualities \cite{ref:Witten:revolution,ref:Vafa--Brennan--Carta,ref:Becker--Becker--Schwarz}.
The eleven-dimensional supergravity theory is the theory with which this thesis is concerned.

This chapter gives an introduction to supersymmetry and eleven-dimensional supergravity.
In this thesis, we are interested in supersymmetric theories in an anti-de Sitter (AdS) spacetime (after compactification).
Still, we start by considering the simpler case of supersymmetry in Minkowski spacetime.
Due to the limited scope of the thesis, the presentation here is incomplete in many ways, although relatively self-contained.
For more thorough introductions to supersymmetry, see for instance \cite{ref:Wess--Bagger,ref:West}.

\section{Supersymmetry in Minkowski spacetime}
In this section, we give an introduction to supersymmetry.
For simplicity, we do this in the setting of a four-dimensional Minkowski spacetime.
From a theoretical perspective, the interest in supersymmetry is motivated by the Coleman--Mandula theorem \cite{ref:Coleman--Mandula}.
This theorem states, under quite general assumptions, that the Lie algebra of a connected symmetry group of the $S$-matrix containing the Poincaré algebra is locally isomorphic to the direct product of the Poincaré algebra and a Lie algebra of internal symmetries.%
\footnote{Note that, since the theorem is concerned with symmetries of the $S$-matrix, it does not apply to spontaneously broken symmetries \cite{ref:Coleman--Mandula}, see \cite{ref:Nesti--Percacci} for a counterexample.}
Essentially, this means that the spacetime symmetries can only be extended with internal symmetries in a trivial way.
One way around this is to not consider a Lie algebra but a \emph{Lie superalgebra}.
A Lie superalgebra is a generalisation of a Lie algebra that allows for ``anticommuting'' generators as well as ordinary ``commuting'' generators.%
\footnote{We have put quotes around (anti-)commuting since the generators may fail to (anti-)commute, as measured by the superbracket.}
Formally, it is a $\ZZ_2$-graded vector space equipped with a bilinear Lie superbracket $[\cdot,\cdot\}$ satisfying \cite{ref:Cheng--Wang}
\begin{subequations}
\label{eq:susy:alg:Lie_superalg}
\begin{gather}
    [X, Y\} = -(-1)^{|X||Y|} [X, Y\},\\
    (-1)^{|X||Z|}[X, [Y, Z\}\} + (-1)^{|Z||Y|}[Z, [X, Y\}\} + (-1)^{|Y||X|}[Y, [Z, X\}\} = 0,
\end{gather}
\end{subequations}
where $X$, $Y$ and $Z$ are elements of the Lie superalgebra which are pure in the grading and $|X|$ denotes the degree ($0$ or $1$) of $X$.
\Cref{eq:susy:alg:Lie_superalg} are the natural graded generalisations of the anticommutative property and the Jacobi identity of the ordinary Lie bracket.

Given an associative superalgebra, a Lie superalgebra can be constructed by defining \cite{ref:Cheng--Wang}
\begin{equation}
    [X, Y\} = XY - (-1)^{|X||Y|} YX.
\end{equation}
This means that, if a Lie superalgebra is represented by linear operators on a vector space, the superbracket corresponds to the anticommutator if both elements are of odd degree and the commutator otherwise.%
\footnote{This is similar to how bosonic creation and annihilation operators \emph{commute} with fermionic creation and annihilation operators.}

The generalisation of the Coleman--Mandula theorem to the Lie superalgebra setting is the Haag--Łopuszański--Sohnius theorem \cite{ref:Haag--Lopuszanski--Sohnius}.
This theorem gives a classification of possible Lie superalgebras generating symmetries of the $S$-matrix for a theory in Minkowski spacetime.
To give an introduction to supersymmetry, we consider, in particular, the super-Poincaré algebra.

\subsection{The super-Poincaré algebra} \label{sec:susy:super-Poincare}
The super-Poincaré algebra (in four dimensions) is a Lie superalgebra with neither central charges nor internal symmetries that can be used to describe supersymmetry in a four-dimensional Minkowski spacetime.
The generators of this algebra are the translations $P_a$, Lorentz generators $L_{ab}$ and supercharges $Q^i_\alpha$ and their conjugates $\bar{Q}_{\dot{\alpha} i} = (Q^i_\alpha)^\dag$.
Here, $\alpha$ is a Weyl-spinor index while $i=1,\, \hdots,\, \N$ where $\N$ is the number of supersymmetries.
The nonvanishing independent superbrackets are \cite{ref:Wess--Bagger}
\begin{subequations}
\label{eq:susy:alg:super-Poincare}
\begin{alignat}{2}
    &[L_{ab}, L^{cd}]
    &&= -2 \tensor{(L_{ab})}{^{[c}_e} \tensor{L}{^{|e|}^{d]}},
    \\
    &[L_{ab}, P^c]
    &&= - \tensor{(L_{ab})}{^c_d} P^d,
    \\
\label{eq:susy:alg:super-Poincare_LQ}
    &[L_{ab}, Q^i_\alpha\}
    &&=  - \tensor{(L_{ab})}{_\alpha^\beta} Q^i_\beta,
    \\
\label{eq:susy:alg:super-Poincare_QQ}
    &\{Q^i_\alpha, \bar{Q}_{\dot{\beta} j}\}
    &&= 2 \delta^i_j \sigma^a_{\alpha \dot\beta} P_a,
\end{alignat}
\end{subequations}
where $\sigma^a$ are the Pauli matrices (see \cref{app:conventions:4d-spinors} for conventions and some identities).%
\footnote{Here, we have written the superbrackets in an almost convention-independent way although \cref{eq:susy:alg:super-Poincare_QQ} reveals that $P_a$ is Hermitian and, hence, identified with $-\i\partial_a$.}
From this, it is easy to see that, if the super-Poincaré algebra is realised as operators on a Hilbert space, the energy is bounded from below.
To calculate the energy, we must fix a direction of time.
This introduces $\delta^{\alpha \dot\beta}$ as an invariant under spatial rotations.
Contracting \cref{eq:susy:alg:super-Poincare_QQ} with $\delta^{\alpha \dot\beta}$ gives
\begin{equation}
    \delta^{\alpha \dot \beta} \{Q^i_\alpha, \bar{Q}_{\dot{\beta} j}\} = 2 \delta^i_j (-2) (-H)
    \quad\implies\quad
    H = \frac{1}{4\N} \delta^{\alpha \dot \beta} \{Q^i_\alpha, \bar{Q}_{\dot{\beta} i}\}.
\end{equation}
From this,
\begin{equation}
    \langle\psi| H |\psi\rangle
    = \frac{1}{4 \N} \delta^{\alpha \dot \beta} \langle\psi| Q^i_\alpha \bar{Q}_{\dot{\beta} i} + \bar{Q}_{\dot{\beta} i} Q^i_\alpha |\psi\rangle \geq 0,
\end{equation}
since
\begin{equation}
    \langle\psi|
    Q^i_\alpha \delta^{\alpha \dot \beta} \bar{Q}_{\dot{\beta} i}
    |\psi\rangle
    =
    \sum_{i,\alpha} \|\bar{Q}_{\dot \alpha i} \psi\|^2,
    \qquad
    \langle\psi|
    \bar{Q}_{\dot{\beta} i} \delta^{\alpha \dot \beta} Q^i_\alpha
    |\psi\rangle
    =
    \sum_{i,\alpha} \|Q^i_\alpha \psi\|^2.
\end{equation}
Note that a global minimum of $H$ is attained if $Q^i_\alpha |\psi\rangle = 0 = \bar{Q}_{\dot{\alpha} i} |\psi\rangle$, in which case $|\psi\rangle$ is a supersymmetric vacuum.
Such a state $|\psi\rangle$ might, however, not exist in which case the supersymmetry is said to be spontaneously broken \cite{ref:Wess--Bagger}.%
\footnote{Note the difference between this and, for instance, the spontaneously broken symmetry in the Standard Model. In the latter, an invariant state exists but does not minimise the energy.}

Since $Q^i_\alpha$ and $\bar Q_{\dot\alpha i}$ are spinor generators, they raise or lower the spin of a state by $1/2$.
To see this explicitly, consider the commutator with $S_z \propto L^{12}$, the spin operator in the $z$-direction.%
\footnote{With the normalisation of the Lorentz algebra from \cref{app:SO_conventions} and the conventional $[S_i, S_j] = \i \epsilon_{ijk} S^k$, the exact relation is $S_i = -\i \epsilon_{ijk} L^{jk}$, which explains the factors in \cref{eq:susy:alg:super-Poincare_SzQ}.}
By \cref{eq:susy:alg:super-Poincare_LQ}
\begin{equation}
\label{eq:susy:alg:super-Poincare_SzQ}
    [S_z, Q^i_\alpha] = - \frac{\i}{2} (\sigma^{12})\indices{_\alpha^\beta} Q^i_\beta,
    \qquad\quad
    [S_z, Q_{\dot\alpha i}] = - \frac{\i}{2} (\bar\sigma^{12})\indices{_{\dot\alpha}^{\dot\beta}} Q_{\dot \beta i}.
\end{equation}
Since $\sigma^{12} = -\i \sigma^z$ and $\bar\sigma^{12} = \i \sigma^z$, $Q^i_1$ and $\bar Q_{\dot 2 i}$ lower while $Q^i_2$ and $\bar Q_{\dot 1 i}$ raise $S_z$ by a half unit.
By the spin-statistics theorem, this implies that the fermionic $Q$ operators interchange bosons and fermions.

\subsection{Supermultiplets} \label{sec:Mink_susy:multiplets}
We now turn to the possible particle contents of supersymmetric theories, with supersymmetry implemented by the super-Poincaré algebra, referred to as supermultiplets.
Particles are identified with the nontrivial irreducible unitary representations of the Poincaré algebra.
Physically, this means that, given a particle, it will not change into another particle under translations, rotations or boosts.
That the representations are required to be irreducible corresponds to the fact that we are interested in elementary particles.
To investigate the particle content, we will therefore consider irreducible representations not of the complete super-Poincaré algebra but, rather, of a subalgebra $\mathfrak{k}$ which leaves the momentum $p^a$ invariant, that is, we consider particles in a specific frame.
The subalgebra $\mathfrak{k}$ includes, $P^a$, $Q^i_\alpha$, $Q_{\dot\alpha i}$ and the subalgebra of the Lorentz algebra leaving $p^a$ invariant.

As explained in \cite{ref:West}, the above procedure induces a unique representation of the complete algebra.
We will only consider supermultiplets consisting of a finite number of particles and will assume that $p^a \neq 0$.%
\footnote{The case $p=0$ corresponds to the Minkowski vacuum since the only finite-dimensional unitary irreducible representation, in this case, is the trivial representation.}
Note that $P^a P_a$ is a Casimir of the superalgebra, whence all particles in a supermultiplet have the same mass.%
\footnote{Since we do not observe superpartners with identical mass in nature, this means that supersymmetry must be spontaneously broken if implemented in nature.}

Before investigating the massive and massless supermultiplets in more detail, we give a final remark.
Consider the fermion parity operator $(-1)^{N_\mathrm{f}}$ acting by $+1$ ($-1$) on bosonic (fermionic) one-particle states in an irreducible finite-dimensional representation of $\mathfrak{k}$ \cite{ref:Wess--Bagger}.
From the above remarks, it follows that $(-1)^{N_\mathrm{f}} Q^i_\alpha = - Q^i_\alpha (-1)^{N_\mathrm{f}}$ and similarly for $\bar Q_{\dot\alpha i}$.
Using this and cyclicity of the trace%
\footnote{This is valid since the trace is over the Hilbert space and matrix elements are complex (not Grassmann) numbers.},
we find that
\begin{equation}
\label{eq:susy:alg:super-Poincare:representation_fermion_parity}
    2 \delta^i_j \sigma^a_{\alpha \dot\beta} p_a \trace (-1)^{N_\mathrm{f}}
    =
    \trace\bigl[
        (-1)^{N_\mathrm{f}} \{Q^i_\alpha, \bar Q_{\dot\beta j}\}
    \bigr]
    = \trace\bigl[
        (-1)^{N_\mathrm{f}} (Q^i_\alpha \bar Q_{\dot\beta j} + \bar Q_{\dot\beta j} Q^i_\alpha)
    \bigr]
    = 0.
\end{equation}
Since $p \neq 0$, this implies that the number of bosonic and fermionic states in the representation are equal.

\subsubsection{Massive supermultiplets}
To get the particle content in the massive case, we follow \cite{ref:Wess--Bagger} and consider the rest frame in which $p^a = (m, 0, 0, 0)^a$.
This corresponds to a choice of a timelike vector, whence $\Spin(3, 1)$ is broken to $\Spin(3)\simeq \SU(2)$ and $\delta^{\alpha \dot\beta}$ becomes invariant.
Thus, the subalgebra of the Lorentz algebra contained in $\mathfrak{k}$ is isomorphic to $\so(3)$.
By defining
\begin{equation}
    a_\alpha^i \coloneqq \frac{1}{\sqrt{2m}} Q_\alpha^i,
    \qquad\quad
    a_{\dot \alpha i}^\dag = \frac{1}{\sqrt{2m}} \bar{Q}_{\dot \alpha i}
\end{equation}
we obtain, from \cref{eq:susy:alg:super-Poincare_QQ}, the canonical anticommutation relations
\begin{equation}
\label{eq:susy:alg:super-Poincare:massive_ccr}
    \{a_\alpha^i, a_{\dot\alpha j}^\dag\} = \delta^i_j \delta_{\alpha \dot\alpha},
    \qquad
    \{a_\alpha^i, a_\beta^j\} = 0,
    \qquad
    \{a_{\dot\alpha i}^\dag, a_{\dot\beta j}^\dag\} = 0.
\end{equation}
Since we consider finite-dimensional representations, there exists some state $|\Omega\rangle$, known as a Clifford vacuum, which is quenched by all annihilation operators, $a_\alpha^i |\Omega\rangle = 0$.
From \cref{eq:susy:alg:super-Poincare_LQ}, we see that, given one such state, there must be a complete $\so(3)$-representation of such states, describing the same particle but with different $S_z$-eigenvalues.
To get an irreducible representation, this $\so(3)$-representation must be irreducible \cite{ref:Wess--Bagger}.

We begin by considering the case when the Clifford vacuum has spin 0, that is, the case in which there is precisely one state $|\Omega\rangle$ such that $a_\alpha^i |\Omega\rangle = 0$.
This is known as the fundamental supermultiplet \cite{ref:Ferrara--et_al}.
Note that it contains $2^{2 \N}$ states since there are $2 \N$ anticommuting creation operators.

\Cref{eq:susy:alg:super-Poincare:massive_ccr} is manifestly $(\SU(2) \times \U(\N))$-invariant.
There is, however, a larger invariance group that can be found by defining
\begin{equation}
\begin{alignedat}{2}
    &\Gamma^i = a_1^i + a_{\dot 1 i}^\dag,\qquad\qquad
    &&\Gamma^{\N+i} = a_2^i + a_{\dot 2 i}^\dag,\qquad\\
    &\Gamma^{2\N+i} = \i (a_1^i - a_{\dot 1 i}^\dag),
    &&\Gamma^{3\N+i} =  \i (a_2^i - a_{\dot 2 i}^\dag).
\end{alignedat}
\end{equation}
These are Hermitian and satisfy $\{\Gamma^r, \Gamma^s\} = 2 \delta^{rs}$, where $r,s=1,\hdots,4\N$, which we recognise as the generators of a Clifford algebra with invariance group $\SO(4\N)$.
The $2^{2\N}$ states transform under the Dirac spinor representation of this $\SO(4\N)$, with the irreducible Weyl spinor representations corresponding to bosons and fermions~\cite{ref:Ferrara--et_al}.

To understand the physical content of the representation, we want to label the particles by their spin.
Thus, we wish to keep the original $\operatorname{SU}(2)$ manifest.
It is, however, convenient to consider a larger subgroup of $\SO(4\N)$ than the originally manifest $\SU(2) \times \U(\N)$, namely $\SU(2) \times \Sp(\N)$ \cite{ref:Wess--Bagger}.%
\footnote{$\Sp(\N)$ is sometimes, for instance in \cite{ref:Wess--Bagger,ref:Ferrara--et_al}, denoted $\operatorname{USp}(2\N)$.}
To this end, define
\begin{equation}
    q_\alpha^i = a_\alpha^i,\qquad\quad
    q_\alpha^{\N+i} = \epsilon_{\alpha\beta} \delta^{\beta \dot{\gamma}} a_{\dot\gamma i}^\dag.
\end{equation}
These satisfy the $(\SU(2)\times\Sp(\N))$-invariant anticommutation relations
\begin{equation}
    \{q_\alpha^m, q_\beta^n\} = -\epsilon_{\alpha \beta} \Lambda^{mn},
    \qquad\quad
    \Lambda^{mn} =
    \begin{pmatrix}
    0 & \1\\
    -\1 & 0
    \end{pmatrix}^{\!\!mn},
\end{equation}
where $m,n =  1,\hdots,2\N$.
When breaking $\SO(4\N)$ to $\SU(2)\times\Sp(\N)$ the Dirac spinor representation decomposes as \cite{ref:Ferrara--et_al}
\begin{equation}
\label{eq:susy:alg:super-Poincare:massive_spin_reps}
    2^{2\N} \rightarrow
    \bigoplus_{k=0}^\N \bigl(\N+1-k,\ d^{(\N)}_k \bigr),
\end{equation}
where the first label, $\N+1-k = 2s+1$, is the dimension of the irreducible spin-$s$ representation of $\operatorname{SU}(2)$ and the second label,
\begin{equation}
    d^{(\N)}_k = \binom{2\N}{k} - \binom{2\N}{k-2},
\end{equation}
is the dimension of the irreducible representation of $\Sp(\N)$ consisting of traceless%
\footnote{The trace is of course taken with the antisymmetric invariant $\Lambda^{mn}$.}
completely antisymmetric tensors $T_{m_1 \hdots m_k}$.
Hence, the number of spin-$s$ particles is~$d^{(\N)}_{\N-2s}$.

The number of particles with each spin can also be calculated directly with combinatorics by analysing the spin in the $z$-direction.
Due to \cref{eq:susy:alg:super-Poincare_SzQ} it is clear that, since we start from an $s=0$ Clifford vacuum, the highest $S_z$ eigenvalue in the supermultiplet is $\N/2$ which occurs with multiplicity $1$.
This implies that there is exactly one spin-$\N/2$ representation with dimension $\N+1$ corresponding to the $k=0$ term in \cref{eq:susy:alg:super-Poincare:massive_spin_reps}.
Similarly, there are $2\N$ states with $s_z = (\N-1)/2$ since we can either use all raising operators, $a_{\dot 1 i}^\dag$, and then one of the lowering operators, $a_{\dot 2 i}^\dag$, ($\N$ choices) or all but one raising operator ($\N$ choices).
Generalising this to $s_z = n/2$ we see, by Vandermonde's identity, that there are
\begin{equation}
    \sum_{k=0}^{\N-n} \binom{\N}{n+k} \binom{\N}{k}
    = \binom{2\N}{\N-n}
\end{equation}
states.
The terms in this sum are interpreted as raising the $s_z$ eigenvalue $n+k$ times and then lowering it $k$ times.
Since there is one $s_z$ state for every irreducible representation with $s \geq s_z$, the multiplicity of the spin-$s$ representation is the number of $s_z=s$ states minus the number of $s_z=s+1$ states.
Hence, the multiplicity of the representation with spin $s=n/2$ is
\begin{equation}
\label{eq:supermultiplet_s=n/2_multiplicity}
    \binom{2\N}{\N-n} - \binom{2\N}{\N-n-2} = d^{(\N)}_{\N-n},
\end{equation}
which is consistent with the above result from the group-theoretic approach.

Now consider the general case of a spin-$s$ Clifford vacuum $|\Omega_s\rangle$.
The $s_z$ eigenvalues can be obtained as all possible sums of one $s_z$ value from the vacuum and one from the creation operators, whence the supermultiplet is given by angular momentum addition with the fundamental supermultiplet.
The dimension of the general supermultiplet is given by $2^{2\N}(2s+1)$.
In \cref{tab:susy:alg:super-Poincare:massive_multiplets}, massive $\N=1$ and $\N=4$ supermultiplets with spin at most $2$ are presented; a complete list can be found in \cite{ref:Wess--Bagger}.
Note that $\N = 4$ is the maximal number of supersymmetries if we require that the spin is at most 2.
\begin{table}[H]
    \centering
    \caption[Massive supermultiplets in four-dimensional Minkowski spacetime.]{Multiplicities of the spin representations, that is, the number of particle types of each spin, for the irreducible massive $\N=1$ and $\N=4$ supermultiplets. $\Omega_s$ specifies the spin, $s$, of the Clifford vacuum of the representation.}
\label{tab:susy:alg:super-Poincare:massive_multiplets}
    \begin{tabular}{l rrrr r}
        \toprule
        & \multicolumn{4}{c}{$\N=1$} & $\N=4$ \\
        \cmidrule(rl){2-5} \cmidrule(rl){6-6}
        Spin & $\Omega_0$ & $\Omega_{1/2}$ & $\Omega_1$ & $\Omega_{3/2}$ & $\Omega_0$ \\ \midrule
        $0$ & 2 & 1 &  &  & 42 \\
        $1/2$ & 1 & 2 & 1 &  & 48 \\
        $1$ &  & 1 & 2 & 1 & 27 \\
        $3/2$ &  &  & 1 & 2 & 8 \\
        $2$ &  &  &  & 1 & 1 \\
        \addlinespace[\aboverulesep] \bottomrule
    \end{tabular}
\end{table}

\subsubsection{Massless supermultiplets}
In the massless case, $p^2 = 0$, the particles have no rest frame.
Instead, again following \cite{ref:Wess--Bagger}, we analyse the situation in the frame in which $p^a=(E, 0, 0, E)^a$.
This corresponds to a choice of a lightlike vector, which breaks $\Spin(3, 1)$ to the double cover of $\operatorname{SE}(2)$ \cite{ref:Wigner}.%
\footnote{$\operatorname{E}(n)$ denotes the isometry group of Euclidean space.}
Since we only consider finite-dimensional representations, they are labelled by their helicity, which coincides with the spin in the $z$-direction due to our choice of coordinates.
From \cref{eq:susy:alg:super-Poincare_QQ} we see that
\begin{equation}
    \{Q_\alpha^i, \bar{Q}_{\dot \beta j}\}
    = 2 \delta^i_j
    \begin{pmatrix}
    2E & 0\\
    0 & 0
    \end{pmatrix}_{\!\!\!\alpha\dot\beta}.
\end{equation}
Since the $\alpha=2$ operators anticommute with everything they act by $0$ on any representation \cite{ref:Wess--Bagger} whence no new states are obtained by applying them.
Thus, we need only introduce $\N$ creation and annihilation operators
\begin{equation}
    a^i = \frac{1}{2\sqrt{E}} Q^i_1,
    \qquad\quad
    a_i^\dag = \frac{1}{2\sqrt{E}} Q_{\dot 1 i},
\end{equation}
where $a^i$ lowers and $a_i^\dag$ raises the helicity by a half unit.
Introducing a Clifford vacuum $|\Omega_\lambda\rangle$ with helicity $\lambda$, we see that we get a total of $2^\N$ states and $\binom{\N}{k}$ states of helicity $\lambda+k/2$.
These supermultiplets are, in general, not CPT-invariant since they are not symmetric around helicity $0$ \cite{ref:Wess--Bagger}.
To create CPT-invariant supermultiplets, two supermultiplets with opposite helicities can be added.
In \cref{tab:susy:alg:super-Poincare:massless_multiplets}, the CPT-invariant $\N=1$ and $\N=8$ supermultiplets with spin at most 2 are presented.
Note that $\N=8$ is the maximal number of supersymmetries if we require the spin to be at most 2.
This assumption is often, but not universally, employed due to no-go theorems for higher spins \cite{ref:Rahman--Taronna}.
\begin{table}[H]
    \centering
    \caption[Massless supermultiplets in four-dimensional Minkowski spacetime.]{Multiplicities of the helicities for the massless CPT-invariant $\N=1$ and $\N=8$ supermultiplets. $\Omega_\lambda$ specifies the helicity, $\lambda$, of the lowest-helicity Clifford vacuum of the representation. The $\N=8$ supermultiplet is irreducible while the $\N=1$ supermultiplets contain two irreducible parts to make them CPT-invariant.}
\label{tab:susy:alg:super-Poincare:massless_multiplets}
    \begin{tabular}{l rrrr r}
        \toprule
        & \multicolumn{4}{c}{$\N=1$} & $\N=8$ \\
        \cmidrule(rl){2-5} \cmidrule(rl){6-6}
        Helicity & $\Omega_{-2}$ & $\Omega_{-3/2}$ & $\Omega_{-1}$ & $\Omega_{-1/2}$ & $\Omega_{-2}$ \\ \midrule
        $2$ & 1 &  &  &         & 1 \\
        $3/2$ & 1 & 1 &  &      & 8 \\
        $1$ &  & 1 & 1 &        & 28 \\
        $1/2$ &  &  & 1 & 1     & 56 \\
        $0$ &  &  &  & 2        & 70 \\
        $-1/2$ &  &  & 1 & 1    & 56 \\
        $-1$ &  & 1 & 1 &       & 28 \\
        $-3/2$ & 1 & 1 &  &     & 8 \\
        $-2$ & 1 &  &  &        & 1 \\
        \addlinespace[\aboverulesep] \bottomrule
    \end{tabular}
\end{table}

\subsection{The Wess--Zumino model} \label{sec:susy:Wess--Zumino}%
Here, we give an example of a supersymmetric theory in Minkowski spacetime known as the Wess--Zumino model \cite{ref:Wess--Zumino:model}.
The theory consists of a massless $\N=1$ supermultiplet with two scalar and two spinor real on-shell degrees of freedom, see \cref{tab:susy:alg:super-Poincare:massless_multiplets}.
We formulate the theory using a Weyl spinor $\psi_\alpha$ and a complex scalar $\phi$.
Note that the spinor has four real off-shell degrees of freedom but only two real on-shell degrees of freedom.
The Lagrangian of the free theory is
\begin{equation}
\label{eq:susy:Wess--Zumino:L}
    \mathcal{L}
    = - \partial_a \phi^\ast \partial^a \phi - \i \bar\psi \bar\sigma^a \partial_a \psi,
\end{equation}
which is real since
\begin{equation}
    \bigl( \i \bar\psi_{\dot\alpha} \bar\sigma^{a \dot \alpha \beta} \partial_a \psi_\beta \bigr)^\ast
    = \i \psi_\alpha \sigma^{a \alpha \dot \beta} \partial_a \bar\psi_{\dot \beta}
    \simeq \i \psi_{\dot\beta} \bar\sigma^{a \dot\beta \alpha} \partial_a \psi_\alpha,
\end{equation}
where, in the step indicated by $\simeq$, we have used integration by parts and disregarded boundary terms.

The supersymmetry transformation can be written as
\begin{equation}
\label{eq:susy:Wess--Zumino:susy_transformation}
    \delta_\xi \phi
    =\sqrt{2} \xi^\alpha \psi_\alpha,
    \qquad
    \delta_\xi \psi_\alpha
    = \sqrt{2} \i \sigma^a_{\alpha \dot\beta} \bar\xi^{\dot\beta} \partial_a \phi,
\end{equation}
where the factor $\sqrt{2}$ is purely conventional.
Note that the transformation parameter $\xi^\alpha$ is fermionic since supersymmetry interchanges bosons and fermions.
The above implies that
\begin{equation}
    \delta_\xi \phi^\ast
    = \sqrt{2} \bar\xi_{\dot\alpha} \bar\psi^{\dot\alpha},
    \qquad\quad
    \delta_\xi \bar\psi^{\dot\alpha}
    = \sqrt{2} \i \bar\sigma^{a \dot\alpha \beta} \xi_\beta \partial_a \phi^\ast.
\end{equation}
From this, we see that the Lagrangian is, indeed, supersymmetric since
\begin{align}
\nonumber
    \delta_\xi \mathcal{L}
    &= -\sqrt{2} \bar\xi \partial_a \bar\psi \partial^a \phi
    - \sqrt{2} \partial_a \phi^\ast \xi \partial^a \psi
    - \sqrt{2} \partial_b \phi^\ast \xi \sigma^b \bar\sigma^a \partial_a \psi
    + \sqrt{2} \bar\psi \bar\sigma^a \sigma^b \bar\xi \partial_a \partial_b \phi
    \simeq\\
    &\simeq -\sqrt{2} \bar\xi \partial_a \bar\psi \partial^a \phi
    - \sqrt{2} \partial_a \phi^\ast \xi \partial^a \psi
    + \sqrt{2} \partial_a \phi^\ast \xi \partial^a \psi
    + \sqrt{2} \partial_a \bar\psi \bar\xi \partial^a \phi
    = 0,
\end{align}
where we have integrated by parts and used properties of the Pauli matrices.

To see the relation to the super-Poincaré algebra, consider the commutator $[\delta_\xi, \delta_\varepsilon]$ of two supersymmetry transformations.
From $\delta_\xi \delta_\varepsilon \phi = 2\i \varepsilon \sigma^a \bar\xi \partial_a \phi$, we immediately find
\begin{equation}
\label{eq:susy:Wess--Zumino:commutator_phi}
    [\delta_\xi, \delta_\varepsilon] \phi
    = (\xi^\alpha \bar\varepsilon^{\dot\beta} - \varepsilon^\alpha \bar\xi^{\dot\beta}) 2 \sigma^a_{\alpha \dot\beta} (-\i \partial_a) \phi.
\end{equation}
Similarly, $\delta_\xi \delta_\varepsilon \psi_\alpha = 2\i \sigma^a_{\alpha \dot\beta} \bar\varepsilon^{\dot\beta} \xi^\gamma \partial_a \psi_\gamma$ whence
\begin{equation}
    [\delta_\xi, \delta_\varepsilon] \psi_\alpha
    = 2\i \sigma^a_{\alpha \dot\beta} \delta^\gamma_\delta \delta^{\dot\beta}_{\dot\epsilon} (\bar\varepsilon^{\dot\epsilon} \xi^\delta - \bar\xi^{\dot\epsilon} \varepsilon^\delta) \partial_a \psi_\gamma.
\end{equation}
Writing the deltas using $\sigma^b_{\delta \dot\epsilon} \bar\sigma_b^{\dot\beta \gamma} = -2 \delta^\gamma_\delta \delta^{\dot\beta}_{\dot\epsilon}$ and then using $\sigma^a \bar\sigma^b = - 2\eta^{ab} - \sigma^b \bar\sigma^a$ this becomes
\begin{equation}
\label{eq:susy:Wess--Zumino:commutator_psi}
    [\delta_\xi, \delta_\varepsilon] \psi
    = (\xi^\alpha \bar\varepsilon^{\dot\beta} - \varepsilon^\alpha \bar\xi^{\dot\beta}) 2 \sigma^a_{\alpha \dot\beta} (-\i \partial_a) \psi
    - \i (\xi \sigma^b \bar\varepsilon - \varepsilon \sigma^b \bar\xi) \sigma_b \bar\sigma^a \partial_a \psi.
\end{equation}
By writing a supersymmetry transformation as $\delta_\xi = \xi Q + \bar\xi \bar Q$ \cite{ref:Wess--Bagger}, we find, from \cref{eq:susy:alg:super-Poincare_QQ},
\begin{equation}
    [\delta_\xi, \delta_\varepsilon]
    = \xi^\alpha \{Q_\alpha, \bar Q_{\dot \beta}\} \bar\varepsilon^{\dot \beta}
    - \varepsilon^\alpha \{Q_\alpha, \bar Q_{\dot \beta}\} \bar\xi^{\dot \beta}
    = (\xi^\alpha \bar\varepsilon^{\dot\beta} - \varepsilon^\alpha \bar\xi^{\dot\beta}) 2 \sigma^a_{\alpha \dot\beta} P_a.
\end{equation}
Identifying $P_a$ with $(-\i\partial_a)$, we see that this agrees with \cref{eq:susy:Wess--Zumino:commutator_phi,eq:susy:Wess--Zumino:commutator_psi} except for the last term in \cref{eq:susy:Wess--Zumino:commutator_psi}.
Note that this term vanishes on-shell since, then, $\psi$ satisfies the Weyl equation $\bar\sigma^a \partial_a \psi = 0$.
Hence, the supersymmetry is said to close on-shell.
This is expected from the representation theory of the super-Poincaré algebra since there is always an equal number of bosonic and fermionic states in a representation.
In the Wess--Zumino model, as presented here, their numbers are equal on-shell but not off-shell.
Thus, we expect that we can make the algebra close off-shell by adding an auxiliary complex scalar field with no on-shell degrees of freedom.
In \cref{sec:superspace:Wess--Zumino}, we will see that this is indeed the case.

\section{Superspace formalism}
In this section, we introduce the formalism of superspace and superfields.
With this formalism, one can construct manifestly supersymmetric theories.
For instance, one can add interaction terms to the Wess--Zumino model without having to check that the Lagrangian is supersymmetric by hand, as we did in \cref{sec:susy:Wess--Zumino}.
Furthermore, it provides insight into the quantum theory by simplifying calculations and explaining seemingly miraculous cancellations in component calculations by keeping the supersymmetry manifest \cite{ref:Grisaru--Siegel--Rocek,ref:Brink--Lindgren--Nilsson,ref:West}.
It is also the language that we will use to construct eleven-dimensional supergravity.
We will not attempt to give a mathematically rigorous presentation of supermanifolds%
\footnote{For such a treatment, see for instance \cite{ref:DeWitt}.}
but provide the tools necessary to formulate supergravity theories.

Superspace%
\footnote{With ``superspace'', we refer to any supermanifold, not necessarily flat superspace.}
is parameterised by supercoordinates $z^M = (x^m, \theta^\mu, \bar\theta_{\dot\mu})^M$ or $z^M = (x^m, \theta^\mu)^M$, where $x^m$ are real and Grassmann-even (bosonic) while $\theta^\mu$ and $\bar\theta_{\dot\mu}$ are Grassmann-odd (fermionic), see \cref{app:Grassmann}.
In dimensions and signatures in which the irreducible spinor(s) are Majorana, $\theta^\mu$ can be taken as real and, then, the supercoordinates are $(x^m, \theta^\mu)$.
For extended supersymmetry, that is, with $\N>1$, one adds additional anticommuting coordinates \cite{ref:West}.

\subsection{Superfields}
Here, we introduce the concept of superfields.
We consider the case of flat superspace with 4 bosonic dimensions and supercoordinates $z^A = (x^a, \theta^\alpha, \bar\theta_{\dot \alpha})^A$, although several aspects naturally generalise to arbitrary dimension.
Similar to how translations and rotations can be implemented as differential operators on ordinary space, we wish to realise the supercharges, $Q_\alpha$ and $\bar Q_{\dot\alpha}$, as differential operators on superspace.
Following \cite{ref:Wess--Bagger}, we define
\begin{equation}
\label{eq:superfields:Q}
    Q_\alpha
    = \partial_\alpha
    - \i \sigma^a_{\alpha \dot\beta} \bar\theta^{\dot\beta} \partial_a,
    \qquad\quad
    \bar Q_{\dot\alpha}
    = -\partial_{\dot\alpha}
    + \i\theta^\beta \sigma^a_{\beta \dot \alpha} \partial_a,
\end{equation}
from which we find the single nonvanishing anticommutator
\begin{equation}
\label{eq:superfields:D_anticommutator}
    \{Q_{\alpha}, \bar Q_{\dot\beta}\}
    = 2\i \sigma^a_{\alpha \dot\beta} \partial_a
    = -2 \sigma^a_{\alpha \dot \beta} P_a.
\end{equation}
Note the difference in sign compared to \cref{eq:susy:alg:super-Poincare_QQ}.
However, by changing coordinates $x^a \mapsto -x^a$, which implies $P_a \mapsto -P_a$, we recover the super-Poincaré algebra as previously defined.

The partial derivate $\bar\partial_{\dot\alpha}$ does not anticommute with $Q_\alpha$.
Hence, we introduce supercovariant derivatives
\begin{equation}
\label{eq:superfields:D}
    D_\alpha = \partial_\alpha + \i \sigma^a_{\alpha \dot\beta} \bar\theta^{\dot\beta} \partial_a,
    \qquad\quad
    \bar D_{\dot \alpha} = - \bar\partial_{\dot\alpha} - \i \theta^\beta \sigma^a_{\beta \dot\alpha} \partial_a,
\end{equation}
satisfying
\begin{subequations}
\begin{alignat}{2}
    &\{D_\alpha, \bar D_{\dot\beta}\} = -2\i \sigma^a_{\alpha \dot\beta} \partial_a,
    \qquad\quad
    &&\{D_\alpha, D_\beta\} = 0,\\
    &\{D_\alpha, Q_{\beta}\} = 0,
    &&\{D_\alpha, \bar Q_{\dot \beta}\} = 0.
\end{alignat}
\end{subequations}
We raise and lower the indices on $D_\alpha$ and $\bar D_{\dot\alpha}$ using the usual convention for spinors, in contrast to how the indices on $\partial_\alpha$ and $\bar\partial^{\dot\alpha}$ are raised and lowered, see \cref{app:Grassmann}.

Now that we have defined the supercharges as differential operators and introduced supercovariant derivatives, we are ready to introduce superfields.
A superfield $F(x, \theta, \bar{\theta})$ is a Grassmann-even function on superspace.
Since $\alpha$ can only take two values, we can form $\theta^2 = \theta^\alpha \theta_\alpha$ but $\theta^2 \theta^\alpha = 0$ and similarly for dotted indices.%
\footnote{Note that $\theta^\alpha \theta^\beta = k \epsilon^{\alpha \beta} \theta^2$ since there is only one antisymmetric combination. By contracting with $\epsilon_{\alpha \beta}$, one finds $k=-1/2$.}
Hence, we may expand $F$ in powers of $\theta$ and $\bar{\theta}$ and only get a finite number of terms
\begin{align}
    \nonumber
    F(x,\theta,\bar{\theta})
    &=
    f(x)
    + \theta \psi(x) + \bar{\theta} \bar{\chi}(x)
    + \theta^2 m(x) + \theta \sigma^a \bar{\theta} v_a(x) + \bar{\theta}^2 n(x)
    \\
\label{eq:susy:superspace:field_expansion}
    &\quad
    + \theta^2 \bar{\theta} \bar{\lambda}(x) + \bar{\theta}^2 \theta \eta(x)
    + \theta^2 \bar{\theta}^2 h(x).
\end{align}
Here, the $x$-dependent expansion-coefficients are referred to as component fields.

The supersymmetry transformation of a superfield is defined as \cite{ref:Wess--Bagger}
\begin{equation}
\label{eq:susy:superspace:field_transformation}
    \delta_\xi F = (\xi Q + \bar\xi \bar Q) F,
\end{equation}
while the supersymmetry transformations of the component fields are defined by
\begin{align}
\label{eq:susy:superspace:component_transformation}
    \nonumber
    \delta_\xi F
    &=
    \delta_\xi f
    + \theta \,\delta_\xi \psi + \bar{\theta} \,\delta_\xi \bar{\chi}
    + \theta^2 \,\delta_\xi m + \theta \sigma^a \bar{\theta} \,\delta_\xi v_a + \bar{\theta}^2 \,\delta_\xi n
    \\
    &\quad
    + \theta^2 \bar{\theta} \,\delta_\xi \bar{\lambda} + \bar{\theta}^2 \theta \,\delta_\xi \eta
    + \theta^2 \bar{\theta}^2 \,\delta_\xi h.
\end{align}
\Cref{eq:susy:superspace:field_transformation} is so important that we do not call functions of $x$, $\theta$ and $\bar\theta$ superfields if they do not obey it (similarly to how the word tensor is used in physics).
Thus, $\partial_\alpha F$ is, for instance, not a superfield since $\{\bar Q^{\dot\beta}, \partial_\alpha\} \neq 0$.
Given two superfields $F$ and $G$, a linear combination of them is, however, a superfield since $Q_\alpha$ and $\bar Q_{\dot\alpha}$ are linear operators.
Similarly, the product $FG$ is a superfield since the supercharges are graded derivations, that is, they satisfy the super-Leibniz's rule.
Also, the supercovariant derivative of a superfield is again a superfield since $D_\alpha$ and $\bar D_{\dot\alpha}$ anticommute with $Q_\alpha$ and $\bar Q_{\dot\alpha}$.

Since the supersymmetry transformation is linear, the space of superfields transform under a representation of the superalgebra.
However, as is clear from \cref{eq:susy:superspace:field_expansion}, an unconstrained superfield contains quite many spacetime fields and the representation is not irreducible \cite{ref:Wess--Bagger}.
To get an irreducible representation, the superfield must be constrained.
To not break Lorentz invariance or supersymmetry, the constraining equations should be Lorentz invariant and respect supersymmetry in the sense that the variation $\delta_\xi F$ also satisfies the constraints \cite{ref:Wess--Bagger}.
If the theory should be kept off-shell, the constraining equations should, furthermore, not imply differential equations for the remaining component fields.
Two possibilities are
\begin{equation}
    \bar D \Phi = 0,
    \qquad\quad
    V^\ast = V,
\end{equation}
where $\Phi$ is called a chiral superfield while $V$ is called a vector superfield.
It turns out that every supersymmetric renormalisable Lagrangian can be written in terms of chiral and vector superfields \cite{ref:Wess--Bagger}.

\clearpage%
\subsection{Revisiting the Wess--Zumino model} \label{sec:superspace:Wess--Zumino}%
Now that we have introduced superfields, we demonstrate the formalism by studying chiral superfields in more detail.
As it will turn out, we will find a superspace formulation of the Wess--Zumino model.
As stated above, a chiral superfield satisfies $\bar D \Phi = 0$.
Naturally, this forces the $\bar\theta$-dependence of the field.
To see this explicitly, define
\begin{equation}
    T
    \coloneqq \exp(-\i\theta\sigma^a \bar\theta \partial_a)
    = 1 - \i \theta \sigma^a \bar\theta \partial_a + \frac{1}{4} \theta^2 \bar\theta^2 \square,
\end{equation}
which satisfies $T \bar D_{\dot\alpha} = -\bar\partial_{\dot\alpha} T$.
With a short calculation, one can verify that $T^{-1} = \exp(\i\theta\sigma^a \bar\theta \partial_a)$, as expected.
Thus, since $T$ is invertible, $\bar D_{\dot\alpha} \Phi = 0$ is equivalent to $\bar\partial_{\dot\alpha} T \Phi = 0$.
Hence, $T\Phi$ can be expanded, in terms of component fields, as
\begin{equation}
\label{eq:superspace:Wess--Zumino:TPhi}
    T\Phi(x, \theta, \bar\theta) = \phi(x) + \sqrt{2} \theta\psi(x) + \theta^2 F(x),
\end{equation}
where the $\sqrt{2}$ might seem arbitrary at this point.
Acting with $T^{-1}$ gives, in agreement with \cite{ref:Wess--Bagger},
\begin{equation}
\label{eq:superspace:Wess--Zumino:Phi}
    \Phi
    = \phi + \sqrt{2} \theta\psi + \theta^2 F
    + \i \theta\sigma^a \bar\theta \partial_a \phi
    + \frac{\i}{\sqrt{2}} \theta^2 \bar\theta \bar\sigma^a \partial_a \psi
    + \frac{1}{4} \theta^2 \bar\theta^2 \square \phi.
\end{equation}

To derive the supersymmetry transformation of the component fields, note that they are obtained as the analogous components in the $\theta$-expansion of $\delta_\xi F$, see \cref{eq:susy:superspace:component_transformation}.
Since $\phi = \Phi|_{\theta=0=\bar\theta}$, this means that $\delta_\xi \phi = \delta_\xi \Phi|_{\theta=0=\bar\theta}$.
Henceforth, we will omit $\theta=0=\bar\theta$ at the evaluation bar and simply write, for instance, $\Phi|$ to save ink.
Note that, acting with $(\xi Q + \bar\xi \bar Q)$ and $(\xi D + \bar\xi \bar D)$ gives the same result when evaluated at $\theta=0=\bar\theta$ since only the $\partial_\alpha$ and $\bar\partial_{\dot\alpha}$ terms survive.
Using this, and that $\Phi$ is chiral, we find
\begin{equation}
    \delta_\xi \phi
    = (\xi Q + \bar\xi \bar Q) \Phi \bigr|
    = (\xi D + \bar\xi \bar D) \Phi \bigr|
    = \xi D \Phi \bigr|
    = \sqrt{2} \xi \psi.
\end{equation}
We recognise this from \cref{eq:susy:Wess--Zumino:susy_transformation}, which explains why we normalised $\psi$ with a factor of $\sqrt{2}$ in \cref{eq:superspace:Wess--Zumino:TPhi}.
Next, we note that $D_\alpha \Phi| = \sqrt{2} \psi_\alpha$ whence
\begin{align}
    \nonumber
    \delta_\xi \psi_\alpha
    &= \frac{1}{\sqrt{2}} D_\alpha (\xi Q + \bar{\xi} \bar Q) \Phi \bigr|
    = \frac{1}{\sqrt{2}} (\xi D + \bar\xi \bar D) D_\alpha \Phi \bigr|
    =\\ \nonumber
    &= \frac{1}{\sqrt{2}} (\xi^\beta D_\beta D_\alpha - \bar\xi^{\dot\beta} \{\bar D_{\dot\beta}, D_\alpha\} ) \Phi \bigr|
    = \frac{1}{\sqrt{2}} (2 \xi^\beta \epsilon_{\alpha\beta} F + 2\i \sigma^a_{\alpha\dot\beta} \bar\xi^{\dot\beta} \partial_a \phi)
    =\\
    &= \sqrt{2} \xi_\alpha F + \sqrt{2} \i \sigma^a_{\alpha\dot\alpha} \bar{\xi}^{\dot\alpha} \partial_a \phi.
\end{align}
Here, it is important that we move $D_\alpha$ to the right of the supercharges before we replace them with covariant derivatives.
This is reminiscent of \cref{eq:susy:Wess--Zumino:susy_transformation} but there is an extra term involving the field $F$.
More on this later.
In the above, we used that $D_\alpha D_\beta \theta^2 = 2 \epsilon_{\beta \alpha}$.
This implies that $F = -1/4\, D^2 \Phi|$, which we now use to compute the supersymmetry transformation of $F$,
\begin{align}
    \nonumber
    \delta_\xi F
    &= -\frac{1}{4} D^2 (\xi Q + \bar{\xi} \bar{Q}) \Phi \bigr|
    = -\frac{1}{4} (\xi D + \bar{\xi} \bar{D}) D^2 \Phi \bigr|
    = - \frac{1}{4} \bar{\xi} \bar{D} D^2 \Phi \bigr|
    =\\ \nonumber
    &= \frac{1}{4} \bar{\xi}^{\dot\alpha} (D^2 \bar{D}_{\dot\alpha} - 2 \{\bar{D}_{\dot{\alpha}}, D_\beta\}  D^\beta ) \Phi \bigr|
    = \sqrt{2} \i \bar{\xi}^{\dot\alpha} \sigma^a_{\beta\dot\alpha} \partial_a \psi^\beta
    =\\
    &= \sqrt{2} \i \bar{\xi} \bar{\sigma}^a \partial_a \psi.
\end{align}
%
% Note the appearance of the Weyl equation in this transformation.

Having found the supersymmetry transformations of the component fields we would like to construct a supersymmetric Lagrangian.
To obtain an ordinary spacetime Lagrangian, $\mathcal{L}$, that does not depend on the Grassmann variables, we integrate over $\theta$ and $\bar\theta$.
For this, we use the Berezin integral, see \cref{app:Grassmann}.
Therefore, we want to know what quantities can be integrated to a supersymmetric Lagrangian.
Note that
\begin{equation}
\label{eq:superspace:Wess--Zumino:nonsaturated_integrals}
    \int \d^2 \theta\, \d^2 \bar\theta\, \partial_\alpha = 0,
    \qquad\quad
    \int \d^2 \theta\, \d^2 \bar\theta\, \bar\partial_{\dot\alpha} = 0,
\end{equation}
since the integral is only nonzero for $\theta^2 \bar\theta^2$-terms.
Hence, any superfield integrated over $\theta$ and $\bar\theta$ transforms into a total $x$-derivative under a supersymmetry transformation.
Thus, the integral of any superfield gives a supersymmetric Lagrangian, disregarding boundary terms.
We can now write a Lagrangian for a chiral superfield $\Phi$ as
\begin{equation}
    \mathcal{L} = \int \d^2 \theta\, \d^2 \bar\theta\, \Phi^\ast \Phi.
\end{equation}
As we will soon see, this is a kinetic Lagrangian, even though it does not contain any explicit derivatives.
Instead, the derivatives will come from \cref{eq:superspace:Wess--Zumino:Phi}.
Note that $\Phi^\ast$ is an antichiral superfield, $D_\alpha \Phi^\ast = 0$, whence $\Phi^\ast \Phi$ is neither chiral nor antichiral.
However, it is a superfield, which is all that matters for $\mathcal{L}$ to be supersymmetric.
To calculate the integrals, we use that
\begin{equation}
\label{eq:superspace:Wess--Zumino:theta_bartheta_integral}
    \int \d^2 \theta\, \d^2 \bar\theta
    \simeq \frac{1}{16} D^2 \bar D^2 \bigr|,
\end{equation}
where the difference, which comes from the $\theta$-term in $\bar D$, is a total $x$-derivative.
Using that $\Phi$ is chiral
\begin{align}
\nonumber
    \int \d^2 \theta\, \d^2 \bar\theta\, \Phi^\ast \Phi
    &\simeq \frac{1}{16} \bigl(
        (D^2 \bar D^2 \Phi^\ast) \Phi
        -2 (D_\alpha \bar D^2 \Phi^\ast) (D^\alpha \Phi)
        + (\bar D^2 \Phi^\ast) (D^2 \Phi)
    \bigr)\bigr|
    =\\ \nonumber
    &= \square \phi^\ast \phi
    + \i \partial_a \bar\psi^{\dot\beta} \bar\sigma^a_{\dot\beta \alpha} \psi^\alpha
    + F^\ast F
    \simeq\\
\label{eq:superspace:Wess--Zumino:L}
    &\simeq
    - \partial_a \phi \partial^a \phi
    - \i \bar\psi \bar\sigma^a \partial_a \psi
    + F^\ast F,
\end{align}
where, in the second step, we have used
\begin{subequations}
\begin{alignat}{2}
    &D_\alpha \bar D^2 \Phi^\ast
    &&= (\bar D^2 D_\alpha + 2 \{D_\alpha, \bar D_{\dot\beta}\} \bar D^{\dot\beta}) \Phi^\ast
    = -4\i \sigma^a_{\alpha \dot\beta} \partial_a \bar D^{\dot \beta} \Phi^\ast,
    \\
    &D^2 \bar D^2 \Phi^\ast
    &&= - 4\i \sigma^a_{\alpha \dot\beta} \partial_a D^\alpha \bar D^{\dot\beta} \Phi^\ast
    = -8 \sigma^a_{\alpha \dot\beta} \bar\sigma^{b\dot\beta\alpha} \partial_a \partial_b \Phi^\ast
    = 16 \square \Phi^\ast.
\end{alignat}
\end{subequations}
Since no derivatives of $F$ enter in $\mathcal{L}$, we say that $F$ is an auxiliary field.
Note that if we impose the Euler--Lagrange equation $F=0$, we recover the Lagrangian and supersymmetry transformations of the Wess--Zumino model in \cref{sec:susy:Wess--Zumino}.
Without imposing $F=0$, the supersymmetry transformations close off-shell due to \cref{eq:superfields:D_anticommutator}.%
\footnote{Due to how we define supersymmetry transformations of component fields, $\delta_\xi \delta_\varepsilon \Phi| = \delta_\varepsilon \delta_\xi \phi$ and similarly for $\psi$ and $F$ \cite{ref:Wess--Bagger}. Thus, $[\delta_\xi, \delta_\varepsilon]$ gives the same result as in \cref{sec:susy:Wess--Zumino} due to the sign difference between \cref{eq:superfields:D_anticommutator,eq:susy:alg:super-Poincare_QQ}.}%
\todo[disable]{}
Thus, we have realised the supersymmetry off-shell by introducing an auxiliary field $F$, as alluded to in \cref{sec:susy:Wess--Zumino}.

\subsubsection{Interactions}
At this point, it might seem like we have not gained much by introducing the superspace formalism.
To illustrate part of the power of the formalism, we consider supersymmetric interactions.
As noted above, the superspace formalism also provides insight into supersymmetric quantum field theories.

To only get renormalisable interactions, we require that the coupling constants have positive momentum dimensions.
For this analysis, we need to define $[\theta]$.
As usual, in momentum dimensions, $[x] = -1$, $[\phi] = 1$, which implies $[\Phi] = 1$.
From this, and $[\psi] = 3/2$, we get $[\theta] = -1/2$ which is consistent with the dimensions in \cref{eq:superfields:Q,eq:superfields:D}.
Hence, if we integrate with respect to both $\theta^2$ and $\bar\theta^2$, the integrand must have dimension $2$ since $[\mathcal{L}]=4$.%
\footnote{Note that $\int \d^2 \theta$ contributes $+1$ to the dimension due to how the integral is defined.}
Such a term can, hence, only have two powers of $\Phi$.
However, from \cref{eq:superspace:Wess--Zumino:theta_bartheta_integral}, we see that any chiral integrand will only contribute with boundary terms.
Similarly, any antichiral integrand will only contribute with boundary terms since the integration over $\theta$ and $\bar\theta$ can be carried out in any order.
Thus, the only possible term of this kind is the kinetic term.

To construct interaction terms, we instead use $\int \d^2 \theta + \mathrm{c.c}$.
Such an integral will, in general, not produce a $\bar\theta$-independent result.
However, if the integrand is chiral, any term containing $\bar\theta$ will be a total $x$-derivative, as seen from \cref{eq:superspace:Wess--Zumino:Phi}, that we disregard.
A term of this kind in the Lagrangian is supersymmetric due to \cref{eq:superspace:Wess--Zumino:nonsaturated_integrals} and the fact that $\delta_\xi \Phi$ can be written without $\bar\partial_{\dot\alpha}$-terms since $\bar D_{\dot\alpha} \Phi = 0$.
Since we only integrate over $\theta$ or $\bar\theta$ in the interaction terms, the integrand must have dimension $3$.
Hence, we write
\begin{equation}
    \mathcal{L}_\mathrm{int.}
    = \int \d^2 \theta \bigl(\lambda \Phi + \frac{m}{2} \Phi^2 + \frac{g}{3} \Phi^3 \bigr)
    + \int \d^2 \bar\theta \bigl(\lambda \Phi^\ast + \frac{m}{2} (\Phi^\ast)^2 + \frac{g}{3} (\Phi^\ast)^3 \bigr),
\end{equation}
where $[\lambda]=2$, $[m]=1$ and $[g]=0$.
Using that $\int \d^2 \theta$ and $-1/4\, D^2|$ differs by a total $x$-derivative, we compute the component field interactions
\begin{subequations}
\begin{alignat}{2}
    &\int \d^2 \theta\, \Phi
    &&\simeq -\frac{1}{4} D^2 \Phi \bigr|
    = F,
    \\ %\nonumber
    &\int \d^2 \theta\, \Phi^2
    &&\simeq %-\frac{1}{4} D^2 \Phi^2 \bigr|
    %=  -\frac{1}{4} D^\alpha (2 \Phi D_\alpha \Phi) \bigr|
    %=
    -\frac{1}{2} (D^\alpha \Phi D_\alpha \Phi + \Phi D^2 \Phi) \bigr|
    %=\\
    %& &&%
    = - \psi^\alpha \psi_\alpha + 2 \phi F,
    \\ %\nonumber
    &\int \d^2 \theta\, \Phi^3
    &&\simeq %-\frac{1}{4} D^2 \Phi^3 \bigr|
    %= -\frac{1}{4} D^\alpha (3 \Phi^2 D_\alpha \Phi) \bigr|
    %=
    -\frac{3}{4} (2 \Phi D^\alpha \Phi D_\alpha \Phi + \Phi^2 D^2 \Phi) \bigr|
    %=\\
    %& &&%
    = -3 \phi \psi^\alpha \psi_\alpha + 3 \phi^2 F.
\end{alignat}
\end{subequations}
Generalising this to multiple chiral superfields $\Phi^i$, we can write the most general renormalisable supersymmetric Lagrangian as
\begin{align}
    \nonumber
    \mathcal{L}
    &=
    \int \d^2 \theta\, \d^2 \bar{\theta}\, \Phi^\ast_i \Phi^i
    + \Bigl[ \int \d^2 \theta\, \bigl( \lambda_i \Phi^i + \frac{1}{2} m_{ij} \Phi^i \Phi^j + \frac{1}{3} g_{ijk} \Phi^i \Phi^j \Phi^k \bigr) + \mathrm{c.c.} \Bigr]
    =\\ \nonumber
    &=
    - \partial^a \phi^\ast_i \partial_a \phi^i
    - \i \bar\psi_i \bar\sigma^a \partial_a \psi^i
    + F^*_i F^i
    +\\
\label{eq:superspace:Wess--Zumino:general_Lagrangian}
    &\quad+ \Bigl[
    \lambda_i F^i
    -\frac{1}{2} m_{ij} \psi^i \psi^j + m_{ij} \phi^i F^j
    - g_{ijk} \phi^i \psi^j \psi^k + g_{ijk} \phi^i \phi^j F^k
    + \mathrm{c.c.} \Bigr],
\end{align}
where $m_{ij}$ and $g_{ijk}$ are completely symmetric \cite{ref:Wess--Bagger}.%
\footnote{An equivalent way of constructing supersymmetric Lagrangians is by picking out a component of a superfield that transforms into a total $x$-derivative under supersymmetry, see \cite{ref:Wess--Bagger}.}
Note that even after adding interactions, $F^i$ are still auxiliary fields without dynamics.
The Euler--Lagrange equation for $F^i$ reads
\begin{equation}
\label{eq:superspace:Wess--Zumino:F_eom_interactions}
    F^*_i + \lambda_i + m_{ij} \phi^j + g_{ijk} \phi^j \phi^k = 0.
\end{equation}
We can now eliminate the auxiliary fields from the theory by inserting the solution into \cref{eq:superspace:Wess--Zumino:general_Lagrangian}.%
\footnote{The dynamics remain the same when eliminating the auxiliary field, as can be proven in general.}
The Lagrangian then becomes
\begin{align}
    \nonumber
    \mathcal{L}
    &=
    - \partial^a \phi^*_i \partial_a \phi^i
    - \i \bar\psi_i \bar\sigma^a \partial_a \psi^i
    - \frac{1}{2} m_{ij} \psi^i \psi^j
    - \frac{1}{2} m^{\ast ij} \bar{\psi}_i \bar{\psi}_j
    +\\
    &\quad
    - g_{ijk} \phi^i \psi^j \psi^k
    - g^{* ijk} \phi^*_i \bar\psi_j \bar\psi_k
    - F_i^\ast(\phi) F^i(\phi).
\end{align}
Here, the last term is a potential term containing powers of $\phi$ from order zero to four.
We recognise the kinetic and mass terms for the scalars and spinors, the Yukawa interactions and the scalar potential from quantum field theory.
Note that, due to the supersymmetry, there are very particular relations between the masses, Yukawa couplings and parameters in the potential.

The last term is the only term where a vacuum expectation value can enter.
From the Lagrangian, we can see explicitly that the conclusions regarding nonnegative energy and spontaneously broken supersymmetry in \cref{sec:susy:super-Poincare} are valid.
If $\langle F^i \rangle = 0$ we get a vacuum with $0$ energy which is supersymmetric since a supersymmetry transformation leaves all fields unchanged.
If, on the other hand, $\langle F^i \rangle \neq 0$, the ground state energy is positive and the supersymmetry is spontaneously broken since $\psi^i$ is not invariant under a supersymmetry transformation.
Spontaneously broken supersymmetry can only be guaranteed if $\lambda_i$, $m_{ij}$ and $g_{ijk}$ are such that there is no solution to \cref{eq:superspace:Wess--Zumino:F_eom_interactions} with $F^i = 0$ but can also be obtained as a metastable vacuum where the energy is only minimised locally \cite{ref:Ellis--Smith--Ross}.

\subsection{Superdifferential forms} \label{sec:superspace:superforms}%
Having introduced the concepts of superspace and superfields, we turn to superdifferential forms, or superforms, which we will use to formulate eleven-dimensional supergravity.
Supersymmetry transformations form a subgroup of the diffeomorphism group of a supermanifold \cite{ref:Wess--Bagger}.
Thus far, we have only considered flat superspace and global, or rigid, supersymmetry transformations.
To be able to formulate supergravity theories that are manifestly invariant under general diffeomorphisms, that is, coordinate transformations, we introduce superdifferential forms.%
\footnote{Note that a diffeomorphism is an \emph{active} coordinate transformation. Any theory can be formulated in a way invariant under passive coordinate transformations, that is, changes of coordinates.}
Superforms are not only useful for formulating supergravity, but supersymmetric Yang--Mills theories as well.
These formulations are super-analogous of Cartan's formulation of general relativity and Yang--Mills theory formulated with differential forms, see \cref{app:bundles_gauge_theory_gravity}.

In the following, we will no longer separate commuting and anticommuting coordinates.
Instead, we work directly with supercoordinates $z^M = (x^m, \theta^\mu, \bar\theta_{\dot\mu})^M$, where $\bar\theta$ may be omitted if we impose a Majorana condition on $\theta$,%
\footnote{This comment naturally applies throughout the rest of this section.}
satisfying
\begin{equation}
    z^M z^N = (-1)^{|M||N|} z^N z^M,
\end{equation}
where $|M|$ is $0$ for $M=m$ and $1$ for $M=\mu,\dot\mu$.
Hence, the coordinates are said to be graded-commutative.
Similarly, we write $\d z^M = (\d x^m, \d \theta^\mu, \d \bar\theta_{\dot\mu})^M$ and $\partial_M = (\partial_m, \partial_\mu, \bar\partial^{\dot\mu})_M$ and introduce a ``graded-anticommutative'' wedge product
\begin{equation}
    \d z^M \wedge \d z^N = - (-1)^{|M||N|} \d z^N \wedge \d z^M.
\end{equation}
A general superdifferential $p$-form can now be written as
\begin{equation}
    \Omega
    = \frac{1}{p!} \d z^{M_1} \wedge \hdots \wedge \d z^{M_p} \Omega_{M_p \hdots M_1}(z)
    = \d z^{M_I} \Omega_{M_I}(z),
\end{equation}
where $M_I$ is a multi-index, $\d z^{M_I} = \d z^{M_1} \wedge \hdots \d z^{M_p}$ and $\Omega_{M_I} = \Omega_{M_p \hdots M_1}/p!$.
Note the order of indices and placement of $\Omega$.
This is purely conventional but will turn out to be practical.
For Grassmann-even superforms, $\Omega_{M_I}$ is Grassmann-odd if the number of spinor indices is odd and Grassmann-even otherwise.%
\footnote{When considering, for instance, a vector superform $V^M = \d z^{N_I} \tensor{V}{_{N_I}^{\! M}}$, this is complemented according to whether $M$ is bosonic or fermionic.}
The wedge product is, of course, extended bilinearly to arbitrary superforms.
It is a straightforward exercise to show that the wedge product is associative, $\Lambda \wedge (\Omega \wedge \Xi) = (\Lambda \wedge \Omega) \wedge \Xi$, and satisfies $\Omega \wedge \Lambda = (-1)^{pq} \Lambda \wedge \Omega$ for a super $p$-form $\Omega$ and super $q$-form $\Lambda$ \cite{ref:Wess--Bagger}, like the wedge product of ordinary differential forms.

The exterior derivative of a superform $\Omega$ is defined as
\begin{equation}
    \d \Omega = \d z^{M_I} \wedge \d z^N \partial_N \Omega_{M_I}
\end{equation}
and is a super $(p+1)$-form where $p$ is the form-degree of $\Omega$.
Note that this differs from the conventional definition in ordinary space since $\d z^N \partial_N$ is to the right of $\d z^{M_I}$.
This implies that
\begin{equation}
\label{eq:superspace:superforms:d_distribution}
    \d(\Omega \wedge \Lambda) = \Omega \wedge \d \Lambda + (-1)^{|\Lambda|} \d \Omega \wedge \Lambda,
\end{equation}
where $|\Lambda|$ is the form-degree of $\Lambda$.
As usual, $\d^2 = 0$ which, together with \cref{eq:superspace:superforms:d_distribution} and $\d F = \d z^M \partial_M F$ for super 0-forms $F$, provides an alternative definition of the exterior derivative \cite{ref:Wess--Bagger}.

\subsubsection{Connection form, covariant derivative and field strength tensor}
In gauge theory, one considers transformations under a gauged structure group $G$.
The structure group is a compact Lie group in case of Yang--Mills theory and the Lorentz group in Cartan's formulation of gravity.
In the superspace setting, we use right-action, so a tensorial super $p$-form $\Omega^i$ transforms under some right-representation $\rho$ of the group, that is, $\Omega'^i = \tensor{\Omega}{^j} \rho(g)\indices{_j^i}$ where $g$ is a group element.
The use of a right-action is motivated by our convention for the exterior derivative.
Henceforth, we will not write the representation explicitly but instead simply write $\Omega'^i = \tensor{\Omega}{^j} g\indices{_j^i}$ or, dropping the indices as well, $\Omega' = \Omega g$.
Note that $\d \Omega'$ is not a tensor, since $\d(\Omega g) = \Omega \wedge \d g + \d \Omega\, g$.
To remedy this, we introduce the Lie algebra-valued (local) connection 1-form $\phi$ and a covariant exterior derivative
\begin{equation}
    \D = \d + \phi,
\end{equation}
which acts on a tensor as $\D \Omega = \d \Omega + \Omega \wedge \phi$.%
\footnote{In the super Yang--Mills context, the connection could be denoted by $A$ in analogy with the conventional notation in ordinary Yang--Mills theory.}
To make $\D \Omega$ a tensor, that is,
\begin{equation}
    \D' \Omega'
    = (\d + \phi') (\Omega g)
    = \d \Omega\, g + \Omega \wedge \d g + \Omega g \wedge \phi'
    = (\d \Omega + \Omega \wedge \phi) g,
\end{equation}
we need
\begin{equation}
    \phi' = g^{-1} \phi g - g^{-1} \d g.
\end{equation}
In this equation, which defines how the connection form transforms, $g^{-1} \phi g$ is the adjoint right action of a Lie group element on a Lie algebra element and, hence, a well-defined element in the Lie algebra.
The second term is also an element in the Lie algebra.

With $T_r$ generators of $\g = \operatorname{Lie}(G)$, the connection 1-form may be written as $\phi = \phi^r \i T_r$ where $\phi^r = \d z^N \phi_N^r$.%
\footnote{Here, we use the convention that a group element is $g=\exp(\i T)$ but keep the $\i$ close to the generator to be able to switch conventions without effort.}
The action on $\Omega$ is then
\begin{equation}
    \Omega \wedge \phi
    = \d z^{M_I} \wedge \d z^N \phi_N^r \Omega_{M_I} \i T_r.
\end{equation}

Next, we define the $\g$-valued field strength, or curvature 2-form,
\begin{equation}
    F
    = F^r\, \i T_r
    = \frac{1}{2} \d z^M \wedge \d z^N F^r_{NM} \i T_r
    \coloneqq \d \phi + \phi \wedge \phi.
\end{equation}
Note that we define the wedge product between Lie algebra-valued forms using the associative product%
\footnote{Given a representation, this corresponds to matrix multiplication.}
of the universal enveloping algebra $\UU(\g)$ whence it is, in general, $\UU(\g)$-valued, not $\g$-valued.
$F$ is, however, Lie algebra-valued as is seen from
\begin{equation}
    \phi \wedge \phi
    = \phi^r \i T_r \wedge \phi^s \i T_s
    = \frac{1}{2} \phi^r \wedge \phi^s [\i T_r, \i T_s]
    \eqqcolon \frac{1}{2} [\phi \wedge \phi],
\end{equation}
where $[\cdot\wedge\cdot]$ is a wedge product defined using the Lie bracket.
Using the definition of $F$, the transformation law of $\phi$ and $0=\d(g^{-1} g) = g^{-1} \d g + \d g^{-1} g$, it is straightforward to show that $F'=g^{-1} F g$, whence $F$ is a tensor carrying the adjoint representation.
Hence, $\phi$ acts on $F$ as
\begin{equation}
    F \wedge_{\ad} \phi
    = [F \wedge \phi]
    = F^r \wedge \phi^s [\i T_r, \i T_s]
    = F \wedge \phi - \phi \wedge F.
\end{equation}
Using the definition of $F$ to calculate $\d F$, this implies
\begin{align}
\nonumber
    \D F
    &= \d(\d \phi + \phi \wedge \phi) + [F \wedge \phi]
    = \phi \wedge \d \phi - \d \phi \wedge \phi + [F \wedge \phi]
    =\\
\label{eq:superspace:superform:BI2}
    &=0,
\end{align}
which is known as the Bianchi identity of the second type \cite{ref:Wess--Bagger}.
The Bianchi identity of the first type is
\begin{align}
\nonumber
    \D^2 \Omega
    &= \D (\d \Omega + \Omega \wedge \phi)
    = (\Omega \wedge \d \phi - \d \Omega \wedge \phi)
    + (\d \Omega \wedge \phi + \Omega \wedge \phi \wedge \phi)
    =\\
\label{eq:superspace:superform:BI1}
    &= \Omega \wedge F.
\end{align}
If we define $\mathcal{F}$ as the operator acting as $\mathcal{F} \Omega = \Omega \wedge F$, the first Bianchi identity reads $\D^2 = \mathcal{F}$ while the second Bianchi identity reads $[\D, \mathcal{F}] = 0$ and follows immediately from the first.%
\footnote{Here, we distinguish the 2-form $F$ from the operator $\mathcal{F}$.
Note, however, that the situation is similar for the connection, where we use $\phi$ for both the operator and 1-form.}
Also, from the definition of the covariant derivate, it is clear that $\D' = g \D g^{-1}$ where the juxtaposition denotes operator composition ($\D$ is not acting on $g^{-1}$) from which it follows that $\mathcal{F}' = g \mathcal{F} g^{-1}$.
This is consistent with the above $F' = g^{-1} F g$ since $\mathcal{F}$ acts from the right, $\mathcal{F}' \Omega = \Omega \wedge g^{-1} F g$.

\subsubsection{Spin connection, vielbeins and torsion}
Thus far, we have used the coordinate frame as a basis for tangent vectors.
Now, we consider another frame, related to the coordinate frame by a local change of basis that is, in general, not induced by a change of coordinates,
\begin{equation}
    E_A = \tensor{E}{_A^M} \partial_M,
    \qquad\quad
    E^A = \d z^M \tensor{E}{_M^A},
\end{equation}
where $\tensor{E}{_A^M} \tensor{E}{_M^B} = \delta_A^B$ and $\tensor{E}{_M^A} \tensor{E}{_A^N} = \delta_M^N$.
The vielbein $E_A$ generalises the concept of vierbeins from four-dimensional spacetime.
We define the vielbeins to transform covariantly under local Lorentz transformations.
Going forward, we use $A,B,C,\hdots$ for Lorentz indices (flat) and $M,N,P,\hdots$ for ``Einstein'' indices (curved).
As in the case of ordinary manifolds, globally defined vielbeins $E_A$ do not exist in general \cite{ref:DeWitt} but only for parallelisable supermanifolds, for instance, flat superspace.
Instead, the vielbeins are defined locally and related by local Lorentz transformations on intersections of patches.

Since we are interested in fields carrying a spin-representation, we need a spin connection $\omega$.
The curvature 2-form of the spin connection is denoted by $R$.
In gravity, $\omega$ is dynamical and the only connection we are concerned with, while in super Yang--Mills, in a fixed background, there is a dynamical Yang--Mills connection while $\omega$ is fixed.
Note that a Lorentz transformation $\tensor{\Lambda}{_A^B}$ does not mix the bosonic and fermionic parts of tensors, that is, $\tensor{\Lambda}{_A^B}$ is only nonzero when both indices are of the same type.
Splitting the index $A=(a, \alpha, \dot\alpha)$ each part transforms under its usual vector or spinor representation.
Hence, the spin connection $\tensor{\omega}{_A^B} = E^C \omega_{C de} (L^{de})\indices{_A^B}$, where $L^{de}$ are the Lorentz generators,%
\footnote{We use the geometrical convention that a group element is $\Lambda=\exp(L)$ for the Lorentz group.}
and the curvature 2-form $\tensor{R}{_A^B}$ are only nonzero when both indices $A\ B$ are either bosonic or fermionic.
Explicitly, the nonzero components of $(L^{de})\indices{_A^B}$ are
\begin{equation}
    (L^{de})\indices{_a^b} = \delta^{[d}_a \eta^{e] b},
    \qquad\quad
    (L^{de})\indices{_\alpha^\beta} = \frac{1}{4} (\Gamma^{de})\indices{_\alpha^\beta},
\end{equation}
whence all Lorentz algebra-valued quantities have this relation between their components.

Given a spin connection and vielbeins, one may define the torsion 2-form as
\begin{equation}
    T^A
    = \D E^A
    = \d E^A + E^B \wedge \tensor{\omega}{_B^A}.
\end{equation}
The torsion transforms covariantly under local Lorentz transformations.
From the Bianchi identity of the first type \cref{eq:superspace:superform:BI1}, we see that
\begin{equation}
\label{eq:superspace:superforms:Ricci_identity}
    \D T^A = E^B \wedge \tensor{R}{_B^A},
\end{equation}
which is known as the Ricci identity (with torsion).

\subsection{Coordinate transformations in superspace}
To conclude this section, we review the transformation laws for various fields under an infinitesimal coordinate transformation%
\footnote{We consider active transformations, that is, diffeomorphisms in superspace. Thus, $z'$ and $z$ describe different points in superspace and we only consider a single coordinate system.}
\begin{equation}
\label{eq:superspace:diffeos:infinitesimal_transformation}
    z^M
    \mapsto z'^M
    = z^M + \xi^M.
\end{equation}
The transformation of a scalar field $\phi(z)$ is, as usual, defined by $\phi'(z') = \phi(z)$.
This is interpreted as moving the field: the value of the moved field, $\phi'$, at the moved point, $z'$, is the same as the value of the original field, $\phi$, at the original point, $z$.
For the infinitesimal transformation in \cref{eq:superspace:diffeos:infinitesimal_transformation},
\begin{equation}
    \delta_\xi \phi = - \xi^M \partial_M \phi,
\end{equation}
where $\delta_\xi \phi \coloneqq (\phi'(z) - \phi(z))|_{\Ordo(\xi)}$ (that is, only terms up to first order in $\xi$ are kept when computing $\delta_\xi \phi$).

Similarly, the transformation of a vector field is defined by $V'^M(z') \partial'_M = V^M(z) \partial_M$ and that of a covector by $\d z'^M U'_M(z') = \d z^M U_M(z)$.
Using the chain rules
\begin{equation}
    \d z'^M = \d z^N \frac{\partial z'^M}{\partial z^N},
    \qquad\quad
    \partial_M = \frac{\partial z'^N}{\partial z^M} \partial'_N,
\end{equation}
one finds
\begin{equation}
    \partial_\xi V^M
    = - \xi^N \partial_N V^M + V^N \partial_N \xi^M,
    \qquad
    \partial_\xi U_M
    = - \xi^N \partial_N U_M - \partial_M \xi^N U_N.
\end{equation}
These equations, which agree with \cite{ref:Wess--Bagger}, can be generalised to arbitrary tensors.
Then, one picks up signs when moving factors through $\d z^M$ and $\partial_M$, for instance
\begin{align}
\nonumber
    \delta_\xi \Omega_{MN}
    &= - \xi^P \partial_P \Omega_{MN}
    - \partial_M \xi^P \Omega_{PN}
    - (-1)^{|M|(|N|+|P|)} \partial_N \xi^P \Omega_{MP}
    =\\
\label{eq:superspace:diffeos:2-form_transformation}
    &= - \xi^P \partial_P \Omega_{MN}
    - 2 \partial_{[M} \xi^P \Omega_{|P|N)},
\end{align}
where the last equality only holds in general for graded-antisymmetric $\Omega_{MN}$, corresponding to a 2-form $\Omega$.
Here $[MN)$ denotes graded antisymmetrisation of $M\, N$, that is, ordinary antisymmetrisation but picking up an extra sign when fermionic indices pass through each other.%
\footnote{Similarly, $(MN]$ denotes graded symmetrisation. The idea behind this notation is that the left (right) symbol indicates what to do with bosonic (fermionic) indices.}

Now, we consider fields that carry some representation of the Lorentz group.%
\footnote{Parts of the following naturally generalise to the case of an arbitrary structure group.}
When transforming a Lorentz vector field $V^A$, we must combine the above with a local Lorentz transformation to ensure that the result is Lorentz covariant.
Therefore, we write
\begin{equation}
    \delta_\xi V^A
    = - \xi^M \partial_M V^A + V^B \tensor{L}{_B^A}
    = - \xi^M \D_M V^A + (-1)^{|M||B|} \xi^M V^B \tensor{\omega}{_M_B^A} + V^B \tensor{L}{_B^A}.
\end{equation}
Demanding Lorentz covariance, that the expression is linear in $\xi$ (there cannot be a constant term since $V'=V$ for $\xi=0$) and linearity in $V$, that is, $\delta_\xi(V^A + U^A) = \delta_\xi V^A + \delta_\xi U^A$, we find $\tensor{L}{_B^A} = - \xi^M \tensor{\omega}{_M_B^A}$.
We now require that all Lorentz tensors transform with this local Lorentz transformation and find that, for a covector $U_A$,
\begin{equation}
    \delta_\xi U_A
    = - \xi^M \partial_M U_A + \xi^M \tensor{\omega}{_M_A^B} U_B
    = - \xi^M \D_M U_A.
\end{equation}
This generalises to multiple indices.

Since $\xi^M$ is a Lorentz scalar, the covariant derivative acts on it by a partial derivative.
Hence, the above rules for variations of fields with only curved indices generalise to tensors with Lorentz indices by replacing all partial derivatives with covariant derivatives.
Thus, the vielbein transforms as
\begin{equation}
\label{eq:superspace:diffeos:delta_E:1}
    \delta_\xi \tensor{E}{_M^A}
    = - \xi^N \D_N \tensor{E}{_M^A}
    - \D_M \xi^N \tensor{E}{_N^A}.
\end{equation}
By noting that
\begin{subequations}
\begin{align}
    &\D_M \xi^A
    = \D_M (\xi^N \tensor{E}{_N^A})
    = \D_M \xi^N \tensor{E}{_N^A}
    + (-1)^{|M||N|} \xi^N \D_M \tensor{E}{_N^A}
    \\
    &\tensor{T}{_M_N^A}
    = 2 \D_{[M} \tensor{E}{_{N)}^A}
    = \D_{M} \tensor{E}{_{N}^A}- (-1)^{|M||N|} \D_N \tensor{E}{_M^A}
\end{align}
\end{subequations}
we see that \cref{eq:superspace:diffeos:delta_E:1} can be written as
\begin{equation}
\label{eq:superspace:diffeos:delta_E:2}
    \delta_\xi \tensor{E}{_M^A}
    = -\D_M \xi^A - \xi^N \tensor{T}{_N_M^A}.
\end{equation}

Lastly, we turn to the transformation of the connection.
Remembering the inhomogeneous term in the Lorentz transformation of the connection,
\begin{align}
\nonumber
    \delta_\xi \tensor{\omega}{_M_A^B}
    &= - \xi^N \partial_N \tensor{\omega}{_M_A^B}
    - \partial_M \xi^N \tensor{\omega}{_N_A^B}
    +\\ \nonumber
    &\quad
    + \tensor{\omega}{_M_A^C} (-\xi^N \tensor{\omega}{_N_C^B})
    - (-\xi^N \tensor{\omega}{_N_A^C}) \tensor{\omega}{_M_C^B}
    - \partial_M (-\xi^N \tensor{\omega}{_N_A^B})
    =\\ \nonumber
    &= - 2 \xi^N \bigl(
        \partial_{[N} \tensor{\omega}{_{M)}_A^B}
        - \tensor{\omega}{_{[N}_{|A|}^C} \tensor{\omega}{_{M)}_C^B}
    \bigr)
    =\\
\label{eq:superspace:diffeos:delta_omega}
    &= - \xi^N \tensor{R}{_N_M_A^B}.
\end{align}

\section{Eleven-dimensional supergravity} \label{sec:sugra}
Eleven-dimensional supergravity, the theory this thesis is mainly concerned with, was first formulated as a spacetime theory \cite{ref:Cremmer--Julia--Scherk} and then later in superspace \cite{ref:Cremmer--Ferrara}.
Here, we give a superspace formulation.
The theory is invariant under the diffeomorphism group of a curved supermanifold, similar to general relativity but in the setting of supermanifolds.
We will construct the theory as a completely geometrical theory in superspace with supercoordinates $z^M = (x^m, \theta^\mu)^M$, where $x$ has $D=11$ real Grassmann-even components and $\theta$ has 32 real Grassmann-odd components.
In \cref{sec:sugra:components}, we summarise the theory in component form with left-action conventions.

Similar to what we did when formulating the Wess--Zumino model in superspace, see \cref{sec:superspace:Wess--Zumino}, we wish to reduce the number of component fields by placing constraints on the superfields.
This time, however, the constraints will imply equations of motion for the remaining component fields and put the theory on-shell.
Having constrained the fields, there are Bianchi identities that are no longer automatically satisfied.
These give relations between the remaining components, including the equations of motions.

After solving the constraints and Bianchi identities, we want the spacetime metric $g_{mn}$ to remain as a physical field.%
\footnote{We will work the vielbein, from which the metric is constructed as $g_{mn} = \eta_{ab} \tensor{e}{_m^a} \tensor{e}{_n^b}$.}
The metric has
\begin{equation}
    \frac{1}{2} (D-1)(D-2) -1 = 44,
\end{equation}
independent on-shell degrees of freedom, since they sit in the traceless symmetric transverse part \cite{ref:Becker--Becker--Schwarz}.
Since spinors in $D=11$ have 32 components and the graviton is massless, we expect a total of $2^{32/4}=256$ on-shell degrees of freedom, by an argument analogous to that in \cref{sec:susy:super-Poincare}.%
\footnote{$256$ degrees of freedom is also the number we expect if maximal $\N=8$ supergravity in $D=4$ should be obtainable via dimensional reduction of eleven-dimensional supergravity.}
Hence, there should be $128$ fermionic degrees of freedom and another $84$ bosonic ones.
The bosonic ones can be obtained from a 3-form $B$ Abelian gauge potential.
Gauge invariance then implies that there are
\begin{equation}
    \frac{1}{3!}(D-2)(D-3)(D-4) = 84
\end{equation}
degrees of freedom since only the transverse directions contribute.
Lastly, we need the fermionic degrees of freedom.
$128$ is precisely the number of on-shell degrees of freedom of a massless spin-$3/2$ field in $D=11$ \cite{ref:Cremmer--Julia--Scherk}, since the tensor product of a transverse vector and spinor is $\vec{9}\otimes\vec{16}\simeq \vec{16}\oplus \vec{128}$ \cite{ref:Becker--Becker--Schwarz}.
Hence, we expect a spin-$3/2$ gravitino $\tensor{\psi}{_m^\alpha}$ associated to some gauge invariance corresponding to local supersymmetry \cite{ref:Becker--Becker--Schwarz}.

In superspace, we have a dynamical vielbein $\tensor{E}{_M^A}$.
The metric will then be obtained form the $\theta = 0$ component of $\tensor{E}{_m^a}$ and we can hope to similarly obtain the gravitino from $\tensor{E}{_m^\alpha}$.
It is, however, not immediately clear how the 3-form $B$ should be obtained from the geometrical quantities in superspace.
Therefore, we introduce a Lorentz scalar 3-form $B$ in superspace.
This might seem to contradict the above claim that the theory is entirely geometrical but, as we will see, this is not the case.

Starting from the vielbein, spin connection and 3-form, we can construct the curvature 2-form $\tensor{R}{_A^B}$, the torsion $T^A$ and the Abelian field strength $H = \D B = \d B$.
These satisfy the Bianchi identities
\begin{equation}
\label{eq:sugra:BI}
    \D T^A = E^B \wedge \tensor{R}{_B^A},
    \qquad
    \D \tensor{R}{_A^B} = 0,
    \qquad
    \D H = 0,
\end{equation}
where the middle equation is the Bianchi identity of the second type, \cref{eq:superspace:superform:BI2}, while the other two are Bianchi identities of the first type, \cref{eq:superspace:superform:BI1}.
According to a theorem due to Dragon \cite{ref:Dragon}, these are not independent equations.
Specifically, $\tensor{R}{_A^B}$ can be expressed in terms of the torsion by using the first identity and the second identity is then automatically satisfied.%
\footnote{$E^B \wedge \D \tensor{R}{_B^A} = 0$ follows directly from Bianchi identities of the first type. One then has to check, using that $\tensor{R}{_A^B}$ is Lie algebra-valued, that this implies $\D \tensor{R}{_A^B} = 0$. The theorem holds in $D>3$ but fails in three dimensions, see for instance \cite{ref:Cederwall--Gran--Nilsson}.}

\subsection{Constraints} \label{sec:sugra:constraints}%
As mentioned above, we constrain the superfields to reduce the number of component fields.
This will put the theory on-shell.
We arrive at the constraints motivated by the field content of the theory and dimensional analysis, similar to \cite{ref:Brink--Howe}.
Due to Dragon's theorem, we do not constrain the curvature 2-form, only the torsion and the 3-form.
Note that, since we expect the gravitino $\tensor{\psi}{_m^\alpha}$ to be related to the $\theta=0$ component of $\tensor{E}{_m^\alpha}$, the corresponding field strength $\tensor{S}{_m_n^\alpha}$ should be related to $\tensor{T}{_m_n^\alpha}$.

For the dimensional analysis, we use mass dimensions, so $[\d z^m] = -1$.
Since, in the superalgebra, the commutator of two supercharges is a translation in spacetime, we need $[\d z^\mu] = -1/2$.
Starting from a superform $\Omega$ with dimension $[\Omega] = n$, we can deduce the dimensions of the components.
For each bosonic index, the dimension of the component is raised by 1 unit to balance the dimension of $\d z^m$.
Similarly, the dimension is raised by a half unit for every fermionic index.
To be able to contract upper with lower indices, the dimension of upper indices must contribute in the opposite way.

As usual, the derivatives have dimensions opposite to those of the coordinates, whence the exterior derivative is dimensionless, $[\D] = 0$.
From their definitions, it is clear that the dimension of the curvature and torsion components come solely from their indices, which we may write as $[R]=[T]=0$.
Since the 3-form, $B$, and its field strength, $H$, are nongeometrical, we cannot derive their dimensions from the above.
However, it will turn out to be reasonable to set $[H_{mnpq}]=1$, which implies $[H]=-3$.

We will constrain the components with flat indices.
Still, we only keep components of $T$ and $H$ corresponding to the field strengths $\tensor{S}{_m_n^\alpha}$ and $H_{mnpq}$ as dynamical.
Apart from that, we allow nondynamical components expressed in terms of $\Gamma$-matrices.
To not introduce dimensionful constants, these components must be dimensionless.
Thus, for $H$, we have the nonzero components
\begin{equation}
\label{eq:sugra:H_nonzero}
    H_{abcd},
    \qquad\quad
    H_{ab \gamma\delta} = 2\i (\Gamma_{ab})\indices{_\gamma_\delta},
\end{equation}
where the second equation determines the normalisation of $H$.%
\footnote{Note that the notation here differs from \cref{chap:sugra_comp,app:conventions:11d-spinors}. Here $\Gamma^a$ denotes the eleven-dimensional $\Gamma$-matrices.}%
\footnotemarksep%
\footnote{$H$ being a real 4-form implies that $H_{ab\gamma\delta}$ is imaginary due to how Grassmann-odd quantities are complex conjugated.}
For $T$, we can also form a nonvanishing component using the dynamical component of $H$, which leads to the nonzero
\begin{equation}
\label{eq:sugra:T_nonzero}
    \tensor{T}{_a_b^\gamma},
    \qquad
    \tensor{T}{_\alpha_\beta^c}
    = 2\i (\Gamma^c)\indices{_\alpha_\beta},
    \qquad
    \tensor{T}{_a_\beta^\gamma}
    = H_{bcde} \bigl(
        k_1 \delta_a^{[b} (\Gamma^{cde]})\indices{_\beta^\gamma}
        + k_2 (\tensor{\Gamma}{_a^{bcde}})\indices{_\beta^\gamma}
    \bigr),
\end{equation}
where $k_1$ and $k_2$ are, as of yet, undetermined dimensionless constants that will be fixed by the Bianchi identities.
Here, $\tensor{T}{_\alpha_\beta^c} = 2\i (\Gamma^c)\indices{_\alpha_\beta}$ is motivated by that flat superspace should be a solution to the theory \cite{ref:Candiello--Kurt}.%
\footnote{Flat superspace has nonvanishing torsion, see for instance \cite{ref:Wess--Bagger,ref:West}.}

Since $\tensor{T}{_a_\beta^\gamma}$ is expressed in terms of the only dynamical component field in $H$, the theory could have been formulated without $H$, something we alluded to above.%
\todo[disable]{}
At this point, it is not clear whether we, in that case, would have to implement additional constraints on $T$ or if it follows from the Bianchi identities that $\tensor{T}{_a_\beta^\gamma}$ can be written in this way.
By a more careful analysis, one can show that the 3-form, $B$, and its field strength, $H$, do not have to be introduced by hand without adding additional constraints \cite{ref:Candiello--Kurt}.
In fact, the theory can be derived, without introducing the 3-form by hand, from the single constraint $\tensor{T}{_\alpha_\beta^c} \sim (\Gamma^c)_{\alpha\beta}$ \cite{ref:Howe}.

\subsubsection{Solution to the Bianchi identities}
To solve the Bianchi identities \cref{eq:sugra:BI} subject to the constraints, we first write them as tensor equations.
For the torsion,
\begin{align}
\nonumber
    &\D T^D
    &&= \frac{1}{2} \D \bigl(E^C \wedge E^B \tensor{T}{_B_C^D}\bigr)
    =\\ \nonumber
    & &&= \frac{1}{2} E^C \wedge E^B \wedge E^A \D_A \tensor{T}{_B_C^D}
    + \frac{1}{2} E^C \wedge T^B \tensor{T}{_B_C^D}
    - \frac{1}{2} T^C \wedge E^B \tensor{T}{_B_C^D}
    =\\ \nonumber
    & &&= \frac{1}{2} E^C \wedge E^B \wedge E^A \D_A \tensor{T}{_B_C^D}
    + E^C \wedge T^B \tensor{T}{_B_C^D}
    =\\
    & &&= \frac{1}{2} E^C \wedge E^B \wedge E^A \D_A \tensor{T}{_B_C^D}
    + \frac{1}{2} E^C \wedge E^B \wedge E^A \tensor{T}{_A_B^E} \tensor{T}{_E_C^D}.
\end{align}
By completely analogous calculations for the curvature 2-form and the 4-form, we find that the Bianchi identities on tensor form are
\begin{subequations}
\label{eq:sugra:tensor_BI}
\begin{alignat}{3}
    &\D_{[A} \tensor{T}{_{BC)}^D}
    &&+ \tensor{T}{_{[AB}^E} \tensor{T}{_{|E|C)}^D}
    &&= \tensor{R}{_{[ABC)}^D},
    \\
\label{eq:sugra:tensor_BI:R}
    &\D_{[A} \tensor{R}{_{BC)}_D^E}
    &&+ \tensor{T}{_{[AB}^F} \tensor{R}{_{|F|C)}_D^E}
    &&= 0,
    \\
    &\D_{[A} H_{BCDE)}
    &&+ 2 \tensor{T}{_{[AB}^F} \tensor{H}{_{|F|CDE)}}
    &&= 0.
\end{alignat}
\end{subequations}
As noted above, the second of these follows from the first due to Dragon's theorem.

To proceed, one splits the equations into all different index types.
Here, we present the results of this analysis, a detailed derivation can be found in \cref{app:sugra_calc} (see also \cite{ref:Brink--Howe}).
One finds that the only dynamical independent component fields of $R$, $H$ and $T$ are the $\theta=0$ components of $R_{abcd}$, $H_{abcd}$ and $\tensor{T}{_a_b^\gamma}$.
The partially undetermined components of the torsion, see \cref{eq:sugra:T_nonzero}, are found to be
\begin{equation}
\label{eq:sugra:constraints:T_k1k2}
    \tensor{T}{_a_\beta^\gamma}
    = - \frac{1}{288} H_{bcde} \bigl(
        8 \delta_a^{[b}  (\tensor{\Gamma}{^{cde]}})\indices{_\beta^\gamma}
        +  (\tensor{\Gamma}{_a^{bcde}})\indices{_\beta^\gamma}
    \bigr).
\end{equation}
The field strengths satisfy Bianchi identities%
\todo[disable]{}
\begin{subequations}
\begin{align}
    & \tensor{R}{_{[abc]}^d} = 0,
    \\
    & \D_{[a} H_{bcde]} = 0,
    \\
    & \D_{[a} \tensor{T}{_{bc]}^\delta}
    - \tensor{T}{_{[ab}^\varepsilon} \tensor{T}{_{c]\varepsilon}^\delta}
    = 0,
\end{align}
\end{subequations}
and equations of motion
\begin{subequations}
\begin{align}
\label{eq:sugra:constraints:eom:R}
    &R_{ab} - \frac{1}{2} \eta_{ab} R = \frac{1}{12} \tensor{H}{_{acde}} \tensor{H}{_b^{cde}} - \frac{1}{96} \eta_{ab} H^2,
    \\
\label{eq:sugra:constraints:eom:H}
    &\D^d H_{dabc} = - \frac{1}{1152} \tensor{\epsilon}{_{abc}^{d_1 d_2 d_3 d_4 e_1 e_2 e_3 e_4}} H_{d_1 d_2 d_3 d_4} H_{e_1 e_2 e_3 e_4},
    \\
\label{eq:sugra:constraints:eom:T}
    &\tensor{T}{_a_b^\beta} (\Gamma^{abc})\indices{_\beta^\alpha} = 0.
\end{align}
\end{subequations}

\clearpage%
\subsection{The spacetime theory} \label{sec:sugra:spacetime}%
Now that we have discussed eleven-dimensional supergravity in superspace, we turn to the component formulation in spacetime.
To distinguish between superfields and spacetime fields, we put a hat on all superfields.
The spacetime fields are defined as the $\theta=0$ components of the corresponding superfields with \emph{curved} form-indices \cite{ref:Brink--Howe}.
This is motivated geometrically since we, at least locally, can embed spacetime in superspace by $x^m \mapsto (x^m, 0)^M$ and the definition then ensures that spacetime fields are tangent to the spacetime in this embedding.
Lie algebra-indices, like $A\, B$ on $\tensor{\omega}{_M_A^B}$, and the flat indices on $E^A$ and $T^A$ are, however, kept flat.
Local supersymmetry corresponds to local translations in the $\theta$-directions in superspace whence invariance under gauged supersymmetry corresponds to invariance under the choice of local embedding.

The first step in converting the superspace theory into a spacetime theory is to impose gauge conditions.
Then, we compute the supersymmetry transformations and, lastly, derive the equations of motions for the spacetime fields.
Note that this section uses right-action conventions even in spacetime.
In \cref{sec:sugra:components}, we switch to left-action conventions and summarise the component formulation of the theory.

\subsubsection{Gauge fixing}
Consider an infinitesimal superspace coordinate transformation $\hat\xi^M(z)$ combined with an infinitesimal local Lorentz transformation $\tensor{\hat L}{_A^B}(z)$.
From \cref{eq:superspace:diffeos:delta_omega}, we find
\begin{equation}
    \delta \tensor{\hat \omega}{_M_A^B}
    = - \hat\xi^N \tensor{\hat R}{_N_M_A^B}
    + \tensor{\hat \omega}{_M_A^C} \tensor{\hat L}{_C^B}
    - \tensor{\hat L}{_A^C} \tensor{\hat \omega}{_M_C^B}
    - \hat \partial_M \tensor{\hat L}{_A^B}.
\end{equation}
Since the theory in spacetime will contain gravity, we keep manifest invariance under arbitrary spacetime coordinate transformations $\xi^m(x)$ and local Lorentz transformations $\tensor{L}{_A^B}(x)$.
For $M=\mu$, we can, however, use the $\theta$-component of $\tensor{\hat L}{_A^B}$ to set the last term to a tensor $\tensor{C}{_\mu_A^B}(x)$ which is arbitrary apart from being Lie algebra-valued in its last two indices.
Hence, we can use that term to gauge away $\smash[b]{\tensor{\hat \omega}{_\mu_A^B}|}$ and then maintain that condition for transformations with arbitrary $\theta=0$ components of $\smash[t]{\hat \xi^M}$ and $\tensor{\hat L}{_A^B}$.%
\footnote{All evaluation bars denote evaluation at $\theta=0$.}

Similarly, from \cref{eq:superspace:diffeos:delta_E:2},
\begin{equation}
\label{eq:sugra:spacetime:transformation:E}
    \delta \tensor{\hat E}{_M^A}
    = - \hat \D_M \hat\xi^A - \hat\xi^N \tensor{\hat T}{_N_M^A} + \tensor{\hat E}{_M^B} \tensor{\hat L}{_B^A}.
\end{equation}
Here, we can use the $\theta$-component of $\hat \xi^A$ to set $\tensor{\hat E}{_\mu^a} | = 0$ and $\tensor{\hat E}{_\mu^\alpha}| = \delta_\mu^\alpha$.
It is not trivial that this is possible.
For an infinitesimal transformation, we can indeed transform $\tensor{\hat E}{_\mu^A}|$ in any desired direction.
This does, however, not imply that we can set the corresponding components to whatever we like.
For instance, we cannot set $\tensor{\hat E}{_\mu^A}|=0$ since that would render $\tensor{\hat E}{_M^A}$ noninvertible.
With the above gauge choice, there is, however, no such singularity and we therefore expect it to be viable.
In the following, we assume this to be the case.
Similar remarks apply to the other gauge conditions.

Lastly, we turn to $\hat B_{MNP}$.
Since only the field strength $\hat H=\hat \d \hat B$ enters the theory, we have an Abelian gauge symmetry
\begin{equation}
    \delta_{\hat \lambda} \hat B = \hat \d \hat \lambda,
    \quad\Longleftrightarrow\quad
    \delta_{\hat \lambda} \hat B_{MNP} = 3 \hat \partial_{[M} \hat\lambda_{NP)},
\end{equation}
that leaves $\hat H$ invariant.
Combining this with an infinitesimal coordinate transformation, see \cref{eq:superspace:diffeos:2-form_transformation}, we get
\begin{align}
\nonumber
    \delta \hat B_{MNP}
    &= - \hat \xi^Q \hat \partial_Q \hat B_{MNP}
    - 3 \hat \partial_{[M} \hat \xi^Q \hat B_{|Q|NP)}
    + 3 \hat \partial_{[M} \hat \lambda_{NP)}
    =\\
\label{eq:sugra:spacetime:B_transformation}
    &= - \hat \xi^Q \hat H_{QMNP}
    - 3 \hat \partial_{[M} (\xi^Q \hat B_{|Q|NP)})
    + 3 \hat \partial_{[M} \hat \lambda_{NP)}.
\end{align}
In this transformation, we use the $\theta$-component of $\hat \lambda_{mn}$ to gauge away $\hat B_{\mu np}|$.

The gauge we have arrived at can be summarised as
\begin{subequations}
\begin{alignat}{4}
    &\tensor{\hat E}{_M^A} \bigr|
    &&=
    \begin{pmatrix}
        \tensor{e}{_m^a}(x) & \tensor{\psi}{_m^\alpha}(x) \\
        0 & \delta_\mu^\alpha
    \end{pmatrix}\indices{_{\!\!\!M}^{A}}
    ,
    \qquad\quad
    &&\tensor{\hat E}{_A^M} \bigr|
    &&=
    \begin{pmatrix}
        \tensor{e}{_a^m}(x) & -\tensor{\psi}{_a^\mu}(x) \\
        0 & \delta_\alpha^\mu
    \end{pmatrix}\indices{_{\!\!\!A}^{M}},
    \\[6pt]
    &\tensor{\hat\omega}{_\mu_A^B}\bigr|
    &&= 0,
    &&\hat B_{\mu n p} \bigr|
    &&= 0.
\end{alignat}
\end{subequations}
Similar gauges are used in, for instance, \cite{ref:Brink--Gell-Mann--Ramond--Schwarz,ref:Wess--Zumino:superspace,ref:Wess--Bagger}.
Note that, on spacetime fields, we convert between curved and flat indices using $\tensor{e}{_m^a}$ and $\delta_\mu^\alpha$, for example, $\tensor{\psi}{_m^\alpha} = \tensor{e}{_m^a} \delta_\mu^\alpha \tensor{\psi}{_a^\mu}$.
In superspace, on the other hand, curved and flat indices are converted using $\tensor{\hat E}{_M^A}$, which illustrates the importance of keeping spacetime fields and $\theta = 0$ components of superfields apart.

Since we have only used the $\theta$-components of $\hat \xi^M$, $\tensor{\hat L}{_A^B}$ and $\hat \lambda_{mn}$, we can still make transformations with $\theta=0$ while maintaining the gauge with $\theta$-components.
From the above, it is clear that $\xi^m(x)$ corresponds to coordinate transformations in spacetime, $\tensor{L}{_A^B}(x)$ local Lorentz transformations and $\lambda_{mn}(x)$ Abelian gauge transformations of $B_{mnp}$.
Since $\xi^\mu(x)$ is fermionic, it must correspond to a gauged supersymmetry transformation.

There are still some transformations that we have not considered, including higher $\theta$-components and $\hat \lambda_{MN}$ with one or two fermionic indices.
However, we are only interested in $\tensor{e}{_m^a}$, $\tensor{\psi}{_m^a}$, $B_{mnp}$ and $\omega_m$ and these are invariant under such transformations.

\subsubsection{Supersymmetry transformations}
The (gauged) supersymmetry transformation is generated by $\xi^\mu(x)$.
Note that $\hat \xi^\alpha| = \delta_\mu^\alpha \xi^\mu = \xi^\alpha$ and $\hat \xi^a| = 0$, since $\hat \xi^A = \hat \xi^M \tensor{\hat E}{_M^A}$.
From \cref{eq:sugra:spacetime:transformation:E}, we find
\begin{equation}
    \delta_\xi \tensor{e}{_m^a}
    = - \xi^\mu \tensor{\hat T}{_\mu_m^a} \bigr|.
\end{equation}
At this point, we need a torsion component with curved indices.
As already mentioned, flat indices in superspace are converted to curved ones using $\tensor{\hat E}{_M^A}$, with sign factors similar to those in \cref{eq:superspace:diffeos:2-form_transformation}.
For the torsion
\begin{gather}
\nonumber
    \frac{1}{2} \hat \d \hat z^N \wedge \hat \d \hat z^M \tensor{\hat T}{_M_N^C}
    = \frac{1}{2} \hat E^B \wedge \hat E^A \tensor{\hat T}{_A_B^C}
    = \frac{1}{2} \hat \d \hat z^N \tensor{\hat E}{_N^B} \wedge \hat \d \hat z^M \tensor{\hat E}{_M^A} \tensor{\hat T}{_A_B^C}
    \\
\label{eq:sugra:spacetime:curved_indices:T}
    \mathllap{\implies\quad}
    \tensor{\hat T}{_M_N^C} = (-1)^{|M|(|N|+|B|)} \tensor{\hat E}{_N^B} \tensor{\hat E}{_M^A} \tensor{\hat T}{_A_B^C}.
\end{gather}
Thus, the torsion component of interest is, see \cref{eq:sugra:T_nonzero},
\begin{equation}
    \tensor{\hat T}{_\mu_m^a} \bigr|
    = - \tensor{\psi}{_m^\beta} \delta_\mu^\alpha \tensor{\hat T}{_\alpha_\beta^a}
    = -2\i (\Gamma^a)_{\mu \beta} \tensor{\psi}{_m^\beta}
\end{equation}
and the supersymmetry transformation of the vielbein
\begin{equation}
\label{eq:sugra:spacetime:susy:e}
    \delta_\xi \tensor{e}{_m^a}
    = 2\i \xi^\alpha (\Gamma^a)_{\alpha \beta} \tensor{\psi}{_m^\beta}.
\end{equation}

For the spin-$3/2$ field, again using \cref{eq:sugra:spacetime:transformation:E},
\begin{equation}
    \delta_\xi \tensor{\psi}{_m^\alpha}
    = - \D_m \xi^\alpha
    - \xi^\nu \tensor{\hat T}{_\nu_m^\alpha} \bigr|.
\end{equation}
By \cref{eq:sugra:spacetime:curved_indices:T,eq:sugra:constraints:T_k1k2}, the relevant torsion component is
\begin{equation}
    \tensor{\hat T}{_\nu_m^\alpha} \bigr|
    = \tensor{e}{_m^a} \delta_\nu^\beta \tensor{\hat T}{_\beta_a^\alpha} \bigl|
    = \frac{1}{288} \tensor{e}{_m^a} \bigl(
        8 \delta_a^{[b_1} (\Gamma^{b_2 b_3 b_4]})\indices{_\nu^\alpha}
        + (\tensor{\Gamma}{_a^{b_1 b_2 b_3 b_4}})\indices{_\nu^\alpha}
    \bigr) \hat H_{b_1 b_2 b_3 b_4}\bigr|.
\end{equation}
To express this in terms of the spacetime fields, we first relate the components of $\hat H$ with flat and curved indices, similar to \cref{eq:sugra:spacetime:curved_indices:T} but for $\hat H$.
We find
\begin{align}
\nonumber
    \hat H_{MNPQ}
    &= (-1)^{(|N|+|B|)|M| + (|P|+|C|)(|M|+|N|) + (|Q|+|D|)(|M|+|N|+|P|)}
    \cdot\\
\label{eq:sugra:spacetime:curved_indices:H}
    &\quad \cdot
    \tensor{\hat E}{_Q^D} \tensor{\hat E}{_P^C} \tensor{\hat E}{_N^B} \tensor{\hat E}{_M^A} \hat H_{ABCD},
\end{align}
whence the spacetime field $H_{mnpq}$ is given by, see \cref{eq:sugra:H_nonzero},
\begin{align}
\nonumber
    H_{mnpq}
    &= \tensor{\hat E}{_q^D} \tensor{\hat E}{_p^C} \tensor{\hat E}{_n^B} \tensor{\hat E}{_m^A} \hat H_{ABCD}\bigr|
    =\\ \nonumber
    &= \tensor{e}{_q^d} \tensor{e}{_p^c} \tensor{e}{_n^b} \tensor{e}{_m^a}
    \hat H_{abcd} \bigr|
    + 6 \tensor{\psi}{_{[q}^\delta} \tensor{\psi}{_p^\gamma} \tensor{e}{_n^b} \tensor{e}{_{m]}^a}
    \bigl(2\i (\Gamma_{ab})_{\gamma \delta} \bigr)
    \\
\label{eq:sugra:spacetime:def:tildeH}
    \mathllap{\implies\quad}
    \hat H_{abcd} \bigr|
    &= H_{abcd}
    + 12\i \tensor{\psi}{_{[a}^\gamma} (\Gamma_{bc})_{|\gamma \delta|} \tensor{\psi}{_{d]}^\delta}
    \eqqcolon \tilde H_{abcd}.
\end{align}
This equation illustrates, explicitly, the importance of distinguishing between $\theta = 0$ components of superfields ($\hat H_{abcd} |$) and spacetime fields ($H_{abcd}$).
Putting the above together, we find the supersymmetry transformation
\begin{align}
\nonumber
    \delta_\xi \tensor{\psi}{_m^\alpha}
    &= - \D_m \xi^\alpha
    - \frac{1}{288} \xi^\beta \tensor{e}{_m^a} \bigl(
        8 \delta_a^{[b_1} (\Gamma^{b_2 b_3 b_4]})\indices{_\beta^\alpha}
        + (\tensor{\Gamma}{_a^{b_1 b_2 b_3 b_4}})\indices{_\beta^\alpha}
    \bigr) \tilde H_{b_1 b_2 b_3 b_4}
    =\\
\label{eq:sugra:spacetime:susy:psi}
    &\eqqcolon - \tilde \D_m \xi^\alpha.
\end{align}

Lastly, we turn to the supersymmetry transformation of $B_{mnp}$.
From \cref{eq:sugra:spacetime:B_transformation},
\begin{equation}
    \delta_\xi B_{mnp} = - \xi^\sigma \hat H_{\sigma mnp} \bigr|.
\end{equation}
The relevant component of $\hat H$ is, by \cref{eq:sugra:spacetime:curved_indices:H,eq:sugra:H_nonzero},
\begin{equation}
    \hat H_{\sigma mnp} \bigr|
    = - 3 \tensor{e}{_{[p}^c} \tensor{e}{_n^b} \tensor{\psi}{_{m]}^\alpha} \delta_\sigma^\delta \hat H_{\delta \alpha b c} \bigr|
    = - 6\i (\Gamma_{[mn})_{|\sigma \alpha|} \tensor{\psi}{_{p]}^\alpha},
\end{equation}
whence the supersymmetry transformation is
\begin{equation}
\label{eq:sugra:spacetime:susy:B}
    \delta_\xi B_{mnp}
    = 6\i \xi^\alpha (\Gamma_{[mn})_{|\alpha \beta|} \tensor{\psi}{_{p]}^\beta}.
\end{equation}

\subsubsection{Equations of motion}
Now that we have found the supersymmetry transformations, we derive the equations of motion for the spacetime fields.
We begin by examining the torsion.
From \cref{eq:sugra:spacetime:curved_indices:T,eq:sugra:T_nonzero},
\begin{subequations}
\begin{align}
    \label{eq:sugra:spacetime:curved_indices:T:mn_a}
    &\tensor{T}{_m_n^a}
    = \tensor{\psi}{_n^\beta} \tensor{\psi}{_m^\alpha} \tensor{\hat T}{_\alpha_\beta^a} \bigr|
    = -2\i \tensor{\psi}{_m^\alpha} (\Gamma^a)_{\alpha\beta} \tensor{\psi}{_n^\beta}
    \\
\label{eq:sugra:spacetime:curved_indices:T:mn_alpha}
    &\tensor{T}{_m_n^\alpha}
    = \tensor{e}{_n^b} \tensor{e}{_m^a} \tensor{\hat T}{_a_b^\alpha}\bigr|
    + 2 \tensor{\psi}{_{[n}^\beta} \tensor{e}{_{m]}^a} \tensor{\hat T}{_a_\beta^\alpha}\bigr|.
\end{align}
\end{subequations}
By the definition of the torsion, $\tensor{T}{_m_n^\alpha} = 2 \hat D_{[m} \tensor{\hat E}{_{n]}^\alpha}| = 2 \D_{[m} \tensor{\psi}{_{n]}^\alpha}$, whence \cref{eq:sugra:spacetime:curved_indices:T:mn_alpha,eq:sugra:constraints:T_k1k2} gives
\begin{align}
\nonumber
    \tensor{\hat T}{_a_b^\alpha}\bigr|
    &= 2 \tensor{e}{_a^m} \tensor{e}{_b^n} \Bigl[
        \D_{[m} \tensor{\psi}{_{n]}^\alpha}
        + \frac{1}{288} \tensor{\psi}{_{[n|}^\beta} \bigl(
            8 \delta_{|m]}^{[p_1} (\Gamma^{p_2 p_3 p_4]})\indices{_\beta^\gamma}
            + (\tensor{\Gamma}{_{|m]}^{p_1 p_2 p_3 p_4}})\indices{_\beta^\gamma}
        \bigr) \tilde H_{p_1 p_2 p_3 p_4}
    \Bigr]
    =\\
\label{eq:sugra:spacetime:hatT_ab^alpha|}
    &= 2 \tensor{e}{_a^m} \tensor{e}{_b^n} \tilde\D_{[m} \tensor{\psi}{_{n]}^\alpha},
\end{align}
with $\tilde\D_m$ as defined in \cref{eq:sugra:spacetime:susy:psi}.
Thus, the equation of motion for $\tensor{T}{_a_b^\alpha}$, \cref{eq:sugra:constraints:eom:T}, gives the equation of motion for $\tensor{\psi}{_m^\alpha}$
\begin{equation}
\label{eq:sugra:spacetime:eom:psi}
    \tilde\D_{m} \tensor{\psi}{_{n}^\beta}\, (\Gamma^{mnp})\indices{_\beta^\alpha}
    = 0.
\end{equation}

Writing the spin connection as $\tensor{\omega}{_m_a^b} = \tensor{\mathring\omega}{_m_a^b} + \tensor{K}{_m_a^b}$, where $\mathring\omega$ is the unique torsion-free spin connection and $\tensor{K}{_m_a^b}$ the contorsion tensor, the definition of the torsion gives $\tensor{T}{_m_n^a} = 2 \tensor{K}{_{[mn]}^a}$.
Since the contorsion is Lie algebra-valued, $K_{abc} = K_{[ab]c} - K_{[ac]b} - K_{[bc]a}$.
Hence, by \cref{eq:sugra:spacetime:curved_indices:T:mn_a},
\begin{equation}
\label{eq:sugra:spacetime:contorsion}
    K_{abc}
    = -\i \tensor{\psi}{_a^\alpha} (\Gamma_c)_{\alpha\beta} \tensor{\psi}{_b^\beta}
    + \i \tensor{\psi}{_a^\alpha} (\Gamma_b)_{\alpha\beta} \tensor{\psi}{_c^\beta}
    + \i \tensor{\psi}{_b^\alpha} (\Gamma_a)_{\alpha\beta} \tensor{\psi}{_c^\beta}
\end{equation}

Turning to the equation of motion for $H$, we need to convert between curved and flat indices on the derivative.
The covariant derivative in spacetime is $\D_m = \hat\D_m |$, where the result is evaluated at $\theta=0$ after $\hat\D_m$ has acted.
In superspace, $\hat\D_M = \tensor{\hat E}{_M^A} \hat \D_A$, whence $\hat\D_m | = \tensor{e}{_m^a} \hat\D_a | + \tensor{\psi}{_m^\alpha} \hat\D_{\alpha} |$.
Thus, using \cref{eq:sugra_calc:BI_H:e:solved,eq:sugra:spacetime:hatT_ab^alpha|},
\begin{align}
\nonumber
    \eta^{ab} \hat\D_a \hat H_{bcde} \bigr|
    &= \eta^{ab} \D_a \hat H_{bcde} \bigr|
    - \eta^{ab} \tensor{\psi}{_a^\alpha} \bigl(
        -12\i \tensor{\hat T}{_{[bc}^\beta}  (\Gamma_{de]})\indices{_\beta_\alpha}
    \bigr) \bigr|
    =\\
    &= \D^b \tilde H_{bcde} + 24 \i \psi^{b\alpha} (\Gamma_{[bc})_{|\alpha\beta|} \tensor{e}{_d^m} \tensor{e}{_{e]}^n} \tilde\D_m \tensor{\psi}{_n^\beta}.
\end{align}
To write the equation of motion with curved indices, we have to move vielbeins through the $\D^a$ derivative.
As explained in \cref{app:Cartan_gravity}, we need to replace the Lorentz connection with an affine connection to be able to do this.
Thus, the equation of motion \cref{eq:sugra:constraints:eom:H} becomes
\begin{equation}
\label{eq:sugra:spacetime:eom:H}
    \nabla^m \tilde H_{mnpq}
    + 24\i \psi^{m \alpha} (\tensor{\Gamma}{_{[m n}})\indices{_{|\alpha\beta|}} \tensor{\tilde\D}{_p} \tensor{\psi}{_{q]}^\beta}
    = - \frac{1}{1152} \tensor{\epsilon}{_{npq}^{r_1 r_2 r_3 r_4}^{s_1 s_2 s_3 s_4}} \tilde H_{r_1 r_2 r_3 r_4} \tilde H_{s_1 s_2 s_3 s_4}.
\end{equation}

Lastly, we turn to the equation of motion for $R$, given in \cref{eq:sugra:constraints:eom:R}.
From the above, it is clear what happens to the right-hand side when switching from superspace to spacetime.
For the left-hand side, we need to relate $R_{mnpq}$ to $\hat R_{abcd}|$.
For arbitrary indices
\begin{equation}
    \tensor{\hat R}{_M_N_C^D}
    = (-1)^{(|N|+|B|)|M|} \tensor{\hat E}{_N^B} \tensor{\hat E}{_M^A} \tensor{\hat R}{_A_B_C^D},
\end{equation}
whence
\begin{equation}
    \hat R_{mncd} \bigr|
    = \tensor{e}{_n^b} \tensor{e}{_m^a} \hat R_{abcd} \bigr|
    + 2 \tensor{e}{_{[n}^b} \tensor{\psi}{_{m]}^\alpha} \hat R_{\alpha bcd} \bigr|
    + \tensor{\psi}{_n^\beta} \tensor{\psi}{_m^\alpha} \hat R_{\alpha \beta cd} \bigr|.
\end{equation}
Using \cref{eq:sugra_calc:BI_T:e:solved,eq:sugra:spacetime:hatT_ab^alpha|} we find that the middle term is
\begin{equation}
    2 \tensor{e}{_{[n}^b} \tensor{\psi}{_{m]}^\alpha} \hat R_{\alpha bcd} \bigr|
    = 4\i \tensor{e}{_c^p} \tensor{e}{_d^q} \tensor{\psi}{_{[m|}^\alpha} \bigl(
        (\Gamma_{[p|})_{\alpha\beta} \tilde\D_{|q]} \tensor{\psi}{_{|n]}^\beta}
        - (\Gamma_{[p|})_{\alpha\beta} \tilde\D_{|n]} \tensor{\psi}{_{|q]}^\beta}
        - (\Gamma_{|n]})_{\alpha\beta} \tilde\D_{[p} \tensor{\psi}{_{q]}^\beta}
    \bigr),
\end{equation}
where, in all terms, the antisymmetrisations are $[m\, n]$ and $[p\, q]$.
By \cref{eq:sugra_calc:BI_T:c:solved,eq:sugra:spacetime:def:tildeH}, the third term is
\begin{equation}
    \tensor{\psi}{_n^\beta} \tensor{\psi}{_m^\alpha} \hat R_{\alpha \beta cd} \bigr|
    =
    \frac{4\i}{288} \tensor{\psi}{_m^\alpha} \bigl(
        24 \tensor*{\delta}{*_{c}^{a_1}_{d}^{a_2}} (\Gamma^{a_3 a_4})_{\alpha\beta}
        + (\tensor{\Gamma}{_c_d^{a_1 a_2 a_3 a_4}})_{\alpha\beta}
    \bigr)
    \tensor{\psi}{_n^\beta} \tilde H_{a_1 a_2 a_3 a_4}.
\end{equation}
Defining $\tilde R_{abcd} = \hat R_{abcd} |$ and collecting the above terms, we find
\begin{align}
\nonumber
    \tilde R_{mnpq}
    &= R_{mnpq}
    -
    \frac{4\i}{288} \tensor{\psi}{_m^\alpha} \bigl(
        24 \tensor*{\delta}{*_{p}^{r_1}_{q}^{r_2}} (\Gamma^{r_3 r_4})_{\alpha\beta}
        + (\tensor{\Gamma}{_p_q^{r_1 r_2 r_3 r_4}})_{\alpha\beta}
    \bigr)
    \tensor{\psi}{_n^\beta} \tilde H_{r_1 r_2 r_3 r_4}
    +\\
\label{eq:sugra:spacetime:def:tildeR}
    &\quad
    -
    4\i \tensor{\psi}{_{[m|}^\alpha} \bigl(
        (\Gamma_{[p|})_{\alpha\beta} \tilde\D_{|q]} \tensor{\psi}{_{|n]}^\beta}
        - (\Gamma_{[p|})_{\alpha\beta} \tilde\D_{|n]} \tensor{\psi}{_{|q]}^\beta}
        - (\Gamma_{|n]})_{\alpha\beta} \tilde\D_{[p} \tensor{\psi}{_{q]}^\beta}
    \bigr).
\end{align}
With this definition, the equation of motion reads
\begin{equation}
\label{eq:sugra:spacetime:eom:R}
    \tilde R_{mn} - \frac{1}{2} g_{mn} \tilde R
    = \frac{1}{12} \tensor{\tilde H}{_{mpqr}} \tensor{\tilde H}{_n^{pqr}}
    - \frac{1}{96} g_{mn} \tilde H^2,
\end{equation}
where $\tilde R_{mn} \coloneqq \tensor{\tilde R}{_p_m_n^p}$ and $\tilde R = \tensor{\tilde R}{_m^m}$.%
\footnote{Recall that we are working in right-action conventions, whence this definition agrees with $R < 0$ for AdS.}

\subsection{Component formulation with left-action} \label{sec:sugra:components}
In this section, we give the most important equations from \cref{sec:sugra:spacetime} but in left-action conventions.
The crucial insight to convert a right-action to a left-action is that $(g_1 g_2)^{-1} = g_2^{-1} g_1^{-1}$.
Thus, given a right-action, we can define a left-action by acting with $g^{-1}$ from the right and, given a right-representation $(v \cdot g)^a = v^b g\indices{_b^a}$, we get a left-representation $(g\cdot v)^a = v^b (g^{-1})\indices{_b^a}$.
Hence, we need to replace all group elements by their inverses and all Lie algebra elements by their negatives when switching conventions.

We also switch conventions for the index order on differential forms.
Here, we employ the usual convention
\begin{equation}
    \Omega = \frac{1}{p!} \d x^{m_1} \wedge \hdots \wedge \d x^{m_p} \Omega_{m_1 \hdots m_p}.
\end{equation}
We demand that the 3-form, $B$, is the same when switching convention.
This implies that we must replace $B_{mnp}$ with $-B_{mnp}$ and $H_{mnpq}$ with $-H_{mnpq}$.%
\footnote{Note that $H_{mnpq} = 4 \partial_{[m} B_{npq]}$ in both conventions due to the difference in the definition of the exterior derivative.}
Since the spin connection is Lie algebra-valued, it should be replaced with its negative.
This implies that the curvature 2-form is unchanged while its components $\tensor{R}{_m_n_a^b}$ change sign.%
\footnote{This is consistent with Lie algebra elements being replaced by their negatives due to the additional change of conventions for differential forms.}
In these conventions, we define the Ricci tensor $\tensor{R}{_m_n} = \tensor{R}{_m_p_n^p}$.
Due to how we defined the Ricci tensor in the other set of conventions, it does not change sign.

In this section we will, moreover, not write out spinor indices explicitly.
Spinors have an implicit subscript index ($\chi$ means $\chi_\alpha$), Dirac conjugated spinors have an implicit superscript index ($\bar\chi$ means $\bar\chi^\alpha$) and $\Gamma$-matrices have their indices in the usual positions ($\Gamma^a$ means $(\Gamma^a)\indices{_\alpha^\beta}$).%
\footnote{Due to how we raise and lower spinor indices in $D=11$, the Majorana condition $\smash{\chi_\alpha = \delta_\alpha^{\dot \beta} \chi^\dagger_{\dot\beta}}$ can be written as $\smash[t]{\bar\chi^\alpha = - \chi^\alpha}$ (see \cref{app:conventions:11d-spinors}).}

From \cref{eq:sugra:spacetime:contorsion} we find that the spin connection, with these conventions, is
\begin{equation}
\label{eq:sugra:components:spin_connection}
    \omega_{abc}
    = \mathring\omega_{abc}
    - \i \bigl(
        \bar\psi_a \Gamma_c \psi_b
        - \bar\psi_a \Gamma_b \psi_c
        - \bar\psi_b \Gamma_a \psi_c
    \bigr),
\end{equation}
where $\mathring\omega_{mab}$ is the torsion-free spin connection.
The Lorentz covariant derivative is $\D_m = \partial_m + \omega_m$ and $\nabla_m$ denotes the associated affine connection, see \cref{app:Cartan_gravity}.
$R_{mnpq}$ denotes the curvature tensor of the spin connection $\omega_m$.

The supersymmetry transformations are, by \cref{eq:sugra:spacetime:susy:e,eq:sugra:spacetime:susy:psi,eq:sugra:spacetime:susy:B},
\begin{subequations}
\label{eq:sugra:components:susy}
\begin{align}
    &\delta_\xi \tensor{e}{_m^a} = -2\i \bar\xi \Gamma^a \psi_m,
    \\
    &\delta_\xi \psi_m = - \tilde\D_m \xi,
    \\
    &\delta_\xi B_{mnp} = 6\i \bar\xi \Gamma_{[mn} \psi_{p]},
\end{align}
\end{subequations}
and equations of motion, by \cref{eq:sugra:spacetime:eom:psi,eq:sugra:spacetime:eom:H,eq:sugra:spacetime:eom:R},
\begin{subequations}
\label{eq:sugra:components:eom}
\begin{align}
\label{eq:sugra:components:eom:psi}
    &\Gamma^{mnp} \tilde\D_n \psi_p = 0,
    \\
    &\nabla^m \tilde H_{mnpq} + 24\i \psi^m \Gamma_{[mn} \tilde\D_{p} \psi_{q]}
    = \frac{1}{1152} \tensor{\epsilon}{_{npq}^{r_1 r_2 r_3 r_4 s_1 s_2 s_3 s_4}} \tilde H_{r_1 r_2 r_3 r_4} \tilde H_{s_1 s_2 s_3 s_4}
    \\
    &\tilde R_{mn} - \frac{1}{2} g_{mn} \tilde R
    = \frac{1}{12} \tensor{\tilde H}{_{mpqr}} \tensor{\tilde H}{_n^{pqr}}
    - \frac{1}{96} g_{mn} \tilde H^2
\end{align}
\end{subequations}
where, by \cref{eq:sugra:spacetime:susy:psi,eq:sugra:spacetime:def:tildeH,eq:sugra:spacetime:def:tildeR},
\begin{subequations}
\begin{align}
    &\tilde\D_m \xi
    = \D_m \xi - \frac{1}{288} \tilde H_{n_1 n_2 n_3 n_4} \bigl(
        8 \delta_m^{[n_1} \Gamma^{n_2 n_3 n_4]}
        - \tensor{\Gamma}{_m^{n_1 n_2 n_3 n_4}}
    \bigr) \xi,
    \\
    &\tilde H_{mnpq} = H_{mnpq} + 12\i \bar\psi_{[m} \Gamma_{np} \psi_{q]},
    \\ \nonumber
    &\tilde R_{mnpq}
    = R_{mnpq}
    -
    \frac{4\i}{288} \tilde H_{r_1 r_2 r_3 r_4} \bar\psi_m \bigl(
        24 \tensor*{\delta}{*_{p}^{r_1}_{q}^{r_2}} \Gamma^{r_3 r_4}
        + \tensor{\Gamma}{_p_q^{r_1 r_2 r_3 r_4}}
    \bigr)
    \psi_n
    + \\
    &\phantom{\tilde R_{mnpq} =}\,
    + 4\i \bar\psi_{[m|} \bigl(
        \Gamma_{[p} \tilde\D_{q]} \psi_{|n]}
        - \Gamma_{[p|} \tilde\D_{|n]} \psi_{|q]}
        - \Gamma_{|n]} \tilde\D_{[p} \psi_{q]}
    \bigr),
\end{align}
\end{subequations}
where, in the last line, the antisymmetrisations are $[m\, n]$ and $[p\, q]$.

\chapter{Supergravity compactifications} \label{chap:sugra_comp}%
If eleven-dimensional supergravity describes reality (at low energies) an obvious question that arises is why we perceive reality as having only four dimensions.
Historically, this question arose as early as 1921 when Kaluza \cite{ref:Kaluza} proposed a unification of gravity and electromagnetism by introducing a fifth dimension.%
\footnote{The theory also contains a dilaton. At the time, this was, however, inconsistently set to zero~\cite{ref:Duff:KK_theory}.}
To obtain Einstein's and Maxwell's field equations, Kaluza assumed, ad hoc, that all fields are independent of the fifth dimension.
This would also explain why we cannot see the fifth dimension since there can be no dynamics in a direction in which everything is constant.
Still, it seems like an unmotivated assumption; if nature is truly five-dimensional, why should all fields be constant in a specific direction?

A more satisfactory explanation was put forward by Klein \cite{ref:Klein:ZPhys} in 1926.
Klein assumed that the fifth dimension is periodic, that is, that the topology of spacetime is that of $\RR^4 \times S^1$.
Then, all fields can be expanded in Fourier series in the periodic coordinate and ordinary gravity and electromagnetism correspond to the zero-modes in the expansion \cite{ref:Duff:KK_theory}.
Klein's idea also explains the quantisation of electric charge, which corresponds to momentum in the periodic dimension and is naturally quantised due to the periodicity \cite{ref:Klein:Nature}.
Assuming that the smallest unit of charge is that of the electron, Klein derived the period of the compact dimension to be of order \SI{e-30}{\metre}.
This also explains why we do not observe five dimensions in experiments since physics at much larger scales would be averaged over the compact dimension.
However, momentum in the periodic dimension also gives the fields masses.
With the above period, these are of the same order as the Planck mass \cite{ref:Duff:KK_theory}, that is, about \num{e22} times the electron mass.

The situation for string and supergravity theories is similar.
If some of the dimensions form a compact manifold, this explains why we only observe four dimensions, provided that the extra dimensions are sufficiently small.
Similar to how electromagnetism arises in Kaluza--Klein theory, isometries of the compact manifold, or internal space, give rise to, possibly non-Abelian, gauge fields in spacetime \cite{ref:Duff--Nilsson--Pope:KK}.
As explained in more detail in \cref{sec:mass_operators} and also analogous to Kaluza--Klein theory, momentum in the internal directions contribute to the mass of spacetime fields.

In this chapter, and the remainder of the thesis, we use slightly different notation and conventions than in \cref{sec:sugra:components}.
Uppercase indices are used as eleven-dimensional spacetime indices, not superspace indices.
In compactifications, Greek indices ($\alpha, \beta, \gamma, \hdots$ and $\mu, \nu, \rho, \hdots$) are used for the resulting spacetime and Latin lowercase ($a, b, c, \hdots$ and $m, n, p, \hdots$) for the internal manifold, that is, the extra dimensions.
Letters from the beginning of the alphabets are used for flat indices and letters from the middle for curved ones.
Spinor indices are not written out.

To distinguish between the components of the eleven-dimensional $\Gamma$-matrices in the internal directions and the $\Gamma$-matrices on the internal manifold, we denote the eleven-dimensional $\Gamma$-matrices by $\hat\Gamma^A$, similar to \cref{app:conventions:11d-spinors}.
Furthermore, we denote the 3-form by $A$ instead of $B$ and its field strength by $F$ instead of $H$.
We also rescale the gravitino $\psi_M \mapsto \psi_M/\sqrt{2}$ and the supersymmetry parameter $\xi \mapsto -\xi/\sqrt{2}$.
Lastly, $x$ denotes coordinates on the spacetime and $y$ on the internal manifold.

With these conventions, the bosonic equations of motion, resulting from setting $\psi_M = 0$ in \cref{eq:sugra:components:spin_connection,eq:sugra:components:eom}, are
\begin{subequations}
\label{eq:sugra_comp:eom}
\begin{align}
\label{eq:sugra_comp:eom:R}
    &R_{MN} - \frac{1}{2} g_{MN} R
    = \frac{1}{12} \tensor{F}{_{MPQR}} \tensor{F}{_N^{PQR}}
    - \frac{1}{96} g_{MN} F^2,
    \\
\label{eq:sugra_comp:eom:F}
    &\nabla^M F_{MNPQ}
    = \frac{1}{1152} \tensor{\epsilon}{_{NPQ}^{R_1 R_2 R_3 R_4}^{S_1 S_2 S_3 S_4}} \tensor{F}{_{R_1 R_2 R_3 R_4}} \tensor{F}{_{S_1 S_2 S_3 S_4}},
\end{align}
\end{subequations}
where the curvature and covariant derivative are those of the torsion-free connection, and the supersymmetry transformations are, by \cref{eq:sugra:components:susy},
\begin{subequations}
\label{eq:sugra_comp:susy}
\begin{align}
    &\delta_\xi \tensor{e}{_M^A}
    = \i \bar\xi \hat\Gamma^A \psi_M,
    \\
    &\delta_\xi A_{MNP}
    = 3\i \bar\xi \hat\Gamma_{[MN} \psi_{P]},
    \\
\label{eq:sugra_comp:susy:psi}
    &\delta_\xi \psi_M
    = \tilde\D_M \xi
    \simeq \D_M \xi
    - \frac{1}{288} F_{NPQR} \bigl(
        8 \delta_M^{[N} \tensor{\hat\Gamma}{^{PQR]}}
        - \tensor{\hat\Gamma}{_M^{NPQR}}
    \bigr) \xi,
\end{align}
\end{subequations}
where, in the step indicated by $\simeq$, we have dropped terms containing $\psi_M$.
These conventions (apart from how the Dirac conjugate is defined), equations of motion and supersymmetry transformations agree with \cite{ref:Becker--Becker--Schwarz}.

\section{Freund--Rubin compactification} \label{sec:Freund--Rubin}
In theories with extra dimensions, we wish to achieve what is known as spontaneous compactification.
In contrast to ad hoc compactification, we do not simply postulate that some dimensions are compact but instead look for stable ground state, or vacuum, solutions to the field equations that describe, at least locally, a product space $\M_d \times \M_{k}$ \cite{ref:Duff--Nilsson--Pope:KK}.
Here, $D = d + k$ is the dimension of the complete reality in the theory, $d$ the dimension of spacetime (after compactification) and $k$ the dimension of the internal manifold.
We will work towards the Freund--Rubin ansatz \cite{ref:Freund--Rubin}, which is a way of achieving spontaneous compactification, but make some more general comments before arriving at the full set of assumptions in the ansatz.

The first assumption we will employ is to assume that the vacuum spacetime $\M_d$ is maximally symmetric.%
\footnote{We may add, as a zeroth assumption, that we assume $\M_d$ and $\M_k$ to be spin manifolds so that spinors can be defined globally. See \cite{ref:Lawson--Michelsohn} for an introduction to spin geometry.}
This is motivated experimentally and is a generally accepted assumption.%
\footnote{A cosmology with a Big Bang singularity is clearly not maximally symmetric. However, this is due to the matter content, not the vacuum.}
The assumption implies that the vacuum expectation values of the various fields can only be constructed from scalars, the metric and the Levi-Civita tensor in the external space.
We do not consider the possibility of topological phases with vanishing vacuum expectation value of the metric.

The $D=11$ spinor decomposes into the tensor product of the spinor in spacetime and the spinor on the internal space when $\Spin(D-1,1)$ is broken to $\Spin(d-1,1)\times\Spin(k)$, see \cref{app:conventions:11d-spinors}.
Hence, a spinor in $D=11$ can be written as a sum of terms on the form $\varepsilon\otimes\eta$, where $\varepsilon$ is an anticommuting spinor in spacetime and $\eta$ a commuting spinor on the internal manifold.%
\footnote{See \cite{ref:Klinker} for a general discussion on the spinor bundle of product manifolds.}
Since a nonzero spinor or vector-spinor in spacetime would break maximal symmetry \cite{ref:Duff--Nilsson--Pope:KK}, we set $\mathring \psi_M = 0$, where the overset circle denotes that it is a vacuum value.
Since $\psi_M$ is the only fermion field in the theory, this clearly solves its equation of motion.
Note that this does not imply that the vacuum expectation value of fermion bilinears vanish, that is, there can be fermion condensates \cite{ref:Duff--Orzalesi}.%
\footnote{Non-vanishing vacuum expectation values of fermion bilinears have consequences for the cosmological constant \cite{ref:Duff--Orzalesi}.}
In the following, we assume that all fermion bilinears vanish as well, which implies that the relevant equations of motion are those in \cref{eq:sugra_comp:eom} and, by \cref{eq:sugra:components:spin_connection}, that the spin connection is torsion-free in the background.

Maximal symmetry forces the the $x$-dependence of the spacetime vacuum metric $\mathring g_{\mu\nu}$ to be either that of Minkowski, de Sitter (dS), or anti-de Sitter (AdS) spacetime, corresponding to $R=0$, $R>0$ and $R<0$, respectively \cite{ref:Duff--Nilsson--Pope:KK}.%
\footnote{Note that these are local considerations, that is, the spacetime is \emph{locally} isometric to Minkowski, dS or AdS.}
In general, there may also be a $y$-dependence, whence we write $\mathring g_{\mu\nu} = f(y) g^{\mathrm{m.s.}}_{\mu\nu}(x)$ where $f>0$ is known as the warp factor and $g^{\mathrm{m.s.}}_{\mu\nu}(x)$ is maximally symmetric.%
\footnote{Here, we require that the spacetime metric is of constant signature $(d-1, 1)$. Signature changing metrics have been discussed in the context of cosmology and quantum gravity, see \cite{ref:Hayward,ref:Ellis,ref:White}.}%
\footnotemarksep%
\footnote{Intuitively, we glue together copies of $\M_d$ with different sizes over $\M_k$. For instance, $S^2$ without the poles is a warped product of a circle (the equator) and a line (a meridian).}
Since a nonvanishing spacetime vector field would break maximal symmetry, the mixed components of the metric vanish, $\mathring g_{\mu n} = 0$.
Also, since the internal components $\mathring g_{mn}$ are spacetime scalars, they must be $x$-independent to not break maximal symmetry.

Consider now the 4-form $F_{MNPQ} = (F_{\mu\nu\rho\sigma}, F_{\mu\nu\rho q}, F_{\mu\nu pq}, F_{\mu npq}, F_{mnpq})$.
Here, maximal symmetry forces $\mathring F_{mnpq}$ to be $x$-independent.
Furthermore, any of the other components can only be nonvanishing if it is a product of the completely antisymmetric $\epsilon$-tensor in spacetime and a tensor on the internal manifold.
Thus, with $d=2,3,4$ we may have nonzero $\mathring F_{\mu\nu pq}$, $\mathring F_{\mu\nu\rho q}$ and $\mathring F_{\mu\nu\rho\sigma}$, respectively.
For other values of $d$, all three of these vanish.

At this point, we make the additional assumption that $\mathring F_{\mu\nu\rho q} = 0 = \mathring F_{\mu\nu pq}$.
Thus, we can set $\mathring F_{\mu\nu\rho\sigma} = -6m \mathring \epsilon_{\mu\nu\rho\sigma}$, with $m = 0$ in $d\neq 4$ where $\mathring \epsilon_{\mu\nu\rho\sigma}$ does not make sense.
Note that, for dimensional reasons, $m$ has dimension mass and it is independent of $y$ due to the Bianchi identity $\partial_{[M} F_{NPQR]} = 0$.
The Bianchi identity also implies $\mathring F_{mnpq} = \mathring F_{mnpq}(y)$, which, as noted above, also follows from maximal symmetry, and $\partial_{[m} F_{npqr]} = 0$.

We now turn to the Einstein equation \cref{eq:sugra_comp:eom:R}.
It is convenient to write the equation for the vacuum values as
\begin{equation}
\label{eq:Freund--Rubin:eom:R}
    \mathring R_{MN}
    = \frac{1}{12} \tensor{\mathring F}{_{MPQR}} \tensor{\mathring F}{_N^{PQR}}
    - \frac{1}{144} \mathring g_{MN} \tensor{\mathring F}{_{PQRS}} \tensor{\mathring F}{^{PQRS}}
\end{equation}
Using that $\mathring F_{\mu\nu\rho\sigma} = -6m \mathring \epsilon_{\mu\nu\rho\sigma}$ and that $\M_d$ is Lorentzian, we find
\begin{equation}
    \tensor{\mathring F}{_{\mu \rho\sigma\lambda}}
    \tensor{\mathring F}{_{\nu}^{\rho\sigma\lambda}}
    = - 6 (6m)^2 \mathring g_{\mu\nu},
    \qquad\quad
    \tensor{\mathring F}{_{\mu\rho\sigma\lambda}}
    \tensor{\mathring F}{^{\mu\rho\sigma\lambda}}
    = -24 (6m)^2.
\end{equation}
Since the vacuum metric is block diagonal, \cref{eq:Freund--Rubin:eom:R} splits into
\begin{subequations}
\label{eq:Freund--Rubin:eom:R:components}
\begin{align}
\label{eq:Freund--Rubin:eom:R:components:spacetime}
    &\mathring R_{\mu\nu}
    = - \frac{1}{144} \mathring g_{\mu\nu} \bigl(
        12^3 m^2
        + \tensor{\mathring F}{_{mnpq}} \tensor{\mathring F}{^{mnpq}}
    \bigr),
    \\
    &\mathring R_{mn}
    = \frac{1}{12} \tensor{\mathring F}{_{m pqr}} \tensor{\mathring F}{_n^{pqr}}
    - \frac{1}{144} \mathring g_{mn} \bigl(
        \tensor{\mathring F}{_{pqrs}} \tensor{\mathring F}{^{pqrs}}
        - 864 m^2
    \bigr),
    \\
    &\mathring R_{\mu n} = 0.
\end{align}
\end{subequations}
By choosing $f(y)$ appropriately, we can make $R_{(d)}^{\mathrm{m.s.}} \in \set{-1,0,1}$, where $R_{(d)}^{\mathrm{m.s.}}$ is the Ricci scalar of the maximally symmetric metric $g_{\mu\nu}^{\mathrm{m.s.}}$.
Contracting \cref{eq:Freund--Rubin:eom:R:components:spacetime} with $\mathring g^{\mu\nu}$, we find, since $\mathring R_{\mu\nu}$ is independent of $f(y)$,
\begin{equation}
\label{eq:Freund--Rubin:warp_analysis}
    \mathring R_{(d)}(y)
    =
    \frac{R_{(d)}^{\mathrm{m.s.}}}{f(y)}
    = - \frac{d}{144} \bigl(
        12^3 m^2
        + \tensor{\mathring F}{_{mnpq}}(y) \tensor{\mathring F}{^{mnpq}}(y)
    \bigr).
\end{equation}
Since $f>0$, $\mathring R_{(d)}(y)$ is of constant sign.
Furthermore, $\mathring R_{(d)}(y)$ is only zero at a point $y$ if both $m = 0$ and $\mathring F_{mnpq}(y) = 0$ and, then, $\mathring F_{mnpq} = 0$ at all points due to $\mathring R_{(d)}(y)$ having constant sign.
Thus, the only Minkowski solution, under the above assumptions, is the zero flux case $\mathring F_{MNPQ} = 0$ with a Ricci flat internal manifold, $\mathring R_{mn} = 0$.
In all other cases, $\mathring R_{(d)} < 0$, the spacetime is AdS and the internal manifold has everywhere positive scalar curvature $\mathring R_{(k)} > 0$.

Now, we assume that $m\neq 0$ and that there is no internal flux, that is, $\mathring F_{mnpq} = 0$.
These are the last assumptions in the Freund--Rubin ansatz.
As noted above, the former forces $d = 4$ and $k = 7$ while the latter implies that $f$ is independent of $y$ by \cref{eq:Freund--Rubin:warp_analysis}.
Under these assumptions, \cref{eq:Freund--Rubin:eom:R:components} immediately gives
\begin{equation}
\label{eq:Freund--Rubin:Einstein_metrics}
    \mathring R_{\mu\nu} = - 12 m^2 \mathring g_{\mu\nu},
    \qquad\quad
    \mathring R_{mn} = 6 m^2 \mathring g_{mn}.
\end{equation}
Thus, the spacetime is $\mathrm{AdS}_4$ and the internal manifold an Einstein manifold with positive scalar curvature.
Hence, assuming that the internal manifold is complete implies, by the Bonnet--Myers theorem \cite{ref:Myers:Bonnet--Myers_thm}, that it is compact and of finite diameter%
\footnote{The diameter of a Riemannian manifold (that is, pseudo-Riemannian with Euclidean signature) is the supremum of all distances.}.
\todo[disable]{}

The last part of \cref{eq:Freund--Rubin:eom:R:components}, $\mathring R_{\mu n} = 0$, is trivially satisfied since the only nonzero Christoffel symbols of the Levi-Civita connection are $\tensor{\mathring\affine}{_\mu^\nu_\rho}(x)$ and $\tensor{\mathring\affine}{_m^n_p}(y)$.
Note that there not being any other nonzero Christoffel symbols implies that the affine connection splits as $\mathring\nabla_M = \mathring\nabla_\mu \oplus \mathring\nabla_m$, where $\mathring\nabla_\mu$ and $\mathring\nabla_m$ are the Levi-Civita connections of $\mathrm{AdS}_4$ and $\M_7$, respectively.
For nontrivial warp factors, this is not generally true.

Note that the equation of motion for $F$, \cref{eq:sugra_comp:eom:F}, is satisfied since the right-hand side immediately vanishes and $\mathring\nabla^\mu \mathring\epsilon_{\mu\nu\rho\sigma} = 0$.
With $\mathring F_{mnpq} \neq 0$, there would have been a nontrivial equation
\begin{equation}
    \mathring\nabla^m \mathring F_{mnpq}
    = \frac{1}{4} \tensor{\mathring \epsilon}{_{npq}^{rstu}} \mathring F_{rstu}.
\end{equation}

\subsubsection{Unbroken supersymmetries}
Thus far, we have not paid much attention to the gravitino, $\psi_M$, as its background value is $0$.
For supersymmetry, it is, however, crucial.
Since the supersymmetry parameter is fermionic, each term in the transformations of the bosonic fields must contain the gravitino, as is also evident from \cref{eq:sugra_comp:susy}.
Thus, all bosonic fields are invariant under supersymmetry transformations in the vacuum.
That the gravitino is also invariant under a supersymmetry transformation in the vacuum is, therefore, equivalent to the corresponding generator being unbroken.

As noted above, the spinor in eleven dimensions decomposes into a tensor product of a four-component spinor in spacetime and an eight-component spinor on $\M_7$.
Explicitly, we write the eleven-dimensional $\Gamma$-matrices as
\begin{equation}
    \hat\Gamma^\alpha = \gamma^\alpha \otimes \1,
    \qquad\quad
    \hat\Gamma^a = - \gamma^5 \otimes \Gamma^a,
\end{equation}
where, as in \cref{app:conventions:11d-spinors} but with slightly different notation, $\gamma^\alpha$ are the four-dimensional $\gamma$-matrices, $\gamma^5 = -\i \epsilon_{\alpha\beta\gamma\delta} \gamma^{\alpha\beta\gamma\delta}/24$ and $\Gamma^a$ are the seven-dimensional $\Gamma$-matrices.

In the Freund--Rubin vacuum, the supersymmetry transformation of the gravitino is $\delta_\xi \psi_M = \tilde\D_M \xi$.
Here, and in the following, we drop the overset circle to reduce clutter; all quantities refer to vacuum values.
Putting $\delta_\xi \psi_M = 0$, we obtain the (generalised) Killing spinor equation
\begin{equation}
\label{eq:Freund--Rubin:Killing_spinor}
    \tilde\D_M \xi
    = \D_M \xi
    + \frac{6m}{288} \epsilon_{\mu\nu\sigma\rho} \bigl(
        8 \delta_M^{[\mu} \tensor{\hat\Gamma}{^{\nu\sigma\rho]}}
        - \tensor{\hat\Gamma}{_M^{\mu\nu\sigma\rho}}
    \bigr) \xi
    = 0
\end{equation}
Since $\M_7$ is compact, $\xi(x,y)$ can be expanded as $\xi(x,y) = \chi^I(x) \otimes \lambda_I(y)$ where $\lambda_I(y)$ is a complete (infinite but countable) set of linearly independent spinors on $\M_7$.
Using $\epsilon_{\mu\nu\rho\sigma} \tensor{\hat\Gamma}{^{\nu\rho\sigma}}
= -6\i \gamma^5 \gamma_\mu \otimes \1$ and that the spin connection in $D=11$ only has nonzero components $\omega_{\mu\alpha\beta}$ and $\omega_{mab}$, the $M=\mu$ part of \cref{eq:Freund--Rubin:Killing_spinor} gives
\begin{equation}
    \bigl( \D_\mu \chi^I - \i m \gamma^5 \gamma_\mu \chi^I \bigr) \otimes \lambda_I
    = 0,
\end{equation}
where, by abuse of notation, we use $\D_\mu$ to denote the covariant derivative in $\mathrm{AdS}_4$.
Since $\lambda_I$ are linearly independent, the parenthesis must vanish for every $I$.
In $\mathrm{AdS}_4$, this equation admits four linearly independent solutions $\varepsilon^i(x)$ for $\chi$, which is the maximal number in $d=4$ \cite{ref:Duff--Nilsson--Pope:KK}.
Due to the linear independence of these, we may reorder the terms in the expansion and write $\xi(x,y) = \varepsilon^i(x) \otimes \eta_i(y)$, which is now a sum of just four terms.
The $M=m$ part of \cref{eq:Freund--Rubin:Killing_spinor} then reads, using $\epsilon_{\mu\nu\rho\sigma} \tensor{\hat\Gamma}{_m^{\mu\nu\rho\sigma}}
= -24\i\, \1 \otimes \Gamma_m$,
\begin{equation}
    \varepsilon^i \bigl( \D_m \eta_i + \i \frac{m}{2} \Gamma_m \eta_i \bigr)
    = 0.
\end{equation}
Again the parenthesis must vanish for every $i$, now due to the linear independence of $\varepsilon^i$.
Hence, the most general solution $\xi$ to $\delta_\xi \psi_M = 0$ is a sum of terms $\varepsilon(x) \otimes \eta(y)$ where
\begin{subequations}
\begin{align}
\label{eq:Freund--Rubin:spacetime_Killing_spinor}
    &\tilde\D_\mu \varepsilon
    \coloneqq \D_\mu \varepsilon - \i m \gamma^5 \gamma_\mu \varepsilon
    = 0,
    \\
\label{eq:Freund--Rubin:internal_Killing_spinor}
    &\tilde\D_m \eta
    \coloneqq \D_m \eta + \i \frac{m}{2} \Gamma_m \eta
    = 0.
\end{align}
\end{subequations}
Since, as noted above, $\mathrm{AdS}_4$ admits four Killing spinors, we get four linearly independent supercharges forming a spinorial supersymmetry generator $Q$ for each linearly independent solution to \cref{eq:Freund--Rubin:internal_Killing_spinor}.
Accordingly, the number of supersymmetries, $\N$, is the number of linearly independent solutions to the Killing spinor equation \cref{eq:Freund--Rubin:internal_Killing_spinor} on the internal space $\M_7$.

Consider now the curvature of the connection $\tilde\D_m$ on the spinor bundle of $\M_7$.
By extending $\tilde\D_m$ to $\tilde\nabla_m$, which acts on spinors as $\tilde\D_m$ and on vectors as $\nabla_m$, and using that $\nabla_m$ is torsion-free, the curvature of $\tilde\D_m$ may be computed as
\begin{align}
\nonumber
    [\tilde\nabla_m, \tilde\nabla_n] \eta
    &= \bigl(
        [\nabla_m, \nabla_n]
        + \i \frac{m}{2} \nabla_m(\Gamma_n)
        - \i \frac{m}{2} \nabla_n(\Gamma_m)
        - \frac{m^2}{4} [\Gamma_m, \Gamma_n]
    \bigr) \eta
    =\\
\label{eq:Freund--Rubin:tildeD_curvature}
    &=
    \frac{1}{4} \bigl(\tensor{R}{_m_n^p^q} - 2m^2 \tensor*{\delta}{*_m^p_n^q} \bigr) \Gamma_{pq} \eta,
\end{align}
where we have also used that $\nabla_m(\Gamma_n) = 0$ and that $\Gamma_{pq}/4$ are the Lorentz generators in the spinor representation.
Recall that the Weyl tensor, in arbitrary dimension $d>2$, is
\begin{equation}
    \tensor{W}{_m_n^p^q}
    = \tensor{R}{_m_n^p^q}
    - \frac{4}{d-2} \tensor{R}{_{[m}^{[p}} \tensor*{\delta}{_{n]}^{q]}}
    + \frac{2}{(d-1)(d-2)} R \tensor*{\delta}{*_m^p_n^q}.
\end{equation}
Thus, for our seven-dimensional Einstein manifold with $R_{mn} = 6m^2 g_{mn}$,
\begin{equation}
\label{eq:Freund--Rubin:Weyl_tensor}
    \tensor{W}{_m_n^p^q}
    = \tensor{R}{_m_n^p^q} - 2m^2 \tensor*{\delta}{*_m^p_n^q},
\end{equation}
which we recognise from \cref{eq:Freund--Rubin:tildeD_curvature}.
Hence, any solution $\eta$ to the Killing spinor equation \cref{eq:Freund--Rubin:internal_Killing_spinor} also satisfies the integrability condition
\begin{equation}
\label{eq:Freund--Rubin:susy_integrability}
    \tensor{W}{_m_n^p^q} \Gamma_{pq} \eta = 0.
\end{equation}

The holonomy of a connection $\hat\nabla$ on a vector bundle $E$, with fibre $F$, at a point $p \in \M$, $\Hol_p(\hat\nabla)$, is defined as the subgroup of $\GL(E_p)$ obtained by all possible parallel transports%
\footnote{Parallel transport is defined by demanding that the covariant derivative along the curve vanishes. See for instance \cite{ref:ONeil,ref:Joyce}.}
around closed loops in $\M$ starting at $p$.
For a proper introduction to holonomy,
see \cite{ref:Joyce}.%
\footnote{As an example, the holonomy of the Levi-Civita connection on the sphere $S^2$ is $\SO(2)$, that is, parallel transport can rotate vectors arbitrarily but not change their lengths nor turn a right-handed pair of vectors into a left-handed pair.}
For an introduction to bundles, see \cref{app:bundles}.
If $\M$ is connected, which we assume for our $\M_7$, the holonomy group is independent of the base point $p$ (up to conjugation) and we simply write $\Hol(\hat\nabla) \subseteq \GL(F)$.
The restricted holonomy group $\Hol^0(\hat\nabla)$, defined as $\Hol(\hat\nabla)$ but restricted to null-homotopic loops, is a connected Lie subgroup of $\GL(F)$ \cite{ref:Joyce};
it is the identity component of $\Hol(\hat\nabla)$.
For a principal connection, the holonomy is a subgroup of the structure group \cite{ref:Joyce}.
In our case with $\tilde\D$, the holonomy group $\Hol(\tilde\D)$ is the subgroup of invertible linear transformations on the space of 8-component spinors obtained by the parallel transport maps defined by $\tilde\D$.

The Lie algebra of the restricted holonomy group, $\hol(\hat\nabla)$, is related to the curvature of the connection.
From pseudo-Riemannian geometry, this seems plausible since the curvature gives the change of a vector when parallel transported around an infinitesimally small parallelogram \cite{ref:Weinberg:GR}.
More precisely, the Lie algebra-valued curvature 2-form take values in $\hol_p(\hat\nabla) \subseteq \gl(E_p)$ \cite{ref:Joyce} and, by the Ambrose--Singer theorem \cite{ref:Ambrose--Singer}, $\hol_p(\hat\nabla)$ is spanned by the curvature 2-form at all points connected to $p$ by piecewise smooth curves, parallel transported to $p$.%
\footnote{It is, of course, important to consider the curvature not only at $p$ since one can have a flat region on a generally curved manifold with nontrivial holonomy.}

Putting the above together, the integrability condition $[\tilde\nabla_m, \tilde\nabla_n] \eta = 0$ implies that the number of unbroken supersymmetries is at most the number of singlets in the decomposition of the spinor when restricting $\so(7)$ to $\hol(\tilde\D)$.
Furthermore, by \cref{eq:Freund--Rubin:tildeD_curvature,eq:Freund--Rubin:Weyl_tensor}, $\hol(\tilde\D)$ is spanned by $\tensor{W}{_{mn}^{ab}} \Sigma_{ab}$, where $\Sigma_{ab}$ are the generators of $\so(7)$ \cite{ref:Duff--Nilsson--Pope:KK}.
Note that $\tilde\D_m \eta = 0$ is stronger than $[\tilde\nabla_m, \tilde\nabla_n] \eta = 0$, so there might be fewer supersymmetries than singlets in the decomposition.
Also, there can be at most eight supersymmetries since spinors on $\M_7$ have eight components and the value of $\eta$ at a point determines its differential at the same point by the Killing spinor equation \cref{eq:Freund--Rubin:internal_Killing_spinor} \cite{ref:Duff--Nilsson--Pope:KK}.

For each vacuum with $m\neq 0$, there is another vacuum obtained by skew-whiffing, that is, reversing the direction of the flux, $m\mapsto -m$.
One can show that, except for $S^7$ with its usual round metric, at most one of the two solutions related by skew-whiffing can admit Killing spinors and, hence, at most one of the solutions can have unbroken supersymmetries \cite{ref:Duff--Nilsson--Pope:KK}.
This is known as the skew-whiffing theorem.

The above considerations are local.
If the space $\M_7$ is not simply connected%
\footnote{A connected manifold is simply connected if all closed loops are null-homotopic. The circle, $S^1$, is not simply connected.},
there may, in addition, be global obstructions to the existence of Killing spinors \cite{ref:Duff--Nilsson--Pope:KK}.%
\todo[disable]{}

\section{Anti-de Sitter, mass operators and supersymmetry} \label{sec:ads_mass_susy}%
A general feature of Kaluza--Klein compactifications is that there are infinite towers of fields obtained by expanding the $D=11$ fields in modes on the internal space \cite{ref:Duff--Nilsson--Pope:KK}.
The masses of these fields are related to certain differential operators on the compactification manifold $\M_7$.
Before we turn to the specific expressions for these, we should define what we mean by mass in $\mathrm{AdS}_4$.
In Minkowski spacetime, the mass is simply defined as (the nonnegative root of) $M^2 = - p_\mu p^\mu$.
For perturbative stability, that is, stability against small field fluctuations, $M^2\geq 0$ is needed.%
\footnote{$M^2 \geq 0$ does not imply perturbative stability since the potential could have nonzero slope.}
In $\mathrm{AdS}_4$, the situation is more complicated since the isometry group is $\SO(3,2)$, which does not contain momentum operators $p^\mu$.
There are two ways forward, one can study the field equations in AdS to try to come up with reasonable definitions of the masses or one can investigate the unitary irreducible representations of $\Spin(3,2)$ to characterise the particles.
Below, we review aspects of both approaches.
As in \cref{sec:Freund--Rubin}, we use conventions in which the cosmological constant is $\Lambda = -12m^2$, that is, the curvature radius is $1/(2m)$ and $R_{\mu\nu} = -12 m^2 g_{\mu\nu}$.

\subsection{Defining mass in AdS}
Following \cite{ref:Duff--Nilsson--Pope:KK}, we define the masses for different spin $s$ through the linear, free field equations
\begin{subequations}
\label{eq:ads_susy:mass_def}
\begin{alignat}{2}
\label{eq:ads_susy:mass_def:0}
    &s = 0\colon\qquad
    &&\Delta_0 \phi - 8m^2 \phi + M^2 \phi = 0,
    \\
\label{eq:ads_susy:mass_def:1/2}
    &s = \frac{1}{2}\colon
    &&\i \gamma^\mu \nabla_\mu \chi - M \gamma^5 \chi = 0,
    \\
    &s = 1\colon
    &&\Delta_1 A_\mu + \nabla_\mu \nabla^\nu A_\nu + M^2 A_\mu = 0,
    \\
    &s = \frac{3}{2}\colon
    && \i \gamma^{\mu\nu\rho} \tilde\nabla_\nu \psi_\rho - M \gamma^5 \gamma^{\mu\nu} \psi_\nu = 0,
    \\
    &s = 2\colon
    && \Delta_\mathrm{L} h_{\mu\nu} + 2 \nabla_{(\mu} \nabla^\rho h_{\nu)\rho} - \nabla_{(\mu} \nabla_{\nu)} \tensor{h}{_\rho^\rho} + 24 m^2 h_{\mu\nu} + M^2 h_{\mu\nu} = 0,
\end{alignat}
\end{subequations}
where $\Delta_p$ is the Hodge--de Rham operator, see \cref{app:Hodge_de_Rham_harmonic_forms},
\begin{equation}
    \tilde\nabla_\mu = \nabla_\mu - \i m \gamma^5 \gamma_\mu,
\end{equation}
as in \cref{eq:Freund--Rubin:spacetime_Killing_spinor}, and $\Delta_\mathrm{L}$ is the Lichnerowicz operator
\begin{equation}
\label{eq:ads_susy:mass:Lichnerowicz}
    \Delta_\mathrm{L} \tensor{h}{_{\mu\nu}}
    \coloneqq - \square \tensor{h}{_{\mu\nu}} - 2 \tensor{R}{_\mu^\rho_\nu^\sigma} \tensor{h}{_\rho_\sigma} + 2 \tensor{R}{_{(\mu}^\rho} \tensor{h}{_{\nu)\rho}}.
\end{equation}
For $s=0,\, 1/2$, these equations with $M = 0$ are Weyl invariant%
\footnote{A Weyl transformation is a local rescaling of the metric $g_{\mu\nu} \mapsto \Omega^2(x) g_{\mu\nu}$ \cite{ref:Karananas--Monin}.}
(if $\phi$ and $\chi$ given proper Weyl weights)
and they can be generalised Weyl-invariantly to arbitrary dimension \cite{ref:Deser--Nepomechie}.%
\footnote{Without transforming $m$, \cref{eq:ads_susy:mass_def:0} with $M = 0$ should be written as $\Delta_0 \phi + R/6\, \phi = 0$ to be Weyl invariant.}
As explained in \cite{ref:Karananas--Monin}, Weyl invariance together with diffeomorphism invariance implies conformal invariance (although conformal invariance and diffeomorphism invariance do not imply Weyl invariance \cite{ref:Karananas--Monin,ref:Wu} as sometimes claimed~\cite{ref:Farnsworth--et_al}).
One can show that \cref{eq:ads_susy:mass_def:0,eq:ads_susy:mass_def:1/2} with $M = 0$ imply that $\phi$ and $\chi$ propagate on the local light cones in AdS by using that $\d s^2 = 0$ is Weyl invariant and that $\mathrm{AdS}_4$ is related, locally, to four-dimensional Minkowski spacetime via a Weyl transformation~\cite{ref:Deser--Nepomechie}.

While for $s < 1$, we may define masslessness by requiring propagation on the local light cones, the situation for $s\geq 1$ is different.
Here, we instead define masslessness by requiring gauge invariance.
In $d=4$ Minkowski space, $p^2 = 0$ implies, via the group theory of the Poincaré group, that the particle only has two states (helicity), see \cref{sec:Mink_susy:multiplets}.
However, to reduce the number of propagating degrees of freedom to two for $s\geq 1$, gauge invariance is needed, whence propagation on the light cone and gauge invariance coincide \cite{ref:Deser--Nepomechie}.
This is not the case in arbitrary spacetimes.
Maxwell theory, that is, $s=1$, is Weyl invariant in precisely $d=4$ and photons, therefore, propagate on the local light cones in $\mathrm{AdS}_4$, by the above argument \cite{ref:Deser--Nepomechie}.
The gauge-invariant $s=3/2,\, 2$ theories are, however, not Weyl invariant in $d=4$ and the fields propagate not only on but also in the interior of the light cone \cite{ref:Deser--Nepomechie}.

Another peculiarity regarding masses in AdS is the bound for perturbative stability.
For $s > 0$, the requirement is $M^2 \geq 0$, just as in Minkowski spacetime \cite{ref:Duff--Nilsson--Pope:KK}.
Remarkably, for $s=0$, only $M^2 \geq -m^2$ is required to avoid exponentially growing modes for small field fluctuations \cite{ref:Breitenlohner--Freedman:short,ref:Breitenlohner--Freedman:long}.
This is known as the Breitenlohner--Freedman bound.
A theory in which all scalar fields satisfy the Breitenlohner--Freedman bound and all $s>0$ fields satisfy $M^2 \geq 0$ is said to be BF stable.
BF stability does not imply perturbative stability since the slope of the potential can be nonvanishing, as signalled by a tadpole.

\subsection{Masses from operators on the internal space} \label{sec:mass_operators}%
As stated above, the masses of the fields in $\mathrm{AdS}_4$ are related to differential operators on the internal space.
We will not give a derivation of the mass operators for the Freund--Rubin ansatz but make some comments on the derivation.
For details, see \cite{ref:Duff--Nilsson--Pope:KK}.
To derive the mass operators, one writes the $D=11$ fields as their background values plus a fluctuation (for instance, $g_{MN}(x,y) = \mathring g_{MN}(x,y) + h_{MN}(x,y)$ for the metric) and derives the linearised field equations from $D=11$ supergravity using the background values of the Freund--Rubin ansatz.
Then, gauge conditions are imposed and the fluctuations are expanded in modes on the internal space, similar to what we did when discussing supersymmetry in \cref{sec:Freund--Rubin}.
Finally, the mass operators are derived by analysing the equations resulting from inserting the expansions in the linearised field equations and comparing with \cref{eq:ads_susy:mass_def}.
The result is presented in \cref{tab:ads_susy:mass_op}.

To make the below results plausible, note that, for instance, $h_{\mu\nu}$ is a scalar on the internal space and will, hence, be expanded in scalar modes.
Since $s=2$ corresponds to the transverse and traceless part of $h_{\mu\nu}$, the $s=2$ mass operator must act on scalar fields on $\M_7$.

The operators appearing in \cref{tab:ads_susy:mass_op} are the Hodge--de Rham operator $\Delta_p$ acting on transverse $p$-forms with $p=0,\, 1,\, 2$; the Lichnerowicz operator $\Delta_\mathrm{L}$ acting on transverse traceless symmetric rank-2 tensors; the Dirac operator $\i\slashed\D$ acting on Majorana spinors and transverse $\Gamma$-traceless Majorana vector-spinors and the operator $Q$ acting on transverse 3-forms (defined in \cref{eq:ads_susy:laplacian:Q_def}).%
\footnote{Transversality means that $\nabla^m Y_{mn} = 0$ et cetera and a vector-spinor is $\Gamma$-traceless if $\Gamma^m \psi_m = 0$.}
At this point, we switch to consider fields with flat indices on the internal space.
In particular, $\i\slashed\D_{3/2}$ acts both on the vector and spinor index of the vector-spinor.
All operators should be interpreted in terms of their eigenvalues.
Note that all of the operators are self-adjoint and respect the transversality and tracelessness conditions, whence there are bases of eigenmodes with real eigenvalues spanning the corresponding function spaces.
We discuss this further in \cref{sec:ads_susy:Laplacian}.

In the derivation of \cref{tab:ads_susy:mass_op}, some special cases arise.
In the table, the subscripts label various towers of fields with the first (second) subscript referring to the top (bottom) sign or, for $\i\slashed\D$ and $Q$, the positive (negative) part of the spectrum of the operator.%
\footnote{Note that the labels of the towers agree with \cite{ref:Duff--Nilsson--Pope:KK} although other conventions differ.}
The eigenvalues $7m^2$ of $\Delta_0$ in $0^+_1$ and $-7m/2$ of $\i\slashed\D_{1/2}$ in $1/2_1$ correspond to singletons \cite{ref:Duff--Nilsson--Pope:KK}.
Singletons have no Poincaré analogue and are topological in the sense that the fields have no degrees of freedom in the bulk, only on the boundary~\cite{ref:Flato--et_al}.
Still, as explained in \cite{ref:Nilsson--Padellaro--Pope}, they must be kept in the theory.
The last exception is the eigenvalue $0$ of $\Delta_0$ in $0^+_1$, which should be omitted from the physical spectrum \cite{ref:Duff--Nilsson--Pope:KK}.
\begin{table}[H]
    \everymath{\displaystyle}
    \setlength{\defaultaddspace}{0.3em}
    \centering
    \caption[Mass operators in Freund--Rubin compactification.]{Mass operators in Freund--Rubin compactification of $D=11$ supergravity. Here, $s$ denotes spin, $p$ parity and $t$ labels the tower. The operators are understood in terms of their eigenvalues. Singletons correspond to $7m^2$ eigenvalues of $\Delta_0$ in $0^+_1$ and $-7m/2$ of $\i\slashed\D_{1/2}$ in $1/2_1$. The eigenvalue $0$ of $\Delta_0$ should be omitted from $0^+_1$.}
\label{tab:ads_susy:mass_op}
    \begin{tabular}{c@{\hphantom{${}_{3,2}$\hskip 1.5em}}l}
        \toprule
        $s_{\mathrlap{t}}^{\mathrlap{p}}$
        & Mass operator
        \\ \midrule
        $2^{\mathrlap{+}}$
        & $\Delta_0$
        \\ \addlinespace
        $\frac{3}{2}_{\mathrlap{1,2}}$
        &
        $-\i\slashed\D_{1/2} + \frac{7m}{2}$
        \\
        $1^{\mathrlap{-}}_{\mathrlap{2,1}}$
        & $\Delta_1 + 12m^2 \pm 6m \sqrt{\Delta_1 + 4m^2}$
        \\ \addlinespace
        $1^{\mathrlap{+}}$
        & $\Delta_2$
        \\ \addlinespace
        $\frac{1}{2}_{\mathrlap{4,1}}$
        & $-\i\slashed\D_{1/2} - \frac{9m}{2}$
        \\ \addlinespace[0.5em]
        $\frac{1}{2}_{\mathrlap{3,2}}$
        & $\i\slashed\D_{3/2} + \frac{3m}{2}$
        \\
        $0^{\mathrlap{+}}_{\mathrlap{3,1}}$
        & $\Delta_0 + 44m^2 \pm 12 m \sqrt{\Delta_0 + 9m^2}$
        \\ \addlinespace
        $0^{\mathrlap{+}}_{\mathrlap{2}}$
        & $\Delta_L - 4m^2$
        \\ \addlinespace
        $0^{\mathrlap{-}}_{\mathrlap{2,1}}$
        & $Q^2 + 6m Q + 8m^2$
        \\ \addlinespace[\aboverulesep] \bottomrule
    \end{tabular}
\end{table}
Note that there is precisely one massless spin-$2$ field, that is, graviton, since $\Delta_0$ has precisely one zero-mode.
This is reassuring.
From \cref{eq:Freund--Rubin:internal_Killing_spinor}, we see that each unbroken symmetry gives a mode of $\i\slashed\D_{1/2}$ with eigenvalue $7m/2$.
As is clear from \cref{tab:ads_susy:mass_op}, this eigenvalue corresponds to massless spin-$3/2$ fields, that is, gravitinos, the gauge fields of gauged supersymmetry \cite{ref:Duff--Nilsson--Pope:KK}.%
\footnote{From \cref{eq:sugra_comp:susy:psi}, we see that the supersymmetry transformation of the gravitino is $\delta_\varepsilon \psi_\mu = \tilde\D_\mu \varepsilon$, analogous to the transformation of the gauge potential in Yang--Mills theory.}
Similarly, unbroken gauge symmetries correspond to $12m^2$ eigenvalues of $\Delta_1$ in $1^-_1$ ($\Delta_1$ has no zero-mode on compact Einstein spaces with positive curvature \cite{ref:Duff--Nilsson--Pope:KK}).
There can be additional unbroken gauge symmetries from zero-modes of $\Delta_2$ in $1^+$.
These come from the Abelian gauge invariance in $D=11$ and all fields are, thus, neutral under them \cite{ref:Duff--Nilsson--Pope:KK}.%
\footnote{This is true of the supergravity fields; the M2 and M5-branes of M-theory, and the corresponding supergravity solutions, are electrically and magnetically charged under this symmetry, respectively.}

Note that the masses of spins $3/2$, $1/2$ and $0^-$ are sensitive to skew-whiffing, $m\mapsto -m$.
Also, by the skew-whiffing theorem, there are no spin-$1/2$ singletons for supersymmetric vacua except for the round $S^7$.
By the same argument, a vacuum with $\N=0$ related to a supersymmetric vacuum by skew-whiffing has exactly as many spin-$1/2$ singletons as there are supersymmetries in the other vacuum.
As we will see below, the only supermultiplet containing singletons is the Dirac singleton supermultiplet.
Thus, there are no spin-$0$ singletons in supersymmetric vacua, except for the round $S^7$, nor their skew-whiffed partners since the $0^+$ spectrum is insensitive to skew-whiffing.
In fact, one can prove that the first nonzero eigenvalue of $\Delta_0$ is at least $7m^2$ with equality only for the round $S^7$ \cite{ref:Duff--Nilsson--Pope:KK}, whence there are never spin-$0$ singletons in any other cases.

From \cref{tab:ads_susy:mass_op}, one can also draw some conclusions regarding BF stability.
Firstly, there cannot be any negative $M^2$ values in $2^+$, $3/2_{1,2}$, $1^-_2$, $1^+$, $1/2_{1,2,3,4}$ nor $0^+_3$ since $\Delta_p$ is nonnegative and $\i\slashed\D$ is Hermitian.
One can prove that $\Delta_1 \geq 12 m^2$ \cite{ref:Duff--Nilsson--Pope:vacuum_stability} which is precisely what is needed to ensure $M^2 \geq 0$ for $1^-_1$.
Since $Q$ has real eigenvalues and $\Delta_0 \geq 0$, the $0^+_1$ and $0^-_{1,2}$ towers satisfy, but might saturate, the Breitenlohner--Freedman bound $M^2 \geq -m^2$.
The only remaining tower is $0_2^+$, for which the BF stability criterion reads $\Delta_\mathrm{L} \geq 3m^2$.
This criterion is not satisfied by all compact Einstein spaces with positive curvature \cite{ref:Duff--Nilsson--Pope:vacuum_stability}.
If the vacuum is supersymmetric, it is perturbatively stable (see below) \cite{ref:Duff--Nilsson--Pope:KK}.
Since the $0_2^+$ spectrum is insensitive to skew-whiffing, $\N=0$ skew-whiffed counterparts of supersymmetric vacua are BF stable.
As remarked above, this does not imply that they are perturbatively stable.

One class of unstable solutions are Riemannian products.
If $\M_7 = \M_{(1)}\times \M_{(2)}$ with the product metric, that is, $g_{mn}$ is block diagonal over the two factors, then there is a mode of $\Delta_\mathrm{L}$ with eigenvalue $0 < 3m^2$ corresponding to one of the factors expanding while the other contracts \cite{ref:Duff--Nilsson--Pope:vacuum_stability}.

\subsection{\texorpdfstring{$\Spin(3,2)$}{Spin(3,2)}-representations and supersymmetry}
Similar to what we described in \cref{sec:Mink_susy:multiplets}, elementary particles in $\mathrm{AdS}_4$ correspond to nontrivial irreducible unitary representations of $\Spin(3,2)$, the double cover of the identity component of the isometry group $\SO(3,2)$ of $\mathrm{AdS}_4$.
We denote the generators of $\Spin(3,2)$ by $M_{\hat\alpha \hat\beta}$ where $\hat\alpha, \hat\beta,\hdots$ are 5-dimensional indices.
$\mathrm{AdS}_4$ is the universal cover of the connected hyperboloid $\smash{\eta_{\hat\alpha \hat\beta} x^{\hat\alpha} x^{\hat\beta} = -1/(2m)^2}$ \cite{ref:Nicolai}.%
\footnote{For example in \cite{ref:Avis_et_al}, AdS is used for the hyperboloid and CAdS for the cover.}

We split the 5-dimensional index as $\hat\alpha = (\alpha, 5)$ in the basis in which $\eta_{\hat\alpha\hat\beta}$ is block-diagonal with blocks $\eta_{\alpha\beta}$ and $-1$.
By considering the Poincaré limit $m\to 0$, one sees that $M_{\alpha\beta}$ and $P_\alpha \propto m M_{\alpha 5}$ are the AdS analogues of the Lorentz and momentum generators of the Poincaré algebra, respectively \cite{ref:Deser--Nepomechie}.
Hence, $M_{05}$ is identified as a dimensionless energy operator.
We consider representations with energy bounded from below, that is, for which there is a smallest eigenvalue $E_0$ of $M_{05}$.
The algebra $\so(3,2)$ is of rank 2 and irreducible representations can, hence, be specified by the eigenvalues of two Casimir operators \cite{ref:Fuchs--Schweigert}.
Equivalently, and more physically relevant, the irreducible representations can be denoted by $D(E_0, s)$ where $E_0$ is the lowest energy eigenvalue in the representation and $s$ the spin%
\footnote{The spin is, as usual, defined via the Casimir of the $\so(3)$ corresponding to spatial rotations.}
of the particle \cite{ref:Nicolai}.
A representation is said to be unitary if there exists an invariant, positive definite scalar product on it.
One can start from $2s+1$ lowest energy states and construct an invariant scalar product by declaring that these are orthonormal and that the generators are (anti-)Hermitian (depending on conventions).
Demanding that the scalar product is positive definite then leads to unitarity bounds on $E_0$ \cite{ref:Nicolai}.%
\footnote{Actually, it is sufficient to demand positive semi-definiteness and then factor out the zero norm states. This leads to multiplet shortening and corresponds to saturating the inequalities in \cref{eq:ads_susy:E0_unitarity}. See \cite{ref:Nicolai} for details.}
The result of this analysis is that
\begin{equation}
\label{eq:ads_susy:E0_unitarity}
    s < 1\colon\quad E_0 \geq s + \frac{1}{2},
    \qquad\qquad
    s \geq 1\colon\quad E_0 \geq s + 1.
\end{equation}
There are some special representations.
These are the massless $D(2, 0)$ and, for all $s$, $D(s+1,s)$ and the Dirac singleton representations $D(1/2, 0)$ and $D(1, 1/2)$ \cite{ref:Nicolai}.

The lowest energy eigenvalue, $E_0$, characterising a representation can be related to the AdS mass $M$ discussed above.
Here, we do not give a derivation but simply quote the result in \cref{tab:ads_susy:E0_M} \cite{ref:Duff--Nilsson--Pope:KK}.
The sign ambiguities for $s<1$ arise from a quadratic equation and are eliminated for $s \geq 1$ by the unitarity bounds.%
\footnote{The different signs correspond to different boundary conditions \cite{ref:Breitenlohner--Freedman:long,ref:Witten:AdS/CFT,ref:Klebanov--Witten}.}
For unitarity and $E_0 \in \RR$, we see that $M^2_{s=0} \geq -m^2$ is required for the plus sign, again the Breitenlohner--Freedman bound, and $-m^2 \leq M^2_{s=0} \leq 3m^2$ for the minus sign.
For the spinors, all real $M_{s=1/2}$ are allowed for the plus sign and $|M_{s=1/2}| \leq m$ for the minus sign.
One can prove that the absolute values of the eigenvalues of $\i\slashed\D_{1/2}$ on $\M_7$ are greater than or equal to $7 |m|/2$ \cite{ref:Duff--Nilsson--Pope:KK}.
Thus, $M_{s=3/2}$ as given in \cref{tab:ads_susy:mass_op} cannot violate the unitarity bound in the table below.
Lastly, for $s=1,\, 2$, unitarity requires $M^2 \geq 0$ in agreement with the above stability criteria.
Note that the unitarity bounds for $s>1$ correspond to masslessness while for $s \leq 1/2$ they correspond to the singleton representations.
\begin{table}[H]
    \everymath{\displaystyle}
    \setlength{\defaultaddspace}{0.6em}
    \centering
    \caption[Relation between energy and mass in $\mathrm{AdS}_4$.]{Relation between energy, $E_0$, and mass, $M$, for various spins, $s$, in $\mathrm{AdS}_4$. The corresponding $\Spin(3,2)$-representations, $D(E_0, s)$, are unitary for $E_0 \geq E_0^\mathrm{min}$.}
\label{tab:ads_susy:E0_M}
    \begin{tabular}{c@{\hskip 2em}l@{\hskip 2em}l}
        \toprule
        $s$ & $E_0$ & $E_0^\mathrm{min}$
        \\ \midrule
        $0$
        & $\frac{3}{2} \pm \frac{1}{2} \sqrt{\frac{M^2}{m^2} + 1}$
        & $\frac{1}{2}$
        \\ \addlinespace
        $\frac{1}{2}$
        & $\frac{3}{2} \pm \frac{1}{2} \Bigl|\frac{M}{m}\Bigr|$
        & $1$
        \\ \addlinespace
        $1$
        & $\frac{3}{2} + \frac{1}{2} \sqrt{\frac{M^2}{m^2} + 1}$
        & $2$
        \\ \addlinespace
        $\frac{3}{2}$
        & $\frac{3}{2} + \frac{1}{2} \Bigl|\frac{M}{m} - 2\Bigr|$
        & $\frac{5}{2}$
        \\ \addlinespace
        $2$
        & $\frac{3}{2} + \frac{1}{2} \sqrt{\frac{M^2}{m^2} + 9}$
        & $3$
        \\ \addlinespace[\aboverulesep] \bottomrule
    \end{tabular}
\end{table}

The above representations $D(E_0, s)$ can be combined into supermultiplets.
The superalgebra in $\mathrm{AdS}_4$ is not the super-Poincaré algebra of \cref{sec:susy:super-Poincare} but the orthosymplectic Lie superalgebra $\osp(\N| 4)$.
This algebra is the graded extension of $\so(\N)$ and $\sp(4, \RR) \simeq \so(3, 2)$ where $\N$ denotes the number of supersymmetries as usual.
Accordingly, there are, apart from the 10 generators $M_{\hat\alpha\hat\beta}$ of $\so(3,2)$, $\N(\N-1)/2$ generators $T^{ij}$ of $\so(\N)$ and $4\N$ supercharges $Q^i$ (with a suppressed Dirac spinor index).
The supersymmetry generators are Majorana and the nonvanishing superbrackets are
\cite{ref:Nicolai}
\begin{subequations}
\begin{align}
    &\{Q^i, Q^j\}
    = - \frac{1}{2} \delta^{ij} \hat\gamma^{\hat\alpha \hat\beta} C^{-1} M_{\hat\alpha \hat\beta}
    + \i C^{-1} T^{ij},
    \\
    &[M_{\hat\alpha \hat\beta}, M^{\hat\gamma \hat\delta}]
    = 4\i \tensor*{\eta}{_{[\hat\alpha}^{[\hat\gamma}} \tensor{M}{_{\hat\beta]}^{\hat\delta]}},
    \\
    &[M_{\hat\alpha \hat\beta}, Q^i] = \frac{\i}{2} \hat\gamma_{\hat\alpha\hat\beta} Q^i,
    \\
    &[T^{ij}, T_{kl}] = 4\i \tensor*{\delta}{^{[i}_{[k}} \tensor{T}{^{j]}_{l]}},
    \\
    &[T_{ij}, Q^k] = 2\i \tensor*{\delta}{^k_{[i}} \tensor{Q}{_{j]}},
\end{align}
\end{subequations}
where $\hat\gamma^{\hat\alpha}$ are the $\gamma$-matrices of $\so(3,2)$, that is, $\hat\gamma^\alpha = \gamma^\alpha$ and $\hat\gamma^5 = \i \gamma^5$, and $C$ is the $\so(3,2)$ charge conjugation matrix, $C \hat\gamma^{\hat\alpha} C^{-1} = + \hat\gamma^{\hat\alpha \transpose}$.%
\footnote{Here, we use the convention with an $\i$ in the exponent when exponentiating to a group element.}
By tracing with $C\hat\gamma_{05}$ and contracting with $\delta_{ij}$ one finds
\begin{equation}
    M_{05} = \frac{1}{2 \N} \delta_{ij} Q^i C\hat\gamma_{05} Q^j \geq 0,
\end{equation}
since $C\hat\gamma_{05}$ is symmetric and positive definite.%
\footnote{Note that $C$ is only well-defined up to a sign. The other sign gives $M_{05} \leq 0$ and we must then associate the energy with $-M_{05}$ if we demand it to be bounded from below rather than above.}
Taken as a relation for the quantum operators, this implies perturbative stability for supersymmetric vacua, that is, vacua annihilated by the supersymmetry generators $Q^i$.

By a method similar to that for the representations $D(E_0, s)$ one can determine the possible unitary irreducible supermultiplets with energy bounded from below, see \cite{ref:Nicolai}.
The results for $\N=1$ were first obtained by Heidenreich \cite{ref:Heidenreich} and for $\N=8$ by Freedman and Nicolai \cite{ref:Freedman--Nicolai}.
We present the results for $\N = 1$ in \cref{tab:ads_susy:supermultiplets}.
\begin{table}[H]
    \everymath{\displaystyle}
    \setlength{\defaultaddspace}{0.6em}
    \centering
    \caption[The unitary $\N=1$ supermultiplets in $\mathrm{AdS}_4$.]{The unitary $\N=1$ supermultiplets in $\mathrm{AdS}_4$ with $\Spin(3,2)$\hyp{}representations ordered decreasingly by spin.}
\label{tab:ads_susy:supermultiplets}
    \begin{tabular}{l l}
        \toprule
        Class & Multiplet name and unitary $\Spin(3,2)$-representations \\ \midrule
        1
        & Dirac singleton
        \\
        & $D\Bigl(1, \frac{1}{2}\Bigr) \oplus D\Bigl(\frac{1}{2}, 0\Bigr)$
        \\ \addlinespace
        2
        & Wess--Zumino supermultiplet for $E_0 > 1$
        \\
        & $D\Bigl(E_0, \frac{1}{2}\Bigr)
        \oplus D\Bigl(E_0 + \frac{1}{2}, 0\Bigr)
        \oplus D\Bigl(E_0 - \frac{1}{2}, 0\Bigr)$
        \\ \addlinespace
        3
        & Massless higher spin supermultiplets for $s \geq 1$
        \\
        & $D(s+1, s)
        \oplus D\Bigl(s+\frac{1}{2}, s-\frac{1}{2} \Bigr)$
        \\ \addlinespace
        4
        & Massive higher spin supermultiplets for $s \geq 1$ and $E_0 > s+1$
        \\
        &$D(E_0, s)
        \oplus D\Bigl(E_0 + \frac{1}{2}, s - \frac{1}{2}\Bigr)
        \oplus D\Bigl(E_0 - \frac{1}{2}, s - \frac{1}{2}\Bigr)
        \oplus D(E_0, s-1)$
        \\ \addlinespace[\aboverulesep] \bottomrule
    \end{tabular}
\end{table}

\subsection{Differential operators and a universal Laplacian} \label{sec:ads_susy:Laplacian}%
Here, we discuss some properties of the operators appearing in \cref{tab:ads_susy:mass_op} and relate them to a universal Laplacian.
We assume that the compact manifold $\M_7$ is without boundary or that the boundary conditions are such that all boundary integrals vanish.
To see where the universal Laplacian comes from, consider first the Hodge--de Rham operator, or Hodge Laplacian,
\begin{equation}
    \Delta_p = \dd \d + \d \dd,
\end{equation}
acting on $p$-forms.
Here, $\d$ is the exterior derivative and $\dd$ the codifferential, see \cref{app:conventions:diff_forms}.
A $p$-form $\alpha$ is transverse if $\dd \alpha = 0$, which follows from the definition $\D^b a_{b a_1 \hdots a_{p-1}} = 0$ of transversality and the component formula for $\dd \alpha$.
Thus, $\Delta_p$ maps transverse $p$-forms to transverse $p$-forms since $\dd \Delta_p \alpha = \dd \d \dd \alpha$.
Also, $\Delta_p$ is manifestly self-adjoint and nonnegative since $\dd$ is the adjoint of $\d$.

By using the definitions of the exterior derivative $\d$ and the codifferential $\dd$ we immediately find
\begin{equation}
\label{eq:ads_susy:laplacian:HdR_1}
    \Delta_p \alpha_{a_1 \hdots a_p}
    = - \square \alpha_{a_1 \hdots a_p}
    - p [\D_{[a_1}, \D^b] \alpha_{|b| a_2 \hdots a_p]},
\end{equation}
where $\D$ is the torsion-free spin connection.
Using the Ricci identity $[\D_a, \D_b] = R_{abcd} \Sigma^{cd}$, this can be written as
\begin{equation}
    \Delta_p \alpha_{a_1 \hdots a_p}
    = - \square \alpha_{a_1 \hdots a_p}
    - p(p-1) \tensor{R}{_{[a_1}^{b_1}_{a_2}^{b_2}} \alpha_{|b_1 b_2| a_3 \hdots a_p]}
    + p \tensor{R}{_{[a_1}^b} \alpha_{|b| a_2 \hdots a_p]},
\end{equation}
which is known as a Weitzenböck identity.
There is another way of writing $\Delta_p$.
For this, note that the second term in \cref{eq:ads_susy:laplacian:HdR_1} can be written as
\begin{equation}
    p [\D_{[a_1}, \D^b] \alpha_{|b|a_2 \hdots a_p]}
    =
    [\D_{c_1}, \D_{c_2}] p \delta^{\, c_1\, c_2}_{[a_1|b|} \tensor{\alpha}{^b_{a_2 \hdots a_p]}}
    =
    [\D_{c_1}, \D_{c_2}] \Sigma^{c_1 c_2} \alpha_{a_1 \hdots a_p}.
\end{equation}
Again using the Ricci identity, we find
\begin{align}
\nonumber
    \Delta_p \alpha_{a_1 \hdots a_p}
    &= -\square \alpha_{a_1 \hdots a_p}
    - [\D_{b_1}, \D_{b_2}] \Sigma^{b_1 b_2} \alpha_{a_1 \hdots a_p}
    =\\
    &= -\square \alpha_{a_1 \hdots a_p}
    - R_{b_1 b_2 c_1 c_2} \Sigma^{c_1 c_2} \Sigma^{b_1 b_2} \alpha_{a_1 \hdots a_p}.
\end{align}
This form of the Laplacian can be generalised to a field carrying any representation of $\Spin(7)$.
Thus, we define
\begin{equation}
\label{eq:ads_susy:laplacian:def}
    \Delta
    \coloneqq
    -\square  -  [\D_{a_1}, \D_{a_2}] \Sigma^{a_1 a_2}
    =
    - \square - R_{a_1 a_2 b_1 b_2} \Sigma^{a_1 a_2} \Sigma^{b_1 b_2}.
\end{equation}
We refer to this as the \emph{universal Laplacian} since it can act on a field carrying any representation of $\Spin(7)$ and, as we will see, is related to all the other Laplacians we are interested in.

Let us show that $\Delta$ is self-adjoint.
To this end, let $Y_A$ be a field carrying any finite-dimensional real representation of $\Spin(7)$.%
\footnote{We restrict to the real case since that is what we are interested in and for notational convenience. What follows is easily generalised to the complex case by replacing the symmetric invariant $\delta^{AB}$ with a Hermitian invariant $\delta^{\bar A B}$.}
Since $\Spin(7)$ is compact, the representation is unitary and there is an invariant symmetric nondegenerate $\delta^{AB}$ with Euclidean signature.
We have an $L^2$ inner product defined by
\begin{equation}
    \langle Y, X \rangle
    = \int \vol\, Y_A \delta^{AB} X_B.
\end{equation}
To see that $\Delta$ is self-adjoint with respect to this inner product, note that
\begingroup
\allowdisplaybreaks
\begin{subequations}
\label{eq:ads_susy:laplacian:self-adjoint_and_pos}
\begin{align}
    &%
    \begin{aligned}[b]
        \langle Y, -\square X \rangle
        &= - \int \vol\, Y_A \delta^{AB} \square X_B
        = + \int \vol\, \D_a Y_A \delta^{ab} \delta^{AB} \D_b X_B
        =\\
        &= - \int \vol\, \square Y_A \delta^{AB} X_B
        = \langle -\square Y, X \rangle,
    \end{aligned}
    \\ &%
    \begin{aligned}[b]
        \langle Y, -\D_a \D_b \Sigma^{ab} X \rangle
        &= - \int \vol\, Y_A (\Sigma^{ab})^{AB} \D_a \D_b X_B
        =\\
        &= + \int \vol\, \D_a Y_A (\Sigma^{ab})^{AB} \D_b X_B
        =\\
        &= - \int \vol\, \D_b \D_a Y_A (\Sigma^{ab})^{AB} X_B
        = \langle -\D_a \D_b \Sigma^{ab} Y, X \rangle,
    \end{aligned}
\end{align}
\end{subequations}
\endgroup
where we have used $\delta^{AB}$ to raise an index on $(\Sigma^{ab})_A{}^B$, that $\delta^{AB}$ and $(\Sigma^{ab})^{AB}$ are invariant tensors and, in the last step, that $(\Sigma^{ab})^{AB}$ is antisymmetric both in $a\, b$ and $A\, B$.
The last statement follows from $\delta^{AB}$ being a symmetric invariant.
Thus, glossing over some mathematical subtleties regarding the distinction between symmetric and self-adjoint unbounded operators on infinite-dimensional Hilbert spaces \cite{ref:Hall:quantum_mechanics}, $\Delta$ is self-adjoint and there is a basis of eigenmodes of $\Delta$ with real eigenvalues.%
\todo[disable]{}

We would also like to show that $\Delta$ respects the various conditions (transversality, tracelessness, et cetera) placed on the fields.
We do this separately in the cases of interest.
Since $\Delta = \Delta_p$ when acting on $p$-forms and $\Delta_p$ respects transversality, $\Delta$ can be restricted to transverse $p$-forms.

Turning to symmetric rank-2 tensors, a short calculation shows that
\begin{equation}
\label{eq:ads_susy:laplacian:Lichnerowicz}
    \Delta h_{ab}
    = \Delta_\mathrm{L} h_{ab}
    \coloneqq
    - \square h_{ab}
    - 2 \tensor{R}{_a^c_b^d} h_{cd}
    + 2 \tensor{R}{_{(a}^c} h_{b)c},
\end{equation}
where $\Delta_\mathrm{L}$ is the Lichnerowicz Laplacian as defined in \cref{eq:ads_susy:mass:Lichnerowicz} but here for the internal space.
By contracting $a$ and $b$ one immediately sees that $\Delta_\mathrm{L}$ maps traceless tensors into traceless tensors.
To show that $\Delta_\mathrm{L}$ respects transversality, that is, that $\D^a h_{ab} = 0$ implies $\D^a \Delta_\mathrm{L} h_{ab} = 0$, first note that
\begin{equation}
\label{eq:ads_susy:laplacian:transv_Lich_term_2}
    \D^a \bigl( \tensor{R}{_{(a}^c} h_{b)c} \bigr)
    = 0,
    \qquad\quad
    \D^a \bigl( \tensor{R}{_a^c_b^d} h_{cd} \bigr)
    = \tensor{R}{^a^c_b^d} \D_a h_{cd}.
\end{equation}
Here, we have used that the manifold is Einstein, $R_{ab} = 6m^2 \delta_{ab}$, which implies that $\D^a R_{abcd} = 0$ by contracting the Bianchi identity $\D_{[a} R_{bc]de} = 0$.
When computing $\D^a \Delta_\mathrm{L} h_{ab}$, the first term in \cref{eq:ads_susy:laplacian:Lichnerowicz} gives
\begin{align}
\nonumber
    \D^a \square h_{ab}
    &= \D^c \D^a \D_c h_{ab}
    + \tensor{R}{^a^c_c^d} \D_d h_{ab}
    + \tensor{R}{^a^c_a^d} \D_c h_{db}
    + \tensor{R}{^a^c_b^d} \D_c h_{ad}
    =\\
\label{eq:ads_susy:laplacian:transv_Lich_term_1}
    &= \D^c \bigl(\tensor{R}{^a_c_a^d} h_{db} + \tensor{R}{^a_c_b^d} h_{ad}  \bigr)
    + \tensor{R}{^a^c_b^d} \D_c h_{ad}
    = 2 \tensor{R}{^a^c_b^d} \D_c h_{ad}.
\end{align}
Since the right-hand sides of \cref{eq:ads_susy:laplacian:transv_Lich_term_2,eq:ads_susy:laplacian:transv_Lich_term_1} cancel when combined as in \cref{eq:ads_susy:laplacian:Lichnerowicz}, $\Delta_\mathrm{L}$ can indeed be restricted to the space of transverse $h_{ab}$.

Next, we turn to the 3-form operator $Q$, defined as
\begin{equation}
\label{eq:ads_susy:laplacian:Q_def}
    Q \coloneqq \hodge \d,
\end{equation}
which maps 3-forms into 3-forms.
If $\alpha$ is a transverse 3-form,
\begin{equation}
    \Delta_3 \alpha
    = \dd \d \alpha
    = \hodge \d \hodge \d \alpha
    = Q^2 \alpha,
\end{equation}
since $\dd$ is acting on the 4-form $\d \alpha$.
This gives the relation between $Q$ and the Laplacian $\Delta$.
In seven dimensions and Euclidean signature, $\hodge^2 = 1$, $\hodge$ is self-adjoint and, thus,
\begin{equation}
    Q^\ast = \dd \hodge = \hodge \d \hodge^2 = Q,
\end{equation}
that is, $Q$ is self-adjoint.
Also, $Q$ maps into transverse 3-forms since $\dd Q = - \hodge \d^2  = 0$.

Turning now to half-integer spins, the relevant operator is the Dirac operator
\begin{equation}
    \i\slashed\D = \i \Gamma^a \D_a.
\end{equation}
The Dirac operator $\i\slashed\D$ maps Majorana spinors to Majorana spinors.
This is easily seen in the basis from \cref{app:octonions:spin7} since the Majorana condition then reduces to a reality condition for each component and $\Gamma_a$ are purely imaginary.
It also maps the space of transverse $\Gamma$-traceless vector-spinors, that is, vector-spinors $\psi_a$ satisfying $\D^a \psi_a = 0 = \Gamma^a \psi_a$, to itself since
\begin{subequations}
\begin{align}
    &\Gamma^a \i\slashed\D \psi_a
    = - \i\slashed\D \Gamma^a \psi_a + 2 \i\D^a \psi_a
    = 0,
    \\
    &\D^a \i\slashed\D \psi_a
    = \i\slashed\D \D^a \psi_a
    + \frac{\i}{4} \tensor{R}{^a_b_c_d} \Gamma^b \Gamma^{cd} \psi_a
    + \i \Gamma^b \tensor{R}{^a_b_a^c} \psi_c
    = 0,
\end{align}
\end{subequations}
where, in the last step, we used that $R_{a[bcd]} = 0$ and that $\M_7$ is Einstein.
Furthermore, $\i\slashed\D$ is self-adjoint.
To see this, let $\psi_A$ and $\chi_A$ be Majorana tensor-spinors, where $A$ is an arbitrary (flat) tensor index and we have suppressed the spinor indices, and note that%
\footnote{The Dirac operator can also be considered in the context of Dirac spinors. In this case, one should use the Dirac conjugate instead of the Majorana conjugate in the $L^2$ inner product. The eigenvalues are the same since the Majorana eigenbasis provides an eigenbasis for the space of Dirac spinors as well.}
\begin{align}
\nonumber
    \langle \psi, \i\slashed\D \chi \rangle
    &= \int \vol\, \psi_A^\transpose C \delta^{AB} \i\slashed\D \chi_B
    = \int \vol\, \i\D^a \psi_A^\transpose \Gamma_a^\transpose C \delta^{AB} \chi_B
    =\\
    &= \int \vol\, (\i\slashed\D \psi_A)^\transpose C \delta^{AB} \chi_B
    = \langle \i\slashed\D \psi, \chi \rangle.
\end{align}
Here, the sign from the integration by parts cancels the sign from $C \Gamma_a = - \Gamma_a^\transpose C$.

Lastly, we investigate the relation between $\i\slashed\D$ and $\Delta$.
The square of the Dirac operator is
\begin{equation}
    (\i\slashed\D)^2
    = - \Gamma^a \Gamma^b \D_a \D_b
    = - \square - \frac{1}{2} \Gamma^{ab} [\D_a, \D_b].
\end{equation}
Acting on $\psi_A$, this gives
\begin{align}
\nonumber
    (\i\slashed\D)^2 \psi_A
    &= -\square \psi_A
    - \frac{1}{2} \Gamma^{a_1 a_2} \tensor{R}{_{a_1 a_2}^{b_1 b_2}}
    \Bigl(\frac{1}{4}\Gamma_{b_1 b_2} \psi_A
    + \tensor{(\Sigma_{b_1 b_2})}{_A^B} \psi_B \Bigr)
    =\\
    &= - \square \psi_A
    + \frac{1}{4} R \psi_A
    - \frac{1}{2} \tensor{R}{_{a_1 a_2}^{b_1 b_2}} \Gamma^{a_1 a_2} \tensor{(\Sigma_{b_1 b_2})}{_A^B} \psi_B.
\end{align}
Acting instead with the Laplacian $\Delta$ from \cref{eq:ads_susy:laplacian:def} on $\psi_A$, we find, after a short calculation,
\begin{align}
\nonumber
    \Delta \psi_A
    &= - \square \psi_A
    + \frac{1}{8} R \psi_A
    - \frac{1}{2} \tensor{R}{_{a_1 a_2}^{b_1 b_2}} \Gamma^{a_1 a_2} \tensor{(\Sigma_{b_1 b_2})}{_A^B} \psi_B
    +\\ \nonumber
    &\eqspace - \tensor{R}{_{a_1 a_2}^{b_1 b_2}} \tensor{(\Sigma^{a_1 a_2})}{_A^B} \tensor{(\Sigma_{b_1 b_2})}{_B^C} \psi_C
    =\\
\label{eq:diff_op:Delta_on_tensor-spinor}
    &= (\i\slashed\D)^2 \psi_A
    - \frac{1}{8} R \psi_A
    - \tensor{R}{_{a_1 a_2}^{b_1 b_2}} \tensor{(\Sigma^{a_1 a_2})}{_A^B} \tensor{(\Sigma_{b_1 b_2})}{_B^C} \psi_C.
\end{align}
Thus, for spinors and vector-spinors, respectively,
\begin{subequations}
\label{eq:diff_op:tensor-spinor_lambda_kappa_relation}
\begin{align}
    &(\i\slashed\D)^2 \psi = \Delta \psi + \frac{21}{4} m^2 \psi,\\
    &(\i\slashed\D)^2 \psi_a = \Delta \psi_a - \frac{3}{4} m^2 \psi_a,
\end{align}
\end{subequations}
where we have used that $R_{ab} = 6 m^2 \delta_{ab}$.
Note that, for Einstein spaces, the eigenvalues of $\Delta$ and $(\i\slashed\D)^2$ on spinors and vector-spinors are related by a constant.
This seems to break down for higher tensor-spinors since the difference between $\Delta$ and $(\i\slashed\D)^2$ will contain contributions from the Weyl tensor from the last term in \cref{eq:diff_op:Delta_on_tensor-spinor}.
However, we are only interested in spinors and vector-spinors, whence this is not a problem.

To conclude, we have seen that all of the operators $\Delta_p$, $\Delta_\mathrm{L}$, $Q$ and $\i\slashed\D$ are self-adjoint and can be restricted to the relevant functions spaces which, thus, have bases of eigenmodes with real eigenvalues.
Moreover, we have found that all of the operators are related to a universal Laplacian $\Delta$ \cref{eq:ads_susy:laplacian:def}
\begin{equation}
\label{eq:ads_susy:laplacian:universal}
    \Delta
    =
    - \square - R_{a_1 a_2 b_1 b_2} \Sigma^{a_1 a_2} \Sigma^{b_1 b_2},
\end{equation}
which differs from $(\i\slashed\D)^2$ only by a constant when acting on spinors and vector-spinors and coincides with $\Delta_p$, $\Delta_\mathrm{L}$ and $Q^2 = \Delta_3$.

\chapter{Homogeneous spaces} \label{chap:coset}%
Homogeneous pseudo-Riemannian%
\footnote{We include all signatures of the metric in ``pseudo-Riemannian''.}
manifolds are a rich, yet particularly simple, set of manifolds.
Due to this, they are often used as the internal space in string and supergravity compactifications.
Examples of homogeneous spaces include Euclidean spaces, spheres, flat tori and hyperbolic spaces in Euclidean signature; Minkowski, de Sitter (dS) and anti-de Sitter (AdS) spaces in Lorentzian signature and super-Minkowski and super-AdS superspaces.

A homogeneous space is, intuitively, a space in which all points are equivalent, or ``look the same'', in some sense appropriate to the setting.
Technically, it is a space $\M$ on which a group $G$ of automorphisms acts transitively and effectively, that is, for every $x,y \in \M$ there is a $g \in G$ such that $gx = y$ and there is no $g\in G$ which acts trivially on $\M$.
That $G$ acts by automorphisms means that it preserves the structure of $\M$ and we require $G$ to act effectively since, otherwise, it is really $G/N$, where $N$ is the kernel of the $G$-action, that acts on $\M$.%
\footnote{When considering spinor fields, we will have reason to loosen the latter requirement slightly.}

Since we are interested in connected oriented pseudo-Riemannian manifolds,
the automorphisms are orientation-preserving isometries of $\M$, that is, $G \subseteq \operatorname{Iso}_+(\M)$.
So, a homogeneous (pseudo-Riemannian) manifold is a manifold on which the isometry group acts transitively.
The isometry group is a Lie group \cite{ref:Myers--Steenrod,ref:Kobayashi}, whence, in the following, we consider Lie groups $G$.

The stabiliser $H_y$ of $y\in \M$, that is, the subgroup of $G$ fixing $y$, is called the isotropy group of $y$.
It is easy to see that the isotropy groups of different points in $\M$ are conjugate subgroups in $G$.
Thus, we often need not distinguish between them and simply write $H$ for \emph{the} isotropy subgroup, which is a closed subgroup of $G$ \cite{ref:ONeil}.%
\footnote{A closed subgroup of a Lie group is a Lie subgroup by Cartan's theorem.}
Note that the isotropy group $H_y$ is a subgroup of $\SO_g(T_y \M)$, where $g$ is the $G$-invariant metric on $\M$.%
\footnote{With $\SO_g(V)$ we mean the orientation-preserving subgroup of $\GL(V)$ that leaves the metric $g$ on $V$ invariant.}
This follows from the fact that an isometry of a connected manifold is determined by its value and differential at a single point \cite{ref:ONeil}.%
\todo[disable]{}
Thus, since an element of $H_y$ fixes $y$, it is completely determined by its differential at $y$.
Since $H_y$ preserves the metric and orientation on $\M$, and particularly on the tangent space at the fixed point, $T_y \M$, there is a natural homomorphism $H_y \to \SO_g(T_y \M)$ and we may view $H_y$ as a subgroup of $\SO_g(T_y \M)$.

$G$ is a principal $H$-bundle over $\M$ \cite{ref:Mickelsson--Niederle}.
For an introduction to fibre bundles, see \cref{app:bundles}.
Similar to the frame bundle, we may view $G$ as the bundle of $H$-frames over $\M$.
To see this, pick an orthonormal frame at a point $o$ and use the pushforward by every element in $g$ to get a set of frames.
This set is naturally in one-to-one correspondence with $G$ since $G$ acts effectively and isometries are determined by their value and differential at a single point.
Due to $G$ preserving the metric of $\M$, the set will only contain orthonormal frames.
Thus, $G$ is a subbundle of the orthonormal frame bundle and we have a reduction of the structure group from $\SO_g(\dim \M)$ to~$H$.

By choosing an origin $o \in \M$ we get a natural map $\phi\from G/H_o \to \M$ by $\phi(g H_o) = g o$.
It is easy to see that $\phi$ is a well-defined bijection and it is, in fact, a diffeomorphism \cite{ref:ONeil}.
The $G$-action on $\M$ is realised on $G/H$ by left-multiplication.
By pulling back the $G$-invariant metric on $\M$ to $G/H$, the two spaces become isometric whence we do not distinguish them in the following.

Instead of starting from a manifold $\M$ and then realising it as a coset space (provided that the isometry group acts transitively), one can start from the groups $G$ and $H$ and construct a homogeneous manifold as $G/H$.
We do this in \cref{coset:geometry}.
Note, however, that there might exist a proper subgroup of $\operatorname{Iso}_+(\M)$ which acts transitively on $\M$ \cite{ref:Arvanitoyeorgos}, in which case $\M$ can be described by different cosets $G/H$.
Also, if we loosen the requirement that $G$ acts effectively, it may be possible to describe $\M$ by additional cosets \cite{ref:Duff--Nilsson--Pope:KK}.
In \cref{coset:analysis}, we discuss harmonic analysis on coset spaces.

% Sections:
\section{Geometry} \label{coset:geometry}
Let $G$ be a Lie group and $H$ a closed subgroup with Lie algebras $\g$ and $\h$, respectively.
We will assume that the algebras are reductive, that is, that there exists an $\operatorname{Ad}(H)$-invariant subspace $\m$ of $\g$ such that $\g = \h \oplus \m$ \cite{ref:ONeil}.
Here, $\operatorname{Ad}$ is the adjoint representation of $G$.
Thus, reductivity means that the adjoint representation of $G$ splits, when restricted to $H$, into a direct sum of the adjoint representation of $H$ and another representation.
The latter is a $H$-representation on $\m \simeq T_o \M$ called the isotropy representation \cite{ref:Arvanitoyeorgos}.
This implies that $[\h, \m] \subseteq \m$ while the converse implication holds for connected $H$ \cite{ref:Arvanitoyeorgos}.
Reductivity is not a very restrictive assumption: every homogeneous space admitting a $G$-invariant metric with Euclidean signature is reductive \cite{ref:Arvanitoyeorgos}.
Since we are, in this thesis, interested in compact manifolds with Euclidean signature to be used in a Kaluza--Klein compactification, we find this assumption acceptable.

In the above, we assumed that the $G$-action on $\M$ is effective.
When constructing a coset space, it is natural to ask what this means for $G$ and $H$.
Since any element $g \in G$ which acts trivially on $\M$ in particular fixes $o$, it follows that $g \in H_o$.
Since an element of $H_o$ is determined by its differential at $o$, that is, its action on $T_o \M$, this implies that the $G$-action is effective if and only if the isotropy representation of $H$ is faithful.

We wish to put a $G$-invariant metric on $G/H$.
By a theorem \cite{ref:ONeil}, $G$-invariant tensor fields on $G/H$ are in one-to-one correspondence with $H$-invariant tensors of the same type on $\m$.
Thus, for a $G$-invariant metric on $G/H$ we need a $H$-invariant symmetric nondegenerate rank-2 tensor.
Note that, if there is only one such tensor (up to a constant factor), the coset space $G/H$, equipped with the $G$-invariant metric, is an Einstein space, since the Ricci-tensor will also be proportional to the invariant \cite{ref:Duff--Nilsson--Pope:KK}.
This happens if the isotropy representation is irreducible \cite{ref:Coquereaux}.

In this section, we use indices $A, B, C,\hdots$ for $\mathfrak{g}$, $a,b,c,\hdots$ for $\m$ and $i,j,k\hdots$ for $\h$.
With $T_A$ generators and $\tensor{f}{_A_B^C}$ structure constants of $\g$, $[\h, \m] \subseteq \m$ and the fact that $\h$ is a subalgebra of $\g$ implies that $\tensor{f}{_i_a^j} = 0$ and $\tensor{f}{_i_j^a} = 0$.
We use $g_{ab}$ to denote the $H$-invariant tensor that defines the $G$-invariant metric on $G/H$ and raise and lower indices $a,b,c\hdots$ using $g_{ab}$ and its inverse $g^{ab}$.
Since, $\ad_\g$ splits into $\ad_\h$ and the isotropy representation when restricted to $\h$, the latter representation is, explicitly, $(T_i)_a{}^b = -f_{ia}{}^b$.
Note that $g_{ab}$ being $\h$-invariant is equivalent to $f_{iab}$ being antisymmetric in $ab$.
Thus, the isotropy algebra is the subalgebra of $\so_g(\m)$ given by
\begin{equation}
\label{eq:coset:geometry:isotropy_embedding}
    T_i = - \tensor{f}{_i_a_b} \Sigma^{ab},
\end{equation}
where $\Sigma_{ab}$ are the generators of $\so_g(\m)$,%
\footnote{See \cref{app:SO_conventions} for conventions regarding the normalisation of the generators of $\so_g(\m)$.}
since the isotropy representation is faithful.

Following \cite{ref:Duff--Nilsson--Pope:KK,ref:Bais--Nicolai--van_Nieuwenhuizen}, we can write a group element close to the identity as
\begin{equation}
\label{eq:coset:geometry:group_element}
    g = \exp(y \cdot T_{(\m)}) \exp(h \cdot T_{(\h)})
\end{equation}
where $(T_{(\h)})_i = T_i$ and $(T_{(\m)})_a = T_a$ are the generators of $\g$ in $\h$ and $\m$, respectively, and $y$ and $h$ are coordinates on $G$.
Note that $y \cdot T_{(\m)}$ and $h \cdot T_{(\h)}$ are simple sums, there are no vielbeins to convert the curved indices of the coordinates to flat indices, like the ones on the generators.%
\footnote{A familiar example is a $\SU(2)$-group element close to the identity, $\exp\bigl(\i \frac{\alpha}{2} \sigma^1 + \i \frac{\beta}{2} \sigma^2 + \i \frac{\gamma}{2} \sigma^3\bigr)$.}
Still, the coordinates are ``curved'' due to the noncommutativity of the generators.
By writing a group element as in \cref{eq:coset:geometry:group_element}, we get a natural representative of each coset $g H$
\begin{equation}
\label{eq:coset:geometry:coset_representative}
    L_y = \exp(y \cdot T_{(\m)}).
\end{equation}
Since $L_y^{-1} \d L_y$ is a $\g$-valued 1-form \cite{ref:Duff--Nilsson--Pope:KK}, we can define 1-forms $e^a$ and $\Omega^i$ by
\begin{equation}
\label{eq:coset:geometry:Ly}
    \tilde\omega
    \coloneqq
    L_y^{-1} \d L_y
    \eqqcolon e^a T_a + \Omega^i T_i.
\end{equation}
Note that $e^a$ are left-invariant by construction.
The $G$-invariant metric on $G/H$ is, in this local coordinate patch, given by
\begin{equation}
    g_{mn} = g_{ab} \tensor{e}{_m^a} \tensor{e}{_n^b},
\end{equation}
where $\tensor{e}{_m^a}$ are the components of $e^a$, that is, $e^a = \d y^m \tensor{e}{_m^a}$.
Here, the left-invariance of $e^a$ ensures that $G$ acts by isometries on $G/H$.
We also see that the metric is invariant under right-multiplication of $L_y$ by $h \in H$ due to $g_{ab}$ being $H$-invariant.
This means that the metric is independent of which representative we choose for a coset, which is needed to globally extend the metric on $\m \simeq T_o \M$ to $G/H$ \cite{ref:Arvanitoyeorgos}.

Note the similarity between $\tilde\omega$ and the Maurer--Cartan form.
In fact, $\tilde\omega$ satisfies the Maurer--Cartan equation
\begin{equation}
    \d \tilde\omega + \tilde\omega \wedge \tilde\omega = 0,
\end{equation}
which follows from differentiating \cref{eq:coset:geometry:Ly} by using
\begin{equation}
    0
    = \d (L_y^{-1} L_y)
    = \d L_y^{-1}\, L_y + L_y^{-1} \d L_y.
\end{equation}
By using $\tilde\omega = e^a T_a + \Omega^i T_i$ and reductivity the Maurer--Cartan equation splits into
\begin{equation}
\label{eq:coset:geometry:Maurer--Cartan}
    \d e^a
    = - \frac{1}{2} e^b \wedge e^c \tensor{f}{_b_c^a}
    - e^b \wedge \Omega^i \tensor{f}{_b_i^a},
    \qquad
    \d \Omega^i
    = - \frac{1}{2} e^a \wedge e^b \tensor{f}{_a_b^i}
    - \frac{1}{2} \Omega^j \wedge \Omega^k \tensor{f}{_j_k^i}.
\end{equation}
From this, we can find an expression for the Levi-Civita spin connection $\omega$, that is, the unique torsion-free spin connection.
Since it is torsion-free, $0 = \d e^a + \tensor{\omega}{^a_b} \wedge e^b$.
From this and \cref{eq:coset:geometry:Maurer--Cartan}, one can read off that
\begin{equation}
    \tensor{\omega}{_{[cb]}^a} = - \frac{1}{2} \tensor{f}{_c_b^a} - \Omega_{[c}^i \tensor{f}{_{|i|b]}^a}.
\end{equation}
Since $\omega_{abc}$ is antisymmetric in its last two indices $\omega_{abc} = \omega_{[ab]c} - \omega_{[ac]b} - \omega_{[bc]a}$, where we have lowered the last index using $g_{ab}$, whence
\begin{equation}
\label{eq:coset:geometry:spin_connection}
    \omega_{abc} = - \frac{1}{2} F_{abc} - \Omega_a^i f_{i bc},
    \qquad
    F_{abc} \coloneqq 2 f_{a [bc]} - f_{bc a}.
\end{equation}
Note that the Jacobi identity and reductivity implies that all nonvanishing parts of $\tensor{f}{_A_B^C}$ ($\tensor{f}{_a_b^c}$, $\tensor{f}{_a_b^k}$, $\tensor{f}{_i_b^c}$ and $\tensor{f}{_i_j^k}$) are $\h$-invariant tensors.
Thus, $F_{abc}$ is also $\h$-invariant since $g_{ab}$ is.

The curvature 2-form is, per definition, $\tensor{\R}{_a^b} = \d \tensor{\omega}{_a^b} + \tensor{\omega}{_a^c} \wedge \tensor{\omega}{_c^b}$.%
\footnote{Note that we use $R_{ab}$ to denote both the curvature 2-form and the Ricci tensor; it should be clear from the context which is being referred to.}
By direct computation, we find from \cref{eq:coset:geometry:Maurer--Cartan,eq:coset:geometry:spin_connection}
\begin{subequations}
\begin{align}
    &\d \tensor{\omega}{_a^b}
    =
    e^d \wedge e^e \Bigl( \frac{1}{4} \tensor{f}{_d_e^c} \tensor{F}{_c_a^b} + \frac{1}{2} \tensor{f}{_d_e^i} \tensor{f}{_i_a^b} \Bigr)
    - \frac{1}{2} e^d \wedge \Omega^i \tensor{f}{_i_d^c} \tensor{F}{_c_a^b}
    + \frac{1}{2} \Omega^j \wedge \Omega^k \tensor{f}{_j_k^i} \tensor{f}{_i_a^b},
    \\
    &\tensor{\omega}{_a^c} \wedge \tensor{\omega}{_c^b}
    =
    \frac{1}{4} e^d \wedge e^e \tensor{F}{_d_a^c} \tensor{F}{_e_c^b}
    + e^d \wedge \Omega^i \Bigl(\frac{1}{2}\tensor{F}{_d_a^c} \tensor{f}{_i_c^b} - \frac{1}{2} \tensor{f}{_i_a^c} \tensor{F}{_d_c^b} \Bigr)
    + \Omega^i \wedge \Omega^j \tensor{f}{_i_a^c} \tensor{f}{_j_c^b}.
\end{align}
\end{subequations}
When adding these, the mixed terms $e^d \wedge \Omega^i$ cancel due to $F_{abc}$ being $\h$-invariant and the $\Omega^i \wedge \Omega^j$ terms cancel since $\tensor{(T_i)}{_a^b} = - \tensor{f}{_i_a^b}$ in the isotropy representation and $T_{[i} T_{j]} = \tensor{f}{_i_j^k} T_k/2$.
Thus,
\begin{equation}
    \tensor{\R}{_a^b}
    = e^d \wedge e^e \Bigl(
        \frac{1}{2} \tensor{f}{_d_e^i} \tensor{f}{_i_a^b}
        + \frac{1}{4} \tensor{f}{_d_e^c} \tensor{F}{_c_a^b}
        + \frac{1}{4} \tensor{F}{_d_a^c} \tensor{F}{_e_c^b}
    \Bigr).
\end{equation}
The curvature 2-form is related to the Riemann tensor by
\begin{equation}
    \tensor{\R}{_a^b} = \frac{1}{2} \tensor{R}{_c_d_a^b} \d x^c \wedge \d x^d,
\end{equation}
whence
\begin{equation}
\label{eq:coset:geometry:Riem}
    \tensor{R}{_c_d_a^b}
    = \tensor{f}{_c_d^i} \tensor{f}{_i_a^b}
    + \frac{1}{2} \tensor{f}{_c_d^e} \tensor{F}{_e_a^b}
    + \frac{1}{2} \tensor{F}{_{[c|a|}^e} \tensor{F}{_{d]}_e^b}.
\end{equation}

Lastly, we mention that there is a special class of $G$-invariant metrics which are particularly simple.
Consider a positive definite $G$-invariant tensor $g_{AB}$.
One can then take $\m$ as the orthogonal complement of $\h$.
This makes $g_{AB}$ block-diagonal on $\h \oplus \m$ and $g_{ab}$, the restriction to $\m$, $H$-invariant.
Such a metric $g_{ab}$ is said to be a normal homogeneous metric on $G/H$ \cite{ref:Arvanitoyeorgos}.
Since $g_{AB}$ is $G$-invariant and block-diagonal, $f_{abc}$ is completely antisymmetric and $F_{abc} = f_{abc}$.
Then, \cref{eq:coset:geometry:Riem} simplifies to
\begin{equation}
\label{eq:coset:geometry:Riem_normal}
    \tensor{R}{_c_d_a^b}
    = \tensor{f}{_c_d^i} \tensor{f}{_i_a^b}
    + \frac{1}{2} \tensor{f}{_c_d^e} \tensor{f}{_e_a^b}
    + \frac{1}{2} \tensor{f}{_{[c|a|}^e} \tensor{f}{_{d]}_e^b},
\end{equation}
which agrees with \cite{ref:Bais--Nicolai--van_Nieuwenhuizen}.%
\todo[disable]{}

\subsubsection{Spin geometry} \label{sec:coset:geometry:spin}
To be able to globally define spinors on $\M=G/H$, it must admit a spin structure, that is, a lift of the structure group from $\SO_g(\m)$ to $\Spin_g(\m)$.%
\footnote{For a proper introduction to spin structures and spin geometry, see \cite{ref:Lawson--Michelsohn}.}
Not all coset spaces $G/H$ admit a spin structure, for instance $\CC\mathbf{P}^2 \simeq \SU(3)/\U(2)$ does not \cite{ref:Mickelsson--Niederle}.
Therefore, like \cite{ref:Mickelsson--Niederle}, we assume that there is a covering group $\bar G$ of $G$ such that the embedding of $H$ in $\SO_g(\m)$ lifts%
\todo[disable]{}
to an embedding of the corresponding cover $\bar H$ of $H$ in $\Spin_g(\m)$.
This implies that $\M \simeq \bar G / \bar H$ where $\bar H$ is a subgroup of $\Spin_g(\m)$.
The $\bar G$-action on $\M$ is not effective since the lift is nontrivial.
However, it is infinitesimally effective in the sense that the isotropy representation of $\bar{\h} = \h$ is faithful.

Analogous to what we saw above, $\bar G$ is a principal $\bar H$-bundle over $\M$ and we have a reduction of the structure group from $\Spin_g(\m)$ to $\bar H$ \cite{ref:Mickelsson--Niederle}.
Correspondingly, the associated vector bundles of the principal bundle of spin frames split into direct sums of vector bundles carrying irreducible $\bar H$-representations.
This means that tensor and spinor fields on $\M$ can be decomposed (globally) into pieces transforming under some representation of $\bar H$.%
\todo[disable]{}

\section{Harmonic analysis} \label{coset:analysis}
In this section, we discuss harmonic analysis on coset spaces, a generalisation of Fourier series and spherical harmonics.
Harmonic analysis on coset spaces is important in Kaluza--Klein compactifications since, if the $(4+k)$-dimensional theory is compactified on a manifold which is locally isometric to $\M_4 \times \M_k$, the fields can be expanded on $\M_k$ using harmonics with spacetime fields as coefficients, yielding a $4$-dimensional theory, see for instance \cite{ref:Duff--Nilsson--Pope:KK,ref:Mickelsson--Niederle}.

We assume that the manifold is spin and that the group of effective isometries lifts as described in \cref{sec:coset:geometry:spin}.
Since we will only be concerned with the lifted groups, we denote the lifted isometry group by $G$ and the lifted isotropy group, which is a subgroup of $\Spin_g(\m)$, by $H$.
Also, since the isometry group of a compact manifold is compact \cite{ref:Kobayashi--Nomizu} and the representation theory of compact Lie groups is particularly well behaved, we restrict our attention to compact $G$.
The relevant theory is based on the Peter--Weyl theorem and the fact that $G$ is a principal $H$-bundle over $G/H$.

\subsubsection{The Peter--Weyl theorem}
Recall that every representation of a compact Lie group $G$ is unitary in the sense that there exists a $G$-invariant positive definite scalar product, that is, $\delta_{\bar P Q}$ is invariant.
This can be seen by Weyl's unitarian trick \cite{ref:Hall:Lie}.
The Peter--Weyl theorem \cite{ref:Peter--Weyl} for compact groups states that an orthogonal basis for $L^2(G)$, that is, complex square-integrable%
\footnote{The integration measure on $G$ is known as the Haar measure. See \cite{ref:Kaniuth--Taylor} for details.}
functions on $G$, is provided by the matrix elements of all irreducible representations.
The orthogonality relation is
\begin{equation}
    \langle \rho^{(\sigma)}\negphantom{{}^{\sigma}}{}\indices{_P^Q}, \rho^{(\tau)}\negphantom{{}^{\tau}}{}\indices{_R^S} \rangle
    \coloneqq \int\limits_G \d g\, \bar\rho^{(\sigma)}(g)\indices{_{\bar P}^{\bar Q}}\, \rho^{(\tau)}(g)\indices{_R^S}
    = \frac{V_G}{\dim \rho^{(\sigma)}} \delta^{{(\sigma)} {(\tau)}} \delta_{\bar P R} \delta^{\bar Q S},
\end{equation}
where $\rho^{(\sigma)}$ and $\rho^{(\tau)}$ are irreducible representations of $G$; ${\sigma}$ and ${\tau}$ label%
\footnote{We think of $\sigma$ and $\tau$ as \emph{labels} and do not employ the Einstein summation convention on them.}
all inequivalent irreducible representations of $G$;%
\footnote{Note that there are infinitely many inequivalent irreducible representations of (nonfinite) compact Lie groups.}
$P,Q,R,S$ are indices for the corresponding representations; bars denote complex conjugation and
\begin{equation}
    \int\limits_G \d g = V_G
\end{equation}
is the volume of $G$.
Thus, $(\rho^{(\sigma)}\negphantom{{}^{\sigma}}{}\indices{_P^Q})_{\sigma,P,Q}$ is an orthogonal basis for $L^2(G)$ and we may expand a function $X \from G \to \CC$ as \cite{ref:Salam--Strathdee}
\begin{equation}
    X(g)
    = \sum_{\sigma} \rho^{(\sigma)}(g)\indices{_P^Q}\, X^{(\sigma)}\negphantom{{}^{\sigma}}{}\indices{_Q^P}.
\end{equation}
Using that $\rho^{(\sigma)}(g)\indices{_P^Q} \bar\rho^{(\sigma)}(g)\indices{_{\bar R}^{\bar S}} \delta_{Q \bar S} = \delta_{P \bar R}$, that is, that the representations are unitary, one finds the coefficients%
\footnote{With $G=\U(1)$ this is ordinary Fourier series.}
\begin{equation}
\label{eq:coset:analysis:Peter--Weyl-coeffs}
    X^{(\sigma)}\negphantom{{}^{\sigma}}{}\indices{_P^Q}
    = \frac{\dim \rho^{(\sigma)}}{V_G} \int\limits_G \d g\, \rho^{(\sigma)}(g^{-1})\indices{_P^Q}\, X(g).
\end{equation}

\subsubsection{Coset harmonics}
Consider now a tensor (or spinor) field on a coset space $G/H$, that is, a section of a vector bundle over $\M$ carrying a particular representation of $\Spin_g(\m)$.
As explained above, these bundles split into direct sums of vector bundles carrying irreducible $H$-representations due to the reduction of the structure group from $\Spin_g(\m)$ to the subgroup $H$.
Thus, we wish to find a basis for vector bundles with $H$ as structure group.
Such bundles are constructed from a vector space $V$, which $H$ acts on by a representation $\rho_H$, and the principal $H$-bundle $G$ via the associated bundle construction, see \cref{app:bundles}, and will be denoted $G\times_{\rho_H}\! V$ \cite{ref:Mickelsson--Niederle}.
A basis can then be constructed by noting that sections of $G\times_{\rho_H}\! V$ are in one-to-one correspondence with $V$-valued functions on $G$ satisfying the equivariance condition \cite{ref:Mickelsson--Niederle}
\begin{equation}
    X_p(gh) = \rho_H(h^{-1})\indices{_p^q} X_q(g).
\end{equation}
Here $p,q$ are indices for the $H$-representation $\rho_H$.
By the Peter--Weyl theorem, each component $X_p(g)$ can be expanded as%
\footnote{Here, we apply the above expansion to $X_p(g^{-1})$ for later convenience.}
\begin{equation}
    X_p(g) = \sum_{\sigma} \rho^{(\sigma)}(g^{-1})\indices{_P^Q} X_p^{(\sigma)}\negphantom{{}^{\sigma}}{}\indices{_Q^P}.
\end{equation}
Imposing the equivariance condition and using that all functions $\rho^{(\sigma)}(g^{-1})\indices{_P^Q}$ are independent, we find that
\begin{equation}
    X_p^{(\sigma)}\negphantom{{}^{\sigma}}{}\indices{_Q^P}
    = \rho_H(h)\indices{_p^q}\,
    X_q^{(\sigma)}\negphantom{{}^{\sigma}}{}\indices{_Q^R}\,
    \rho^{(\sigma)}(h^{-1})\indices{_R^P}.
\end{equation}
Hence, for each fixed $\sigma$ and $Q$, $X_p^{(\sigma)}\negphantom{{}^{\sigma}}{}\indices{_Q^P}$ is an intertwiner between $\rho_H$ and the restriction $\rho^{(\sigma)}|_H$ of the $G$-representation $\rho^{(\sigma)}$ to $H$.
By Shur's lemma, the only such intertwiners are linear combinations of projections from the restricted $G$-representation to subrepresentations equivalent to $\rho_H$.
Thus, we can write $X_p^{(\sigma)}\negphantom{{}^{\sigma}}{}\indices{_Q^P} = X^{(\sigma) \xi}_Q P_\xi^{(\sigma)}\negphantom{{}^{\sigma}}{}\indices{_p^P}$, where $P_\xi^{(\sigma)}$ is the projection onto the $\xi$'th subrepresentation of $\rho^{(\sigma)}|_H$ that is equivalent to $\rho_H$, and
\begin{equation}
\label{eq:coset:analysis:harmonic_expansion}
    X_p(g) = \sum_{\sigma} \rho^{(\sigma)}(g^{-1})\indices{_{p\xi}^Q} X^{(\sigma)\xi}_Q.
\end{equation}
In this expansion, we refer to the basis functions $\rho^{(\sigma)}(g^{-1})\indices{_{p\xi}^Q}$ as harmonics on the coset.

There is a left $G$-action on $X_p$, defined by $\tilde g \cdot X_p(g) = X_p(\tilde g^{-1} g)$.
This $G$\hyp{}representation is said to be induced from the $H$-representation $\rho_H$ \cite{ref:Mickelsson--Niederle,ref:Kaniuth--Taylor}.
By identifying $X_p(g)$ with the coefficients $X^{(\sigma) \xi}_Q$ via the above expansion, we see that the induced representation splits into a direct sum of irreducible $G$-representations, each $X^{(\sigma) \xi}_Q$, for fixed $\sigma$ and $\xi$, transforming under $\rho^{(\sigma)}$.
That the multiplicity of $\rho^{(\sigma)}$ in the induced representation coincides with the multiplicity of $\rho_H$ in $\rho^{(\sigma)}|_H$ is known as Frobenius reciprocity \cite{ref:Kaniuth--Taylor}.

Lastly, we expand on the link between sections of $G\times_{\rho_H}\! V$, that is, fields carrying the $H$-representation $\rho_H$, and $H$-equivariant $V$-valued functions on $G$.
This is explained in more detail in \cref{app:bundles}.
Given a local trivialisation of $G$, considered as a principal $H$-bundle over $\M$, we get local embeddings $\psi^\alpha\from U^\alpha \to G$, where $\set{U^\alpha}_\alpha$ are coordinate charts on $\M$ \cite{ref:Mickelsson--Niederle}.
Locally, a section of $G\times_{\rho_H}\! V$ is equivalent to a $V$-valued function on $\M$ and the section corresponding to $X_p(g)$ is simply given by
\begin{equation}
    Y_p^\alpha(y) = X_p(\psi^\alpha(y)).
\end{equation}
The $H$-equivariance of $X_p$ is needed to ensure that $Y_p$ defines a global section of $G\times_{\rho_H}\! V$.
In the above chart, with local embedding given by $y \mapsto L_y$, the expansion \cref{eq:coset:analysis:harmonic_expansion} reads
\begin{equation}
    Y_p (y)
    = \sum_{\sigma} \rho^{(\sigma)}(L_y^{-1})\indices{_{p\xi}^Q} X^{(\sigma)\xi}_Q,
\end{equation}
which agrees with \cite{ref:Salam--Strathdee} apart from irrelevant normalisation of the coefficients.%
\footnote{A well-known example is the expansion of a scalar field on $S^2\simeq \SO(3)/\SO(2) \simeq \Spin(3)/\Spin(2)$ in terms of spherical harmonics.}

\section{The coset master equation} \label{sec:coset:master_eq}
In this section, we discuss what we will refer to as \emph{the coset master equation}, which we will use to compute the eigenvalue spectrum of the squashed $S^7$ in \cref{chap:squashed_spectrum}.
The equation is based on the fact that the tensor and spinor fields can be expanded in terms of harmonics that come from the irreducible representations of $G$, as described in the preceding section.
For reasons explained below, we restrict our attention to normal homogeneous metrics, so that $F_{abc} = f_{abc}$ is completely antisymmetric, and compact Euclidean manifolds $\M$.

As explained above, $G$ is a principal $H$-bundle over $\M$.
There is a natural principal $H$-connection on this bundle induced by the splitting $\g = \h \oplus \m$ \cite{ref:Mickelsson--Niederle}.
In our local coordinates, the $H$-connection is given by \cite{ref:Bais--Nicolai--van_Nieuwenhuizen}%
\todo[disable]{}
\begin{equation}
\label{eq:coset:master_eq:H-derivative}
    \Dg_m = \partial_m + \Omega_m^i T_i.
\end{equation}
Let $\rho$ be any representation of $G$.
Then, by \cref{eq:coset:geometry:Ly},
\begin{equation}
\label{eq:coset:master_eq}
    \Dg_a \rho(L_y^{-1})\indices{_P^{(Q)}}
    = - (T_a)\indices{_P^R} \rho(L_y^{-1})\indices{_R^{(Q)}},
\end{equation}
where $\Dg$ only acts on the first index of $\rho(L_y^{-1})$, which we indicate by the parentheses around $Q$.
This is what we will refer to as the coset master equation.
From \cref{eq:coset:geometry:isotropy_embedding,eq:coset:geometry:spin_connection}, we see that the torsion-free spin connection $\D_a$ is related to the principal $H$-connection $\Dg_a$ in \cref{eq:coset:master_eq:H-derivative} by
\begin{equation}
\label{eq:coset:master_eq:H-derivative_spin}
    \Dg_a = \D_a + \frac{1}{2} f_{abc} \Sigma^{bc},
\end{equation}
since $F_{abc} = f_{abc}$ in the normal homogeneous case.
Note, however, that this relation only is valid when $\Dg_a$ acts on a tensor carrying a $\Spin_g(\m)$-representation since, otherwise, the right-hand side is not defined.
In particular, it cannot be used directly in \cref{eq:coset:master_eq}.

Since $T_a$ are the generators of $\m$, which are not block-diagonal over the irreducible $H$-representations in $\rho|_{H}$, $T_a$ cannot act as matrices on the harmonics.
However, if multiple generators are combined to an element in the universal enveloping algebra, $\UU(\g)$, which is block-diagonal over the $H$-representations, the corresponding analogue of \cref{eq:coset:master_eq} can be restricted to any particular $H$-representation and then applies to the harmonics.
A short calculation shows that%
\footnote{In \cref{eq:coset:master_eq:T_D_relation}, $\Dg_b$ acts not only on the first index of $\rho(L_y^{-1})$ but also on the $a$-index.}
\begin{align}
\nonumber
    T_a T_b \rho(L_{y}^{-1})
    &= -T_a (\partial_b + \Omega_b^i T_i) \rho(L_{y}^{-1})
    =\\ \nonumber
    &= -(\partial_b + \Omega_b^i T_i) T_a \rho(L_{y}^{-1})
    - \Omega_b^i \tensor{f}{_a_i^c} T_c \rho(L_{y}^{-1})
    =\\ \nonumber
    &= (\partial_b + \Omega_b^i T_i) \Dg_a \rho(L_{y}^{-1})
    - \Omega_b^i \tensor{f}{_i_a^c} \Dg_c \rho(L_{y}^{-1})
    =\\
\label{eq:coset:master_eq:T_D_relation}
    &= \Dg_b \Dg_a \rho(L_{y}^{-1}).
\end{align}
Now we will make use of the assumption that the metric is normal homogeneous, that is, that $g_{ab}$ comes from the restriction of a $G$-invariant $g_{AB}$.
Then,
\begin{equation}
    g^{ab} T_a T_b
    %= g^{AB} T_A T_B - g^{ij} T_i T_j
    = - \C_{\g} + \C_{\h},
    \qquad\quad
    \C_{\g} \coloneqq -g^{AB} T_A T_B,
    \qquad
    \C_{\h} \coloneqq - g^{ij} T_i T_j,
\end{equation}
where $\C_{\g}$ and $\C_{\h}$ are quadratic Casimir invariants of $\g$ and $\h$, respectively.%
\footnote{For semisimple $\g$, $\C_\g$ is some linear combination of the quadratic Casimirs of the simple constituent Lie algebras, see \cref{app:quadratic_casimirs}.}%
\footnotemarksep%
\footnote{The normalisation here might not be conventional for concrete cases.}
Here, $\C_{\g}$ acts by a constant on any particular $G$-representation $\rho$ and $\C_{\h}$ acts by a constant on every irreducible part of $\rho|_H$.
Thus,
\begin{equation}
\label{eq:coset:master_eq:quadratic}
    (\C_{\g} - \C_{\h}) Y
    = -\check\square Y,
\end{equation}
where $Y$ is a field (with suppressed index) carrying a representation of $\Spin_g(\m)$ and $\check\square \coloneqq g^{ab} \Dg_a \Dg_b$.
We will refer to this equation as the quadratic master equation.%
\footnote{This has been referred to as the squared coset master equation in, for instance, \cite{ref:Ekhammar--Nilsson}. Since only the operators are squared, we use ``quadratic'' instead.}
As noted above, the field $Y$ splits into irreducible $H$-components each carrying an induced $G$-representation.
$\C_{\h}$ acts by a constant on each irreducible $H$-component and can thus be implemented as a matrix acting on the spin-index of $Y$ while $\C_{\g}$ acts by a constant on each irreducible $G$-representation in the decompositions of the induced representations and cannot be implemented as a matrix.

As we saw in \cref{sec:ads_susy:Laplacian}, the mass spectrum of a Freund--Rubin compactification is related to the eigenvalue spectrum of a universal Laplacian \cref{eq:ads_susy:laplacian:universal}
\begin{equation}
\label{eq:coset:master_eq:Laplacian}
    \Delta Y = \kappa^2 Y,
    \qquad\quad
    \Delta \coloneqq -\square - R_{abcd} \Sigma^{ab} \Sigma^{cd}.
\end{equation}
Since $\M$ is compact and Euclidean, both $\Spin_g(\M)$ and $G$ are compact, the finite-dimensional representations of $\Spin_g(\M)$ are unitary and the fields carrying such representations form a unitary $G$-representation, with respect to the appropriate $L^2$ inner product, which is, thus, completely reducible.
Hence, the eigenmodes of $\Delta$ fall into irreducible representations of $G$ since $\Delta$ is manifestly invariant under isometries.

To be able to use \cref{eq:coset:master_eq:quadratic} to compute the eigenvalues of $\Delta$, we wish to relate $\square$ and $\DDg$.
Using \cref{eq:coset:master_eq:H-derivative_spin}, that $g_{ab}$ and $f_{ab}{}^c$ are $H$-invariant and that $\rho(\Sigma_{ab})$ is an $\SO_g(\m)$-invariant for any representation $\rho$, we find
\begin{align}
\nonumber
    \DDg
    &= \Dg^a \Bigl(\D_a + \frac{1}{2} f_{abc} \Sigma^{bc} \Bigr)
    = \square
    + \frac{1}{2} f^{ade} \Sigma_{de} \D_a
    + \frac{1}{2} f_{abc} \Sigma^{bc} \Dg^a
    =\\
    &= \square
    + f_{abc} \Sigma^{bc} \Dg^a
    - \frac{1}{4} f_{abc} f^{ade} \Sigma^{bc} \Sigma_{de}.
\end{align}
Combining this with \cref{eq:coset:master_eq:quadratic,eq:coset:master_eq:Laplacian} gives
\begin{equation}
\label{eq:coset:master_eq:almost_done}
    f_{abc} \Sigma^{bc} \Dg^a
    = \Delta - \C_\g + \C_\h + R_{abcd} \Sigma^{ab} \Sigma^{cd} + \frac{1}{4} f_{abc} f^{ade} \Sigma^{bc} \Sigma_{de}.
\end{equation}
We can simplify \cref{eq:coset:master_eq:almost_done} a bit further.
For this, note that, by \cref{eq:coset:geometry:Riem_normal},
\begin{equation}
    R\indices{_{abcd}} \Sigma\indices{^{ab}} \Sigma\indices{^{cd}}
    = f\indices{_{iab}} f\indices{^{icd}} \Sigma\indices{^{ab}} \Sigma\indices{_{cd}}
    + \frac{1}{2} f\indices{_{abc}} f\indices{^{a}_{de}} \Sigma\indices{^{bc}} \Sigma\indices{^{de}}
    - \frac{1}{2} f\indices{_{abd}} f\indices{^a_{ce}} \Sigma\indices{^b^c} \Sigma\indices{^d^e},
\end{equation}
whence, by \cref{eq:coset:geometry:isotropy_embedding},%
\footnote{We also use the fact that it does not matter whether $\Sigma_{ab}$ acts on $\Sigma_{cd}$, as in $\rho(\Sigma_{ab} \cdot \Sigma^{cd}) = 2 \delta^{[c|e}_{\hphantom{[}a\hphantom{|}b} \rho(\tensor{\Sigma}{_e^{|d]}}) + \rho(\Sigma_{ab}) \rho(\Sigma^{cd})$, or is $\UU(\so_g(\m))$-multiplied by $\Sigma_{cd}$, as in $\rho(\Sigma_{ab} \circ \Sigma^{cd}) = \rho(\Sigma_{ab}) \rho(\Sigma^{cd})$, as long $\{ab\}\,\{cd\}$ are symmetrised (as in, for instance, $R_{ab cd} = R_{cd ab}$).}
\begin{equation}
\label{eq:coset:master_eq:done}
    f\indices{_{abc}} \Sigma\indices{^{bc}} \Dg\indices{^a}
    = \Delta
    - \C_\g
    + \frac{3}{4} f\indices{_{abc}} f\indices{^{a}_{de}} \Sigma\indices{^{bc}} \Sigma\indices{^{de}}
    - \frac{1}{2} f\indices{_{abd}} f\indices{^a_{ce}} \Sigma\indices{^b^c} \Sigma\indices{^d^e}.
\end{equation}
Remarkably, $\C_\h$ from $\DDg$ was cancelled by the first term in \cref{eq:coset:geometry:Riem_normal}.
As mentioned above, $f_{abc}$ is an $\h$-invariant and, hence, $H^0$-invariant, where $H^0$ is the identity component of $H$.
There may, however, be a larger group $\tilde H$ leaving $f_{abc}$ and $g_{ab}$ invariant, such that $H^0 \subseteq \tilde H \subseteq \Spin_g(\m)$.
This can lead to significant simplifications, as we will see explicitly for the squashed seven-sphere in \cref{chap:squashed_spectrum}.
At this point, it is, however, not clear that $\C_\h$ will not re-enter in the calculation from $\DDg$ or $[\Dg_a, \Dg_b]$.
Note that the last two terms in \cref{eq:coset:master_eq:done} can, for any $\Spin_g(\m)$-representation $\rho$, be expressed in terms of projection operators that project onto the $\tilde H$-irreducible parts of $\rho|_{\tilde H}$, by Shur's lemma, since $f_{abc}$ is $\tilde H$-invariant.

For symmetric spaces%
\footnote{A space is said to be (locally) symmetric if there exists, for each $y\in\M$, a (local) isometry that fixes $y$ and reverses all geodesics through $y$ \cite{ref:Arvanitoyeorgos}. By a theorem due to Cartan, a space is locally symmetric if and only if $\D_a R_{bcde} = 0$ \cite{ref:Arvanitoyeorgos}.}%
, which have $f_{ab}{}^c = 0$ \cite{ref:Duff--Nilsson--Pope:KK}, this reduces the problem of finding the eigenvalues of $\Delta$ to the problem of decomposing the induced $G$-representation into irreducible $G$-representations on which $\C_\g$ is just a number.
However, the case we are ultimately interested in, the squashed seven-sphere, is not a symmetric space.

\subsubsection{Curvature and torsion of $\Dg$}
Lastly, we give some properties of $\Dg$.
We have already seen, in \cref{eq:coset:master_eq:H-derivative_spin}, that
\begin{equation}
    \Dg = \d + \check\omega,
    \qquad\quad
    \check\omega = \omega + \check\kappa,
    \qquad\quad
    \check\kappa_{abc} = \frac{1}{2} f_{abc},
\end{equation}
where $\check\kappa$ is the Lie algebra-valued contorsion 1-form of $\Dg$.
The torsion of this spin connection is, as usual, defined by $\check T^a = \Dg e^a = \d e^a + \check\omega^a{}_b \wedge e^b$.
Since $\D$ is torsion-free,
\begin{equation}
    \check T^a = \check\kappa\indices{^a_b} \wedge e^b,
    \qquad\quad
    \tensor{\check T}{_a_b^c}
    = - 2 \check\kappa\indices{_{[ab]}^c}
    = - f_{abc}.
\end{equation}
The Lie algebra-valued curvature 2-form is, per definition $\check R = \d \check\omega + \check\omega \wedge \check\omega$.
This is related to the curvature 2-form $R$ of the torsion-free spin connection by \cref{eq:gauge_Cartan:arbitrary_R}
\begin{equation}
    \check R = R + \Dg \check\kappa - \check\kappa \wedge \check\kappa,
    \quad\
    \check R\indices{_c_d_a^b}
    = R\indices{_c_d_a^b}
    + 2 \Dg_{[c} \check\kappa\indices{_{d]}_a^b}
    + \check T\indices{_c_d^e} \check\kappa\indices{_e_a^b}
    - 2 \check\kappa\indices{_{[c|}_a^e} \check\kappa\indices{_{|d]}_e^b},
\end{equation}
where the index expression follows from $\Dg (e^c \check\kappa\indices{_c_a^b}) = \check T^c \check\kappa\indices{_c_a^b} - e^c \wedge \Dg \check\kappa\indices{_c_a^b}$.
Using \cref{eq:coset:geometry:Riem_normal} and that $\Dg_a f_{bcd} = 0$ since $f_{abc}$ is $\h$-invariant, this simplifies to
\begin{equation}
    \check R_{abcd}
    = \tensor{f}{_a_b^i} \tensor{f}{_i_c_d}
    = (T^i)_{ab} (T_i)_{cd}.
\end{equation}
The Ricci identity $\Dg^2 \Omega = \check R \wedge \Omega$ can thus be written as
\begin{equation}
\label{eq:coset:master_eq:Ricci_id}
    [\Dg_a, \Dg_b]
    = \check R\indices{_a_b^c^d} \Sigma_{cd}
    - \check T\indices{_a_b^c} \Dg_c
    = (T^i)_{ab} T_i + \tensor{f}{_a_b^c} \Dg_c,
\end{equation}
since $\Dg^2 \Omega
= \D (e^a \D_a \Omega)
= T^a \wedge \D_a \Omega - e^a \wedge e^b \Dg_b \Dg_a \Omega$.
Note that we raise and lower $\h$-indices using $g_{ij}$.

\chapter{Squashed sphere geometry} \label{chap:squashed_geometry}%
In this chapter, we study the geometry of the manifold on which we will compactify eleven-dimensional supergravity: the squashed seven-sphere.
With squashing, we mean a smooth deformation of a homogeneous manifold, that is, a deformation of the metric (the topology is unchanged), that keeps the manifold homogeneous.
Although homogeneity should be preserved, the isometry group may change when squashing.
For instance, when we squash the round $S^7$, part of the $\SO(8)$ isometry is broken.
Note that the existence of a squashing deformation is nontrivial.
For instance, one cannot squash $S^2$ \cite{ref:Duff--Nilsson--Pope:KK}.

Below, we present two constructions of the squashed seven-sphere.
First, it is realised as a nontrivial principal $\SU(2)$-bundle over $S^4$ and, second, as a coset space (the subscripts are explained below)
\begin{equation}
    \frac{\Sp(2)\times\Sp(1)_C}{\Sp(1)_{A}\times\Sp(1)_{B+C}}.
\end{equation}
We also discuss the relation between these constructions and an isometric embedding in the quaternionic projective space $\HH\mathbf{P}^2$.

\section{Squashed \texorpdfstring{$S^7$}{S7} as a principal bundle} \label{sec:squashed_geometry:fibration}
This construction is based on the fact that $S^7$ can be realised as a principal $\SU(2)$-bundle over $S^4$ and starts from the fact that the group of unit quaternions, which is isomorphic to $\SU(2)$, has the topology of $S^3$.
Let $U$ be a unit quaternion, parametrised by Euler angles as \cite{ref:Duff--Nilsson--Pope:KK}
\begin{equation}
  U = \e^{\k \phi/2} \e^{\i \theta/2} \e^{\k \psi/2},
\end{equation}
where $\i$, $\j$ and $\k$ are the imaginary units of $\HH$.
Consider the Lie algebra-valued 1-form
\begin{equation}
\label{eq:squashed_geometry:SU2_form_from_unit_quaternion}
    \sigma \coloneqq
    2 U^{-1} \d U
    = \i \sigma^1 + \j \sigma^2 + \k \sigma^3
    = h_i \sigma^i,
\end{equation}
which is proportional to the Maurer--Cartan form.
Here, $\sigma^i$ are left-invariant 1-forms and $h^i= (\i,\j,\k)^i$.%
\footnote{The explicit expressions for $\sigma^i$ in terms of the Euler angles can be found in \cite{ref:Duff--Nilsson--Pope:KK}.}
Using $0 = \d(U^{-1} U) = \d U^{-1}\: U + U^{-1} \d U$ and $h_i h_j = - \delta_{ij} + \tensor{\epsilon}{_i_j^k} h_k$ we immediately find
\begin{equation}
\label{eq:squashed_geometry:SU2_form}
    \d \sigma
    % = - 2 U^{-1} \d U \wedge U^{-1} \d U
    = - \frac{1}{2} \sigma \wedge \sigma
    = -\frac{1}{2} h_i \tensor{\epsilon}{^i_j_k} \sigma^j \wedge \sigma^k,
\end{equation}
which is essentially the Maurer--Cartan equation.
Because of \cref{eq:squashed_geometry:SU2_form}, we say that the 1-forms $\sigma^i$ satisfy the $\su(2)$ algebra.

The quaternionic left-invariant 1-form $\sigma$ and the invariant $\delta_{ij}$ allows us to construct a metric on $S^3$ as%
\footnote{Note that, by writing the metric like this, we have chosen a length unit. Thus, we work in a dimensionless unit system.}
\begin{equation}
    \d s^2(S^3)
    = \| \sigma \|^2
    = \sigma_i \sigma^i.
\end{equation}
In fact, this is the standard metric on $S^3$ up to a constant conformal factor \cite{ref:Duff--Nilsson--Pope:KK}.
That $\sigma$ is left-invariant, that is, invariant under $U\mapsto aU$ where $a\in\SU(2)$, is immediate from \cref{eq:squashed_geometry:SU2_form_from_unit_quaternion}.
The metric $\|\sigma\|^2$ is, however, right-invariant as well, that is, invariant under $U\mapsto Ua$, since then $\sigma \mapsto a^{-1}\sigma a$.
Thus, the metric is said to be bi-invariant.

From the above metric on $S^3$, the metric on $S^4$ can be written as \cite{ref:Duff--Nilsson--Pope:KK}
\begin{equation}
\label{eq:squashed_geometry:ds2_S4}
    \d s^2(S^4)
    = \d \mu^2 + \frac{1}{4} \sin^2{\mu}\: \|\tilde\Sigma\|^2,
\end{equation}
where $0<\mu<\pi$ and $\tilde\Sigma$ is a quaternionic left-invariant 1-form satisfying the $\su(2)$ algebra.
This construction uses that $S^4$ without the north and south pole is diffeomorphic to $(0, \pi) \times S^3$ and the $\sin^2\mu$ factor gives the three-spheres their correct sizes, smaller close to the poles and larger closer to the ``equator''.%
\footnote{The construction is analogous to glueing together circles of various sizes along a semicircle to make a sphere.}
The coordinate patch described by these coordinates is a warped product space (compare to \cref{sec:Freund--Rubin}).

Now that we have briefly discussed the metrics on $\SU(2)\simeq S^3$ and $S^4$ we turn to the real case of interest, that is, an $\SU(2)$-bundle over $S^4$.
We now have two $S^3$ manifolds and use one real coordinate, $\mu$, and two unit quaternions $\tilde U, \tilde V$.
Hence, there are two independent $\su(2)$ algebras and we need two sets of imaginary units, $h_i^{\tilde U}$ and $h_i^{\tilde V}$.
Let $\tilde\sigma = h_i^{\tilde U} \tilde\sigma^i$ and $\tilde\Sigma = h_i^{\tilde V} \tilde\Sigma^i$ be the $\su(2)$-forms, constructed as in \cref{eq:squashed_geometry:SU2_form_from_unit_quaternion}, corresponding to the two unit quaternion coordinates.
A metric can then be written as
\begin{equation}
\label{eq:squashed_geometry:squashed_fibration_metric}
    \d s^2
    = \d \mu^2 + \frac{1}{4} \sin^2{\mu}\: \|\tilde\Sigma\|^2
    + \lambda^2 \|\tilde\sigma - A\|^2,
\end{equation}
where $A = h_i^{\tilde U} A^i$ is a Yang--Mills $\SU(2)$ gauge potential.
With $A = 0$ this would just be $S^4 \times S^3$, with $\lambda$ determining the relative size of the factors, but if the potential describes a topologically nontrivial instanton, the topology of the bundle is affected \cite{ref:Duff--Nilsson--Pope:KK}.
In particular, with
\begin{equation}
    A^i = \cos^2 \frac{\mu}{2}\: \tilde\Sigma^i,
\end{equation}
the topology is that of $S^7$ \cite{ref:Duff--Nilsson--Pope:KK}.
Note, however, that the topology of the chart covered by our coordinates is still that of $(0,\pi)\times S^3\times S^3$.
The parameter $\lambda$ in \cref{eq:squashed_geometry:squashed_fibration_metric} will be referred to as the squashing parameter.

\subsubsection{Rewriting the metric}
Before computing the Riemann tensor, we rewrite the metric in \cref{eq:squashed_geometry:squashed_fibration_metric} as \cite{ref:Duff--Nilsson--Pope:KK}
\begin{equation}
\label{eq:squashed_geometry:squashed_S7_HP2_metric}
    \d s^2 = \d \mu^2 + \frac{1}{4} \sin^2{\mu}\: \|\varpi\|^2
    + \frac{1}{4} \lambda^2 \| \nu  + \cos\mu\: \varpi \|^2,
\end{equation}
where
\begin{equation}
\label{eq:squashed_geometry:nu_omgea_def}
    \nu^i \coloneqq \sigma^i + \Sigma^i,
    \qquad
    \varpi^i \coloneqq \sigma^i - \Sigma^i,
\end{equation}
where $\sigma$ and $\Sigma$ are quaternionic left-invariant 1-forms satisfying the $\su(2)$ algebra related to two unit quaternions $U$ and $V$, respectively, as in \cref{eq:squashed_geometry:SU2_form_from_unit_quaternion}.
This form of the metric comes from an isometric embedding of the squashed $S^7$ in the quaternionic projective space $\HH\mathbf{P}^2$ \cite{ref:Duff--Nilsson--Pope:KK}.
In the isometric embedding, only $0<\lambda^2\leq 1$ is possible \cite{ref:Duff--Nilsson--Pope:KK}, although there seems to be no such upper bound on $\lambda^2$ in \cref{eq:squashed_geometry:squashed_fibration_metric}.

Note that we could not have written \cref{eq:squashed_geometry:nu_omgea_def} as is without indices since $\sigma$ contains $h_i^{U}$ while $\Sigma$ contains $h_i^{V}$.
To remedy this, we indicate which set of imaginary units is being used with a superscript as $\sigma^{U} = h_i^{U} \sigma^i$, $\sigma^{V} = h_i^{V} \sigma^i$, $\varpi^{V} = \sigma^{V} - \Sigma^{V}$ and so on.
Since $\|\sigma\|=\sigma_i \sigma^i$ regardless of which set of unit quaternions is being used, we need not worry about this in the metrics \cref{eq:squashed_geometry:squashed_fibration_metric,eq:squashed_geometry:squashed_S7_HP2_metric}.
The relation between the two constructions is
\begin{equation}
\label{eq:squashed_geometry:fibre_embedding_relation}
    \tilde\sigma^{V} = -V \Sigma^V V^{-1},
    \qquad \tilde\Sigma^{V} = V \varpi^{V} V^{-1}.
\end{equation}
To see this, first note that the first and middle terms of the metrics \cref{eq:squashed_geometry:squashed_fibration_metric,eq:squashed_geometry:squashed_S7_HP2_metric} are equal since $V$ is a unit quaternion.
That the last terms are also equal follows from
\begin{align}
\nonumber
    \bigl\|\tilde\sigma - \cos^2\frac{\mu}{2}\, \tilde\Sigma\bigr\|^2
    &= \bigl\|\Sigma + \cos^2\frac{\mu}{2}\, \varpi\bigr\|^2
    = \bigl\|\Sigma + \frac{1}{2} (\sigma - \Sigma) + \frac{1}{2} \cos\mu\:\varpi \bigr\|^2
    =\\
    &= \frac{1}{4} \| \nu + \cos\mu\:\varpi \|^2.
\end{align}
Furthermore, that $\tilde\sigma$ from \cref{eq:squashed_geometry:fibre_embedding_relation} satisfies the $\su(2)$ algebra is seen from
\begin{align}
\nonumber
    \d \tilde\sigma
    &= -\d V \wedge \Sigma V^{-1}
    - V \d \Sigma V^{-1}
    + V \Sigma \wedge \d V^{-1}
    =\\ \nonumber
    &= -\frac{1}{2} V \Sigma \wedge \Sigma V^{-1}
    + \frac{1}{2} V \Sigma \wedge \Sigma V^{-1}
    - \frac{1}{2} V \Sigma \wedge \Sigma V^{-1}
    = -\frac{1}{2} \tilde\sigma \wedge \tilde\sigma,
\end{align}
where we have dropped the superscript but $h_i^V$ are the only unit quaternions appearing.%
\todo[disable]{}
Similarly,
\begin{align}
\nonumber
    \d \tilde\Sigma
    &= \d V \wedge \varpi V^{-1}
    + V \d \varpi V^{-1}
    - V \varpi \wedge \d V^{-1}
    =\\ \nonumber
    &= \frac{1}{2} V \Sigma \wedge (\sigma - \Sigma) V^{-1}
    - \frac{1}{2} V (\sigma \wedge \sigma - \Sigma \wedge \Sigma) V^{-1}
    + \frac{1}{2} V (\sigma - \Sigma) \wedge \Sigma V^{-1}
    =\\ \nonumber
    &= -\frac{1}{2} V (\sigma - \Sigma)\wedge(\sigma - \Sigma) V^{-1}
    = - \frac{1}{2} \tilde\Sigma \wedge \tilde\Sigma.
\end{align}

\subsubsection{The spin connection and curvature}
Now, we derive expressions for the spin connection, Riemann tensor, Ricci tensor and curvature scalar, starting from the metric \cref{eq:squashed_geometry:squashed_S7_HP2_metric}
\begin{equation}
    \d s^2 = \d \mu^2 + \frac{1}{4} \sin^2{\mu}\: \|\varpi\|^2
    + \frac{1}{4} \lambda^2 \| \nu  + \cos\mu\: \varpi \|^2.
\end{equation}
From this metric, we see that an orthonormal frame is provided by
\begin{equation}
\label{eq:squashed_geometry:fibration:vielbein}
    e^{\hati} = \frac{1}{2} \lambda \bigl(\nu^i + \cos\mu\: \varpi^i\bigr),
    \qquad
    e^0 = \d \mu,
    \qquad
    e^i = \frac{1}{2} \sin{\mu}\: \varpi^i,
\end{equation}
where we have split the seven-dimensional index as $a= (\hati, 0, i)$.
Note that, from a covariant perspective, the index on the first $e$ should be $i$ due to the right-hand side.
However, in the index split, we need to distinguish between $i$ and $\hati$.
Therefore, the notation is not completely covariant and, to avoid confusion, we will only use indices $\hati, \hatj, \hat k,\hdots$ on seven-dimensional objects and not on the $\SU(2)$-invariants $\epsilon_{ijk}$ and $\delta_{ij}$.

The torsion-free spin connection can be determined from
\begin{equation}
\label{eq:squashed_geometry:fibration:torsion}
    0 = T^a
    \coloneqq \D e^a
    = \d e^a + \tensor{\omega}{^a_b} \wedge e^b.
\end{equation}
To determine $\omega$, we first have to compute $\d e^a$.
To this end, we compute $\d \nu^i$ and $\d \varpi^i$ using
\begin{subequations}
\begin{alignat}{2}
    &\nu^i \wedge \nu^j + \varpi^i \wedge \varpi^j
    &&= 2 \bigl( \sigma^i\wedge\sigma^j + \Sigma^i\wedge\Sigma^j \bigr),
    \\
    &\nu^i \wedge \varpi^j + \varpi^i \wedge \nu^j
    &&= 2 \bigl( \sigma^i \wedge \sigma^j - \Sigma^i \wedge \Sigma^j \bigr),
\end{alignat}
\end{subequations}
and express the results in terms of $e^a$ by inverting \cref{eq:squashed_geometry:fibration:vielbein},
\begin{equation}
    \varpi^i = \frac{2}{\sin\mu} e^i,
    \qquad
    \nu^i = - 2 \cot\mu\: e^i + \frac{2}{\lambda} e^{\hati}.
\end{equation}
Using also that $\sigma$ and $\Sigma$ satisfy the $\su(2)$ algebra, we find
\begingroup
\allowdisplaybreaks
\begin{subequations}
\begin{align}
    &%
    \begin{aligned}[b]
        \d \nu^i
        &= - \frac{1}{2} \tensor{\epsilon}{^i_j_k} \bigl( \sigma^j \wedge \sigma^k + \Sigma^j \wedge \Sigma^k \bigr)
        = - \frac{1}{4} \tensor{\epsilon}{^i_j_k} \bigl( \nu^j \wedge \nu^k + \varpi^j \wedge \varpi^k \bigr)
        =\\
        % &= - \tensor{\epsilon}{^i_j_k} \Bigl[
        %     \bigl(\cot\mu\: e^j - \frac{1}{\lambda} e^{\hatj} \bigr)
        %     \wedge
        %     \bigl(\cot\mu\: e^k - \frac{1}{\lambda} e^{\hat k} \bigr)
        %     + \frac{1}{\sin^2\mu} e^j \wedge e^k
        % \Bigr]
        % =\\
        &=
        - \frac{1 + \cos^2\mu}{\sin^2\mu} \tensor{\epsilon}{^i_j_k} e^j \wedge e^k
        + \frac{2}{\lambda} \cot\mu\: \tensor{\epsilon}{^i_j_k} e^j \wedge e^{\hat k}
        - \frac{1}{\lambda^2} \tensor{\epsilon}{^i_j_k} e^{\hatj} \wedge e^{\hat k}
    \end{aligned}
    \\[2pt] &%
    \begin{aligned}[b]
        \d \varpi^i
        &= - \frac{1}{2} \tensor{\epsilon}{^i_j_k} \bigl( \sigma^j \wedge \sigma^k + \Sigma^j \wedge \Sigma^k \bigr)
        = - \frac{1}{2} \tensor{\epsilon}{^i_j_k} \nu^j \wedge \varpi^k
        =\\
        % &= 2 \tensor{\epsilon}{^i_j_k} \bigl( \cot\mu\: e^j - \frac{1}{\lambda} e^{\hatj} \bigr) \wedge \frac{1}{\sin\mu} e^k
        % =\\
        &= 2 \frac{\cos\mu}{\sin^2\mu} \tensor{\epsilon}{^i_j_k} e^j \wedge e^k
        - \frac{2}{\lambda \sin\mu} \tensor{\epsilon}{^i_j_k} e^j \wedge e^{\hat k}.
    \end{aligned}
\end{align}
\end{subequations}
\endgroup
Thus, since $\d \mu = e^0$,
\begin{subequations}
\label{eq:squashed_geometry:fibration:de}
\begin{alignat}{2}
    &\d e^0 &&= 0,\\
    &\d e^i
    &&= \cot\mu\: e^0 \wedge e^i
    - \frac{1}{\lambda} \tensor{\epsilon}{^i_j_k} e^j \wedge e^{\hat k}
    + \cot\mu\: \tensor{\epsilon}{^i_j_k} e^j \wedge e^k,\\
    &\d e^{\hati}
    &&= - \lambda e^0 \wedge e^{i}
    - \frac{\lambda}{2} \tensor{\epsilon}{^i_j_k} e^j \wedge e^k
    - \frac{1}{2\lambda} \tensor{\epsilon}{^i_j_k} e^{\hatj} \wedge e^{\hat k}.
\end{alignat}
\end{subequations}
Reading off $\omega_{[ab]c}$ from \cref{eq:squashed_geometry:fibration:torsion} and \cref{eq:squashed_geometry:fibration:de}, using the standard trick $\omega_{abc} = \omega_{[ab]c} - \omega_{[ac]b} - \omega_{[bc]a}$, we find
\begin{subequations}
\label{eq:squashed_geometry:fibration:spin_connection}
\begin{alignat}{2}
    &\tensor{\omega}{_0^i}
    &&= -\cot\mu\: e^i
    + \frac{\lambda}{2} e^{\hati},
    \\
    &\tensor{\omega}{_0^{\hati}}
    &&= \frac{\lambda}{2} e^i,
    \\
    &\tensor{\omega}{_i_j}
    &&= \cot\mu\: \tensor{\epsilon}{_i_j_k} e^k
    + \Bigl(\frac{\lambda}{2} - \frac{1}{\lambda} \Bigr)
    \tensor{\epsilon}{_i_j_k} e^{\hat k},
    \\
    &\tensor{\omega}{_{\hati}_{\hatj}}
    &&= - \frac{1}{2\lambda} \tensor{\epsilon}{_i_j_k} e^{\hat k},
    \\
    &\tensor{\omega}{_i_{\hatj}}
    &&= -\frac{\lambda}{2} \delta_{ij} e^0
    - \frac{\lambda}{2} \tensor{\epsilon}{_i_j_k} e^k.
\end{alignat}
\end{subequations}
The curvature 2-form and Riemann tensor are defined by
\begin{equation}
    \tensor{\R}{_a^b}
    = \frac{1}{2} \tensor{R}{_c_d_a^b} e^c \wedge e^d
    = \d \tensor{\omega}{_a^b} + \tensor{\omega}{_a^c} \wedge \tensor{\omega}{_c^b}.
\end{equation}
Using \cref{eq:squashed_geometry:fibration:spin_connection,eq:squashed_geometry:fibration:de}, we find
\begin{subequations}
\label{eq:squashed_geometry:fibration:Riem}
\begin{alignat}{2}
    &\tensor{\R}{_0^i}
    &&= \Bigl(1 - \frac{3}{4} \lambda^2 \Bigr) e^0 \wedge e^i
    + \frac{1}{4} (1 - \lambda^2) \tensor{\epsilon}{^i_j_k} e^{\hatj} \wedge e^{\hat k},
    \\
    &\tensor{\R}{_0^{\hati}}
    &&= \frac{\lambda^2}{4} e^0 \wedge e^{\hati}
    - \frac{1}{4} (1 - \lambda^2) \tensor{\epsilon}{^i_j_k} e^j \wedge e^{\hat k},
    \\
    &\tensor{\R}{^i^j}
    &&= \Bigl(1- \frac{3}{4} \lambda^2 \Bigr) e^i \wedge e^j
    + \frac{1}{2} (1 - \lambda^2) e^{\hati} \wedge e^{\hatj},
    \\
    &\tensor{\R}{^{\hati}^{\hatj}}
    &&= \frac{1}{2} (1-\lambda^2) \tensor{\epsilon}{^i^j_k} e^0 \wedge e^k
    + \frac{1}{2} (1-\lambda^2) e^i \wedge e^j
    + \frac{1}{4 \lambda^2} e^{\hati} \wedge e^{\hatj},
    \\ \nonumber
    & \tensor{\R}{^i^{\hatj}}
    &&= - \frac{1}{4} (1-\lambda^2) \tensor{\epsilon}{^i^j_k} e^0 \wedge e^{\hat k}
    + \frac{1}{4} (1-\lambda^2) e^{\hati} \wedge e^j
    + \frac{\lambda^2}{4} e^i \wedge e^{\hat j}
    +\\
    & &&\eqspace + \frac{1}{4} (1-\lambda^2) \delta^{ij} \delta_{k \ell} e^k \wedge e^{\hat\ell}.
\end{alignat}
\end{subequations}
From this, it follows that the nonzero components of the Ricci tensor, $R_{ab}=\tensor{R}{_a_c_b^c}$, are
\begin{equation}
\label{eq:squashed_geometry:fibration:Ricci}
    R_{00} = 3 \Bigl(1 - \frac{\lambda^2}{2}\Bigr),
    \qquad
    R_{ij} = 3 \Bigl(1 - \frac{\lambda^2}{2}\Bigr) \delta_{ij},
    \qquad
    R_{\hati \hatj} = \Bigl(\lambda^2 + \frac{1}{2\lambda^2}\Bigr) \delta_{i j}.
\end{equation}
Finally, the curvature scalar is
\begin{equation}
    R = \frac{3}{2} \Bigl(8 - 2\lambda^2 + \frac{1}{\lambda^2} \Bigr).
\end{equation}
Note that, for sufficiently large $\lambda^2$, the curvature scalar is negative.

The Ricci tensor is diagonal in the basis we have chosen.
In particular, we see that the manifold is Einstein if and only if $\lambda^2 = 1$ or $\lambda^2 = 1/5$.
The $\lambda^2 = 1$ solution corresponds to the ordinary round $S^7$ \cite{ref:Duff--Nilsson--Pope:KK} while $\lambda^2 = 1/5$ corresponds to what we will call the Einstein-squashed or simply \emph{the} squashed seven-sphere.

\section{Coset construction with arbitrary squashing} \label{sec:squashed_geometry:coset}%
As mentioned above, the squashed seven-sphere can be isometrically embedded in $\HH\mathbf{P}^2$.
More precisely, it can be realised as a distance-sphere, that is, as all points at a fixed distance from an origin, in $\HH\mathbf{P}^2$ for squashing parameters in the range $0<\lambda^2\leq 1$ \cite{ref:Duff--Nilsson--Pope:KK}.
This realisation provides insight into the isometry group of the squashed sphere.
In suitable inhomogeneous coordinates on $\HH\mathbf{P}^2$, one finds that left-multiplication by quaternionic unitary $2\times 2$ matrices and right-multiplication by unit quaternions leave both the metric of $\HH\mathbf{P}^2$ and the embedding equation invariant \cite{ref:Duff--Nilsson--Pope:KK}.
Thus, the isometry group of the squashed sphere contains $\Sp(2)\cdot\Sp(1)$ as a subgroup, where $\Sp(n)\simeq\U(n, \HH)$ is the compact real form of $\Sp(2n, \CC)$, isomorphic to the quaternionic unitary group.%
\footnote{Here $\Sp(2)\cdot\Sp(1)=\Sp(2)\times\Sp(1)/\ZZ_2$ where $\ZZ_2$ is the diagonal subgroup of the centre. This comes from the fact that left-multiplication by $\diag(-1,-1)$ and right-multiplication by $-1$ are equivalent.}
Note that, by well-known exceptional isomorphisms $\Sp(2)\simeq \Spin(5)$ and $\Sp(1)\simeq \Spin(3)$.
The group $\Sp(2)\cdot\Sp(1) \subset \SO(8)$ acts transitively and effectively on the squashed sphere, whence the latter is a homogeneous space.

In this section, we use the theory from \cref{coset:geometry} to construct the squashed $S^7$, with arbitrary squashing parameter, as a coset.
As prescribed in \cref{coset:geometry}, we work with spin groups, that is, $\Sp(2)\times\Sp(1) \subset \Spin(8)$, since the manifold is spin, although we will almost exclusively be concerned with the Lie algebras.

If we denote the (lifted) group of isometries by $G = \Sp(2)\times\Sp(1)_C$ and break $\Sp(2)$ to $\Sp(1)_A\times \Sp(1)_B \simeq \Spin(4)$ (corresponding to fixing a $\SO(5)$-vector), the isotropy subgroup of $G$ is $H = \Sp(1)_A \times \Sp(1)_{B+C}$, where $\Sp(1)_{B+C}$ denotes the diagonal subgroup of $\Sp(1)_B\times \Sp(1)_C$.
Again, this is seen from the embedding in $\HH\mathbf{P}^2$ \cite{ref:Duff--Nilsson--Pope:KK}.
Thus, the squashed seven-sphere, with any squashing parameter $0<\lambda^2\leq 1$, is isometric to the coset space
\begin{equation}
    \frac{G}{H}
    = \frac{\Sp(2)\times\Sp(1)_C}{\Sp(1)_A \times \Sp(1)_{B+C}},
\end{equation}
with an appropriate metric.
We will now demonstrate this in detail and find that this is the case even for $\lambda^2 > 1$.
The construction is similar to that of \cite{ref:Bais--Nicolai--van_Nieuwenhuizen} but we generalise it to an arbitrary squashing parameter.

\subsubsection{The metric}
As explained in \cref{coset:geometry}, we need an $\h$-invariant symmetric tensor $g_{ab}$ to construct the metric on the coset ($g_{ab}$ is the metric with flat indices).
By using, for instance, \cite{ref:LieART,ref:McKay--Patera} one finds that the $\so(7)$-representation $\vec{7}^{\odot 2}\simeq \vec{1}\oplus\vec{27}$ contains two $\h$-singlets.%
\footnote{The decomposition of the relevant $\so(7)$-representations can also be found in \cite{ref:Nilsson--Padellaro--Pope}.}
Two is also the number of simple factors in $\g$ whence there are two $\g$-invariants $g^{(1)}_{AB}$ and $g^{(2)}_{AB}$ corresponding to the two quadratic Casimirs of $\g$.
Hence, all $G$-invariant metrics on the coset are of normal homogeneous form.%
\footnote{This is not entirely true in the strict sense of \cref{coset:geometry} since the $\g$-invariant may not be of Euclidean signature. Also, there can be exceptions in degenerate cases.}
To get a metric on the coset, we, therefore, start by finding the invariants $g^{(n)}_{AB}$.
To this end, we compute all commutators and the Cartan--Killing metric of $\g$ and then relate the latter to the invariants via the Casimirs of $\sp(2)$ and $\sp(1)_C$.

To make everything explicit but not lose generality, we work in a faithful representation.
Recall that $\gamma$-matrices of $\so(5)$ can be constructed by joining $\gamma^5$ to the $\gamma$-matrices of $\so(4)$.
We use the tensor product of the spinor representation of $\so(5)\simeq\sp(2)$ and the 2-dimensional spinor representation of $\so(3)\simeq\sp(1)_C$.
Thus, the generators can be written as
\begin{gather}
\nonumber
    T^{(A)}_i
    = - \frac{\i}{2}
    \begin{pmatrix}
        \sigma_i & 0 \\
        0 & 0
    \end{pmatrix}
    \otimes \1_2,
    \quad\
    T^{(B)}_i
    = - \frac{\i}{2}
    \begin{pmatrix}
        0 & 0 \\
        0 & \sigma_i
    \end{pmatrix}
    \otimes \1_2,
    \quad\
    T^{(C)}_i
    = - \frac{\i}{2} \1_4 \otimes \sigma_i
    \\[4pt]
    T^{(Q)}_0
    = \frac{1}{2}
    \begin{pmatrix}
        0 & -\1_2 \\
        \1_2 & 0
    \end{pmatrix}
    \otimes \1_2,
    \qquad
    T^{(Q)}_i
    = - \frac{\i}{2}
    \begin{pmatrix}
        0 & \sigma_i \\
        \sigma_i & 0
    \end{pmatrix}
    \otimes \1_2,
\end{gather}
where the labels $A, B, C$ indicate which $\sp(1)$ algebra the generators belong to, $T^{(Q)}$ are the remaining generators, $\1_n$ is the $n\times n$ unit matrix and $\sigma_i$ are the Pauli matrices.
To see this, note that the $\gamma$-matrices of $\so(4)$ can be obtained from those of $\so(3,1)$ by multiplying $\gamma^0$ by $\i$.
The generators of $\so(3,1)$, which are proportional to $\gamma^{\alpha\beta}$, are block-diagonal in the Weyl-basis of \cref{app:conventions:4d-spinors}.
The blocks in these six generators are $\i\sigma^i$ and by appropriate linear combinations, $T^{(A)}_i$ and $T^{(B)}_i$ can be obtained.%
\footnote{Note that we use the convention in which a group element is $g=\exp(T)$.}
This explicitly demonstrates the well-known exceptional isomorphisms $\so(4)\simeq \su(2)\oplus\su(2)$.
The last generators of $\so(5)$ are proportional to $\gamma^{\alpha 5}$ and are the ones denoted $T^{(Q)}$ above.
The normalisations of the three commuting $\sp(1)$ algebras are such that $[T_i, T_j] = \tensor{\epsilon}{_i_j^k} T_k$.

Apart from the $\sp(1)$ commutation relations, there are nonvanishing Lie brackets between $T^{(Q)}$ and $T^{(A),(B),(Q)}$.
By straightforward computation
\begin{subequations}
\label{eq:squashed_geometry:coset:Q_commutators}
\begin{alignat}{2}
    &\bigl[T^{(A)}_i, T^{(Q)}_0\bigl] = -\frac{1}{2} T^{(Q)}_i,
    \qquad\quad
    &&\bigl[T^{(A)}_i, T^{(Q)}_j\bigr] = \frac{1}{2} \delta_{ij} T^{(Q)}_0 + \frac{1}{2} \tensor{\epsilon}{_i_j^k} T^{(Q)}_k,
    \\
    &\bigl[T^{(B)}_i, T^{(Q)}_0\bigl] = \frac{1}{2} T^{(Q)}_i,
    &&\bigl[T^{(B)}_i, T^{(Q)}_j\bigr] = -\frac{1}{2} \delta_{ij} T^{(Q)}_0 + \frac{1}{2} \tensor{\epsilon}{_i_j^k} T^{(Q)}_k,
    \\
    &\bigl[T^{(Q)}_i, T^{(Q)}_0\bigr] = T^{(A)}_i - T^{(B)}_i,
    \qquad
    &&\bigl[T^{(Q)}_i, T^{(Q)}_j\bigr] = \tensor{\epsilon}{_i_j^k} \bigl(T^{(A)}_k + T^{(B)}_k\bigr).
\end{alignat}
\end{subequations}
From these commutation relations, it is easy to see that the Cartan--Killing metric $\kappa_{AB} = f_{AC}{}^D f_{BD}{}^C$, where $A,B,C,\hdots$ are $\g$-indices, is block diagonal in our basis.
For instance, the $(A),(Q)$-block vanishes since $T^{(A)}$ only has nonvanishing brackets $[T^{(A)}, T^{(A)}] \sim T^{(A)}$ and $[T^{(A)}, T^{(Q)}] \sim T^{(Q)}$ while $[T^{(Q)}, T^{(A)}]$ has no $T^{(A)}$ part and $[T^{(Q)}, T^{(Q)}]$ has no $T^{(Q)}$ part.
A short calculation gives,
\begin{equation}
\label{eq:squashed_geometry:coset:Cartan--Killing}
    \kappa_{AB}
    = \diag(-3\cdot \1_3, -3\cdot \1_3, -2\cdot \1_3, -6\cdot \1_4)_{AB},
\end{equation}
where we have ordered the generators as $(T^{(A)}, T^{(B)}, T^{(C)}, T^{(Q)})$.

We define the quadratic Casimir of $\g$ by
\begin{equation}
    \C_\g = 6 \kappa^{AB} T_A T_B.
\end{equation}
Since $\g = \sp(2) \oplus \sp(1)_C$, $\C_\g$ is a linear combination of $\C_{\sp(2)}$ and $\C_{\sp(1)_C}$.
To find the coefficients, we compute $\ad_\g(\C_\g)$.
From the definition, we immediately find that $\ad_\g(\C_\g)$ is $6\cdot \1_3$ on the $(C)$-block, that is, the $\sp(1)_C$ part.
After a short calculation, we find that $\ad_\g(\C_\g)$ is $6\cdot \1_3$ on the $(A)$-block as well, whence it is $6\cdot\1_{10}$ on the $\sp(2)$ part by Shur's lemma.
With the normalisation of the Casimirs from \cref{app:quadratic_casimirs}, we conclude that
\begin{equation}
    \C_\g = 2\, \C_{\sp(2)} + 3\, \C_{\sp(1)_C},
\end{equation}
since $C_{\sp(2)}(\ad_{\sp(2)}) = 3$ and $C_{\sp(1)}(\ad_{\sp(1)}) = 2$ and $\ad_\g=\ad_{\sp(2)}\oplus\ad_{\sp(1)_C}$.%
\footnote{We use $\C$ to denote the Casimir operators and $C$ for its eigenvalues.}

The Casimirs $\C_{\sp(2)}$ and $\C_{\sp(1)_C}$ can be written as
\begin{equation}
    \C_{\sp(1)_C} = -g_{(1)}^{AB} T_A T_B,
    \qquad\quad
    \C_{\sp(2)} = -g_{(2)}^{AB} T_A T_B,
\end{equation}
where $g_{(1)}^{AB} = \diag(0, 0, \1_3, 0)^{AB}$ and $g_{(2)}^{AB} = \diag(\1_3, \1_3, 0, \1_4/2)^{AB}$ are $\g$-invariant tensors.
From these, we can form the invariants
\begin{equation}
\label{eq:squashed_geometry:coset:g-invariants}
    g^{(1)}_{AB} = \diag(0,\, 0,\, \1_3,\, 0)_{AB},
    \qquad
    g^{(2)}_{AB} = \diag(\1_3,\, \1_3,\, 0,\, 2\cdot \1_4)_{AB}.
\end{equation}
Note that $g^{(n)}_{AB}$ is not the inverse of $g_{(n)}^{AB}$.
Rather, $P^{(n)}{}_{\mkern-8mu A}^{\, C} = g^{(n)}_{AB} g_{(n)}^{BC}$, for $n = 1,2$, are the projection operators onto $\sp(1)_C$ and $\sp(2)$, respectively.
Since there are two simple factors in $\g$, these span the space of $\g$-invariants $g_{AB}$.

We now write the $\g$-invariant that we will use to define the metric on the coset as
\begin{equation}
\label{eq:squashed_geometry:coset:g-metric}
    g_{AB} = \frac{1}{2 \sin(\theta)} \bigl( \cos\theta\, g^{(1)}_{AB} + \sin\theta\, g^{(2)}_{AB} \bigr),
\end{equation}
where $\theta$ is referred to as the squashing angle for reasons that will soon become apparent.
Apart from an overall constant factor, which is irrelevant for the geometry, this is an arbitrary $\g$-invariant except that we have to exclude $\theta = 0$ due to the prefactor.
We need only consider half a revolution for the squashing angle since $\theta\mapsto\theta + \pi$ leaves $g_{AB}$ invariant.
At this point, we thus have two relevant regions $0<\theta<\pi/2$ and $\pi/2 < \theta < \pi$ in which $g_{AB}$ is of signature $(13,0)$, that is, Euclidean, and $(10,3)$, respectively, as well as the midpoint $\theta = \pi/2$ in which $g_{AB}$ is degenerate.
As we will see, this will change when we go over to the metric on the coset.

Following \cref{coset:geometry}, we now wish to split $\g$ into a direct sum $\h\oplus\m$ such that $[\h, \m] \subseteq \m$ and $g_{AB}$ is block-diagonal over the terms.
At the same time, we will switch to a basis in which $g_{ab} = \delta_{ab}$, where $g_{ab}$ is the restriction of $g_{AB}$ to $\m$ (we use $a,b,c\hdots$ for $\m$-indices).
To this end, we write the generators of $\g$ as
\begin{subequations}
\label{eq:squashed_geometry:coset:reductive_basis}
\begin{alignat}{3}
    &\h\colon\qquad
    && T^{(A)}_i,\qquad
    && T^{(B+C)}_i = T^{(B)}_i + T^{(C)}_i,
    \\
    &\m\colon\qquad
    && T^{(Q)}_\alpha,\qquad
    && T^{(T)}_i = f(\theta) T^{(B)}_i - \tan\theta\, f(\theta) T^{(C)}_i,
\end{alignat}
\end{subequations}
where
\begin{equation}
    f(\theta) = \sqrt{\frac{2}{1 + \tan{\theta}}}.
\end{equation}
Note that $T^{(B+C)}_i$ generate $\sp(1)_{B+C}$, the diagonal subalgebra of $\sp(1)_B \oplus \sp(1)_C$.
Here, we need to exclude the region in which $\tan\theta \leq -1$ and $\theta = \pi/2$ where $\tan\theta$ diverges.
In the new basis for $\g$,
\begin{equation}
\label{eq:squashed_geometry:coset:g-metric_ii}
    g_{AB} = \diag(\1_3/2,\, (1+\cot\theta) \1_3/2,\, \1_7)_{AB},
\end{equation}
where the index $A$ is split as $A=(R, a)$ where $R$ is an $\h$-index and $a$ an $\m$-index.
The index $R$ is further split as $R=(r, \dot r)$ corresponding to the terms in $\h = \sp(1)_A \oplus \sp(1)_{B+C}$.
The metric in \cref{eq:squashed_geometry:coset:g-metric_ii} is computed from \cref{eq:squashed_geometry:coset:g-invariants,eq:squashed_geometry:coset:g-metric} by matrix multiplication, $g'_{AB} = M_A{}^C M_B{}^D g_{CD}$, where $M_A{}^B$ is the matrix relating the two bases, $T'_A = M_A{}^{B} T_B$.
Note that the split $\h\oplus\m$ is reductive since $g_{AB}$ is $\g$-invariant which implies that $f_{AB}{}^{D} g_{DC}$ is completely antisymmetric whence $f_{S a}{}^T = 0$ follows from $g_{AB}$ being invertible and $\h$ being a subalgebra of $\g$.

Before we turn to the Riemann and Ricci tensors, some comments about the range of values of $\theta$.
Due to $\tan\theta < -1$ being excluded and $\theta \sim \theta + \pi$, it is convenient to consider $\theta$ in the interval $-\pi/4 < \theta < \pi/2$.
The quotient between the coefficients of $T^{(B)}$ and $T^{(C)}$ in $T^{(T)}$ are determined by requiring $g_{AB}$ to be diagonal in the new basis while the factor $f(\theta)$ and the prefactor in \cref{eq:squashed_geometry:coset:g-metric} ensures $g_{ab} = \delta_{ab}$.
Since there are two $\h$-invariants $g_{ab}^{(1),(2)}$, we conclude that $-\pi/4<\theta<\pi/2$ corresponds to linear combinations of these such that the result has signature $(7,0)$, that is, Euclidean, while $\pi/2<\theta<3\pi/4$ corresponds to signature $(4,3)$.
Similarly, the points $\theta = -\pi/4,\, \pi/2$ corresponds to degenerate linear combinations.
This situation is precisely what one expects from there being two $\h$-invariants $g^{(1),(2)}_{ab}$.
Recall, however, that we also had to exclude $\theta = 0$.
Without the diverging prefactor in \cref{eq:squashed_geometry:coset:g-metric}, $g_{AB}$ would only have rank 3 for $\theta=0$.
Clearly, such a $g_{AB}$ cannot be restricted to a nondegenerate metric on $\h$.
However, it seems like $\theta=0$ will not be problematic when we forget about $g_{AB}$ and only consider the coset since the divergences sit in $g_{RS}$, not $g_{ab}$, and there is no singularity in the basis in \cref{eq:squashed_geometry:coset:reductive_basis} at $\theta = 0$.
This agrees with the above remark that the $\h$-metric is degenerate for $\theta=-\pi/4,\,\pi/2$, not $\theta=0$.
Note that the coset is Euclidean even when $g_{AB}$ is non-Euclidean (but nondegenerate) as long as $T^{(B+C)}_i$ are time-like.
The problematic $\theta=-\pi/4$ corresponds to light-like $T^{(B+C)}_i$.

\subsubsection{The Riemann and Ricci tensors}
To compute the Riemann tensor, we use \cref{eq:coset:geometry:Riem_normal}.
Thus, we first need to compute the structure constants in the basis in \cref{eq:squashed_geometry:coset:reductive_basis}.
We use the index split $A=(R, a)=(r,\dot r, a)$ described above and further split $a$ as $a=(\hati, 0, i)$, where $T_{\hati} = T^{(T)}_{\hati}$, $T_0 = T^{(Q)}_0$ and $T_i = T^{(Q)}_i$.

Since $T^{(C)}_i$ commute with everything else, all commutators are easily computed using the $\sp(1)$ commutation relations and \cref{eq:squashed_geometry:coset:Q_commutators}.
We find that the nonvanishing components are%
\begingroup
\allowdisplaybreaks
\begin{subequations}
\label{eq:squashed_geometry:coset:structure_constants}
\begin{gather}
    \begin{alignedat}{2}
        &\tensor{f}{_r_s^t}
        = \tensor{\epsilon}{_r_s^t},
        \qquad\qquad
        &&\tensor{f}{_{\dot r}_{\dot s}^{\dot t}}
        = \tensor{\epsilon}{_{\dot r}_{\dot s}^{\dot t}},
    \end{alignedat}
    \\[4pt]
    \begin{alignedat}{3}
        &\tensor{f}{_r_0^k}
        = - \frac{1}{2} \delta_r^k,
        \qquad\quad
        &&\tensor{f}{_r_j^0}
        = \frac{1}{2} \delta_{rj},
        \qquad\quad
        &&\tensor{f}{_r_j^k}
        = \frac{1}{2} \tensor{\epsilon}{_r_j^k},
        \\
        &\tensor{f}{_{\dot r}_0^k}
        = \frac{1}{2} \delta_{\dot r}^k,
        \qquad\quad
        &&\tensor{f}{_{\dot r}_j^0}
        = - \frac{1}{2} \delta_{\dot r j},
        \qquad\quad
        &&\tensor{f}{_{\dot r}_j^k}
        = \frac{1}{2} \tensor{\epsilon}{_{\dot r}_j^k},
        \\
        & &&\tensor{f}{_{\dot r}_{\hatj}^{\hat k}}
        = \tensor{\epsilon}{_{\dot r}_{\hatj}^{\hat k}},
        &&
    \end{alignedat}
    \\[4pt]
    \begin{alignedat}{3}
        &\tensor{f}{_i_0^t}
        = \delta_i^t,
        \qquad
        &&\tensor{f}{_i_0^{\dot t}}
        = - \frac{1}{1+\cot\theta} \delta_i^{\dot t},
        \qquad
        &&\tensor{f}{_i_0^{\hat k}}
        = - \frac{f(\theta)}{2} \delta_i^{\hat k},
        \\
        &\tensor{f}{_i_j^t}
        = \tensor{\epsilon}{_i_j^t},
        \qquad
        &&\tensor{f}{_i_j^{\dot t}}
        = \frac{1}{1+\cot\theta} \tensor{\epsilon}{_i_j^{\dot t}},
        \qquad
        &&\tensor{f}{_i_j^{\hat k}}
        = \frac{f(\theta)}{2} \tensor{\epsilon}{_i_j^{\hat k}},
        \\
        &\tensor{f}{_{\hati}_0^k}
        = \frac{f(\theta)}{2} \delta_{\hati}^k,
        \qquad
        &&\tensor{f}{_{\hati}_{j}^0}
        = - \frac{f(\theta)}{2} \delta_{\hati j},
        \qquad
        &&\tensor{f}{_{\hati}_j^{k}}
        = \frac{f(\theta)}{2} \tensor{\epsilon}{_{\hati j}^k},
        \\
        &
        &&\tensor{f}{_{\hati \hatj}^{\dot t}}
        = \frac{2}{1+\cot\theta} \tensor{\epsilon}{_{\hati \hatj}^{\dot t}},
        \qquad
        &&\tensor{f}{_{\hati \hatj}^{\hat k}}
        = f(\theta)(1-\tan\theta) \tensor{\epsilon}{_{\hati \hatj}^{\hat k}},
    \end{alignedat}
\end{gather}
\end{subequations}
\endgroup
where we have grouped the components based on whether they come from $[\h, \h]$, $[\h, \m]$ or $[\m, \m]$, in that order.
Here, some of the $\theta$-dependence comes from
\begin{equation}
    T^{(B)}_i
    = \frac{\tan\theta}{1+\tan\theta} T^{(B+C)}_i
    + \frac{f(\theta)^{-1}}{1+\tan\theta} T^{(T)}_i,
    \quad\!
    T^{(C)}_i = \frac{1}{1+\tan\theta} T^{(B+C)}_i - \frac{f(\theta)^{-1}}{1+\tan\theta} T^{(T)}_i.
\end{equation}

Computing the Riemann tensor using \cref{eq:coset:geometry:Riem_normal}, we find
\begin{subequations}
\label{eq:squashed_geometry:coset:Riem}
\begin{alignat}{2}
    &\tensor{\R}{_0^i}
    &&= \frac{5+8\tan\theta}{8(1+\tan\theta)} e^0 \wedge e^i
    + \frac{1+2\tan\theta}{8(1+ \tan\theta)} \tensor{\epsilon}{^i_{jk}} e^{\hatj} \wedge e^{\hat k},
    \\
    &\tensor{\R}{_0^{\hati}}
    &&= \frac{1}{8(1+\tan\theta)} e^0 \wedge e^{\hati}
    - \frac{1+2\tan\theta}{8(1+\tan\theta)} \tensor{\epsilon}{^{i}_{j k}} e^{\hatj} \wedge e^k,
    \\
    &\tensor{\R}{^{ij}}
    &&= \frac{5+8\tan\theta}{8(1+\tan\theta)} e^i \wedge e^j
    + \frac{1+2\tan\theta}{4(1+\tan\theta)} e^{\hati} \wedge e^{\hatj},
    \\
    &\tensor{\R}{^{\hati \hatj}}
    &&= \frac{1+2\tan\theta}{4(1+\tan\theta)} \tensor{\epsilon}{^i^j_k} e^0 \wedge e^k
    + \frac{1+2\tan\theta}{4(1+\tan\theta)} e^i \wedge e^j
    + \frac{1+\tan\theta}{2} e^{\hati} \wedge e^{\hatj},
    \\\nonumber
    &\tensor{\R}{^i^{\hatj}}
    &&= - \frac{1+2\tan\theta}{8(1+\tan\theta)} \tensor{\epsilon}{^i^j_k} e^0 \wedge e^{\hat k}
    + \frac{1+2\tan\theta}{8(1+\tan\theta)} e^{\hati} \wedge e^j
    + \frac{1}{8(1+\tan\theta)} e^i \wedge e^{\hatj}
    +{}\\
    & &&\eqspace
    + \frac{1+2\tan\theta}{8(1+\tan\theta)} \delta^{ij} \delta_{k \hat\ell} e^k \wedge e^{\hat\ell},
\end{alignat}
\end{subequations}
where we, as in \cref{sec:squashed_geometry:fibration}, have dropped hats on indices on the $\su(2)$-invariants and it is understood that, for instance, $i$ and $\hati$ take the same value when they appear in the same equation.
This result agrees with \cref{eq:squashed_geometry:fibration:Riem} after identifying the relation between the squashing parameter $\lambda$ and the squashing angle $\theta$ as
\begin{equation}
    \lambda^2 = \frac{1}{2(1+\tan\theta)}.
\end{equation}
The squashing parameter $\lambda^2$ takes all values in the interval $(0, \infty)$ and decreases monotonically for $\theta\in (-\pi/4, \pi/2)$.
The values $\lambda^2 = 1,\, 1/5$ for which the coset manifold is Einstein, correspond to $\tan\theta = -1/2$ and $\tan\theta = 3/2$, respectively.
Hence, the round metric is not normal homogeneous in the strict sense since $g_{AB}$ has indefinite signature for $\tan\theta = -1/2$.
The Einstein-squashed sphere, on the other hand, is not only normal homogeneous but standard homogeneous, that is, $g_{AB} \propto \kappa_{AB}$ for $\tan\theta=3/2$.
More specifically, $g_{AB} = - \kappa_{AB}/6$ in the Einstein-squashed case.
Note that the two Einstein metrics are separated by $\theta = 0$, corresponding to the only $G$-invariant metric on the coset that cannot be obtained from an invariant $g_{AB}$.%
\todo[disable]{}
Lastly, note that, for the Einstein-squashed $S^7$,
\begin{equation}
    \tensor{f}{_a_b^c}
    = - \frac{1}{\sqrt{5}} \tensor{a}{_a_b^c},
\end{equation}
where $a_{abc}$ are the octonion structure constants from \cref{app:octonions}.
That $f_{abc}$ is proportional to $a_{abc}$ only happens for the Einstein-squashed sphere since it depends crucially on $(1-\tan\theta)=-1/2$.

\chapter{Eigenvalue spectra of the squashed seven-sphere} \label{chap:squashed_spectrum}%
In this chapter, we derive the eigenvalue spectra of the universal Laplacian of the squashed seven-sphere.
We will only consider the Einstein-squashed sphere and, henceforth, refer to it simply as the squashed $S^7$.
We will consider eigenmodes of the Laplacian, $\Delta$, from \cref{sec:ads_susy:Laplacian} and use \cref{eq:coset:master_eq:done} to determine the possible eigenvalues.
As mentioned in \cref{sec:outline}, the spectra of all operators we consider except $\i\slashed\D_{3/2}$ are already known \cite{ref:Nilsson--Pope,ref:Yamagishi,ref:Duff--Nilsson--Pope:KK,ref:Ekhammar--Nilsson}.

On a high level, the derivation goes as follows.
First, assume that we have an eigenmode of $\Delta$ with some eigenvalue $\kappa^2$.
Since, as explained in \cref{sec:coset:master_eq}, the eigenmodes of $\Delta$ fall into irreducible representations of $G=\Sp(2)\times\Sp(1)_C$, $\C_\g$ can be replaced by its corresponding eigenvalue $C_\g$ on the relevant representation.
We will not investigate which irreducible $G$-representations occur in the $G$-representation induced by the relevant irreducible $H$-representations.
It has, however, been done using Young tableaux techniques \cite{ref:Nilsson--Padellaro--Pope}.
Replacing $\Delta$ by $\kappa^2$ and $\C_\g$ by $C_\g$, we are left with a linear map that acts on the spin-index of the mode in the right-hand side of \cref{eq:coset:master_eq:done}.
In the left-hand side, however, we have a first-order differential operator.
To get rid of this, we will use various techniques, such as squaring it, acting with projection operators and combinations thereof.
This will, eventually, result in a polynomial equation for $\kappa^2$.

In general, there can be false roots, that is, solutions to the polynomial equation that are not actual eigenvalues of $\Delta$.
We will not deal with this in the current chapter.
Note, however, that, as long as we do not introduce any assumptions in the derivation, as we will not, $\kappa^2$ being an eigenvalue of $\Delta$ \emph{implies} that it is a root of the polynomial.
Thus, although we can get false roots, we cannot miss any eigenvalues.

In \cref{chap:squashed_geometry}, we used a dimensionless unit system such that, for Einstein-squashing, \cref{eq:squashed_geometry:fibration:Ricci}
\begin{equation}
    R_{ab} = \frac{27}{10} \delta_{ab},
    \qquad\quad
    R = \frac{189}{10}.
\end{equation}
To connect this to \cref{chap:sugra_comp}, in which the internal manifold has $R_{ab} = 6m^2 \delta_{ab}$, we see that the dimensionless system results from setting
\begin{equation}
\label{eq:squashed_spectrum:m}
    m^2 = \frac{9}{20}.
\end{equation}
Here, we will mostly continue to use the dimensionless system for convenience.
Note, however, that there is a sign choice in \cref{eq:squashed_spectrum:m} which is relevant for skew-whiffing and the number of Killing spinors.
In this chapter, we only concern ourselves with operators on the squashed seven-sphere.
The sign choice will become important in \cref{chap:masses_and_susy}, where we investigate the number of unbroken supersymmetries and masses in the compactifications of eleven-dimensional supergravity on the squashed seven-sphere.

We begin by concretising some details from \cref{sec:coset:master_eq} for the case of interest.
Note that the $\g=\sp(2)\oplus\sp(1)_C$ invariant is $g_{AB} = -\kappa_{AB}/6$ for the Einstein-squashed $S^7$.
The metric with flat indices is $\delta_{ab}$ in the basis from \cref{sec:squashed_geometry:coset}.
Thus, comparing the normalisations of the quadratic Casimirs $\C_\g$ and $\C_\h$ in \cref{sec:coset:master_eq,app:quadratic_casimirs} gives the quadratic master equation%
\footnote{Referred to as the squared coset master equation in, for instance, \cite{ref:Ekhammar--Nilsson}.}
\cref{eq:coset:master_eq:quadratic}
\begin{equation}
\label{eq:squashed_spectrum:quadratic_master}
    -\DDg Z = (\C_\g - \C_\h) Z.
\end{equation}
Next, the relevant structure constants are, as remarked in \cref{sec:squashed_geometry:coset},
\begin{equation}
    f_{abc} = - \frac{1}{\sqrt{5}} a_{abc},
\end{equation}
where $a_{abc}$ are the octonion structure constants from \cref{app:octonions}.
Hence, the covariant $H$-derivative is \cref{eq:coset:master_eq:H-derivative_spin}
\begin{equation}
\label{eq:squashed_spectrum:Dg}
    \Dg_a = \D_a - \frac{1}{2\sqrt{5}} a_{abc} \Sigma^{bc}.
\end{equation}
As explained in \cref{app:octonions}, the largest group that leave $a_{abc}$ and $\delta_{ab}$ invariant is $G_2 \eqqcolon \tilde H \supset H$.
Therefore, $G_2$ will play an important role in the derivation of the spectrum.
In \cref{tab:squashed_spectrum:SO7_to_G2}, the decompositions of the relevant $\Spin(7)$-representations restricted to $G_2$ are given.
These can be found by using \cite{ref:LieART} or \cite{ref:McKay--Patera} or by looking in \cite{ref:Nilsson--Padellaro--Pope}.
Note that, since $H$ is a subgroup of $G_2$,
\begin{equation}
    \Dg_a a_{bcd} = 0,
    \qquad\quad
    \Dg_a c_{bcde} = 0,
\end{equation}
where $c = \hodge a$, see \cref{app:octonions}.
For this reason, it will be convenient to work with $\Dg_a$ instead of $\D_a$.
\begin{table}[H]
    \everymath{\displaystyle}
    \centering
    \caption[Decomposition of $\Spin(7)$-representations when restricted to $G_2$.]{Decomposition of irreducible $\Spin(7)$-representations when restricted to $G_2$. $\vec{1}$, $\vec{7}$, $\vec{21}$ and $\vec{35}$ are $p$-forms for $p=0,\, 1,\, 2,\, 3$; $\vec{27}$ is traceless symmetric rank-2 tensors; $\vec{8}$ spinors and $\vec{48}$ vector-spinors. Each irreducible representation is specified both using its dimension (in bold) and its Dynkin labels.}
\label{tab:squashed_spectrum:SO7_to_G2}
    \begin{tabular}{l@{$\hphantom{\ \to \ }$}l}
        \toprule
        $\Spin(7)$ irrep.$\mathrlap{\ \to}$& $G_2$ rep.
        \\ \midrule
        $\vec{1}=(0,0,0)$ & $\vec{1}=(0,0)$
        \\
        $\vec{7}=(1,0,0)$ & $\vec{7}=(1,0)$
        \\
        $\vec{8}=(0,0,1)$ & $\vec{1}\oplus\vec{7}=(0,0)\oplus(1,0)$
        \\
        $\vec{21}=(0,1,0)$ & $\vec{7}\oplus\vec{14}=(1,0)\oplus(0,1)$
        \\
        $\vec{27}=(2,0,0)$ & $\vec{27}=(2,0)$
        \\
        $\vec{35}=(0,0,2)$ & $\vec{1}\oplus\vec{7}\oplus\vec{27}=(0,0)\oplus(1,0)\oplus(2,0)$
        \\
        $\vec{48}=(1,0,1)$ & $\vec{7}\oplus\vec{14}\oplus\vec{27}=(1,0)\oplus(0,1)\oplus(2,0)$
        \\ \addlinespace[\aboverulesep] \bottomrule
    \end{tabular}
\end{table}
In the coming sections, the projection operators onto the various irreducible $G_2$-representations in the $\Spin(7)$-representations will appear.
We denote these by $P_{n}$, where $n$ is the dimension of the $G_2$-representation and the $\Spin(7)$-representation is understood from the context or index structure.
Of immediate interest are the projection operators $P_7$ and $P_{14}$ from $\vec{21}=\vec{7}^{\wedge 2}$, the adjoint representation of $\Spin(7)$, to $\vec{7}$ and $\vec{14}$, respectively.
Since $a_{abc}$ is a $G_2$-invariant, it can be viewed as an intertwiner between $\vec{21}$ and $\vec{7}$.
Thus, the projection operator $P_7\from\vec{21}\to\vec{21}$ that projects onto the $\vec{7}\subset\vec{21}$ is proportional to $a_{a_1 a_2 c} a^{c b_1 b_2}$.
Working out the normalisation through $P_7{}^2 = P_7$ gives%
\footnote{In this chapter, we make heavy use of the octonion structure constant identities from \cref{app:octonions:structure_constant_identities}. We will, for the most part, not give references to these equations when using them.}
\begin{equation}
\label{eq:squashed_spectrum:P7}
    (P_7)\indices{_{a_1 a_2}^{b_1 b_2}}
    = \frac{1}{6} \tensor{a}{_{a_1 a_2}^c} \tensor{a}{_c^{b_1 b_2}}
    = \frac{1}{3} \delta_{a_1 a_2}^{b_1 b_2} + \frac{1}{6} \tensor{c}{_{a_1 a_2}^{b_1 b_2}}.
\end{equation}
Since $P_7 + P_{14} = \1_{\vec{21}}$, it immediately follows that
\begin{equation}
\label{eq:squashed_spectrum:P14}
    (P_{14})\indices{_{a_1 a_2}^{b_1 b_2}}
    = \frac{2}{3} \delta_{a_1 a_2}^{b_1 b_2} - \frac{1}{6} \tensor{c}{_{a_1 a_2}^{b_1 b_2}}.
\end{equation}

The Weyl tensor of the squashed sphere is, by \cref{eq:Freund--Rubin:Weyl_tensor,eq:coset:geometry:isotropy_embedding,eq:coset:geometry:Riem_normal}
\begin{align}
\nonumber
    \tensor{W}{_a_b^c^d}
    &= \tensor{R}{_a_b^c^d} - \frac{9}{10} \delta_{ab}^{cd}
    = (T^i)_{ab} (T_i)^{cd}
    + \frac{1}{10} \tensor{a}{_a_b^e} \tensor{a}{_e^c^d}
    + \frac{1}{10} \tensor{a}{_{[a}^{ce}} \tensor{a}{_{b] e}^d}
    - \frac{9}{10} \delta_{ab}^{cd}
    =\\
    &= (T^i)_{ab} (T_i)^{cd} - \frac{6}{5} (P_{14})\indices{_a_b^c^d}.
\end{align}
This implies that, using the Casimirs from \cref{app:quadratic_casimirs},
\begin{equation}
\label{eq:squashed_spectrum:Weyl_operator}
    W_{abcd} \Sigma^{ab} \Sigma^{cd}
    = \frac{6}{5}\, \C_{\g_2} - \C_\h.
\end{equation}
Since $H\subset G_2$, $a_{abc} (T_i)^{bc} = 0$.
This is seen by noting that $a_{abc}$ can be interpreted as an intertwiner from $\vec{21}$ to $\vec{7}$ while $(T_i)^{bc}$ can be interpreted as an intertwiner from $\vec{21}$ to $\ad_\h \subset \ad_{\g_2} = \vec{14}$.
Similarly, $a^{abc} (P_{14})_{bc}{}^{de} = 0$, whence also $a^{abc} W_{bc}{}^{de} = 0$.

The Ricci identity \cref{eq:coset:master_eq:Ricci_id} reads
\begin{equation}
\label{eq:squashed_spectrum:Ricci}
    [\Dg_a, \Dg_b]
    = (T^i)_{ab} T_i - \frac{1}{\sqrt{5}} \tensor{a}{_a_b^c} \Dg_c
    = \Bigl(\tensor{W}{_a_b^c^d} + \frac{6}{5} (P_{14})\indices{_a_b^c^d} \Bigr) \Sigma_{cd}
    - \frac{1}{\sqrt{5}} \tensor{a}{_a_b^c} \Dg_c.
\end{equation}
An important special case of this, which follows immediately from the above remark, is
\begin{equation}
\label{eq:squashed_spectrum:Ricci_special}
    \tensor{a}{_a^b^c}\, \Dg_b \Dg_c
    = - \frac{3}{\sqrt{5}} \Dg_a.
\end{equation}

Lastly, since $a_{a[b}{}^d a^a{}_{c]}{}^e = -3 \delta_{bc}^{de} + 6 (P_{14})_{bc}{}^{de}$, \cref{eq:coset:master_eq:done} becomes
\begin{equation}
\label{eq:squashed_spectrum:rewritten_master}
    - \frac{1}{\sqrt{5}} a_{abc} \Sigma^{ab} \Dg^c
    = \Delta - \C_\g - \frac{6}{5}\, \C_{\so(7)} + \frac{3}{2}\, \C_{\g_2}.
\end{equation}
This is the equation we will use to derive the operator spectrum of the squashed seven-sphere.

\section{0-forms} \label{sec:squashed_spectrum:0-form}%
The scalars, or 0-forms, are trivial since all Casimirs vanish and $\rho_{\vec{1}}(\Sigma^{ab}) = 0$.
For a 0-form satisfying $\Delta Y = \kappa^2 Y$ and $\C_\g Y = C_\g Y$, \cref{eq:squashed_spectrum:rewritten_master} immediately gives
\begin{equation}
    \kappa^2 = C_\g.
\end{equation}
The $G$-representation induced by the scalar $H$-representation, that is, the $G$\hyp{}representation carried by scalar fields on the squashed $S^7$, contains precisely one copy of each irreducible $G$-representation $(p,q;r)$ with $p=r$ \cite{ref:Nilsson--Padellaro--Pope}.
Thus, we know exactly which values of $C_\g$ are possible.
In this case, there are no false roots.

\section{1-forms} \label{sec:squashed_spectrum:1-form}
We now turn to transverse 1-forms, $Y_a$.
Transversality means that $\D^a Y_a = 0$ which is equivalent to $\Dg^a Y_a = 0$.
Since $C_{\so(7)}(\vec{7}) = 3$ and $C_{\g_2}(\vec{7}) = 2$, a 1-form eigenmode of $\Delta$ satisfies, by \cref{eq:squashed_spectrum:rewritten_master},
\begin{equation}
\label{eq:squashed_spectrum:1-form:Dop}
    \Dop_1 Y_a
    \coloneqq \tensor{a}{_a^b^c} \Dg_c Y_b
    = - \sqrt{5} \Bigl(\kappa^2 - C_\g - \frac{3}{5} \Bigr) Y_a.
\end{equation}
Squaring the operator $\Dop_1$ gives
\begin{align}
\nonumber
    \Dop_1{}^2 Y_a
    &= \tensor{a}{_a^{b_1 b_2}} \Dg_{b_2} \bigl( \tensor{a}{_{b_1}^{c_1 c_2}} \Dg_{c_2} Y_{c_1} \bigr)
    = - \DDg Y_a + \Dg^b \Dg_a Y_b + \tensor{c}{_a^{b_1 b_2 b_3}} \Dg_{b_1} \Dg_{b_2} Y_{b_3}
    =\\
\label{eq:squashed_spectrum:1-form:Dop^2}
    &= \Bigl( C_\g - \C_\h + \frac{24}{5} - \frac{1}{\sqrt{5}} \Dop_1 \Bigr) Y_a
    + \Dg_a \Dg^b Y_b,
\end{align}
where we have used that the Ricci identity \cref{eq:squashed_spectrum:Ricci} for 1-forms,
\begin{equation}
    [\Dg_a, \Dg_b] Y_c
    = \Bigl(\tensor{W}{_a_b_c^d} + \frac{6}{5} \tensor{(P_{14})}{_a_b_c^d}\Bigr) Y_d
    - \frac{1}{\sqrt{5}} a_{abd} \Dg^d Y_c,
\end{equation}
implies
\begin{subequations}
\label{eq:squashed_spectrum:1-form:asym_G2-derivatives}
\begin{align}
    &[\Dg^b, \Dg_a] Y_b
    = \frac{12}{5} Y_a
    + \frac{1}{\sqrt{5}} \Dop_1 Y_a,\\
    &\tensor{c}{_a^{b_1 b_2 b_3}} \Dg_{b_1} \Dg_{b_2} Y_{b_3}
    = \frac{12}{5} Y_a
    - \frac{2}{\sqrt{5}} \Dop_1 Y_a,
\end{align}
\end{subequations}
since the Weyl tensor is traceless and $W_{[abc]d} = 0$.
Note that $\C_\h$ re-entered the calculation.
However, \cref{eq:squashed_spectrum:Weyl_operator} applied to 1-forms shows that
\begin{equation}
    \tensor{W}{_a^b_{\, b}^c} Y_c = \Bigl(\frac{12}{5} - \C_\h\Bigr) Y_a.
\end{equation}
Since the Weyl tensor is traceless, $\C_\h Y_a = 12/5\, Y_a$.
Using this, \cref{eq:squashed_spectrum:1-form:Dop}, $\Dg^a Y_a = 0$ and $Y_a \neq 0$, \cref{eq:squashed_spectrum:1-form:Dop^2} gives
\begin{equation}
    C_\g + \frac{12}{5} - \Bigl(C_\g - \kappa^2 + \frac{3}{5}\Bigr) = 5 \Bigl(C_\g - \kappa^2 + \frac{3}{5}\Bigr)^2,
\end{equation}
with solutions
\begin{equation}
    \kappa^2 = C_\g + \frac{7}{10} \pm \frac{1}{\sqrt{5}} \sqrt{C_\g + \frac{49}{20}}.
\end{equation}
Note that, from this calculation, we cannot determine whether both solutions occur as eigenvalues of $\Delta_1$.

\section{2-forms} \label{sec:squashed_spectrum:2-form}%
Let $Y_{ab}$ be a transverse eigenmode of the Hodge--de Rham operator, satisfying $\Delta Y_{ab} = \kappa^2 Y_{ab}$, $\C_\g Y_{ab} = C_\g Y_{ab}$ and $\D^a Y_{ab} = 0$.
Since $C_{\so(7)}(\vec{21}) = 5$, $C_{\g_2}(\vec{7}) = 2$ and $C_{\g_2}(\vec{14})=4$, the rewritten quadratic master equation \cref{eq:squashed_spectrum:rewritten_master} becomes%
\footnote{We put brackets around the 2 in $\Dop_{[2]}$ to indicate the antisymmetrisation. $\Dop_{(2)}$ will be defined similarly but with symmetrisation.}
\begin{equation}
\label{eq:squashed_spectrum:2-form:master}
    \Dop_{[2]} Y_{a_1 a_2}
    \coloneqq \tensor{a}{_{[a_1|}^{bc}} \Dg_c Y_{b|a_2]}
    = \frac{2}{\sqrt{5}} \bigl(C_\g - \kappa^2 + 3 P_7 \bigr) Y_{a_1 a_2}.
\end{equation}
The transversality condition $\D^a Y_{ab} = 0$ can be written as
\begin{equation}
    \Dg^b Y_{ba} = \frac{1}{2\sqrt{5}} \tensor{a}{_a^{bc}} Y_{bc}.
\end{equation}

Define another differential operator $\tilde\Dop_{[2]}$ by
\begin{equation}
    \tilde\Dop_{[2]} Y_{a_1 a_2}
    \coloneqq \tensor{a}{_{[a_1}^{b_1 b_2}} \Dg_{a_2]} Y_{b_1 b_2}.
\end{equation}
By using the definitions, the transversality condition and the projection operator in \cref{eq:squashed_spectrum:P7}, we find
\begin{equation}
    \tensor{c}{_{a_1 a_2}^{b_1 b_2}} \Dop_{[2]} Y_{b_1 b_2}
    = \frac{3}{\sqrt{5}} P_7 Y_{a_1 a_2}
    + 2 \Dop_{[2]} Y_{a_1 a_2}
    - 2 \tilde \Dop_{[2]} Y_{a_1 a_2}.
\end{equation}
Taking the appropriate linear combination with $\Dop_{[2]} Y_{a_1 a_2}$ to get $P_7 \Dop_{[2]} Y_{a_1 a_2}$ in the left-hand side and rearranging gives
\begin{equation}
\label{eq:squashed_spectrum:2-form:Dop_relations}
    \tilde\Dop_{[2]} Y
    = 2 \Dop_{[2]} Y - 3 P_7 \Dop_{[2]} Y + \frac{3}{2\sqrt{5}} P_7 Y.
\end{equation}
Note that, if $Y$ is an eigenmode of $\Delta$ with vanishing $\vec{7}$-part, $\Dop_{[2]} Y = 0$ since the other terms in this equation trivially vanish.
This depends on $Y$ being transverse since we used that in the derivation.
We will \emph{not} assume that $Y$ has vanishing $\vec{7}$ part.
However, this remark will prove useful later.

Using \cref{eq:squashed_spectrum:2-form:master} to write the right-hand side of \cref{eq:squashed_spectrum:2-form:Dop_relations} in terms of $Y$ and $P_7 Y$ gives
\begin{equation}
\label{eq:squashed_spectrum:2-form:tilde_Dop_Y}
    \tilde\Dop_{[2]} Y
    = \frac{\sqrt{5}}{2} \Bigl[2 \bigl(C_\g - \kappa^2 \bigr) - 3\Bigl(C_\g - \kappa^2 + \frac{4}{5}\Bigr) P_7 \Bigr] Y.
\end{equation}
Define $Y_a = a_a{}^{bc} Y_{bc}$.
We then immediately see that $a_a{}^{b_1 b_2} \tilde\Dop_{[2]} Y_{b_1 b_2} = \Dop_1 Y_a$, with $\Dop_1$ as defined in \cref{eq:squashed_spectrum:1-form:Dop}.
\Cref{eq:squashed_spectrum:2-form:tilde_Dop_Y} thus implies
\begin{equation}
\label{eq:squashed_spectrum:2-form:Dop1}
    \Dop_1 Y_a = - \frac{\sqrt{5}}{2} \Bigl(C_\g - \kappa^2 + \frac{12}{5}\Bigr) Y_a.
\end{equation}
The situation is now very similar to that in \cref{sec:squashed_spectrum:1-form}.
The only differences are that $Y_a$ might not be transverse and can be 0.
If $Y_a = 0$, then $\tilde\Dop_{[2]} Y_{ab} = \Dg_b Y_a = 0$ and, by \cref{eq:squashed_spectrum:2-form:tilde_Dop_Y}, $\kappa^2 = C_\g$, since $Y_{ab} \neq 0$.
Going forward, we hence assume $Y_a \neq 0$.

To handle that $Y_a$ might not be transverse, we contract \cref{eq:squashed_spectrum:2-form:Dop1} with $\Dg^a$ and use \cref{eq:squashed_spectrum:Ricci_special} to find
\begin{equation}
    \Bigl(C_\g - \kappa^2 + \frac{18}{5}\Bigr) \Dg^a Y_a = 0.
\end{equation}
Hence, either $\kappa^2 = C_\g + 18/5$ or $\Dg^a Y_a = 0$.
If $\Dg^a Y_a = 0$, the calculation in \cref{sec:squashed_spectrum:1-form} can be reused and gives \cref{eq:squashed_spectrum:2-form:kappa^2:c} below.
Thus, the possibilities are
\begin{subequations}
\begin{align}
    &\kappa^2 = C_\g,\\
    &\kappa^2 = C_\g + \frac{18}{5},\\
\label{eq:squashed_spectrum:2-form:kappa^2:c}
    &\kappa^2 = C_\g + \frac{11}{5} \pm \frac{2}{\sqrt{5}} \sqrt{C_\g + \frac{49}{20}},
\end{align}
\end{subequations}
where the first one applies to modes with $P_7 Y = 0$, the second applies to modes with $\Dg^a Y_a \neq 0$ and the third possibility applies to modes with $Y_a \neq 0$ but $\Dg^a Y_a = 0$.
As in \cref{sec:squashed_spectrum:1-form}, we cannot, at this point, say whether all of these occur as eigenvalues of $\Delta_2$.

\section{Symmetric rank-2 tensors} \label{sec:squashed_spectrum:2-rank_sym}%
Now we turn to the eigenvalues of the Lichnerowicz operator.
Let $X_{a_1 a_2}$ be a transverse symmetric traceless rank-2 eigenmode of $\Delta$, satisfying $\Delta X = \kappa^2 X$ and $\D^a X_{ab} = 0$.
Since $C_{\so(7)}(\vec{27}) = 7$ and $C_{\g_2}(\vec{27}) = 14/3$, \cref{eq:squashed_spectrum:rewritten_master} becomes
\begin{equation}
\label{eq:squashed_spectrum:2-rank_sym:rewritten_master}
    \Dop_{(2)} X_{a_1 a_2}
    \coloneqq \tensor{a}{_{(a_1|}^{b_1 b_2}} \Dg_{b_2} X_{b_1 |a_2)}
    = \frac{\sqrt{5}}{2} \Bigl(C_\g - \kappa^2 + \frac{7}{5}\Bigr) X_{a_1 a_2}
    \eqqcolon k X_{a_1 a_2}.
\end{equation}
The transversality condition in terms of $\Dg_a$ reads
\begin{equation}
    \Dg^a X_{ab} = 0.
\end{equation}
Now define, for any rank-2 tensor $Z_{a_1 a_2}$,%
\footnote{We put the free indices outside parentheses around $\Dop_2 Z$ to indicate that $(\Dop_2 Z)_{ab}$ might not have the same symmetries as $Z_{ab}$. Note that, as usual, we drop the parentheses and write, for instance, $\D_a Y_b$ instead of $(\D_a Y)_b$ when there is no risk of confusion.}
\begin{equation}
    (\Dop_2 Z)_{a_1 a_2}
    \coloneqq \tensor{a}{_{a_1}^{b_1 b_2}} \Dg_{b_2} Z_{b_1 a_2}.
\end{equation}
An arbitrary rank-2 tensor consists of a trace, $Z=Z_a{}^a$, a traceless symmetric part, $Z_{(ab)} - \delta_{ab} Z/7$, and an antisymmetric part, $Z_{[ab]}$.
The corresponding $\Spin(7)$-representations are $\vec{1}$, $\vec{27}$ and $\vec{21}$, respectively, where the last one splits into $\vec{7}\oplus\vec{14}$ when restricted to $G_2$.

Consider $(\Dop_2 X)_{ab}$.
The scalar part of this, that is, the trace, vanishes since $X_{ab}$ is symmetric and $a_{abc}$ antisymmetric.
By contracting $(\Dop_2 X)_{b_1 b_2}$ with $a^{a b_1 b_2}$, one sees immediately that the $\vec{7}$ part vanishes as well due to $X$ being traceless, symmetric and transverse.
Define
\begin{equation}
\label{eq:squashed_spectrum:2-rank_sym:Y_def}
    Y_{a_1 a_2}
    \coloneqq (\Dop_{[2]} X)_{a_1 a_2}
    = (\Dop_2 X)_{a_1 a_2 } - k X_{a_1 a_2},
\end{equation}
which is a 2-form with vanishing $\vec{7}$-part by the above remarks and \cref{eq:squashed_spectrum:2-rank_sym:rewritten_master}.
Using \cref{eq:squashed_spectrum:Ricci_special}, we see that
\begin{equation}
    \Dg^a (\Dop_2 X)_{ab}
    = \tensor{a}{_a^{c_1 c_2}} \Dg^a \Dg_{c_2} X_{c_1 b}
    = 0,
\end{equation}
by transversality of $X_{ab}$.
Thus, $Y_{ab}$ is a transverse 2-form.

We will now compute $\Dop_2{}^2 X$.
To this end, note that \cref{eq:squashed_spectrum:Weyl_operator}, applied to a traceless symmetric rank-2 tensor, gives
\begin{equation}
    2 \tensor{W}{_{a_1}^{b_1}_{a_2}^{b_2}} X_{b_1 b_2}
    = \Bigl( \frac{28}{5} - \C_\h \Bigr) X_{a_1 a_2}.
\end{equation}
Using this, the symmetry properties of the Weyl tensor, the projection operators \cref{,eq:squashed_spectrum:P7,eq:squashed_spectrum:P14} and the Ricci identity \cref{eq:squashed_spectrum:Ricci}, we find
\begin{subequations}
\label{eq:squashed_spectrum:2-rank:sym:antisym_Ds}
\begin{align}
    &[\Dg^b, \Dg_{a_1}] X_{b a_2}
    = \frac{1}{2} \C_\h X_{a_1 a_2} + \frac{1}{\sqrt{5}} (\Dop_2 X)_{a_1 a_2},
    \\
    &\tensor{c}{_{a_1}^{b_1 b_2 b_3}} \Dg_{b_1} \Dg_{b_2} X_{b_3 a_2}
    = \frac{1}{2} \C_\h X_{a_1 a_2} - \frac{2}{\sqrt{5}} (\Dop_2 X)_{a_1 a_2}.
\end{align}
\end{subequations}
Squaring $\Dop_{2}$, using the above, properties of $X_{ab}$ and the quadratic master equation \cref{eq:squashed_spectrum:quadratic_master}, we get
\begin{align}
\nonumber
    (\Dop_{2}{}^2 X)_{a_1 a_2}
    &= - \DDg X_{a_1 a_2}
    + \Dg^b \Dg_{a_1} X_{b a_2}
    + \tensor{c}{_{a_1}^{b_1 b_2 b_3}} \Dg_{b_1} \Dg_{b_2} X_{b_3 a_2}
    =\\
\label{eq:squashed_spectrum:2-rank_sym:Dop_2^2}
    &= C_\g X_{a_1 a_2} - \frac{1}{\sqrt{5}} (\Dop_{2} X)_{a_1 a_2}.
\end{align}
Remarkably, the $\C_\h$ from the quadratic master equation was cancelled by the two halves in \cref{eq:squashed_spectrum:2-rank:sym:antisym_Ds}.

Combining \cref{eq:squashed_spectrum:2-rank_sym:Y_def,eq:squashed_spectrum:2-rank_sym:Dop_2^2}, we see that
\begin{equation}
\label{eq:squashed_spectrum:final_eq}
    k^2 X_{a_1 a_2} + k Y_{a_1 a_2} + (\Dop_2 Y)_{a_1 a_2}
    = \Bigl(C_\g - \frac{k}{\sqrt{5}}\Bigr) X_{a_1 a_2} - \frac{1}{\sqrt{5}} Y_{a_1 a_2}.
\end{equation}
The antisymmetric part of this is
\begin{equation}
    \Dop_{[2]} Y_{a_1 a_2}
    = \tensor{a}{_{[a_1|}^{b_1 b_2}} \Dg_{b_2} Y_{b_1 |a_2]}
    = - \Bigl(k + \frac{1}{\sqrt{5}}\Bigr) Y_{a_1 a_2}
    = - \frac{\sqrt{5}}{2} \Bigl( C_\g - \kappa^2 + \frac{9}{5} \Bigl) Y_{a_1 a_2}.
\end{equation}
Since $Y_{ab}$ is a transverse 2-form with vanishing $\vec{7}$-part, we see from \cref{eq:squashed_spectrum:2-form:Dop_relations} that $\Dop_{[2]} Y = 0$.
Hence, either $Y_{ab} = 0$ or $\kappa^2 = C_\g + 9/5$.
If $Y_{ab} = 0$, the symmetric part of \cref{eq:squashed_spectrum:final_eq} is
\begin{equation}
    \Bigl(k^2 + \frac{k}{\sqrt{5}} -C_\g \Bigr) X_{a_1 a_2}
    = 0.
\end{equation}
Note that this is not true if $Y_{a_1 a_2} \neq 0$ since $(\Dop_2 Y)_{a_1 a_2}$ can have a symmetric part even though $Y_{a_1 a_2}$ is antisymmetric.
Inserting $k$ from \cref{eq:squashed_spectrum:2-rank_sym:rewritten_master} and solving for $\kappa^2$ gives \cref{eq:squashed_spectrum:2-rank_sym:kappa^2:b} below, since $X_{ab} \neq 0$.
Thus, we have arrived at
\begin{subequations}
\begin{align}
    &\kappa^2 = C_\g + \frac{9}{5},\\
\label{eq:squashed_spectrum:2-rank_sym:kappa^2:b}
    &\kappa^2 = C_\g + \frac{8}{5} \pm \frac{2}{\sqrt{5}} \sqrt{C_\g + \frac{1}{20}},
\end{align}
\end{subequations}
where the top row applies when $P_{14} (\Dop_{[2]} X)_{ab} \neq 0$ and the bottom one otherwise.

\section{3-forms} \label{sec:squashed_spectrum:3-form}%
A 3-form $Y_{abc}$ belongs to the irreducible representation $\vec{35}$ of $\Spin(7)$, which splits into $\vec{1}\oplus\vec{7}\oplus\vec{27}$ when restricted to $G_2$.
Computing the Casimir eigenvalues on the relevant representations using \cref{tab:Casimirs:conventions}, the rewritten quadratic master equation \cref{eq:squashed_spectrum:rewritten_master} becomes
\begin{equation}
\label{eq:squashed_spectrum:3-form:rewritten_master}
    \frac{3}{\sqrt{5}} \tensor{a}{_{[a_1|}^{b_1 b_2}} \Dg_{b_2} Y_{b_1 |a_2 a_3]}
    = \Bigl(C_\g - \kappa^2 + \frac{36}{5} - 3 P_7 - 7 P_{27}\Bigr) Y_{a_1 a_2 a_3}.
\end{equation}
for a transverse mode $Y_{abc}$ of $\Delta$ with eigenvalue $\kappa^2$.
In \cref{sec:squashed_spectrum:2-form}, we found a trick that made the calculation very short.
Here, we will use essentially the same method but be a bit more systematic.
We begin by analysing the irreducible parts of $Y_{abc}$ and deriving expressions for the projection operators.
Then, we analyse what implications the transversality of $Y_{abc}$ has for the irreducible components.
Lastly, we compute the possible eigenvalues.

\subsubsection{\texorpdfstring{$G_2$}{G2}-components and projection operators}
The $G_2$-scalar in $Y_{abc}$ is, of course,
\begin{equation}
    Y \coloneqq a^{b_1 b_2 b_3} Y_{b_1 b_2 b_3},
\end{equation}
whence $P_1$ is proportional to $a_{a_1 a_2 a_3} a^{b_1 b_2 b_3}$.
Using $P_1{}^2 = P_1$ to determine the constant of proportionality gives
\begin{equation}
    (P_1)\indices{_{a_1 a_2 a_3}^{b_1 b_2 b_3}}
    = \frac{1}{42} a_{a_1 a_2 a_3} a^{b_1 b_2 b_3},
    \qquad
    P_1 Y_{a_1 a_2 a_3} = \frac{1}{42} a_{a_1 a_2 a_3} Y.
\end{equation}

Similarly, the $\vec{7}$-part is
\begin{equation}
    Y_a \coloneqq
    - \tensor{c}{_a^{b_1 b_2 b_3}} Y_{b_1 b_2 b_3}.
\end{equation}
By $P_7{}^2 = P_7$ we find that
\begin{subequations}
\begin{align}
    &%
    \begin{aligned}[b]
        (P_7)\indices{_{a_1 a_2 a_3}^{b_1 b_2 b_3}}
        &= - \frac{1}{24} \tensor{c}{_{a_1 a_2 a_3}^c} \tensor{c}{_c^{b_1 b_2 b_3}}
        =\\
        &= \frac{1}{4} \tensor*{\delta}{*_{a_1}^{b_1}_{a_2}^{b_2}_{a_3}^{b_3}}
        - \frac{1}{24} a_{a_1 a_2 a_3} a^{b_1 b_2 b_3}
        + \frac{3}{8} \tensor{c}{_{[a_1 a_2}^{[b_1 b_2}} \delta_{a_3]}^{b_3]},
    \end{aligned}
    \\
    &%
    \begin{aligned}
        P_7 Y_{a_1 a_2 a_3}
        = \frac{1}{24} \tensor{c}{_{a_1 a_2 a_3}^c} Y_c.
    \end{aligned}
\end{align}
\end{subequations}

Lastly, the $\vec{27}$-part is
\begin{equation}
    X_{a_1 a_2}
    \coloneqq \tensor{a}{_{(a_1}^{b_1 b_2}} \tensor{Y}{_{a_2) b_1 b_2}}
    - \frac{1}{7} \delta_{a_1 a_2} Y.
\end{equation}
For this to be true,
\begin{align}
\nonumber
    (P_{27})\indices{_{a_1 a_2 a_3}^{b_1 b_2 b_3}}
    &\coloneqq (\1 - P_1 - P_{7})\indices{_{a_1 a_2 a_3}^{b_1 b_2 b_3}}
    =\\
    &= \frac{3}{4} \tensor*{\delta}{*_{a_1}^{b_1}_{a_2}^{b_2}_{a_3}^{b_3}}
    + \frac{1}{56} a_{a_1 a_2 a_3} a^{b_1 b_2 b_3}
    - \frac{3}{8} \tensor{c}{_{[a_1 a_2}^{[b_1 b_2}} \delta_{a_3]}^{b_3]}
\end{align}
has to be proportional to
\begin{align}
\nonumber
    &\bigl(\tensor*{\delta}{^{(b_1}_{[a_1}} \tensor{a}{^{b_2)}}_{a_2 a_3]} - \frac{1}{7} \delta^{b_1 b_2} a_{a_1 a_2 a_3} \bigr)
    \bigl(\delta_{(b_1}^{[c_1} \tensor{a}{_{b_2)}}^{c_2 c_3]} - \frac{1}{7} \delta_{b_1 b_2} a^{c_1 c_2 c_3} \bigr)
    =\\
    &= \tensor*{\delta}{*_{a_1}^{c_1}_{a_2}^{c_2}_{a_3}^{c_3}}
    - \frac{1}{7} a_{a_1 a_2 a_3} a^{c_1 c_2 c_3}
    + \frac{1}{2} \tensor{c}{_{[a_1 a_2}^{[c_1 c_2}} \delta_{a_3]}^{c_3]}
    + \frac{1}{2} \tensor{a}{_{[a_1 a_2}^{[c_1}} \tensor{a}{_{a_3]}^{c_2 c_3]}}.
\end{align}
Indeed, by \cref{eq:octonions:structure_constant_identity_asym_aa}, we see that the latter is $4/3\, P_{27}$, whence
\begin{equation}
    P_{27} Y_{a_1 a_2 a_3}
    = \frac{3}{4} \tensor{a}{_{[a_1 a_2}^b} \tensor{X}{_{a_3] b}},
\end{equation}
since $X_{ab}$ is traceless.

Since $\1 = P_1 + P_7 + P_{27}$, we can write $Y_{abc}$ in terms of the irreducible components as
\begin{equation}
\label{eq:squashed_spectrum:3-form:Y_parts}
    Y_{a_1 a_2 a_3}
    = \frac{1}{42} \tensor{a}{_{a_1 a_2 a_3}} Y
    + \frac{1}{24} \tensor{c}{_{a_1 a_2 a_3}^c} \tensor{Y}{_c}
    + \frac{3}{4} \tensor{a}{_{[a_1 a_2}^b} \tensor{X}{_{a_3] b}}.
\end{equation}
We have seen that $\vec{1}\oplus\vec{27}$ sits in the symmetric part of
\begin{equation}
\label{eq:squashed_spectrum:3-form:Z}
    Z_{a_1 a_2} \coloneqq \tensor{a}{_{a_1}^{b_1 b_2}} \tensor{Y}{_{a_2 b_1 b_2}}.
\end{equation}
Since there is no $\vec{14}$ in $\vec{7}^{\wedge 3} = \vec{35}$, we can immediately say that $P_{14} Z_{a_1 a_2} = 0$.
This can easily be verified by a direct computation as well.
The $\vec{7}$-part of $Z_{a_1 a_2}$ has to vanish or be proportional to $Y_a$ by the representation theory.
A direct calculation immediately shows
\begin{equation}
    Y_a = \tensor{a}{_a^{b_1 b_2}} \tensor{Z}{_{b_1 b_2}}.
\end{equation}
Thus, $Z_{a_1 a_2}$ is a rank-2 tensor with vanishing $\vec{14}$-components, containing all irreducible components of $Y_{a_1 a_2 a_3}$.
Its irreducible components are
\begin{equation}
\label{eq:squashed_spectrum:3-form:Z_parts}
    Y = \tensor{Z}{^b_b},
    \qquad
    X_{ab} = Z_{(ab)} - \frac{1}{7} \delta_{ab} Y,
    \qquad
    Y_a = \tensor{a}{_a^{b_1 b_2}} \tensor{Z}{_{b_1 b_2}}.
\end{equation}
Note that, since $Z_{ab}$ has no $\vec{14}$-part,
\begin{equation}
\label{eq:squashed_spectrum:3-form:Z_asym}
    Z_{[a_1 a_2]} = \frac{1}{6} \tensor{a}{_{a_1 a_2}^b} Y_b.
\end{equation}

\subsubsection{Transversality}
The transversality condition $\D^a Y_{abc}$ can be written using $\Dg_a$ as
\begin{equation}
    \Dg^b Y_{b a_1 a_2}
    = \frac{1}{\sqrt{5}} \tensor{a}{_{[a_1}^{b_1 b_2}} Y_{a_2] b_1 b_2}.
\end{equation}
In terms of the irreducible components, this reads
\begin{align}
\nonumber
    \frac{1}{6 \sqrt{5}} \tensor{a}{_{a_1 a_2}^b} Y_b
    &=
    \frac{1}{42} \tensor{a}{_{a_1 a_2}^b} \Dg_b Y
    + \frac{1}{24} \tensor{c}{_{a_1 a_2}^{b_1 b_2}} \Dg_{b_1} Y_{b_2}
    +\\
    &%
    + \frac{1}{4} \tensor{a}{_{a_1 a_2}^{b_1}} \Dg^{b_2} X_{b_1 b_2}
    - \frac{1}{2} \tensor{a}{_{[a_1|}^{b_1 b_2}} \Dg_{b_1} X_{b_2 |a_2]},
\end{align}
where we have used \cref{eq:squashed_spectrum:3-form:Y_parts,eq:squashed_spectrum:3-form:Z_asym}.
This equation contains both a $\vec{7}$-part and a $\vec{14}$-part.
The $\vec{7}$-part can be obtained by contracting with $a_{abc}$ and the $\vec{14}$-part by projecting with $P_{14}$.
They are
\begin{subequations}
\label{eq:squashed_spectrum:3-form:transv_parts}
\begin{align}
\label{eq:squashed_spectrum:3-form:transv_7-part}
    &\frac{1}{\sqrt{5}} Y_a
    = \Dg^b X_{ba}
    + \frac{1}{7} \Dg_a Y
    + \frac{1}{6} \tensor{a}{_a^{b_1 b_2}} \Dg_{b_1} Y_{b_2},
    \\
    &P_{14} \bigl( \Dg_{a_1} Y_{a_2} + 6 \tensor{a}{_{[a_1|}^{b_1 b_2}} \Dg_{b_1} X_{b_2 |a_2]} \bigr)
    = 0.
\end{align}
\end{subequations}

\subsubsection{Computing the eigenvalues}
\Cref{eq:squashed_spectrum:3-form:rewritten_master} is a 3-form equation and hence contains a $\vec{1}$, $\vec{7}$ and $\vec{27}$ part.
We begin by analysing the former two and then turn to the last one.
The scalar part of \cref{eq:squashed_spectrum:3-form:rewritten_master}, obtained by contracting with $a^{b_1 b_2 b_3}$, is
\begin{equation}
\label{eq:suqashed_spectrum:3-form:quad_fund_eq_0-part}
    \frac{3}{\sqrt{5}} \Dg^b Y_b
    = \Bigl( C_\g - \kappa^2 + \frac{36}{5} \Bigr) Y,
\end{equation}
while the vector part, obtained by contracting with $\tensor{c}{_a^{b_1 b_2 b_3}}$, is
\begin{equation}
    \frac{4}{7} \Dg_a Y
    - \frac{1}{6} \tensor{a}{_a^{b_1 b_2}} \Dg_{b_1} Y_{b_2}
    - 3 \Dg^b X_{ba}
    = - \Bigl(C_\g - \kappa^2 + \frac{21}{5} \Bigr) Y_a.
\end{equation}
Using \cref{eq:squashed_spectrum:3-form:transv_7-part} to eliminate $\Dg^b X_{ba}$ from the latter gives
\begin{equation}
\label{eq:squashed_spectrum:3-form:quad_fund_eq_7-part_after_transv}
    \tensor{a}{_a^{b_1 b_2}} \Dg_{b_2} Y_{b_1}
    - 3 \Dg_a Y
    = \sqrt{5} \Bigl(C_\g - \kappa^2 + \frac{12}{5}\Bigr) Y_a.
\end{equation}
Contracting this with $\Dg^a$ and using \cref{eq:suqashed_spectrum:3-form:quad_fund_eq_0-part} to eliminate $\Dg^a Y_a$, we find
\begin{equation}
    \frac{9}{5} C_\g Y
    = \Bigl(C_\g - \kappa^2 + \frac{9}{5}\Bigr) \Bigl(C_\g - \kappa^2 + \frac{36}{5}\Bigr) Y.
\end{equation}
Hence, either $Y = 0$ or
\begin{equation}
    \kappa^2 = C_\g + \frac{9}{2} \pm \frac{3}{\sqrt{5}} \sqrt{C_\g + \frac{81}{20}}.
\end{equation}

If $Y = 0$, \cref{eq:suqashed_spectrum:3-form:quad_fund_eq_0-part,eq:squashed_spectrum:3-form:quad_fund_eq_7-part_after_transv} implies that $Y_a$ is a transverse 1-form satisfying
\begin{equation}
    \tensor{a}{_a^{b_1 b_2}} \Dg_{b_2} Y_{b_1}
    = \sqrt{5} \Bigl(C_\g - \kappa^2 + \frac{12}{5}\Bigr) Y_a,
\end{equation}
Apart from numerical constants, the situation is identical to that in \cref{sec:squashed_spectrum:1-form}.
Reusing that calculation, we find
\begin{equation}
    \Bigl[
        5 \Bigl(C_\g - \kappa^2 + \frac{12}{5}\Bigr)^2
        + \Bigl(C_\g - \kappa^2 + \frac{12}{5}\Bigr)
        - \Bigl(C_\g + \frac{12}{5} \Bigr)
    \Bigr]
    Y_a = 0.
\end{equation}
Thus, either $Y_a = 0$ as well or
\begin{equation}
    \kappa^2 = C_\g + \frac{5}{2} \pm \frac{1}{\sqrt{5}} \sqrt{C_\g + \frac{49}{20}}.
\end{equation}

The only case remaining is when both $Y = 0$ and $Y_a = 0$, that is, when $Y_{abc}$ only has a nonvanishing $\vec{27}$-part. From \cref{eq:squashed_spectrum:3-form:Z,eq:squashed_spectrum:3-form:Z_parts,eq:squashed_spectrum:3-form:Y_parts}, we see that
\begin{equation}
\label{eq:squashed_spectrum:3-form:X_form_Y_only_27}
    X_{a_1 a_2} = \tensor{a}{_{a_1}^{b_1 b_2}} Y_{a_2 b_1 b_2},
    \qquad
    Y_{a_1 a_2 a_3} = \frac{3}{4} \tensor{a}{_{[a_1 a_2}^b} X_{a_3] b}.
\end{equation}
The $\vec{7}$ and $\vec{14}$ parts of the transversality condition, \cref{eq:squashed_spectrum:3-form:transv_parts}, become
\begin{equation}
\label{eq:squashed_spectrum:3-form:tranv_only_27}
    \Dg^a X_{ab} = 0,
    \qquad
    P_{14} (\Dop_{[2]} X)_{ab} = 0.
\end{equation}
The rewritten quadratic master equation \cref{eq:squashed_spectrum:3-form:rewritten_master}, which now only has a nontrivial $\vec{27}$-part by \cref{eq:suqashed_spectrum:3-form:quad_fund_eq_0-part,eq:squashed_spectrum:3-form:quad_fund_eq_7-part_after_transv}, becomes
\begin{equation}
    3 \tensor{a}{_{[a_1|}^{b_1 b_2}} \Dg_{b_2} Y_{b_1| a_2 a_3]}
    = \sqrt{5} \Bigl(C_\g - \kappa^2 + \frac{1}{5} \Bigr) Y_{a_1 a_2 a_3}.
\end{equation}
Contracting this as in the first half of \cref{eq:squashed_spectrum:3-form:X_form_Y_only_27} gives
\begin{align}
\nonumber
    \sqrt{5} \Bigl(C_\g - \kappa^2 + \frac{1}{5} \Bigr) X_{a_1 a_2}
    &= \tensor{a}{_{a_1}^{b_1 b_2}} \bigl(
        \tensor{a}{_{a_2}^{c_1 c_2}} \Dg_{c_2} Y_{c_1 b_1 b_2}
        + 2 \tensor{a}{_{b_1}^{c_1 c_2}} \Dg_{c_2} Y_{c_1 b_2 a_2}
    \bigr)
    =\\ \nonumber
    &= \tensor{a}{_{a_2}^{c_1 c_2}} \Dg_{c_2} X_{c_1 a_1}
    + 2 \tensor{c}{_{a_1}^{c_2 c_1 b_2}} \Dg_{c_2} Y_{c_1 b_2 a_2}
    =\\
\label{eq:squashed_spectrum:3-form:27-part}
    &= - \tensor{a}{_{a_1}^{c_1 c_2}} \Dg_{c_2} X_{c_1 a_2}
    = - (\Dop_2 X)_{a_1 a_2},
\end{align}
where we, in the second to last step, have used
\begin{equation}
    2 \tensor{c}{_{a_1}^{b c_1 c_2}} Y_{c_1 c_2 a_2}
    = \frac{3}{2} \tensor{c}{_{a_1}^{b c_1 c_2}} \tensor{a}{_{[c_1 c_2}^d} X_{a_2] d}
    = \tensor{a}{_{a_1}^{bd}} \tensor{X}{_{d a_2}}
    + \tensor{a}{_{a_2}^{bd}} \tensor{X}{_{d a_1}}
    + \tensor{a}{_{a_1 a_2}^{d}} \tensor{X}{_d^b}.
\end{equation}
From \cref{eq:squashed_spectrum:3-form:tranv_only_27,eq:squashed_spectrum:3-form:27-part}, we see that the situation is the same as that in \cref{sec:squashed_spectrum:2-rank_sym}, apart from numerical constants and the extra piece of information $P_{14} (\Dop_{[2]} X_{ab}) = 0$.
Hence, we can reuse that calculation but only get the two eigenvalues corresponding to \cref{eq:squashed_spectrum:2-rank_sym:kappa^2:b}.
This gives the eigenvalues in \cref{eq:squashed_spectrum:3-form:kappa^2:c} below.
To conclude, we have arrived at the eigenvalues
\begin{subequations}
\begin{align}
    &\kappa^2 = C_\g + \frac{9}{2} \pm \frac{3}{\sqrt{5}} \sqrt{C_\g + \frac{81}{20}},
    \\
    &\kappa^2 = C_\g + \frac{5}{2} \pm \frac{1}{\sqrt{5}} \sqrt{C_\g + \frac{49}{20}},
    \\
\label{eq:squashed_spectrum:3-form:kappa^2:c}
    &\kappa^2 = C_\g + \frac{1}{10} \pm \frac{1}{\sqrt{5}} \sqrt{C_\g + \frac{1}{20}},
\end{align}
\end{subequations}
where the first line applies to modes with a nonzero $\vec{1}$-part, the second one to modes with vanishing $\vec{1}$-part but nonzero $\vec{7}$-part and the last line applies to modes with only a nonvanishing $\vec{27}$-part.
Again, the list exhausts all possibilities but may contain false roots.

\section{Spinors} \label{sec:squashed_spectrum:spinor}%
Now that we have dealt with all tensorial representations, we turn to the spinorial ones, starting with the spinors.
Thus, consider a spinor satisfying $\Delta \psi = \kappa^2 \psi$.%
\footnote{Recall that, for Einstein manifolds, $(\i\slashed\D)^2$ only differs from $\Delta$ by a constant when acting on spinors and vector-spinor.}
The spinor representation $\vec{8}$ of $\Spin(7)$ splits into $\vec{1}\oplus\vec{7}$ when restricted to $G_2$.
Here, the $\vec{1}$ is the $G_2$-invariant spinor $\eta$ from \cref{app:octonions}.
Since $\h \subset \g_2$, $\Dg_a \eta = 0$.
With the normalisation $\bar\eta \eta = 1$, the projection operators are
\begin{equation}
    P_1 = \eta \bar\eta,
    \qquad\quad
    P_7 = \Gamma^a \eta \bar\eta \Gamma_a,
\end{equation}
where $P_1 + P_7 = \1$ by the Fierz identity \cref{eq:octonions:Fierz}.
Hence, we define
\begin{equation}
    Y \coloneqq \bar\eta \psi,
    \qquad\quad
    Y_a = -\i \bar\eta \Gamma_a \psi,
\end{equation}
so that
\begin{equation}
    \psi = Y \eta + \i Y_a \Gamma^a \eta.
\end{equation}

Since $C_{\so(7)}(\vec{8}) = 21/8$, the rewritten quadratic master equation \cref{eq:squashed_spectrum:rewritten_master} becomes
\begin{equation}
\label{eq:squashed_sphere:spinor:rewritten_master}
    \frac{1}{4\sqrt{5}} a_{abc} \Gamma^{ab} \Dg^c \psi
    = \Bigl(C_\g - \kappa^2 + \frac{63}{20} - 3 P_7 \Bigr) \psi.
\end{equation}
Recall from \cref{app:octonions} that
\begin{equation}
    a_{abc} = \i \bar\eta \Gamma_{abc} \eta,
    \qquad\quad
    c_{abcd} = - \bar\eta \Gamma_{abcd} \eta,
\end{equation}
while $\bar\eta \Gamma_a \eta = 0 = \bar\eta \Gamma_{ab} \eta$ since $\Gamma_a$ and $\Gamma_{ab}$ are antisymmetric.
Using this, we find that the scalar and vector parts of \cref{eq:squashed_sphere:spinor:rewritten_master} are
\begin{subequations}
\begin{align}
    &\frac{3}{2\sqrt{5}} \Dg^a Y_a
    = \Bigl(C_\g - \kappa^2 + \frac{63}{20} \Bigl) Y,
    \\
    &\frac{3}{2\sqrt{5}} \Dg_a Y
    + \frac{1}{2\sqrt{5}} \tensor{a}{_a^{b_1 b_2}} \Dg_{b_2} Y_{b_1}
    = -\Bigl(C_\g - \kappa^2 + \frac{3}{20} \Bigl) Y_a,
\end{align}
\end{subequations}
respectively.
Contracting the latter with $\Dg^a$ gives, by using \cref{eq:squashed_spectrum:Ricci_special} and then eliminating $\Dg^a Y_a$ using the scalar equation,
\begin{equation}
    \frac{9}{20} C_\g Y
    = \frac{3}{2\sqrt{5}} \Bigl(C_\g - \kappa^2 + \frac{9}{20} \Bigl) \Dg^a Y_a
    = \Bigl(C_\g - \kappa^2 + \frac{9}{20} \Bigl) \Bigl(C_\g - \kappa^2 + \frac{63}{20} \Bigl) Y.
\end{equation}
Thus, either $\kappa^2$ is given by \cref{eq:squashed_spectrum:kappa^2:a} below or $Y = 0$.
In the latter case, $\Dg^a Y_a = 0$, that is, $Y_a$ is a transverse 1-form, by the scalar equation and the vector equation becomes
\begin{equation}
    \Dop_1 Y_a = - 2 \sqrt{5} \Bigl(C_\g - \kappa^2 + \frac{3}{20} \Bigl) Y_a.
\end{equation}
This situation is identical to that in \cref{sec:squashed_spectrum:1-form} apart from numerical constants.
Reusing that calculation gives \cref{eq:squashed_spectrum:kappa^2:b}.
Thus,
\begin{subequations}
\begin{align}
\label{eq:squashed_spectrum:kappa^2:a}
    &\kappa^2 = C_\g + \frac{9}{5} \pm \frac{3}{2\sqrt{5}} \sqrt{C_\g + \frac{81}{20}},
    \\
\label{eq:squashed_spectrum:kappa^2:b}
    &\kappa^2 = C_\g + \frac{1}{10} \pm \frac{1}{2\sqrt{5}} \sqrt{C_\g + \frac{49}{20}},
\end{align}
\end{subequations}
where the top row applies to modes with nonzero scalar part, $\bar\eta \psi \neq 0$, and the second otherwise.

\section{Vector-spinors} \label{sec:squashed_spectrum:vector-spinor}%
Lastly, we turn to transverse $\Gamma$-traceless vector-spinors.
These carry the representation $\vec{48}$ of $\Spin(7)$ which splits into $\vec{7} \oplus\vec{14} \oplus\vec{27}$ when restricted to $G_2$.
We consider an eigenmode $\psi_a$ of $\Delta$ with eigenvalue $\kappa^2$.
The transversality and $\Gamma$-tracelessness conditions are $\D^a \psi_a = 0$ and $\Gamma^a \psi_a = 0$, respectively.
The $\so(7)$-Casimir is $C_{\so(7)}(\vec{48}) = 49/8$, whence the rewritten quadratic master equation, \cref{eq:squashed_spectrum:rewritten_master}, becomes
\begin{equation}
\label{eq:squashed_spectrum:vector-spinor:rewritten_master}
    \tensor{a}{_a^b^c} \Dg_c \psi_b
    + \frac{1}{4} \tensor{a}{_d_e^c} \Gamma^{de} \Dg_c \psi_a
    = \sqrt{5} \Bigl(C_\g - \kappa^2 + \frac{147}{20} - 3 P_7 - 6 P_{14} - 7 P_{27} \Bigr) \psi_a.
\end{equation}
Before attempting to find $\kappa^2$ from this, we analyse the irreducible components and the transversality condition.

\subsubsection{Irreducible \texorpdfstring{$G_2$}{G2}-components}
Note that $\vec{7} \oplus \vec{14} \oplus \vec{27}$ fits in $\vec{7}^{\otimes 2}$, that is, a rank-2 tensor.
Since we are used to working with rank-2 tensors from \cref{sec:squashed_spectrum:2-form,sec:squashed_spectrum:2-rank_sym,sec:squashed_spectrum:3-form}, we will, here too, translate the problem into one involving a rank-2 tensor.
The $\vec{7}$-part of $\psi_a$ is
\begin{equation}
    Y_a = \i \bar\eta \psi_a,
\end{equation}
where, as in \cref{sec:squashed_spectrum:spinor}, $\eta$ is the $G_2$-invariant spinor.
By the Fierz identity \cref{eq:octonions:Fierz},
\begin{equation}
\label{eq:squashed_spectrum:vector-spinor:psi_Y_Z}
    \psi_a = -\i Y_a \eta + \Gamma^b Z_{ba} \eta,
    \qquad\quad
    Z_{ab} \coloneqq \bar\eta \Gamma_a \psi_b.
\end{equation}
From $\Gamma^a \psi_a = 0$, it immediately follows that $\tensor{Z}{_a^a} = 0$, that is, the $\vec{1}$-part of $Z_{ab}$ is $0$.
The $\vec{14}$ and $\vec{27}$ parts of $\psi_a$ must be the corresponding parts in $Z_{ab}$ since they clearly do not sit in $Y_a$.
By the same representation theory, the $\vec{7}$ part of $Z_{ab}$ is either $0$ or proportional to $Y_a$.
Using the Fierz identity \cref{eq:octonions:Fierz}, $\Gamma^a \psi_a = 0$ and $a_{abc} = \i \bar\eta \Gamma_{abc} \eta$, we find
\begin{equation}
    a^{abc} Z_{bc}
    = \i \bar\eta \Gamma^{abc} \eta \bar\eta \Gamma_b \psi_c
    = -\i \bar\eta \Gamma^{ac} \psi_c
    = Y^a.
\end{equation}
Hence, $Z_{ab}$ contains all three irreducible components of $\psi_a$.
We define
\begin{equation}
    X_{ab} \coloneqq Z_{(ab)},
    \qquad
    Y_{ab} \coloneqq Z_{[ab]},
    \qquad
    Y^7_{ab} \coloneqq P_7 Y_{ab},
    \qquad
    Y^{14}_{ab} \coloneqq P_{14} Y_{ab}.
\end{equation}

\subsubsection{Transversality}
In terms of $\Dg_a$, the transversality condition is
\begin{equation}
    \Dg^a \psi_a
    = - \frac{1}{8\sqrt{5}} \tensor{a}{^a_b_c} \Gamma^{bc} \psi_a.
\end{equation}
Picking out the $\vec{1}$ and $\vec{7}$ parts of this, by contracting with $\bar\eta$ and $\bar\eta \Gamma_a$, respectively,
\begin{equation}
\label{eq:squashed_spectrum:vector-spinor:transversality}
    \Dg^a Y_a = 0,
    \qquad\quad
    \Dg^b Z_{ab}
    = \Dg^b X_{ba} - \Dg^b Y_{ba}
    = \frac{1}{\sqrt{5}} Y_a,
\end{equation}
where we have used
\begin{equation}
\label{eq:squashed_spectrum:vector-spinor:useful_contraction}
    \bar\eta \Gamma_a \Gamma^{cd} \psi_b
    = - \tensor{a}{_a^c^d} Y_b
    - \tensor{c}{_a^c^d^e} Z_{eb}
    + \delta_a^c \tensor{Z}{^d_b}
    - \delta_a^d \tensor{Z}{^c_b},
\end{equation}
which follows from \cref{eq:squashed_spectrum:vector-spinor:psi_Y_Z}.

\subsubsection{Computing the eigenvalues}
To convert \cref{eq:squashed_spectrum:vector-spinor:rewritten_master} into an equation for $Z_{ab}$, we contract with $\bar\eta \Gamma_a$.
Again using \cref{eq:squashed_spectrum:vector-spinor:useful_contraction}, we find
\begin{equation}
\label{eq:squashed_spectrum:vector-spinor:Z_quad_fund_eq}
    -\frac{3}{2\sqrt{5}} \Dg_a Y_b
    + \frac{1}{\sqrt{5}} \tensor{a}{_b^{c_1 c_2}} \Dg_{c_2} Z_{a c_1}
    - \frac{1}{2\sqrt{5}} \tensor{a}{_a^{c_1 c_2}} \Dg_{c_2} Z_{c_1 b}
    = \Bigl(C_\g - \kappa^2 + \frac{7}{20} + P_{14} + 4 P_{7} \Bigr) Z_{ab}.
\end{equation}
Note that
\begin{equation}
\label{eq:squashed_spectrum:vector-spinor:DY7_DopY_relation}
    \Dg^b Y^7_{ba}
    = \frac{1}{6} \tensor{a}{_a^c^b} \Dg_b Y_c
    = \frac{1}{6} \Dop_1 Y_a,
\end{equation}
and that \cref{eq:squashed_spectrum:vector-spinor:transversality} relates the divergences of $X_{ab}$ and $Y_{ab}$ to $Y_a$.
We can get two more relations involving divergences by contracting \cref{eq:squashed_spectrum:vector-spinor:Z_quad_fund_eq} with $\Dg^a$ and $\Dg^b$.
This should suffice to relate $\Dop_1 Y_a \propto \Dg^b Y^7_{ba}$ to $Y_a$ and we will then be in a situation similar to that in \cref{sec:squashed_spectrum:1-form} since $Y_a$ is transverse by \cref{eq:squashed_spectrum:vector-spinor:transversality}.
First, contracting \cref{eq:squashed_spectrum:vector-spinor:Z_quad_fund_eq} with $\Dg^b$ and using the transversality condition, $\Dg^b X_{ba}$ and $\Dg^b Y^{14}_{ba}$ can be expressed as
\begin{subequations}
\label{eq:squashed_spectrum:vector-spinor:divergences}
\begin{alignat}{2}
    &\Dg^b Y^{14}_{ba}
    &&= -3 \Dg^b Y^7_{ba}
    + \frac{\sqrt{5}}{3} \Bigl(C_\g - \kappa^2 + \frac{81}{20} \Bigr) Y_a,
    \\
    &\Dg^b X_{ba}
    &&= - 2 \Dg^b Y^7_{ba}
    + \frac{\sqrt{5}}{3} \Bigl(C_\g - \kappa^2 + \frac{93}{20} \Bigr) Y_a,
\end{alignat}
\end{subequations}
where we have used
\begin{subequations}
\begin{align}
    &\Dg^b \Dg_a Y_b
    = \frac{12}{5} Y_a + \frac{6}{\sqrt{5}} \Dg^b Y^7_{ba},
    \\
    &\tensor{a}{_b^{c_1 c_2}} \Dg^b \Dg_{c_2} Z_{a c_1}
    = \frac{3}{5} Y_a,
    \\
    &\tensor{a}{_a^{c_1 c_2}} \Dg^b \Dg_{c_2} Z_{c_1 b}
    = \frac{6}{5} Y_a
    + \frac{1}{\sqrt{5}} \bigl( \Dg^b X_{ba}
    + 3 \Dg^b Y^{14}_{ba}
    + 3 \Dg^b Y^{7}_{ba} \bigr).
\end{align}
\end{subequations}
\todo[disable]{}%

Now contract \cref{eq:squashed_spectrum:vector-spinor:Z_quad_fund_eq} with $\Dg^a$.
The first and last terms in the left-hand of \cref{eq:squashed_spectrum:vector-spinor:Z_quad_fund_eq} side give
\begin{equation}
    -\frac{3}{2\sqrt{5}} \Dg^a \Dg_a Y_b
    = \frac{3}{2\sqrt{5}} \Bigl(C_\g - \frac{12}{5}\Bigr) Y_b,
    \qquad
    -\frac{1}{2\sqrt{5}} \tensor{a}{_a^{cd}} \Dg^a \Dg_d Z_{cb}
    = -\frac{3}{10} \Dg^c Z_{cb},
\end{equation}
respectively.
Note that $\Dg^c Z_{cb}$ can be expressed in terms of $Y_b$ and $\Dg^a Y^7_{ab}$ by \cref{eq:squashed_spectrum:vector-spinor:divergences}.
The right-hand side of \cref{eq:squashed_spectrum:vector-spinor:Z_quad_fund_eq} contracted with $\Dg^a$ can similarly be expressed in terms of $Y_b$ and $\Dg^a Y^7_{ab}$.
Lastly, the middle term in the left-hand side of \cref{eq:squashed_spectrum:vector-spinor:Z_quad_fund_eq} is proportional to
\begin{equation}
    \tensor{a}{_b^{cd}} \Dg^a \Dg_d Z_{ac}
    = \tensor{a}{_b^{cd}} \Dg_d \Dg^a Z_{ac}
    + \tensor{a}{_b^{cd}} [\Dg^a, \Dg_d] Z_{ac}.
\end{equation}
Here, the commutator term is
\begin{equation}
    \tensor{a}{_b^{cd}} [\Dg^a, \Dg_d] Z_{ac}
    = - Y_b + \frac{4}{\sqrt{5}} \Dg^c Y^7_{cb} - \frac{2}{\sqrt{5}} \Dg^c Y^{14}_{cb},
\end{equation}
while the first term is, by \cref{eq:squashed_spectrum:vector-spinor:divergences,eq:squashed_spectrum:vector-spinor:DY7_DopY_relation},
\begin{equation}
    \tensor{a}{_b^{cd}} \Dg_d \Dg^a Z_{ac}
    = -4 \tensor{a}{_b^{cd}} \Dg_d \Dg^a Y^7_{ac} + 4\sqrt{5} \Bigl(C_\g - \kappa^2 + \frac{87}{20} \Bigr) \Dg^a Y^7_{ab}.
\end{equation}
In this expression, the first term in the right-hand side is
\begin{equation}
    \tensor{a}{_b^{cd}} \Dg_d \Dg^a Y^7_{ac}
    = \frac{1}{6} \tensor{a}{_b^{cd}} \tensor{a}{_c^e^f} \Dg_d \Dg_f Y_e
    = \frac{1}{6} \Bigl(C_\g + \frac{12}{5}\Bigr) Y_b - \frac{1}{\sqrt{5}} \Dg^c Y^7_{cb}.
\end{equation}
Putting the above together, we find
\begin{equation}
    \Bigl(C_\g - \kappa^2 + \frac{109}{40}\Bigr) \Dop_1 Y_a
    = \frac{\sqrt{5}}{2} \Bigl[
        \Bigl(C_\g - \kappa^2 + \frac{57}{20}\Bigr)^2
        - \frac{1}{4} \Bigl(C_\g + \frac{12}{5} \Bigr)
    \Bigr] Y_a.
\end{equation}
The situation is similar to that in \cref{sec:squashed_spectrum:1-form}.
Using \cref{eq:squashed_spectrum:1-form:Dop^2}, we get a fourth-order equation in $\kappa^2$ with solutions
\begin{equation}
    \kappa^2 = C_\g + \frac{14}{5} \pm \frac{1}{2\sqrt{5}} \sqrt{C_\g + \frac{49}{20}},
    \qquad
    \kappa^2 = C_\g + \frac{31}{10} \pm \frac{5}{2\sqrt{5}} \sqrt{C_\g + \frac{49}{20}}.
\end{equation}

If $Y_a = 0$, both $X_{ab}$ and $Y^{14}_{ab}$ are divergence-free by \cref{eq:squashed_spectrum:vector-spinor:divergences}.
The symmetric and antisymmetric parts of \cref{eq:squashed_spectrum:vector-spinor:Z_quad_fund_eq} become
\begin{subequations}
\label{eq:squashed_spectrum:vector-spinor:separated}
\begin{alignat}{3}
    &\frac{1}{2\sqrt{5}} \Dop_{(2)} X_{a b}
    &&- \frac{3}{2\sqrt{5}} (\Dop_{(2)} Y)_{a b}
    &&= \Bigl(C_\g - \kappa^2 + \frac{7}{20}\Bigr) X_{ab},
    \\
\label{eq:squashed_spectrum:vector-spinor:separated_asym}
    -&\frac{3}{2\sqrt{5}} (\Dop_{[2]} X)_{a b}
    &&+ \frac{1}{2\sqrt{5}} \Dop_{[2]} Y_{a b}
    &&= \Bigl(C_\g - \kappa^2 + \frac{27}{20}\Bigr) Y_{ab}.
\end{alignat}
\end{subequations}
From \cref{sec:squashed_spectrum:2-form}, we know that $\Dop_{[2]} Y_{ab} = 0$ since $Y_{ab}$ is divergence-free with vanishing $\vec{7}$-part.
Using this, the above equations can be combined into
\begin{equation}
\label{eq:squashed_spectrum:vector-spinor:combined_X_Y}
    (\Dop_2 X)_{ab} - 3 (\Dop_2 Y)_{ab}
    = 2\sqrt{5} \Bigl(C_\g - \kappa^2 + \frac{7}{20}\Bigr) X_{ab}
    - \frac{2\sqrt{5}}{3} \Bigl(C_\g - \kappa^2 + \frac{27}{20}\Bigr) Y_{ab}.
\end{equation}
Recall from \cref{sec:squashed_spectrum:2-rank_sym}, that \cref{eq:squashed_spectrum:2-rank_sym:Dop_2^2}
\begin{equation}
    (\Dop_2{}^2 X)_{ab} = C_\g X_{ab} - \frac{1}{\sqrt{5}} (\Dop_2 X)_{ab},
\end{equation}
for transverse traceless symmetric rank-2 tensors $X_{ab}$.
A short calculation shows that this holds for transverse 2-forms as well.
Thus, acting with $\Dop_2$ on \cref{eq:squashed_spectrum:vector-spinor:combined_X_Y} and antisymmetrising the free indices,
\begin{equation}
    C_\g Y_{ab}
    = -\frac{2\sqrt{5}}{3} \Bigl(C_\g - \kappa^2 + \frac{9}{20}\Bigr) (\Dop_{[2]} X)_{a b}.
\end{equation}
Eliminating $(\Dop_{[2]} X)_{a b}$ with \cref{eq:squashed_spectrum:vector-spinor:separated_asym}, we find that
\begin{equation}
    C_\g Y_{ab}
    = \frac{20}{9} \Bigl(C_\g - \kappa^2 + \frac{9}{20}\Bigr) \Bigl(C_\g - \kappa^2 + \frac{27}{20}\Bigr) Y_{ab}
\end{equation}
whence either $Y_{ab} = 0$ or
\begin{equation}
    \kappa^2 = C_\g + \frac{9}{10} \pm \frac{3}{2\sqrt{5}} \sqrt{C_\g + \frac{9}{20}}.
\end{equation}
In the case $Y_{ab} = 0$, \cref{eq:squashed_spectrum:vector-spinor:separated} gives us the situation in \cref{sec:squashed_spectrum:2-rank_sym} but with the extra information $(\Dop_{[2]} X)_{ab} = 0$.
As in \cref{sec:squashed_spectrum:3-form}, this implies that we only get two eigenvalues, given in \cref{eq:squashed_spectrum:vector-spinor:kappa^2:d} below.
The possible eigenvalues are, therefore,
\begin{subequations}
\begin{align}
    &\kappa^2 = C_\g + \frac{14}{5} \pm \frac{1}{2\sqrt{5}} \sqrt{C_\g + \frac{49}{20}},
    \\
    &\kappa^2 = C_\g + \frac{31}{10} \pm \frac{5}{2\sqrt{5}} \sqrt{C_\g + \frac{49}{20}},
    \\
    &\kappa^2 = C_\g + \frac{9}{10} \pm \frac{3}{2\sqrt{5}} \sqrt{C_\g + \frac{9}{20}},
    \\
\label{eq:squashed_spectrum:vector-spinor:kappa^2:d}
    &\kappa^2 = C_\g + \frac{2}{5} \pm \frac{1}{2\sqrt{5}} \sqrt{C_\g + \frac{1}{20}}.
\end{align}
\end{subequations}
Here, the top two lines apply to modes with nonzero $\vec{7}$-part, the third one to modes with vanishing $\vec{7}$-part but nonzero $\vec{14}$-part and the last line to modes with only a $\vec{27}$-part.

\section{Spectrum summary}
In \cref{tab:squashed_spectrum:operator_eigenvalues}, we give a summary of the eigenvalues found in \cref{sec:squashed_spectrum:0-form,sec:squashed_spectrum:1-form,sec:squashed_spectrum:2-form,sec:squashed_spectrum:2-rank_sym,sec:squashed_spectrum:3-form,sec:squashed_spectrum:spinor,sec:squashed_spectrum:vector-spinor}.
Recall from \cref{sec:ads_susy:Laplacian} that the Laplacian $\Delta$ is the Hodge--de Rham operator $\Delta_p$ when acting on $p$-forms, the Lichnerowicz operator $\Delta_\mathrm{L}$ when acting on traceless symmetric rank-2 tensors and related to $(\i\slashed\D)^2$ through
\begin{equation}
    (\i\slashed\D_{1/2})^2 = \Delta + \frac{189}{80},
    \qquad\quad
    (\i\slashed\D_{3/2})^2 = \Delta - \frac{27}{80},
\end{equation}
in the dimensionless system in which $m^2 = 9/20$, when acting on spinors and vector-spinors, respectively.
For 3-forms, the operator directly related to the $\mathrm{AdS}_4$ masses is $Q$, related to $\Delta_3$ by $Q^2 = \Delta_3$ on transverse 3-forms.
The possible eigenvalues of $Q$ are thus $\pm \kappa_3$, where $\kappa_3{}^2$ are the eigenvalues of $\Delta_3$.
Note that the eigenvalues of $\Delta_3$ are perfect squares, which we have used to simplify the expressions for the eigenvalues of $Q$.
This applies to $(\i\slashed\D)^2$ as well.
We have switched back to the dimensionful system by inserting appropriate powers of $20 m^2/9$.
Due to the limitations of the method we have used, all eigenvalues of the operators should be present in the table but some of the listed possibilities might not be eigenvalues.%
\footnote{When comparing the spinor eigenvalues to, for instance, \cite{ref:Duff--Nilsson--Pope:KK}, note that what is referred to as the Dirac operator in \cite{ref:Duff--Nilsson--Pope:KK} is $-\i\slashed\D$ in our conventions.}
\begin{table}[H]
    \everymath{\displaystyle}
    \setlength{\defaultaddspace}{0.75em}
    \centering
    \caption[Summary of eigenvalues of differential operators on the squashed $S^7$.]{Summary of eigenvalues of differential operators on the squashed $S^7$ in conventions in which $R_{ab} = 6 m^2 \delta_{ab}$. Note that we have not proven that the listed eigenvalues exist but rather that all eigenvalues are in the list. In \cref{chap:masses_and_susy}, we find that some roots do not fit into supermultiplets and, hence, are false. These are indicated by parentheses on $\pm$ or $\mp$ although there are possibly some exceptions, see \cref{eq:mass_susy:N=1:exceptions}. Note that the sign in front of the square root in the eigenvalues of the linear operators determine whether the eigenvalue belongs to the positive or negative part of the spectrum.}
\label{tab:squashed_spectrum:operator_eigenvalues}
    \begin{tabular}{lll}
        \toprule
        Operator & Possible eigenvalues &
        \\ \midrule
        $\Delta_0$
        & $\frac{m^2}{9} 20 C_\g$
        & \\ \addlinespace
        $\Delta_1$
        & $\frac{m^2}{9} \Bigl(20 C_\g + 14 \pm 2 \sqrt{20 C_\g + 49}\Bigr)$
        & \\ \addlinespace
        $\Delta_2$
        & $\frac{m^2}{9} 20 C_\g\hfil \frac{m^2}{9}\Bigl(20 C_\g + 72\Bigr)$
        & $\frac{m^2}{9} \Bigl(20 C_\g + 44 \pm 4 \sqrt{20 C_\g + 49}\Bigr)$
        \\ \addlinespace
        $\Delta_\mathrm{L}$
        & $\frac{m^2}{9} \Bigl(20 C_\g + 36\Bigr)$
        & $\frac{m^2}{9} \Bigl(20 C_\g + 32 \pm 4\sqrt{20 C_\g + 1}\Bigr)$
        \\ \addlinespace
        $Q$
        & $\varmp \frac{|m|}{3} \Bigl(1 \pm \sqrt{20 C_\g + 1} \Bigr)$
        & $\varpm \frac{|m|}{3} \Bigl(1 \pm \sqrt{20 C_\g + 49} \Bigr)$
        \\ \addlinespace
        & $\varpm \frac{|m|}{3} \Bigl(3 \pm \sqrt{20 C_\g + 81} \Bigr)$
        & \\ \addlinespace
        $\i\slashed\D_{1/2}$
        & $\varmp \frac{|m|}{3} \Bigl(\frac{1}{2} \pm \sqrt{20 C_\g + 49} \Bigr)$
        & $\varpm \frac{|m|}{3} \Bigl(\frac{3}{2} \pm \sqrt{20 C_\g + 81} \Bigr)$
        \\ \addlinespace
        $\i\slashed\D_{3/2}$
        & $\varpm \frac{|m|}{3} \Bigl(\frac{1}{2} \pm \sqrt{20 C_\g + 1} \Bigr)$
        & $\varmp \frac{|m|}{3} \Bigl(\frac{3}{2} \pm \sqrt{20 C_\g + 9} \Bigr)$
        \\ \addlinespace
        & $\varpm \frac{|m|}{3} \Bigl(\frac{1}{2} \pm \sqrt{20 C_\g + 49} \Bigr)$
        & $\varpm \frac{|m|}{3} \Bigl(\frac{5}{2} \pm \sqrt{20 C_\g + 49} \Bigr)$
        \\ \addlinespace[\aboverulesep] \bottomrule
    \end{tabular}
\end{table}

\chapter{Mass spectrum and supermultiplets} \label{chap:masses_and_susy}
Having found the eigenvalue spectra of the various operators on the squashed seven-sphere, see \cref{tab:squashed_spectrum:operator_eigenvalues}, we return to eleven-dimensional supergravity.
Since the squashed seven-sphere is an Einstein space with positive curvature, the background with $\mathrm{AdS}_4$ as the spacetime and the squashed seven-sphere as the internal manifold is a solution to the field equations in the Freund--Rubin ansatz.
There are actually two solutions related by reversing the direction of the flux, that is, by skew-whiffing, $m \mapsto -m$.
As remarked in \cref{sec:Freund--Rubin}, at most one of these can have unbroken supersymmetry.
Here, we begin by demonstrating this explicitly for the two squashed seven-sphere vacua and then turn to the mass spectrum and supermultiplets of the $\N = 1$ vacuum.

Recall from \cref{eq:squashed_spectrum:m} that there is a sign choice when relating the dimensionless and dimensionful unit systems.
This comes from the fact that one can let $m\mapsto -m$ in the Freund--Rubin ansatz to obtain another solution to the field equations.
However, the geometry and eigenvalue spectra of the squashed seven-sphere is independent of the flux-direction in the compactification.
We, therefore, use the relation
\begin{equation}
    |m| = \frac{3}{2\sqrt{5}}
\end{equation}
and insert appropriate powers of $2\sqrt{5} |m|/3$ to translate from the dimensionless to the dimensionful system.
The solution with $m>0$ will be referred to as the left-squashed vacuum and the one with $m<0$ as the right-squashed vacuum.

\section{Unbroken supersymmetry}
Recall that the number of unbroken supersymmetries is given by the number of linearly independent Killing spinors $\eta$, satisfying \cref{eq:Freund--Rubin:internal_Killing_spinor}
\begin{equation}
\label{eq:mass_susy:Killing_spinor}
    \tilde\D_a \eta
    \coloneqq \D_a \eta + \i \frac{m}{2} \Gamma_a \eta
    = 0.
\end{equation}
To investigate whether there are any unbroken supersymmetries we start by considering the holonomy of $\tilde\D_m$ and the integrability condition \cref{eq:Freund--Rubin:susy_integrability}
\begin{equation}
\label{eq:mass_susy:unbroken_integrability}
    W_{ab} \eta
    \coloneqq
    \tensor{W}{_a_b^c^d} \Gamma_{cd} \eta
    = 0.
\end{equation}
By \cref{eq:Freund--Rubin:Weyl_tensor}, the Weyl tensor is given by $W_{ab}{}^{cd} = R_{ab}{}^{cd} - 2m^2 \delta_{ab}^{cd}$.
With the Riemann tensor of the squashed seven-sphere from \cref{eq:squashed_geometry:fibration:Riem}, we find
\begin{subequations}
\begin{alignat}{2}
    &W_{0i}
    &&= \frac{8}{9} m^2 \bigl(
        2 \Gamma_{0i}
        + \epsilon_{ijk} \Gamma^{\hatj \hat k}
    \bigr),
    \\
    &W_{0 \hati}
    &&= -\frac{8}{9} m^2 \bigl(
        2 \Gamma_{0 \hati}
        + \epsilon_{ijk} \Gamma^{j \hat k}
    \bigr),
    \\
    &W_{ij}
    &&= \frac{16}{9} m^2 \bigl(
        \Gamma_{ij}
        + \Gamma_{\hati \hatj}
    \bigr),
    \\
    &W_{\hati \hatj}
    &&= \frac{16}{9} m^2 \bigl(
        2 \Gamma_{\hati \hatj}
        + \Gamma_{ij}
        + \epsilon_{ijk} \Gamma^{0k}
    \bigr),
    \\
    & W_{i \hatj}
    &&= -\frac{8}{9} m^2 \bigl(
        2 \Gamma_{i \hatj}
        + \Gamma_{j \hati}
        + \epsilon_{ijk} \Gamma^{0 \hat k}
        - \delta_{ij} \Gamma_k{}^{\hat k}
    \bigr),
\end{alignat}
\end{subequations}
where we have used the index split $a=(\hati, 0, i)$ and that $\lambda^2 = 1/5$ for the Einstein-squashed sphere.
Here, we see that $W_{ab}$ are linear combinations of the generators of $G_2$ from \cref{eq:octonions:G2:0i,eq:octonions:G2:ij,eq:octonions:G2:0hati,eq:octonions:G2:ihatj}.%
\footnote{Specifically, $W_{0i} \propto T_{0i}$, $W_{0 \hati} \propto T_{0 \hati}$, $W_{ij} \propto T_{ij}$, $W_{\hati \hatj} \propto T_{ij} + \epsilon_{ijk} T^{0 k}/2$ and $W_{i \hatj} \propto T_{i \hatj} + \epsilon_{ijk} T^{0 \hat k}$.}
Thus, the holonomy of $\tilde\D_m$ is $G_2$ \cite{ref:Duff--Nilsson--Pope:KK} and there is exactly one linearly independent solution to \cref{eq:mass_susy:unbroken_integrability}, namely the $G_2$-invariant $\eta$.

To check whether $\eta$ is a Killing spinor, we have to consider \cref{eq:mass_susy:Killing_spinor} and not only the integrability condition.
Recall from \cref{sec:squashed_spectrum:spinor} that $\Dg_a \eta = 0$ since $\h \subset \g_2$, where
\begin{equation}
    \Dg_a = \D_a - \frac{|m|}{3} a_{abc} \Sigma^{bc},
\end{equation}
by \cref{eq:squashed_spectrum:Dg}.
Using that the $\Gamma$-matrices can be represented by octonion multiplication, $\Gamma_a = -\i L_{o_a}$, as described in \cref{app:octonions:spin7}, and that the $G_2$ invariant then is identified with the real unit $o_{\hat 0} = 1\in\OO$, we find
\begin{equation}
    0
    = \Dg^a \eta
    = \D^a o_{\hat 0}
    + \frac{|m|}{12} a^{abc} o_b (o_c o_{\hat 0})
    = \D^a o_{\hat 0}
    + \frac{|m|}{2} o^a
    = \D^a \eta + \i \frac{|m|}{2} \Gamma^a \eta.
\end{equation}
The right-hand side is $\tilde\D^a \eta$ for $m>0$ and differs from $\tilde\D^a \eta$ by $\i m \Gamma^a \eta \neq 0$ for $m<0$.
Hence, $\tilde\D_a \eta = 0$ only for $m>0$.
Note that $S^7$ is simply connected, whence there are no global obstructions from the nonrestricted holonomy group.
Thus, we conclude that the left-squashed vacuum has one unbroken supersymmetry, $\N=1$, while the right-squashed vacuum has none, $\N=0$, \cite{ref:Duff--Nilsson--Pope:KK}.

\section{The left-squashed \texorpdfstring{$\N=1$}{N=1} vacuum} \label{sec:N=1}
We proceed by analysing the left-squashed vacuum with one unbroken supersymmetry.
The possible particle masses, presented in \cref{tab:mass_susy:N=1:masses}, are calculated from the possible eigenvalues in \cref{tab:squashed_spectrum:operator_eigenvalues} and the mass operators in \cref{tab:ads_susy:mass_op}.
Then, when calculating the dimensionless energy $E_0$, we restrict to $G$-representations with sufficiently large quadratic Casimir $C_\g$ to be able to simplify expressions like $|-10+\sqrt{\smash[b]{20 C_\g + 49}}|$, in which the absolute value can be dropped if the root is larger than $10$.
The strongest restriction needed for such simplifications is $C_\g \geq 243/20$.
Additionally, there is a choice of sign in $E_0$ for $s=0,\, 1/2$ for small values of $C_\g$.
To be able to, unambiguously, choose the plus sign we further restrict our attention to $C_\g > 99/4$.
We refer to this as the asymptotic part of the spectrum.
The resulting possibilities for $E_0$ are presented in \cref{tab:mass_susy:N=1:E0}.
\begin{table}[H]
    \everymath{\displaystyle}
    \setlength{\defaultaddspace}{0.3em}
    \centering
    \caption[Possible particle masses in the left-squashed vacuum.]%
    {Possible particle masses ($M^2$ for bosons, $M$ for fermions) in the left-squashed vacuum based on \cref{tab:ads_susy:mass_op} and the eigenvalues in \cref{tab:squashed_spectrum:operator_eigenvalues}. To get a mass (squared) one value should be picked from each pair of braces. When there are multiple braces in an expression, the same position must be chosen in all of them. Each column in a pair of braces corresponds to one eigenvalue-expression from \cref{tab:squashed_spectrum:operator_eigenvalues}, while the rows correspond to different signs in the expressions. The ordering is the same as in \cref{tab:squashed_spectrum:operator_eigenvalues}. When applicable, the first (second) subscript corresponds to the top (bottom) sign.}
\label{tab:mass_susy:N=1:masses}
    \begin{tabular}{c@{\hphantom{${}_{3,2}$\hskip 1.5em}}l}
    \toprule
    $s_{\mathrlap{t}}^{\mathrlap{p}}$
    & Possible masses
    \\ \midrule
% Spin 2+
    $2^{\mathrlap{+}}$
    & $\frac{m^2}{9} 20 C_\g$
    \\ \addlinespace
% Spin 3/2
    $\frac{3}{2}_{\mathrlap{1,2}}$
    &
    $\frac{m}{3}\left(
        \begin{Bmatrix}
            11 & 9\\
            10 & 12
        \end{Bmatrix}
        \mp \sqrt{
            20 C_\g +
            \begin{Bmatrix}
                49 & 81\\
                49 & 81
            \end{Bmatrix}
        }
    \right)$
    \\ \addlinespace
    % $\frac{3}{2}_{\mathrlap{2}}$
    % &
    % $\frac{m}{3}\left(
    %     \begin{Bmatrix}
    %         11 & 9\\
    %         10 & 12
    %     \end{Bmatrix}
    %     + \sqrt{
    %         20 C_\g +
    %         \begin{Bmatrix}
    %             49 & 81\\
    %             49 & 81
    %         \end{Bmatrix}
    %     }
    % \right)$
    % \\ \addlinespace
% Spin 1-
    $1^{\mathrlap{-}}_{\mathrlap{1}}$
    &
    $\frac{m^2}{9}\left(
        20 C_\g +
        \begin{Bmatrix}
            104 \\
            140
        \end{Bmatrix}
        -
        \begin{Bmatrix}
            16 \\
            20
        \end{Bmatrix}
        \sqrt{20 C_\g + 49}
    \right)$
    \\ \addlinespace
    $1^{\mathrlap{-}}_{\mathrlap{2}}$
    &
    $\frac{m^2}{9}\left(
        20 C_\g +
        \begin{Bmatrix}
            140 \\
            104
        \end{Bmatrix}
        +
        \begin{Bmatrix}
            20 \\
            16
        \end{Bmatrix}
        \sqrt{20 C_\g + 49}
    \right)$
    \\ \addlinespace
% Spin 1+
    $1^{\mathrlap{+}}$
    &
    $\frac{m^2}{9}\left(
        20 C_\g +
        \begin{Bmatrix}
            0 & 72 & 44 \\
              &    & 44
        \end{Bmatrix}
        +
        \begin{Bmatrix}
            0 & 0 & +4 \\
              &   & -4
        \end{Bmatrix}
        \sqrt{20 C_\g + 49}
    \right)$
    \\ \addlinespace
% Spin 1/2
    $\frac{1}{2}_{\mathrlap{1,4}}$
    &
    $\frac{m}{3}\left(
        -%
        \begin{Bmatrix}
            13 & 15\\
            14 & 12
        \end{Bmatrix}
        \pm \sqrt{
            20 C_\g +
            \begin{Bmatrix}
                49 & 81\\
                49 & 81
            \end{Bmatrix}
        }
    \right)$
    \\ \addlinespace
    % $\frac{1}{2}_{\mathrlap{4}}$
    % &
    % $\frac{m}{3}\left(
    %     -%
    %     \begin{Bmatrix}
    %         13 & 15\\
    %         14 & 12
    %     \end{Bmatrix}
    %     - \sqrt{
    %         20 C_\g +
    %         \begin{Bmatrix}
    %             49 & 81\\
    %             49 & 81
    %         \end{Bmatrix}
    %     }
    % \right)$
    % \\ \addlinespace
    $\frac{1}{2}_{\mathrlap{2,3}}$
    &
    $\frac{m}{3}\left(
        \begin{Bmatrix}
            5 & 3 & 5 & 7\\
            4 & 6 & 4 & 2
        \end{Bmatrix}
        \mp \sqrt{
            20 C_\g +
            \begin{Bmatrix}
                1 & 9 & 49 & 49\\
                1 & 9 & 49 & 49
            \end{Bmatrix}
        }
    \right)$
    \\ \addlinespace
    % $\frac{1}{2}_{\mathrlap{3}}$
    % &
    % $\frac{m}{3}\left(
    %     \begin{Bmatrix}
    %         5 & 3 & 5 & 7\\
    %         4 & 6 & 4 & 2
    %     \end{Bmatrix}
    %     + \sqrt{
    %         20 C_\g +
    %         \begin{Bmatrix}
    %             1 & 9 & 49 & 49\\
    %             1 & 9 & 49 & 49
    %         \end{Bmatrix}
    %     }
    % \right)$
    % \\ \addlinespace
% Spin 0+
    $0^{\mathrlap{+}}_{\mathrlap{1}}$
    & $\frac{m^2}{9} \Bigl(20 C_\g + 396 - 36 \sqrt{20 C_\g + 81} \Bigr)$
    \\ \addlinespace
    $0^{\mathrlap{+}}_{\mathrlap{3}}$
    & $\frac{m^2}{9} \Bigl(20 C_\g + 396 + 36 \sqrt{20 C_\g + 81} \Bigr)$
    \\ \addlinespace
    $0^{\mathrlap{+}}_{\mathrlap{2}}$
    &
    $\frac{m^2}{9}\left(
        20 C_\g -
        \begin{Bmatrix}
            0 & 4 \\
              & 4
        \end{Bmatrix}
        +
        \begin{Bmatrix}
            0 & +4 \\
              & -4
        \end{Bmatrix}
        \sqrt{20 C_\g + 1}
    \right)$
    \\ \addlinespace
% Spin 0-
    $0^{\mathrlap{-}}_{\mathrlap{1,2}}$
    &
    $\frac{m^2}{9}\left(
        20 C_\g +
        \begin{Bmatrix}
            56 & 140 & 216\\
            92 & 104 & 108
        \end{Bmatrix}
        \mp
        \begin{Bmatrix}
            16 & 20 & 24\\
            20 & 16 & 12
        \end{Bmatrix}
        \sqrt{
            20 C_\g +
            \begin{Bmatrix}
                1 & 49 & 81\\
                1 & 49 & 81
            \end{Bmatrix}
        }
    \right)$
    % \\ \addlinespace
    % $0^{\mathrlap{-}}_{\mathrlap{2}}$
    % &
    % $\frac{m^2}{9}\left(
    %     20 C_\g +
    %     \begin{Bmatrix}
    %         56 & 140 & 216\\
    %         92 & 104 & 108
    %     \end{Bmatrix}
    %     +
    %     \begin{Bmatrix}
    %         16 & 20 & 24\\
    %         20 & 16 & 12
    %     \end{Bmatrix}
    %     \sqrt{
    %         20 C_\g +
    %         \begin{Bmatrix}
    %             1 & 49 & 81\\
    %             1 & 49 & 81
    %         \end{Bmatrix}
    %     }
    % \right)$
    \\ \addlinespace[\aboverulesep] \bottomrule
\end{tabular}

\end{table}
\begin{table}[H]
    \everymath{\displaystyle}
    \setlength{\defaultaddspace}{0.3em}
    \centering
    \caption[Possible asymptotic values of $E_0$ in the left-squashed vacuum.]%
    {Possible asymptotic ($C_\g > 99/4$) values of $E_0$ in the left-squashed vacuum based on \cref{tab:ads_susy:E0_M} with notation and masses from \cref{tab:mass_susy:N=1:masses}.}
\label{tab:mass_susy:N=1:E0}
    \begin{tabular}{c@{\hphantom{${}_{3,2}$\hskip 1.5em}}l}
    \toprule
    $s_{\mathrlap{t}}^{\mathrlap{p}}$
    & Possible values of $E_0$
    \\ \midrule
% Spin 2+
    $2^{\mathrlap{+}}$
    & $\frac{1}{6} \Bigl(9 + \sqrt{20 C_\g + 81}\Bigr)$
    \\ \addlinespace
% Spin 3/2
    $\frac{3}{2}_{\mathrlap{1}}$
    &
    $\frac{1}{6}\left(
        \begin{Bmatrix}
            4 & 6\\
            5 & 3
        \end{Bmatrix}
        + \sqrt{
            20 C_\g +
            \begin{Bmatrix}
                49 & 81\\
                49 & 81
            \end{Bmatrix}
        }
    \right)$
    \\ \addlinespace
    $\frac{3}{2}_{\mathrlap{2}}$
    &
    $\frac{1}{6}\left(
        \begin{Bmatrix}
            14 & 12\\
            13 & 15
        \end{Bmatrix}
        + \sqrt{
            20 C_\g +
            \begin{Bmatrix}
                49 & 81\\
                49 & 81
            \end{Bmatrix}
        }
    \right)$
    \\ \addlinespace
% Spin 1-
    $1^{\mathrlap{-}}_{\mathrlap{1}}$
    &
    $\frac{1}{6}\left(
        \begin{Bmatrix}
            +1\\
            -1
        \end{Bmatrix}
        + \sqrt{20 C_\g + 49}
    \right)$
    \\ \addlinespace
    $1^{\mathrlap{-}}_{\mathrlap{2}}$
    &
    $\frac{1}{6}\left(
        \begin{Bmatrix}
            19\\
            17
        \end{Bmatrix}
        + \sqrt{20 C_\g + 49}
    \right)$
    \\ \addlinespace
% Spin 1+
    $1^{\mathrlap{+}}$
    &
    $\frac{1}{6}\left(
        \begin{Bmatrix}
            9 & 9 & 11 \\
              &   & 7
        \end{Bmatrix}
        + \sqrt{
            20 C_\g +
            \begin{Bmatrix}
                9 & 81 & 49 \\
                  &    & 49
            \end{Bmatrix}
        }
    \right)$
    \\ \addlinespace
% Spin 1/2
    $\frac{1}{2}_{\mathrlap{1}}$
    &
    $\frac{1}{6}\left(
        -%
        \begin{Bmatrix}
            4 & 6\\
            5 & 3
        \end{Bmatrix}
        + \sqrt{
            20 C_\g +
            \begin{Bmatrix}
                49 & 81\\
                49 & 81
            \end{Bmatrix}
        }
    \right)$
    \\ \addlinespace
    $\frac{1}{2}_{\mathrlap{4}}$
    &
    $\frac{1}{6}\left(
        \begin{Bmatrix}
            22 & 24\\
            23 & 21
        \end{Bmatrix}
        + \sqrt{
            20 C_\g +
            \begin{Bmatrix}
                49 & 81\\
                49 & 81
            \end{Bmatrix}
        }
    \right)$
    \\ \addlinespace
    $\frac{1}{2}_{\mathrlap{2}}$
    &
    $\frac{1}{6}\left(
        \begin{Bmatrix}
            4 & 6 & 4 & 2\\
            5 & 3 & 5 & 7
        \end{Bmatrix}
        + \sqrt{
            20 C_\g +
            \begin{Bmatrix}
                1 & 9 & 49 & 49\\
                1 & 9 & 49 & 49
            \end{Bmatrix}
        }
    \right)$
    \\ \addlinespace
    $\frac{1}{2}_{\mathrlap{3}}$
    &
    $\frac{1}{6}\left(
        \begin{Bmatrix}
            14 & 12 & 14 & 16\\
            13 & 15 & 13 & 11
        \end{Bmatrix}
        + \sqrt{
            20 C_\g +
            \begin{Bmatrix}
                1 & 9 & 49 & 49\\
                1 & 9 & 49 & 49
            \end{Bmatrix}
        }
    \right)$
% Spin 0+
    \\ \addlinespace
    $0^{\mathrlap{+}}_{\mathrlap{1,3}}$
    & $\frac{1}{6} \Bigl((9 \mp 18) + \sqrt{20 C_\g + 81}\Bigr)$
    \\ \addlinespace
    % $0^{\mathrlap{+}}_{\mathrlap{3}}$
    % & $\frac{1}{6} \Bigl(27 + \sqrt{20 C_\g + 81}\Bigr)$
    % \\ \addlinespace
    $0^{\mathrlap{+}}_{\mathrlap{2}}$
    &
    $\frac{1}{6}\left(
        \begin{Bmatrix}
            9 & 11 \\
              & 7
        \end{Bmatrix}
        + \sqrt{
            20 C_\g +
            \begin{Bmatrix}
                9 & 1 \\
                  & 1
            \end{Bmatrix}
        }
    \right)$
    \\ \addlinespace
% Spin 0-
    $0^{\mathrlap{-}}_{\mathrlap{1}}$
    &
    $\frac{1}{6}\left(
        \begin{Bmatrix}
            +1 & -1 & -3 \\
            -1 & +1 & +3
        \end{Bmatrix}
        + \sqrt{
            20 C_\g +
            \begin{Bmatrix}
                1 & 49 & 81 \\
                1 & 49 & 81
            \end{Bmatrix}
        }
    \right)$
    \\ \addlinespace
    $0^{\mathrlap{-}}_{\mathrlap{2}}$
    &
    $\frac{1}{6}\left(
        \begin{Bmatrix}
            17 & 19 & 21 \\
            19 & 17 & 15
        \end{Bmatrix}
        + \sqrt{
            20 C_\g +
            \begin{Bmatrix}
                1 & 49 & 81 \\
                1 & 49 & 81
            \end{Bmatrix}
        }
    \right)$
    \\ \addlinespace[\aboverulesep] \bottomrule
\end{tabular}

\end{table}
Now that we have the possible values of $E_0$ (for large $C_\g$), we investigate how they fit into the $\N = 1$ supermultiplets from \cref{tab:ads_susy:supermultiplets}.
Since the supersymmetry generators are $G$-singlets, all fields of a supermultiplet must carry the same irreducible $G$-representation \cite{ref:Nilsson--Padellaro--Pope}.
That the irreducible pieces of the induced $G$-representations fit exactly into supermultiplets has been verified in \cite{ref:Nilsson--Padellaro--Pope}.
The irreducible $G$-representations are labelled by Dynkin labels $(p,q;r)$.
In \cite{ref:Nilsson--Padellaro--Pope}, they also found that, for sufficiently large and fixed $p,q$, there is one $s=2$, six $s=3/2$, six $s=1^-$ and eight $s=1^+$ massive higher spin supermultiplets and 14 Wess--Zumino supermultiplets.

For sufficiently large Dynkin labels, the multiplicities of the irreducible $G$\hyp{}representations appearing in the harmonics are independent of the exact values of $p$ and $q$ \cite{ref:Nilsson--Padellaro--Pope}.
The $G$-representations can be arranged as in \cref{tab:mass_susy:N=1:G_multiplicities}.
If an eigenvalue expression from \cref{tab:squashed_spectrum:operator_eigenvalues} is assigned to a box in one of these diagrams, there is a family of eigenmodes of the corresponding operator with eigenvalues given by this expression.
Each such family has a constant difference between $r$ and $p$ and contains exactly one irreducible $G$-representation $(p,q;r)$ of eigenmodes for each sufficiently large pair $(p,q)$.
If one eigenvalue expression is assigned to $n$ different boxes, we say that it has multiplicity $n$.
Note that this does not imply that there are multiple irreducible $G$-representations with identical eigenvalue since two boxes with the same eigenvalue expression can have different $r-p$.
\begin{table}[H]
    \everymath{\displaystyle}
    \ytableausetup{boxsize=1.2em,aligntableaux=bottom}
    \newcommand{\halfboxsize}{0.6em} % should be half of the above
    \centering
    \caption[Multiplicities of irreducible $G$-representations.]%
    {Multiplicities, found in \cite{ref:Nilsson--Padellaro--Pope}, of irreducible $G$-representations $(p,q;r)$ in the $G$-representation induced by the various $\Spin(7)$-representations (in bold). For large values of the Dynkin labels, the multiplicities are independent of $p$ and $q$ and are given, for the indicated value of $r$ and $\Spin(7)$-representation, by the number of boxes in a row. To give the complete content of the induced representations, not only for large $p$ and $q$, each box can be replaced by a diagram that specifies which values of $p$ and $q$ that do not appear. Such diagrams can be found in \cite{ref:Nilsson--Padellaro--Pope}. By the below supermultiplet analysis, eigenvalue expressions can be assigned to the boxes marked by a cross without ambiguity apart from permutations of the columns.}
\label{tab:mass_susy:N=1:G_multiplicities}
    \begingroup
\renewcommand{\symbol}{\times}
\begin{tabular}{l lll ll}
    \toprule \addlinespace[0.5em]
    $\mathrlap{
    \begin{ytableau}
        \none[\mathrlap{\kern-\halfboxsize r=p+4}]\\
        \none[\mathrlap{\kern-\halfboxsize r=p+2}]\\
        \none[\mathrlap{\kern-\halfboxsize r=p}]\\
        \none[\mathrlap{\kern-\halfboxsize r=p-2}]\\
        \none[\mathrlap{\kern-\halfboxsize r=p-4}]\\
    \end{ytableau}
    }\hphantom{r=p\pm4}$
    &
    \begin{ytableau}
        \none[\vec{1}]\\
        {\symbol} \\
        \none\\
        \none
    \end{ytableau}
    &
    \begin{ytableau}
        \none[\vec{7}]\\
        {\symbol}&{\symbol} \\
        {\symbol}&{\symbol} \\
        {\symbol}&{\symbol} \\
        \none
    \end{ytableau}
    &
    \begin{ytableau}
        \none[\vec{8}]\\
        {\symbol}&{\symbol} \\
        {\symbol}&{\symbol}&{\symbol}&{\symbol} \\
        {\symbol}&{\symbol} \\
        \none
    \end{ytableau}
    &
    \begin{ytableau}
        {}      &\none[\ \vec{21}]\\
        {}&{}&{\symbol}&{\symbol} \\
        {}&{}&{\symbol}&{\symbol} &{\symbol} \\
        {}&{}&{\symbol}&{\symbol} \\
        {}
    \end{ytableau}
    &
    \begin{ytableau}
        {}&{}&{}    &\none[\ \vec{27}]\\
        {}&{}&{}&{} \\
        {}&{}&{}&{} &{}&{} \\
        {}&{}&{}&{} \\
        {}&{}&{}
    \end{ytableau}
    \\ \addlinespace
    $\mathrlap{
    \begin{ytableau}
        \none[\mathrlap{\kern-\halfboxsize r=p+4}]\\
        \none[\mathrlap{\kern-\halfboxsize r=p+2}]\\
        \none[\mathrlap{\kern-\halfboxsize r=p}]\\
        \none[\mathrlap{\kern-\halfboxsize r=p-2}]\\
        \none[\mathrlap{\kern-\halfboxsize r=p-4}]\\
    \end{ytableau}
    }\hphantom{r=p\pm4}$
    &
    \multicolumn{3}{l}{
    \begin{ytableau}
        {}&{}     &\none[\ \vec{35}]\\
        {}&{}&{\symbol}&{\symbol} \\
        {}&{}&{\symbol}&{\symbol} &{}&{}&{\symbol}&{\symbol} \\
        {}&{}&{\symbol}&{\symbol} \\
        {}&{}
    \end{ytableau}
    }
    &
    \multicolumn{2}{l}{
    \begin{ytableau}
        {}&{}&{}&{}           &\none[\ \vec{48}]\\
        {}&{}&{}&{} &{}&{}&{\symbol}&{\symbol} &{\symbol}&{\symbol} \\
        {}&{}&{}&{} &{}&{}&{\symbol}&{\symbol} &{\symbol}&{\symbol}&{}&{} \\
        {}&{}&{}&{} &{}&{}&{\symbol}&{\symbol} &{\symbol}&{\symbol} \\
        {}&{}&{}&{}
    \end{ytableau}
    }
    \\ \addlinespace[\aboverulesep] \bottomrule
\end{tabular}
\endgroup

\end{table}

The massive higher spin supermultiplet families are presented in \cref{tab:mass_susy:N=1:supermultiplets:massive} and the Wess--Zumino supermultiplet families in \cref{tab:mass_susy:N=1:supermultiplets:WZ}.
Each family of fields in a family of supermultiplets, for instance the $2^+$ fields in the $s=2$ massive higher spin supermultiplet family, has a specific eigenvalue expression.
Analogous to the multiplicities of the eigenvalue expressions, there is a multiplicity associated with each supermultiplet family specifying the number of supermultiplets in that family for fixed $(p,q)$ but arbitrary $r$.
By the above, the multiplicity is independent of the exact values of $p,q$ as long as they are sufficiently large.

To deduce some of these multiplicities, we use that there is only one eigenvalue expression for $\Delta_0$ and it has multiplicity one.
Furthermore, the multiplicities of the eigenvalue expressions of $\i\slashed\D_{1/2}$ are known from \cite{ref:Nilsson--Pope}, where the spinor eigenmodes were constructed explicitly from the scalar eigenmodes.
Still, we cannot deduce the multiplicity of the $1^+$ supermultiplet family and two of the Wess--Zumino supermultiplet families.

The multiplicities found in this way are consistent with \cref{tab:mass_susy:N=1:G_multiplicities}.
Assuming that we have not missed anything in our calculations (see below), this implies that the question mark in \cref{tab:mass_susy:N=1:supermultiplets:massive} should be 8 and that the two question marks in \cref{tab:mass_susy:N=1:supermultiplets:WZ} should add to 12.
Except for the possibility that one of the latter two is 0, it follows that the only false roots in the asymptotic part of \cref{tab:squashed_spectrum:operator_eigenvalues} are those marked with parentheses.
\begin{table}[H]
    \everymath{\displaystyle}
    \centering
    \caption[Asymptotic massive higher spin supermultiplet families.]%
    {Spins, parities and towers of fields in asymptotic massive higher spin supermultiplet families that can be formed from the values of $E_0$ in \cref{tab:mass_susy:N=1:E0}. A bracket $[r,c]$ indicates which row, $r$, and column, $c$, are used in the braces in \cref{tab:mass_susy:N=1:E0}. The multiplicity of the supermultiplet family is denoted by $n$. The case $n=3$ applies only when $p,q$ are sufficiently large, see \cite{ref:Nilsson--Padellaro--Pope,ref:Nilsson--Pope}.}
\label{tab:mass_susy:N=1:supermultiplets:massive}
    \begin{tabular}{rllll}
        \toprule
        $n$
        & \multicolumn{4}{l}{Massive higher spin supermultiplets}
        \\ \midrule
        1
        & $2^+$
        & $\frac{3}{2}_{2}[1,2]$
        & $\frac{3}{2}_{1}[1,2]$
        & $1^+[1, 2]$
        \\ \addlinespace
        3
        & $\frac{3}{2}_{1}[1,1]$
        & $1^+[2,3]$
        & $1^-_1[1,1]$
        & $\frac{1}{2}_{2}[1,3]$
        \\ \addlinespace
        3
        & $\frac{3}{2}_{2}[1,1]$
        & $1^-_2[1,2]$
        & $1^+[1,3]$
        & $\frac{1}{2}_{3}[1,3]$
        \\ \addlinespace
        3
        & $1^-_1[2,1]$
        & $\frac{1}{2}_{2}[1,4]$
        & $\frac{1}{2}_{1}[1,1]$
        & $0^-_1[1,2]$
        \\ \addlinespace
        3
        & $1^-_2[1,1]$
        & $\frac{1}{2}_{4}[1,1]$
        & $\frac{1}{2}_{3}[1,4]$
        & $0^-_2[1,2]$
        \\ \addlinespace
        ?
        & $1^+[1,1]$
        & $\frac{1}{2}_{3}[1,2]$
        & $\frac{1}{2}_{2}[1,2]$
        & $0^+_2[1,1]$
        \\ \addlinespace[\aboverulesep] \bottomrule
    \end{tabular}
\end{table}
\begin{table}[H]
    \everymath{\displaystyle}
    \centering
    \caption[Asymptotic Wess--Zumino supermultiplet families.]%
    {Spins, parities and towers of fields in asymptotic Wess--Zumino supermultiplet families, with notation as in \cref{tab:mass_susy:N=1:supermultiplets:massive}, that can be formed from the values of $E_0$ in \cref{tab:mass_susy:N=1:E0}.}
\label{tab:mass_susy:N=1:supermultiplets:WZ}
    \begin{tabular}{rlll}
        \toprule
        $n$
        & \multicolumn{3}{l}{Wess--Zumino supermultiplets}
        \\ \midrule
        1
        & $\frac{1}{2}_{1}[1,2]$
        & $0^-_1[1,3]$
        & $0^+_1$
        \\ \addlinespace
        1
        & $\frac{1}{2}_{4}[1,2]$
        & $0^+_3$
        & $0^-_2[1,3]$
        \\ \addlinespace
        ?
        & $\frac{1}{2}_{2}[1,1]$
        & $0^+_2[2,2]$
        & $0^-_1[1,1]$
        \\ \addlinespace
        ?
        & $\frac{1}{2}_{3}[1,1]$
        & $0^-_2[1,1]$
        & $0^+_2[1,2]$
        \\ \addlinespace[\aboverulesep] \bottomrule
    \end{tabular}
\end{table}

The first thing to notice in \cref{tab:mass_susy:N=1:supermultiplets:massive,tab:mass_susy:N=1:supermultiplets:WZ} is which values of $E_0$ are being used and which eigenvalues these correspond to.
We see that, in the towers corresponding to linear operators on the squashed $S^7$, that is, the spinorial fields and pseudoscalars, only the top rows in the braces in \cref{tab:mass_susy:N=1:E0} fit into supermultiplets.
The bottom rows in the braces correspond to the signs in parentheses in \cref{tab:squashed_spectrum:operator_eigenvalues}, which thus are false roots.
We also note that all possibilities in the towers corresponding to quadratic operators on the squashed $S^7$ fit into a supermultiplet.
This implies that every eigenvalue in the asymptotic part of the spectrum of the Laplacian $\Delta$ that we found in \cref{chap:squashed_spectrum} fits into a supermultiplet.
Furthermore, every value that fits into a supermultiplet fits only into one supermultiplet.

There is a reason to believe that we have not found all operator eigenvalues.
This indicates that there might be subtleties in the approach used in \cref{chap:squashed_spectrum} that could imply that we have missed eigenvalues in several calculations.
Since all eigenvalues of $\Delta$ that we did find fit into supermultiplets, any issue would likely be systematic.

The reason to believe that we might have missed something is the following.
In the case of a spinor on the squashed $S^7$, that is, the representation $\vec{8}$ of $\Spin(7)$, \cite{ref:Nilsson--Pope} found that each of the four columns in the corresponding diagram in \cref{tab:mass_susy:N=1:G_multiplicities} can be assigned a single eigenvalue expression, see \cite{ref:Duff--Nilsson--Pope:KK}.%
\footnote{The eigenvalue expressions are the ones of $\i\slashed\D_{1/2}$ without parentheses in \cref{tab:squashed_spectrum:operator_eigenvalues}.}%
\footnotemarksep%
\footnote{The heights of these columns were used to deduce some multiplicities in \cref{tab:mass_susy:N=1:supermultiplets:massive,tab:mass_susy:N=1:supermultiplets:WZ}.}
If one assumes that this continues to hold for all $\Spin(7)$-representations and that no two columns belonging to the same $\Spin(7)$-representations have the same eigenvalue expressions, the question mark in \cref{tab:mass_susy:N=1:supermultiplets:massive} can only be 3 or 5, not 8.
Similar remarks apply to the question marks in \cref{tab:mass_susy:N=1:supermultiplets:WZ}.
Thus, if our results are complete, some eigenvalue expressions must be assigned to multiple columns and there are unexplained degeneracies.
In this case, the multiplicities can, however, be consistently and essentially unambiguously assigned to the remaining boxes in the table without violating the assumption that each column should have a single eigenvalue expression.

Lastly, we turn to the low end of the spectrum.
When $C_\g \leq 99/4$ there are possibilities to choose the negative sign in the expressions for $E_0$ in \cref{tab:ads_susy:E0_M} for $s=0$ and $s=1/2$.
Also, as explained above, some simplifications of the expressions in \cref{tab:mass_susy:N=1:E0} are only valid in the asymptotic part of the spectrum.
However, since $p$, $q$ and $r$ take nonnegative integer values, there are only a finite number of special cases.
To proceed with the analysis we have to assume that we have found all eigenvalues.
In the low part of the spectrum, it seems like not only the supermultiplets that are part of one of the families in \cref{tab:mass_susy:N=1:supermultiplets:massive,tab:mass_susy:N=1:supermultiplets:WZ} are possible.
In some of these, there are even pseudoscalars of the type $0^-_{1,2}[2,c]$, that is, with a sign within parenthesis in \cref{tab:squashed_spectrum:operator_eigenvalues}.

The majority of the special cases occur for $p=q=r=0$.
Some of these are, however, easily excluded.
Firstly, the $G$-singlet, of course, gives a $H$-singlet when restricted to $H$.
Thus, $p=q=r=0$ only occurs in the $G$-representation induced by the $H$-singlet, by Frobenius reciprocity.
The $H$-singlet only occurs in the $G_2$-representations $\vec{1}$ and $\vec{27}$ \cite{ref:Nilsson--Padellaro--Pope}, whence there are two $G$-singlet 3-form modes.
Both of these are transverse~\cite{ref:Nilsson--Padellaro--Pope}.
One of them is $Y_{abc} = a_{abc}$.
From the Killing spinor equation \cref{eq:mass_susy:Killing_spinor}, it immediately follows that $\nabla_a a_{bcd} = m c_{abcd}$.
Thus, the 3-form $a_{abc}$ is transverse and has $Q$-eigenvalue $4m$.
This is a $[r,c]=[1,3]$ eigenvalue of $Q$ and hence not an exception to the parentheses in the eigenvalue table.

By orthogonality, the other $G$-singlet 3-form mode has only a $\vec{27}$-part.
The analysis in \cref{sec:squashed_spectrum:3-form} implies that the eigenvalue has $[r,c] = [r, 1]$.
Since $S^7$ has Betti number $b_3 = 0$, there are no 0-eigenvalues of $Q$, whence the possibilities are $\pm 2m/3$.
Thus, the $\mathrm{AdS}_4$ field is of type $0^-_1[1,1]$ or $0^-_2[2,1]$.
However, no supermultiplet can be formed using $0^-_2[2,1]$, even for small $p$ and $q$.
Thus, the only remaining possibility for the second $G$-singlet 3-form is $Q Y = -2m/3\, Y$.
The corresponding field belongs to $0^-_1[1,1]$ which is not an exception from the asymptotic case.

The remaining special cases that fit into supermultiplets are
\begin{subequations}
\label{eq:mass_susy:N=1:exceptions}
\begin{alignat}2
    &0^-_1[2,2]\colon \quad
    &&(p,q;r)=(1,0;1),
    \\
    &0^-_1[2,1]\colon \quad
    &&(p,q;r)\in\set{(0,0;4),(0,3;0),(2,1;2),(0,1;0)},
    \\
    &0^-_2[2,3]\colon \quad
    &&(p,q;r)\in\set{(0,0;4),(0,3;0),(2,1;2)}.
\end{alignat}
\end{subequations}
We have neither confirmed nor excluded the existence of modes with the corresponding eigenvalues.

In the low part of the spectrum, there is also the possibility of massless supermultiplets and Dirac singletons.
By analysing the possibilities and using the above remark regarding the occurrence of $G$-singlets in the induced representations, we find that there is one massless $2^+$ supermultiplet for $(p,q;r)=(0,0;0)$, two massless $1^-$ supermultiplets with $(p,q;r)\in\set{(2,0;0),(0,0;2)}$ and no Dirac singletons.
This is precisely what is expected since there is always exactly one massless spin 2 particle, the graviton, and the massless $1^-$ fields correspond to Killing vector fields that generate isometries of $\M_7$ \cite{ref:Duff--Nilsson--Pope:KK}.%
\todo[disable]{}
These massless supermultiplets are multiplet-shortened special cases of the $2^+$ and $1^-_1$ supermultiplets in \cref{tab:mass_susy:N=1:supermultiplets:massive}.

\chapter{Conclusions} \label{chap:conclusions}%
We have studied M-theory, or rather its low-energy limit, eleven-dimensional supergravity, compactified on the squashed seven-sphere, motivated by the AdS instability swampland conjecture.
There are two vacua, the left-squashed $\N=1$ vacuum and the right-squashed $\N=0$ vacuum, related by skew-whiffing.
By the aforementioned conjecture, the $\N=0$ vacuum should be unstable.
One possible instability is related to a tadpole and, in the dual conformal field theory, a global singlet marginal operator (GSMO).
To investigate whether such an instability can occur, considerable parts of the mass spectrum of the theory are needed, which we thus aimed to derive.
We realised that all mass operators in the Freund--Rubin ansatz are related to a universal Laplacian, \cref{eq:ads_susy:laplacian:universal}, which, in particular, enabled significant simplifications by relating Weyl tensor terms to group invariants, \cref{eq:squashed_spectrum:Weyl_operator}.
This is the main advancement of the thesis compared to \cite{ref:Ekhammar--Nilsson,ref:Aspman}.

We have found possible eigenvalue spectra of all operators of interest on the squashed seven-sphere.
By requiring consistency with supersymmetry, some false roots of the first-order operators were excluded and the asymptotic part of the eigenvalue spectra, including the multiplicities of the eigenvalue expressions, could be almost completely determined.
From the perspective of the spectrum of irreducible isometry representations, derived in \cite{ref:Nilsson--Padellaro--Pope}, our results indicate that there are degeneracies that we have not been able to explain.
This could be taken as evidence that our results are incomplete.
That would, however, require that we have missed some eigenvalues of the Laplacian $\Delta$ in such a way that those that we did find still fit into supermultiplets.
We hope to address this in \cite{ref:Karlsson--Nilsson}.

As explained in \cref{sec:GSMO}, GSMO-related instabilities may occur when there is a gauge singlet field, possibly composite, with $E_0 = 3$.
Thus, as in \cite{ref:Murugan}, one has to look for fields that can be combined into a gauge singlet scalar composite such that the energies of the elementary fields add up to $3$.
From the unitarity bounds on $E_0$ in \cref{tab:ads_susy:E0_M}, we see that $s \geq 1$ is immediately excluded by requiring that the field is a spacetime scalar.
Two spin-$1/2$ fields can be combined with up to two scalar fields to form a composite scalar that, a priori, could have $E_0 = 3$.
Also, there is the possibility of using only spin-$0$ fields.
With the bound $E_0 \geq 1/2$ there could be as many as six scalar fields in the composite.

As discussed in \cref{sec:ads_mass_susy}, the unitarity bounds for spins $0$ and $1/2$ correspond to singletons and can only arise in $0^+_1$ and $1/2_1$.
This implies, by the skew-whiffing theorem \cite{ref:Duff--Nilsson--Pope:KK}, that there are no singletons in the left-squashed vacuum, consistent with what we found in \cref{sec:N=1}, and only a single spin-$1/2$ singleton in the right-squashed vacuum.
Thus, there can actually be at most one scalar in a field dual to a GSMO containing two spin-$1/2$ fields and at most five when there are only scalars.

For large values of the quadratic Casimir $C_\g$, the possible energies $E_0$ grow approximately as $\sqrt{\smash[b]{C_\g}}$.
The number of scalars with $E_0 < 3$ and spinors with $E_0 < 3/2$ is therefore finite.
Hence, there are only finitely many combinations that could possibly produce $E_0 = 3$.
Without paying attention to which $G$-representations appear in the various towers, there seems to be plenty of candidates.
However, many of these might be easily excluded by a more careful analysis.
A problem that remains is that there are masses that are small enough that both signs in the expressions for $E_0$ are viable.
If any of these have a multiplicity of at least two, the corresponding $E_0$ values add to 3 if different signs are used.
To settle this, more work is needed.
Note also that the presence of a GSMO does not imply that there is an instability.
Similarly, if it turns out that there are no GSMO-related instabilities, other types of instabilities would have to be considered to strengthen or weaken the AdS instability swampland conjecture.

The swampland program aims to distinguish low-energy effective theories that are consistent when coupled to gravity from those that are not.
As we have seen, swampland criteria can have significant implications for low-energy physics and cosmology, including the role of de Sitter space in string theory.
Thus, the swampland program can bring string theory closer to experiment.
As long as there is no complete, nonperturbative description of M-theory and the stringy swampland conjectures remain unproven, the question of whether such experiments test string/M-theory or only the conjectures remains open.
Still, it is possible to investigate which conjectures are physically implemented in the observable part of the universe.
This could hopefully stimulate further theoretical developments.

% APPENDICES
\cleardoublepage
\appendix
\chapter{Conventions and representations} \label{app:conventions}
In this appendix, we present some conventions and notation used throughout the thesis.
We will always work in natural units in which $c = \hbar = 1$.
For the metric, we use the mostly-plus signature.
Although most equations are covariant and valid in any basis, we use the basis (if nothing else is specified) in which $\eta_{\alpha\beta} = \diag(-1,+1,\hdots,+1)_{\alpha\beta}$ when a basis is needed.

When symmetrising and antisymmetrising tensors, we employ the weight-one
definitions and use parenthesis and bracket notation, respectively.
For instance,
\begin{subequations}
\begin{align}
    u^{(\mu} v^{\nu)}
    &= \frac{1}{2} (u^\mu v^\nu + u^\nu v^\mu),
    \\
    u^{[\mu} v^{\nu]}
    &= \frac{1}{2} (u^\mu v^\nu - u^\nu v^\mu).
\end{align}
\end{subequations}
More generally,
\begin{subequations}
\begin{align}
    T^{(\mu_1 \hdots \mu_n)}
    &= \frac{1}{n!}
    \sum_{\sigma \in S_n}
    T^{\mu_{\sigma^{-1}(1)} \hdots \mu_{\sigma^{-1}(n)}},
    \\
    T^{[\mu_1 \hdots \mu_n]}
    &= \frac{1}{n!}
    \sum_{\sigma \in S_n} \sign{\sigma} \:
    T^{\mu_{\sigma^{-1}(1)} \hdots \mu_{\sigma^{-1}(n)}}.
\end{align}
\end{subequations}
In this notation, we define the generalised Kronecker delta as
\begin{equation}
    \delta_{\nu_1}^{\mu_1}{}_{\hdots}^{\hdots}{}_{\nu_p}^{\mu_p}
    \coloneqq \delta_{[\nu_1}^{\mu_1}\hdots \delta_{\nu_p]}^{\mu_p}.
\end{equation}

In the superspace setting, we use $(\hdots]$ and $[\hdots)$ to denote graded symmetrisation and graded antisymmetrisation of indices, respectively.
The grading means that there is an additional sign when fermionic indices pass through each other.

\section{Representations and index notation}
We use index notation and employ Einstein's summation convention throughout the text.
When elements of a vector space%
\footnote{Here we use ``vector'' in the general sense, not in the sense of an $\SO$-vector, and, in the following, $\alpha, \beta, \hdots$ are not spinor indices but indices for an arbitrary vector space.}
are denoted with lower indices (for instance $v_\alpha$), dual vectors (covectors) are denoted with upper indices (for instance $u^\alpha$).
If the vector space carries a (left-)representation of some group $G$, the dual vector space carries the dual representation and a group element $g$ acts like
\begin{equation}
    g\cdot v_\alpha = g\indices{_\alpha^\beta} v_\beta,
    \qquad\quad
    g \cdot u^\alpha = u^\beta (g^{-1})\indices{_\beta^\alpha},
\end{equation}
where we, by abuse of notation, use the same symbol for the group element and its representation.
Thus, the dual representation is given by $g\indices{^\alpha_\beta} = (g^{-1})\indices{_\beta^\alpha}$.%
\footnote{If one raises and lowers indices using an invertible invariant $M^{\alpha\beta}$, one must define $g\indices{^\alpha_\delta} \coloneqq M^{\alpha\beta} g\indices{_\beta^\gamma} M_{\gamma\delta}$ for $g\indices{^\alpha_\beta}$ to be the dual representation.}
Note that right-multiplication by the inverse, $g^{-1}$, is a left-representation since $g_1^{-1} g_2^{-1} = (g_2 g_1)^{-1}$.
Also, we use the word representation to refer not only to the actual representation but the representation space (module) as well.

Suppose that the vector space is complex.
The complex conjugate vector, which is an element of the complex conjugate vector space and whose coordinates are the complex conjugates of the coordinates of the original vector, is then denoted by $(v_\alpha)^\ast = \bar v_{\bar\alpha}$.
The complex conjugate vector space and its dual carry representations of $G$ and a group element $g$ acts like
\begin{equation}
    g \cdot \bar v_{\bar\alpha}
    = \bar g\indices{_{\bar\alpha}^{\bar\beta}} \bar v_{\bar\beta}
    \qquad\quad
    g \cdot \bar u^{\bar\alpha}
    = \bar u^{\bar\beta} (\bar g^{-1})\indices{_{\bar\beta}^{\bar\alpha}}.
\end{equation}
These transformation rules are summarised in \cref{tab:conventions:transformation_laws}.
\begin{table}[H]
    \centering
    \setlength{\defaultaddspace}{0.3em}
    \caption[Group and Lie algebra transformation laws.]{Transformation of a vector $v_\alpha$, a dual vector $v^\alpha$, a complex conjugated vector $\bar{v}_{\bar \alpha}$ and a complex conjugated dual vector $\bar{v}_{\bar \alpha}$ under a group element $g$ and a Lie algebra element $T$. Here, we use the convention $g = \exp(T)$ without an $\i$ in the exponent. To switch to the convention with an $\i$ in the exponent, let $T \mapsto \i T$ ($\bar T \mapsto -\i \bar T$).}
\label{tab:conventions:transformation_laws}
    \begin{tabular}{l ll}
        \toprule
        Quantity & Finite &Infinitesimal \\ \midrule
        $v_\alpha$
        & $g\cdot v_\alpha = g\indices{_\alpha^\beta} v_\beta$
        & $\delta_T v_\alpha = T\indices{_\alpha^\beta} v_\beta$
        \\ \addlinespace
        $v^\alpha$
        & $g \cdot v^\alpha = v^\beta (g^{-1})\indices{_\beta^\alpha}$
        & $\delta_T v^\alpha = v^\beta (-T\indices{_\beta^\alpha})$
        \\ \addlinespace
        $\bar v_{\bar\alpha}$
        & $g \cdot \bar v_{\bar\alpha}
        = \bar g\indices{_{\bar\alpha}^{\bar\beta}} \bar v_{\bar\beta}$
        & $\delta_T \bar v_{\bar\alpha} = \bar T\indices{_{\bar\alpha}^{\bar\beta}} \bar v_{\bar\beta}$
        \\ \addlinespace
        $\bar v^{\bar\alpha}$
        & $g \cdot \bar v^{\bar\alpha} = \bar v^{\bar\beta} (\bar g^{-1})\indices{_{\bar\beta}^{\bar\alpha}}$ $\quad$
        & $\delta_T \bar v^{\bar\alpha} = \bar v^{\bar\beta} (-\bar T\indices{_{\bar\beta}^{\bar\alpha}})$
        \\ \addlinespace[\aboverulesep] \bottomrule
    \end{tabular}
\end{table}

\section{The Lorentz group and special orthogonal groups} \label{app:SO_conventions}%
The Lorentz group%
\footnote{We use ``Lorentz group'', a bit carelessly, to refer to the identity component $\SO^+(d,1)$ and its double cover $\Spin(d,1)$ as well.}
in $d+1$ spacetime dimensions, denoted $\SO(d, 1)$, is defined through the $(d+1)$-dimensional \emph{vector} representation consisting of matrices with unit determinant that leave $\eta_{ab}$ invariant.
Therefore, we use $\eta_{ab}$ and its inverse to raise and lower vector indices as usual.
We define the generators $L^{ab}$ of the corresponding Lie algebra representation by
\begin{equation}
\label{eq:conventions:Lorentz_generators}
    (L^{ab})_{cd}
    = \tensor*{\delta}{^a_c^b_d}.
\end{equation}
Note that we use the geometrical convention that a group element is $\Lambda = \exp{L}$, without an $\i$ in the exponent.
With this normalisation of the generators, the Lie bracket reads
\begin{equation}
\label{eq:conventions:Lorentz_bracket}
    [L^{ab}, L_{cd}]
    = -2 \delta\indices*{^{[a}_{[c}} L\indices{^{b]}_{d]}}.
\end{equation}

The Lorentz group is, of course, the special case of Lorentzian signature of the more general special orthogonal group $\SO(p,q)$ with arbitrary signature.
We use the same normalisation of the generators in the general case, that is, \cref{eq:conventions:Lorentz_generators,eq:conventions:Lorentz_bracket} are still valid.
Note, however, that the generators of the vector representation $(L^{ab})\indices{^c_d}$ differ depending on the signature since indices are raised and lowered using the metric of the corresponding signature.

\section{Quadratic Casimirs} \label{app:quadratic_casimirs}%
The Casimir operators of a finite-dimensional semisimple Lie algebra $\g$ are special elements of the centre of the universal enveloping algebra, $\mathcal{Z}(\UU(\g))$.
They are elements of the form
\begin{equation}
    \C_\g^{(n)} = t^{a_1 \hdots a_n} T_{a_1} \hdots T_{a_n}
\end{equation}
that commute with all elements in $\g$ \cite{ref:Fuchs--Schweigert}.
Due to the relation $[T_a, T_b] = f_{ab}{}^c T_c$, one can, without loss of generality, take $t^{a_1 \hdots a_n}$ to be completely symmetric.
It is easy to see that $\C_\g^{(n)}$ commutes with all of $\g$ if and only if $t^{a_1 \hdots a_n}$ is an invariant tensor.
One can show that, for an algebra of rank $r$, there are precisely $r$ algebraically independent Casimir operators, which together with $\1$ generate the centre of $\UU(\g)$ via multiplication and linear combinations \cite{ref:Fuchs--Schweigert}.
Since $\C_\g^{(n)}$ commutes with all of $\g$, it acts on an irreducible representation $\rho$ by a constant, $\rho(\C_\g^{(n)}) = C_\g^{(n)}(\rho) \cdot \1$.
The eigenvalues $C_\g^{(n)}$ of the algebraically independent Casimirs can be used to uniquely specify an irreducible representation \cite{ref:Fuchs--Schweigert}, but for practical purposes we use Dynkin labels for this.

In this thesis, we will only be concerned with quadratic Casimirs, that is, the above $t$ is a symmetric rank-2 tensor.
A canonical choice is thus $t^{ab} = \kappa^{ab}$, where $\kappa_{ab}$ is the Cartan--Killing metric of $\g$.
For a semisimple $\g$ consisting of multiple simple Lie algebras, there is, however, one quadratic Casimir corresponding to each simple Lie algebra.
Still, we call the Casimir corresponding to the canonical $t^{ab} = \kappa^{ab}$ \emph{the} quadratic Casimir of $\g$.
The most well-known example of a quadratic Casimir is perhaps $J^2$ of $\so(3)$, the square of the angular momentum.

We are interested in the quadratic Casimirs of four Lie algebras: $\g = \sp(2)\oplus\sp(1)_C$, $\h = \sp(1)_A\oplus\sp(1)_{B+C}$, $\so(7)$ and $\g_2$.
See \cref{sec:squashed_geometry:coset} for an explanation of the subscripts.
The normalisations of the Casimirs of the relevant simple Lie algebras are given in \cref{tab:Casimirs:conventions} and agree with \cite{ref:LieART}.
As mentioned in \cref{sec:squashed_geometry:coset}, we define the $G$-Casimir by
\begin{equation}
\label{eq:Casimir:CG_def}
    \C_\g = 6 \kappa^{AB} T_A T_B,
\end{equation}
which implies $\ad_\g(\C_\g) = 6\cdot \1$.
As described above, $\g$ has two independent quadratic Casimirs corresponding to $\sp(2)$ and $\sp(1)$.
Thus, $\C_\g$ is a linear combination of these.
Since $\ad_\g \simeq \ad_{\sp(2)} \oplus \ad_{\sp(1)}$ we find
\begin{equation}
    C_\g(p,q;r) = 2 C_{\sp(2)}(p,q) + 3 C_{\sp(1)}(r),
\end{equation}
where $(p,q;r)$ are the Dynkin labels of $\g$, see \cite{ref:Nilsson--Padellaro--Pope}.

Since we are considering $\h$ as a subalgebra of $\g$, it will be convenient to use the restriction $\kappa_{RS}$ of the Cartan--Killing metric $\kappa_{AB}$ of $\g$ and normalise the Casimir as
\begin{equation}
\label{eq:Casimir:CH_def}
    \C_\h = 6 \kappa^{RS} T_R T_S.
\end{equation}
There is another independent quadratic Casimir of $\h$, since $\h$ consists of two simple Lie algebras.
It is, however, only this one we will be using.%
\footnote{For arbitrary squashing parameter, the relevant Casimir would be obtained by restricting $g_{AB}$ from \cref{sec:squashed_geometry:coset}, rather than $\kappa_{AB}$, to $\h$.}
With the index-split $R=(r,\dot r)$ for $\h = \sp(1)_A\oplus\sp(1)_{B+C}$ we see from \cref{eq:squashed_geometry:coset:Cartan--Killing} that $\kappa_{rs} = -3 \delta_{rs}$ and $\kappa_{\dot r \dot s} = -5 \delta_{\dot r \dot s}$,%
\footnote{Recall that $\kappa_{AB}$ is $-3$ on $\sp(1)_{A,B}$ and $-2$ on $\sp(1)_C$ which implies that it is $-5$ on $\sp(1)_{B+C}$.}
whence $C_\h$ is proportional to $C_{\sp(1)_A}/3 + C_{\sp(1)_{A+B}}/5$.
To determine the constant of proportionality, we compute $\rho_{\vec{7}}|_\h(\C_\h)=12/5 \cdot \1$, where $\rho_{\vec{7}}|_\h$ is the restriction of the vector representation $\rho_{\vec{7}}$ of $\so(7)$ to $\h$, using \cref{eq:coset:geometry:isotropy_embedding,eq:squashed_geometry:coset:structure_constants}.
Comparing this with the above formula, using that $\vec{7} \to (1,1) \oplus (0,2)$ when restricting to $\h$,%
\footnote{This decomposition is immediate from the structure of the generators presented in \cref{sec:squashed_geometry:coset}. It can also be found in \cite{ref:Nilsson--Padellaro--Pope}.}
we find
\begin{equation}
    C_\h(p,q) = 2 C_{\sp(1)_A}(p) + \frac{6}{5} C_{\sp(1)_{B+C}}(q).
\end{equation}

Turning to $\C_{\so(7)}$, we define%
\footnote{We use the analogous normalisation for $\so(5)\simeq\sp(2)$ but not $\so(3)\simeq\sp(1)$. For the latter, we use the conventional $S^2=s(s+1)$.}
\begin{equation}
\label{eq:Casimir:SO7}
    \C_{\so(7)} = - \tensor*{\delta}{*_{a_1}^{b_1}_{a_2}^{b_2}} \Sigma^{a_1 a_2} \Sigma_{b_1 b_2}.
\end{equation}
Computing $\rho_{\vec{7}}(\C_{\so(7)})$, where $\rho_{\vec{7}}$ is the $\vec{7}=(1,0,0)$-representation of $\so(7)$, we find that this agrees with the normalisation in \cref{tab:Casimirs:conventions}.

Lastly, the generators of $\g_2$ are
\begin{equation}
    T^{(\g_2)}_{a_1 a_2} = \tensor{(P_{14})}{_{a_1 a_2}^{b_1 b_2}} \Sigma_{b_1 b_2},
\end{equation}
where $P_{14}$ is the $\g_2$-projector onto $\vec{14}$ in $\vec{7}^{\wedge 2} \simeq \vec{7} \oplus \vec{14}$, since $\vec{14}=\ad_{\g_2}$.
Hence, $\C_{\g_2}$ is proportional to $\tensor{(P_{14})}{_{a_1 a_2}^{b_1 b_2}} \Sigma^{a_1 a_2} \Sigma_{b_1 b_2}$.
Using the $\vec{7}=(1,0)$-representation we find that the normalisation that agrees with \cref{tab:Casimirs:conventions} is
\begin{equation}
\label{eq:Casimir:CG2_from_generators}
    \C_{\g_2} = - \tensor{(P_{14})}{_{a_1 a_2}^{b_1 b_2}} \Sigma^{a_1 a_2} \Sigma_{b_1 b_2}.
\end{equation}
\begin{table}[H]
    \everymath{\displaystyle}
    \centering
    \caption[Conventions for quadratic Casimir operators.]{Casimir eigenvalues of the four simple Lie algebras of interest in terms of Dynkin labels. In the rightmost column, the values on the adjoint representations are given for convenience.}
\label{tab:Casimirs:conventions}
    \begin{tabular}{l@{$\hphantom{{}=}\negphantom{=}$}l@{\hskip 2em}l}
        \toprule
        Casimir & & $C(\ad)$
        \\ \midrule
        $C_{\so(7)}(p,q,r)$ & $=\frac{1}{2} p (p+2 q+r+5)+q (q+r+4)+\frac{3}{8} r (r+6)$ & 5
        \\ \addlinespace
        $C_{\g_2}(p,q)$ & $=\frac{1}{3} p (p+3 q+5)+q (q+3)$ & 4
        \\ \addlinespace
        $C_{\sp(2)}(p,q)$ & $=\frac{1}{4} p (p+2 q+4)+\frac{1}{2} q (q+3)$ & 3
        \\ \addlinespace
        $C_{\sp(1)}(p)$ & $=\frac{1}{4}p(p+2)$ & 2
        \\ \addlinespace[\aboverulesep] \bottomrule
    \end{tabular}
\end{table}

\chapter{Spinors} \label{app:spinors}%
The vector representation of $\so(r,s)=\operatorname{Lie}(\SO(r,s))$ is not the most elementary $\so(r,s)$-representation in the sense that there is another representation, the spinor representation, which cannot be constructed from it but can be used to construct it.%
\footnote{Note that with signature $(r, s)$, we mean that the metric has $r$ positive and $s$ negative eigenvalues in any basis.}
At the group level, the spinor representations are only projective representations of $\SO(r, s)$ but ordinary representations of $\Spin(r, s)$, the double cover of $\SO(r, s)$.
The Dirac spinor representation can be defined in terms of an irreducible complex representation of the Clifford algebra $\cliff_{r,s}$ \cite{ref:Lawson--Michelsohn}, that is, $\Gamma$-matrices $\Gamma^a$ satisfying
\begin{equation}
\label{eq:spinors:Gamma_anticommutator}
    \{\Gamma^a, \Gamma^b\} = 2 \eta^{ab} \1.
\end{equation}
Since this implies that
\begin{equation}
    [\Gamma^{ab}, \Gamma_{cd}] = -8 \delta^{[a}_{[c} \tensor{\Gamma}{^{b]}_{d]}},
\end{equation}
where $\Gamma^{a_1 \hdots a_n} \coloneqq \Gamma^{[a_1} \hdots \Gamma^{a_n]}$, we get a representation $S$ of $\so(r,s)$ by
\begin{equation}
    S(L^{ab}) = \frac{1}{4} \Gamma^{ab}.
\end{equation}
This is the Dirac spinor representation.
Many properties of spinors depend on the dimension, $d=r+s$, modulo eight%
\footnote{This is related to Bott periodicity \cite{ref:Lawson--Michelsohn}.}
and signature.
Here, we focus on $\Spin(3,1)$ and $\Spin(10, 1)$ but begin with some general remarks.
$\Spin(7)$, which is also of importance for this thesis, is treated in \cref{app:octonions}.
Furthermore, all considerations in this appendix are local.
To be able to define spinor fields globally on a manifold, a spin structure is needed \cite{ref:Lawson--Michelsohn}.%
\todo[disable]{}

Arbitrary products of $\Gamma$-matrices can be computed by combinatorics.
This is most easily seen in a basis in which $\eta = \diag(-1,\dots,-1, 1, \dots, 1)$ but is valid in any basis, as long as the results are written in a basis-independent way, and follows directly from \cref{eq:spinors:Gamma_anticommutator}.
The simplification comes from the fact that, in this basis, $\Gamma^a$ squares to $\pm \1$ and $\Gamma^{a_1 \dots a_n} = \Gamma^{a_1} \dots \Gamma^{a_n}$ as long as the indices are distinct.
As an example, consider $\Gamma^{ab} \Gamma_{cd}$.
Here, either all indices are distinct (1 possibility); one of $a$, $b$ coincides with one of $c$, $d$ (4 possibilities) or both of $a$, $b$ coincide with one of $c$, $d$ each (2 possibilities).
Thus,
\begin{equation}
    \Gamma^{ab} \Gamma_{cd}
    = \tensor{\Gamma}{^a^b_c_d}
    - 4 \delta^{[a}_{[c} \tensor{\Gamma}{^{b]}_{d]}}
    - 2 \delta^{ab}_{cd} \1,
\end{equation}
where the signs come from anticommuting $\Gamma$-matrices with distinct indices.
This can be generalised to cases with contracted indices and more than two factors.
For instance, $\Gamma^a \Gamma_a = d \1$ and
\begin{equation}
    \Gamma_b \Gamma^{a_1\hdots a_n} \Gamma^b
    = (-1)^n (d - 2n) \Gamma^{a_1\hdots a_n}.
\end{equation}
In the latter, there are $d-n$ values of $b$ that do not coincide with any of the $a$'s and $n$ values of $b$ that coincide with one of the $a$'s.
These identities can be used to see that
\begin{equation}
    d \trace \Gamma^{a_1\hdots a_n}
    = \trace(\Gamma^b \Gamma_b \Gamma^{a_1\hdots a_n})
    = \trace(\Gamma_b \Gamma^{a_1\hdots a_n} \Gamma^b)
    = (-1)^n (d - 2n) \trace \Gamma^{a_1\hdots a_n},
\end{equation}
which implies that $\trace \Gamma^{a_1\hdots a_n} = 0$ if $n\neq 0$ and, for odd $d$, $n\neq d$.

\section{Spinors in arbitrary even dimension}
Let us, for a moment, consider the case of even dimension, $d=r+s=2k$.
The dimension of the Dirac spinor representation is $2^k$ \cite{ref:Lawson--Michelsohn}.
A basis of $\operatorname{End}(V_S, V_S)$, that is, $2^k \times 2^k$ matrices after a choice of basis for the Dirac representation space $V_S$, is provided by $\1, \Gamma^{a_1}, \hdots, \Gamma^{a_1 \hdots a_d}$ when restricted to antisymmetrically independent index combinations.
To see this, consider
\begin{equation}
    X = \sum_{i=0}^d x_{a_1\hdots a_i} \Gamma^{a_1 \hdots a_i},
\end{equation}
where all $x_{a_1\hdots a_i}$ are completely antisymmetric.
From the above remarks on how to compute products and traces of $\Gamma$-matrices, we see that $x_{a_1 \hdots a_i} \propto \trace(X \Gamma_{a_1\hdots a_i})$.
Thus, $X=0$ implies that all the $x$'s are 0, whence we have established linear independence.
By counting, we now see that there are $2^d = 2^k\cdot 2^k$ linearly independent components of $X$, which is precisely the dimension of $\operatorname{End}(V_S, V_S)$.

\subsubsection{Chirality and Weyl spinors}
Since there is only one antisymmetrically independent index combination with $d$ indices, we may define $\gamma$ by%
\footnote{In $d=4$, this is commonly denoted by $\gamma^5$.}%
\begin{equation}
    \gamma \epsilon^{a_1\hdots a_d}
    = i^{k+s} \Gamma^{a_1\hdots a_d}.
\end{equation}
It is easy to see that $\gamma^2 = \1$ and $\{\gamma, \Gamma^a\} = 0$.
Note that $(\Gamma^a)_A{}^B$, where $A,B$ are Dirac spinor indices, is a $\so(r,s)$-invariant, as seen from
\begin{equation}
    L_{ab} \cdot (\Gamma^c)\indices{_A^B}
    = \frac{1}{4} (\Gamma_{ab} \Gamma^c)\indices{_A^B}
    - \frac{1}{4} (\Gamma^c \Gamma_{ab})\indices{_A^B}
    + \delta_{ab}^{cd} (\Gamma_d)\indices{_A^B}
    = 0.
\end{equation}
Thus, $\gamma$ is also an invariant and the Dirac spinor representation is reducible by Shur's lemma since $\gamma$ is not proportional to $\1$.
We may form projection operators
\begin{equation}
    P_\pm = \frac{\1 \pm \gamma}{2},
\end{equation}
projecting onto the invariant subspaces with $\gamma$-eigenvalue $\pm 1$, respectively.
These smaller dimensional representations are known as the left and right-handed Weyl spinor representations.
Note that, if $\gamma \Psi = \pm \Psi$, then $\gamma \Gamma^a \Psi = \mp \Gamma^a \Psi$, that is, $\Gamma^a$ interchanges left and right-handed spinor.
Thus, writing a Dirac spinor as $\Psi_A = (\psi_\alpha, \chi^{\tilde\alpha})_A$, where $\psi$ and $\chi$ are Weyl spinors of the two kinds, this implies that
\begin{equation}
    (\Gamma^a)\indices{_A^B}
    =
    \begin{pmatrix}
        0 & (\Gamma^a)_{\alpha \tilde\beta} \\
        (\Gamma^a)^{\tilde\alpha \beta} & 0
    \end{pmatrix}\indices{_{{\mkern-6mu}A}^{B}},
\end{equation}
where we have introduced chiral $\Gamma$-matrices.
In this basis, we see explicitly that the Dirac spinor representation of $\so(r,s)$ is reducible since $\Gamma^{ab}$ is block-diagonal.
For even $d$, one can prove that there is only one irreducible representation of the Clifford algebra and that the Weyl spinor representations of $\so(r,s)$ are irreducible~\cite{ref:Lawson--Michelsohn}.

\subsubsection{Invariant tensors and Majorana spinors}
There are a few other invariants apart from $\gamma$.
To see this, note that we can construct a new set of matrices satisfying \cref{eq:spinors:Gamma_anticommutator} by taking the negative, complex conjugate, transpose or any combination thereof of all matrices $\Gamma^a$.
Consider first the negative.
Since there is only one inequivalent irreducible Clifford algebra representation, there exists a matrix $M_A{}^B$ such that $-\Gamma^a = M \Gamma^{a} M^{-1}$, relating the two representations.
Note that this equation is linear in $M$, which is seen by writing it as $M \Gamma^a = - \Gamma^a M$.
The space of solutions is one-dimensional by Shur's lemma.
From the above, we see that $M = \gamma$ provides a canonical choice of $M$.

Similarly, there are two one-dimensional spaces of solutions $(B_\pm)_{\bar A}{}^B$ to
\begin{equation}
\label{eq:spinors:B}
    B_\pm \Gamma^a = \pm (\Gamma^a)^\ast B_\pm.
\end{equation}
These two spaces are related by $B_- = B_+ \gamma$.
Taking the complex conjugate of \cref{eq:spinors:B}, we find that $(B_\pm^{\ast})^{-1}$ are also solutions whence $(B_\pm^{\ast})^{-1} = z_\pm B_\pm$ for some $z_\pm \in \CC$.
Complex conjugating this relation, we find $z_\pm \in \RR$.
Since the left and right-handed spinor representations are inequivalent,%
\footnote{Note that $\gamma \propto \epsilon_{a_1 b_1 \hdots a_k b_k} \Gamma^{a_1 b_1} \hdots \Gamma^{a_k b_k}$ is a $k$'th-order Casimir with different eigenvalue on the two Weyl representations.}
$B_\pm$ are either block diagonal or block antidiagonal in the Weyl basis, $\gamma = \diag(\1, -\1)$.
With $B_- = B_+ \gamma$, we find that $z_- = \pm z_+$ in the two cases, respectively.
By rescaling $B_\pm$, $z_\pm$ is rescaled by a positive real number whence we may scale $B_\pm$ such that $z_\pm \in \set{-1, +1}$ while maintaining $B_- = B_+ \gamma$.

Due to the index structure of $B$, we may ask whether there are any solutions to $\psi = B^{-1} \psi^\ast$.
To this end, define the antilinear (conjugate-linear) map $R\from V_S \to V_S$ by $\psi \mapsto B^{-1} \psi^\ast$.
Squaring this gives
\begin{equation}
    R^2(\psi)
    = R(B^{-1} \psi^\ast)
    = B^{-1} (B^\ast)^{-1} \psi
    = z \psi,
\end{equation}
where $z=\pm 1$ after the above rescaling of $B$.
Consider first $z = +1$ so that $R^2$ is the identity map.
We claim that there is a basis in which $B$ coincides numerically with the unit matrix, which motivates the notation $B_{\bar A}{}^{B} = \delta_{\bar A}^{B}$.
To see this, pick any $\psi$ and let $\chi = R(\psi)$.
Suppose that $\psi$ and $\chi$ are linearly dependent so that $\chi = c \psi$ for some constant $c$.
By antilinearity and $R^2 = \1$, $\psi = R(\chi) = \bar c \chi = |c|^2 \psi$ whence $c=\e^{\i \theta}$.
Now, with $\psi' = \e^{\i \varphi} \psi$, $R(\psi') = \e^{\i(\theta - 2 \varphi)} \psi'$.
Thus, for an appropriate choice of $\varphi$, we get $R(\psi') = \psi'$.
Taking $\psi'$ as our first basis vector, $B$ becomes block diagonal with a $1$ in the upper left corner.
If, on the other hand, $\psi$ and $\chi$ are linearly independent, we find
\begin{equation}
    R(\psi + \chi) = \chi + \psi,
    \qquad
    R(\i\psi - \i\chi) = -\i\chi + \i\psi.
\end{equation}
In this case, we get two basis vectors, $\psi+\chi$ and $\i\psi - \i\chi$, which are mapped to themselves by $R$.
Proceeding by induction, we conclude that there is a basis such that $B$ is diagonal with $1$s on the diagonal.
Since $B \Gamma^a = \pm (\Gamma^a)^\ast B$, the $\Gamma$-matrices are real ($+$) or imaginary ($-$) in this basis.
In both cases, the generators of $\so(p,q)$ are real.
Furthermore, $\psi = R(\psi)$ reduces to the condition that the components of $\psi$ are real.
This is known as a Majorana basis and the reality condition, $\psi = R(\psi)$, is the Majorana condition.
If $B$ is block diagonal in the Weyl basis, we can define Majorana--Weyl spinors by requiring both $\gamma \psi = \pm \psi$ and $R(\psi) = \psi$.
If $B$ is block antidiagonal, the Majorana condition gives a relation between the left and right-handed components and there are no Majorana--Weyl spinors.
In the above, $B$ is only well-defined up to a global complex phase.
If $B$ is redefined by $B\mapsto \e^{\i \theta} B$, the real subspace defined by the Majorana condition is rotated by $\theta/2$.

In the case $z=-1$, there is clearly no solution to $\psi = R(\psi)$ whence it is not possible to define Majorana spinors.
However, given two Dirac spinors $\psi$ and $\chi$, one may impose the relation $\chi = R(\psi)$, known as the symplectic Majorana condition.%
\footnote{When considering Grassmann-odd spinors, there is another generalisation of the Majorana condition, the graded Majorana condition, which can be imposed on a single Dirac spinor \cite{ref:Kleppe--Wainwright}.}
In this case, $R$ can be used to turn $V_S$ into a quaternionic space by defining $\j \psi = R(\psi)$.%
\footnote{Since $R$ is antilinear, $\i \j = - \j \i$.}
This has applications in extended supersymmetry, see for instance \cite{ref:VanProeyen}.

There are also invariants $(A_\pm)^{\bar A B}$ and $(C_\pm)^{AB}$ satisfying
\begin{equation}
\label{eq:spinors:AC}
    A_\pm \Gamma^a = \pm (\Gamma^a)^\dagger A_\pm,
    \qquad\quad
    C_\pm \Gamma^a = \pm (\Gamma^a)^\transpose C_\pm.
\end{equation}
Note that these equations are linear in $A$ and $C$ whence there are one-dimensional spaces of solutions.
Again, the different signs can be related by $A_- = A_+ \gamma$ and $C_- = C_+ \gamma$.
There is also a relation $A \propto (B^{-1})^\transpose C$, as seen from the index structure or a straightforward calculation.
By defining the Dirac conjugate $\bar\psi = \psi^\dagger A$ and the Majorana conjugate $\tilde\psi = \psi^\transpose C$, we find that the relation $A = (B^{-1})^\transpose C$ is necessary for the Majorana condition $\psi = B^{-1} \psi^\ast$ to take its usual form $\bar\psi = \tilde\psi$.

It is easy to see that $A_\pm^\dagger$ also satisfies \cref{eq:spinors:AC}, whence $A_\pm^\dagger \propto A_\pm$.
When $A$ is rescaled, the phase of the proportionality constant changes.
Thus, we can choose $A$ Hermitian, which makes it well-defined up to a real constant factor.
In the case we are dealing with Majorana spinors, this is not always convenient.
In that case, instead note that $B^\transpose C^\ast B = z_C C_\pm$ by an analogous argument.
With the above normalisation $(B^\ast)^{-1} = \pm B$ (with the plus sign in the Majorana case), it follows that $|z_C| = 1$.
By rescaling $C$, we set $z_C = 1$.
In the Majorana basis, $B=\1$, this means that $C$ is real.
Insisting that $A = (B^{-1})^\transpose C$, we find $A^\dagger = (B^{-1})^\transpose C^\transpose$.
By transposing the equation for $C$ in \cref{eq:spinors:AC}, it is easy to see that $C^\transpose = \pm C$ by using that the solution space is one-dimensional.
Thus, $A$ is (anti-)Hermitian when $C$ is (anti-)symmetric.
Again, $A$ and $C$ are only well-defined up to a real constant after imposing these conditions.

\section{Spinors in arbitrary odd dimension}
Consider now $d=2k+1$.
We may construct a representation of the Clifford algebra by using the $\Gamma$-matrices from $d=2k$ and taking a multiple of $\gamma$ as the last $\Gamma$-matrix.
By adding $\pm \gamma$ to a set of $\Gamma$-matrices with signature $(r,s)$ we end up with signature $(r+1, s)$.
To instead get $(r, s+1)$, $\pm \i \gamma$ should be used as the last $\Gamma$-matrix.
Since there is only one inequivalent representation of $\cliff_{r,s}$ but a sign choice for the last $\Gamma$-matrix when we go up in dimension, there seems to be two inequivalent representations of $\cliff_{r+1,s}$ and $\cliff_{r,s+1}$.
To see that the sign choice really gives inequivalent representations of the $d=2k+1$ Clifford algebras, note that
\begin{equation}
\label{eq:spinors:inequivalent_in_odd_d}
    \Gamma^{d} = c \gamma
    \quad\implies\quad
    \Gamma^{a_1 \hdots a_d}
    = c (-\i)^{k+s} \epsilon^{a_1 \hdots a_d} \1.
\end{equation}
Thus, $c$, which can take two values once the signature is fixed, distinguishes the two representations and they are indeed inequivalent.
These representations are irreducible \cite{ref:Lawson--Michelsohn}.
The inequivalent Clifford representations are related by $\Gamma^a \mapsto -\Gamma^a$.
Note that this implies that the $\so$-representations generated by $\Gamma^{ab}/4$ are the same.
Since there is no longer an invariant $\gamma$, the situation is reversed compared to even dimensions;
in odd dimensions, there are two inequivalent irreducible Clifford algebra representations but only one irreducible spinor representation, the Dirac representation.

Due to \cref{eq:spinors:inequivalent_in_odd_d}, $\Gamma^{a_1 \hdots a_n}$ and $\Gamma^{a_{n+1} \hdots a_d}$ are not linearly independent but related by a contraction with $\epsilon^{a_1 \hdots a_d}$.
Thus, a basis for $\operatorname{End}(V_S, V_S)$ is provided by $\1, \Gamma^{a_1}, \hdots, \Gamma^{a_1 \hdots a_k}$ where, in contrast to the even-dimensional case, there are at most $k$ indices.
Linear independence is proved analogously to the case of even dimension.
Counting the antisymmetrically independent index combinations, we find that the span of $\1, \Gamma^{a_1}, \hdots, \Gamma^{a_1 \hdots a_k}$ is $2^{2k}$-dimensional, which agrees with the dimension of $\operatorname{End}(V_S, V_S)$.

In odd dimension, there are only half as many invariants as in even dimension.
More specifically, only one of the signs in each of $A_\pm$, $B_\pm$ and $C_\pm$ is viable.
Since these invariants intertwine representations of the Clifford algebra, it is easy to see which of them exists by using that the inequivalent representations are distinguished by the sign of $\Gamma^{a_1 \hdots a_d}$.
In \cref{app:conventions:11d-spinors}, we demonstrate this explicitly for $C_\pm$ in $d=11$.

\section{Spinors in four dimensions} \label{app:conventions:4d-spinors}%
In four dimensions, a Dirac spinor, $\Psi_A$, has four components and consists of a left-handed and a right-handed irreducible Weyl spinor, $\psi_\alpha$ and $\bar\chi^{\dot\alpha}$.%
\footnote{Here, we use a dot instead of a bar on complex conjugated indices, as is common in Van der Waerden notation.}

\subsubsection{A chiral basis}
In the chiral (Weyl) basis, $\Psi_A = (\psi_\alpha, \bar\chi^{\dot\alpha})_A$ and (note the factor of $\i$)
\begin{equation}
\label{eq:conventions:4d-spinors:gamma}
    (\gamma^a)\indices{_A^B}
    =
    \i
    \begin{pmatrix}
        0 & \sigma^a_{\alpha \dot\beta} \\
        \bar\sigma^{a \dot\alpha \beta} & 0
    \end{pmatrix}\indices{_{{\mkern-6mu}A}^{B}},
\end{equation}
where, in the basis we choose and with index structure $\sigma^a_{\alpha \dot\beta}$, the Pauli matrices are
\begin{equation}
\label{eq:conventions:4d-spinors:Pauli}
    \sigma^0
    = \begin{pmatrix}
        -1 & 0 \\
        0 & -1
    \end{pmatrix},
    \quad
    \sigma^1
    = \begin{pmatrix}
        0 & 1 \\
        1 & 0
    \end{pmatrix},
    \quad
    \sigma^2
    = \begin{pmatrix}
        0 & -\i \\
        \i & 0
    \end{pmatrix},
    \quad
    \sigma^3
    = \begin{pmatrix}
        1 & 0 \\
        0 & -1
    \end{pmatrix}.
\end{equation}

In four spacetime dimensions, there is a useful exceptional isomorphism $\Spin(3, 1) \simeq \SL(2, \CC)$.
This means that the Weyl spinor representation of $\Spin(3, 1)$ is the defining representation of $\SL(2, \CC)$, whence the antisymmetric tensor $\epsilon_{\alpha\beta}$ is invariant.
This isomorphism is indicated by the fact that there is a one-to-one correspondence between real-valued vectors $v_a$ and Hermitian matrices $V \coloneqq v_a \sigma^a$.
Given $V$, we can construct another Hermitian matrix $V'$ by
\begin{equation}
\label{eq:conventions:4d-spinors:SL2C_transformation}
    V' = \Lambda\, V \Lambda^\dag
\end{equation}
In this transformation, a global phase of $\Lambda$ is irrelevant whence we can demand $\det{\Lambda} \in \RR_{\geq 0}$.
Since $\det{V} = -v^2$, the transformations that preserve $v^2$, or equivalently $\eta_{ab}$, are precisely those with $\det{\Lambda} = 1$, that is, $\Lambda \in \SL(2, \CC)$.
However, $\Lambda = -\1$ is not effective on $V$, whence $\SL(2, \CC)$ is a double cover of $\SO(3,1)$.%
\footnote{Note, however, that $\Lambda = -\1$ is effective in $\psi' = \Lambda \psi$, in perfect agreement with $\Spin(3,1)$ being the double cover of $\SO(3,1)$.}

The above also explains the index structure $\smash[b]{\sigma^a_{\alpha \dot\beta}}$, which is needed for \cref{eq:conventions:4d-spinors:SL2C_transformation} to make sense in index-notation, $V'_{\alpha \dot\beta} = \tensor{\Lambda}{_\alpha^\gamma} \tensor{\bar\Lambda}{_{\dot\beta}^{\dot\delta}} V_{\gamma \dot\delta}$.
This being a Lorentz transformation, that is, $V' = v'_a \sigma^a$ where $v'_a = \tensor{\Lambda}{_a^b} v_b$, further implies that $\sigma^a_{\alpha \dot\beta}$ is an invariant tensor under $\Spin(3,1)$.

As already explained,
\begin{equation}
    \epsilon^{\alpha\beta} =
    \begin{pmatrix}
        0 & 1\\
        -1 & 0
    \end{pmatrix}\indices{^{{\mkern-6mu}\alpha\beta}},
    \qquad\quad
    \epsilon_{\alpha\beta} =
    \begin{pmatrix}
        0 & -1\\
        1 & 0
    \end{pmatrix}\indices{_{{\mkern-6mu}\alpha\beta}},
\end{equation}
are invariant tensors due to $\det \Lambda = 1$ in the spinor representation.
Therefore, we can use them to raise and lower spinor indices.%
\footnote{Dotted indices are raised and lowered using the complex conjugates $\bar\epsilon^{\dot\alpha \dot\beta}$ and $\bar\epsilon_{\dot\alpha \dot\beta}$, which, in our basis, coincide numerically with $\epsilon^{\alpha \beta}$ and $\epsilon_{\alpha \beta}$.}
We do this by left-multiplication, that is, it is always the rightmost index of $\epsilon$ which is contracted with the quantity whose index is being raised or lowered ($\psi^\alpha = \epsilon^{\alpha \beta} \psi_\beta$).
Note that $\epsilon^{\alpha\gamma} \epsilon^{\beta\delta} \epsilon_{\gamma\delta} = \epsilon^{\beta\alpha} \neq \epsilon^{\alpha\beta}$, whence we cannot raise or lower indices on the $\epsilon$-tensors themselves.
This peculiarity is not a problem since two contracted $\epsilon$-tensors can always be written with a Kronecker delta.
Furthermore, these conventions imply that
\begin{equation}
    \psi_\alpha \chi^\alpha
    = \psi_\alpha \epsilon^{\alpha \beta} \chi_\beta
    = - \psi^\beta \chi_\beta.
\end{equation}
Because of this, we need a convention for how to place the indices when switching between index notation and index-free notation.
We use the convention that undotted indices are contracted up-down, while dotted indices are contracted down-up, that is,
\begin{equation}
    \psi \chi
    = \psi^\alpha \chi_\alpha
    = \chi^\alpha \psi_\alpha
    = \chi \psi,
    \qquad\quad
    \bar\psi \bar\chi
    = \bar\psi_{\dot\alpha} \bar\chi^{\dot\alpha}
    = \bar\chi_{\dot\alpha} \bar\psi^{\dot\alpha}
    = \bar\chi \bar\psi,
\end{equation}
for anticommuting (Grassmann-odd) spinors.
Due to how complex conjugation is defined on Grassmann numbers, see \cref{app:Grassmann}, this implies that $(\psi \chi)^\ast = \bar\psi \bar\chi$.

The complex conjugated Pauli matrices $\bar\sigma^{a \dot\alpha \beta}$ are obtained from $\sigma^a_{\alpha \dot\beta}$ by complex conjugation and raising the indices.
Numerically, in our basis and with the above index structure, they are $\bar\sigma^a = (-\1, -\sigma^i)^a$.
In index notation, the statement that $\sigma^a$ are Hermitian reads $\sigma^a_{\alpha \dot\beta} = \bar\sigma^a_{\dot\beta \alpha}$.

One can show that
\begin{equation}
\label{eq:conventions:4d-spinors:sigma_algebra}
    \sigma^{(a}_{\alpha \dot\beta} \bar\sigma^{b) \dot\beta \gamma}
    = - \eta^{ab} \delta_\alpha^\gamma,
    \qquad\quad
    \bar\sigma^{(a| \dot\alpha \beta} \sigma^{|b)}_{\beta \dot\gamma}
    = - \eta^{ab} \delta^{\dot\alpha}_{\dot\beta},
\end{equation}
where the latter is obtained by complex conjugation of the former.
This is equivalent to the Dirac algebra $\{\gamma^a, \gamma^b\} = 2 \eta^{ab}$, with $\gamma^a$ from \cref{eq:conventions:4d-spinors:gamma}.
Using \cref{eq:conventions:4d-spinors:sigma_algebra}, we find $\sigma^a_{\alpha \dot\beta} \bar\sigma^{b \dot\beta \alpha} = -2\eta^{ab}$ since the left-hand side is symmetric in $a\, b$ due to $\sigma^a$ being Hermitian.
We can also derive a type of Fierz identity by writing $\smash[b]{V_{\alpha \dot\beta} = v_a \sigma^a_{\alpha \dot\beta}}$.
Contracting with $\bar\sigma^{b \dot\beta \alpha}$, we find $-2 v^b = \smash[t]{V_{\alpha \dot\beta} \bar\sigma^{b \dot\beta \alpha}}$ and hence
\begin{equation}
\label{eq:conventions:4d-spinors:sigma_Fierz}
    \sigma^a_{\alpha \dot\beta} \bar\sigma_a^{\dot\delta \gamma}
    =
    -2 \delta^\gamma_\alpha \delta^{\dot\delta}_{\dot\beta}.
\end{equation}

Now define $(\sigma^{ab})\indices{_\alpha^\beta} = \sigma^{[a}_{\alpha \dot\gamma} \bar\sigma^{b] \dot\gamma \beta}$.
Using \cref{eq:conventions:4d-spinors:sigma_algebra}, it is straightforward to show that $[\sigma^{ab}, \sigma_{cd}] = 8 \delta^{[a}_{[c} \tensor{\sigma}{^{b]}_{d]}}$, whence $-\sigma^{ab}/4$ are the Lorentz generators in the left-handed spinor representation.
Due to the subtleties when raising and lowering spinor indices, $(\bar\sigma^{ab})\indices{^{\dot\alpha}_{\dot\beta}} = - \bar\sigma^{[a|\dot\alpha \gamma} \sigma^{|b]}_{\gamma \dot\beta}$ and $[\bar\sigma^{ab}, \bar\sigma_{cd}] = 8 \delta^{[a}_{[c} \tensor{\bar\sigma}{^{b]}_{d]}}$.
Hence, $\bar\sigma^{ab}/4$ are the Lorentz generators in the right-handed spinor representation, which is consistent with \cref{tab:conventions:transformation_laws} since $(\sigma^{ab})_{\alpha\beta}$ and $(\bar\sigma^{ab})_{\dot\alpha\dot\beta}$ are symmetric in $\alpha\, \beta$ ($\dot\alpha\, \dot\beta$).%
\footnote{Note that, due to how we raise and lower indices, $-(L^{ab})\indices{^\alpha_\beta}$, rather than $(L^{ab})\indices{^\alpha_\beta}$, are the generators of the dual of the left-handed spinor representation.}
This is precisely what is needed for $\gamma^{ab}/4$ to be the Lorentz generators in the Dirac spinor representation.

We may define the invariant
\begin{equation}
\label{eq:conventions:4d-spinors:gamma5}
    \gamma^5 = -\frac{\i}{4!} \epsilon_{abcd} \gamma^{abcd},
\end{equation}
with the properties $(\gamma^5)^2 = \1$ and $\{\gamma^5, \gamma^a\} = 0$.%
\footnote{In the context of $\Spin(3,1)$, the superscript $5$ is not an index.}
In the Weyl basis,
\begin{equation}
    (\gamma^5)\indices{_A^B}
    =
    \begin{pmatrix}
        \delta_\alpha^\beta & 0 \\
        0 & -\delta^{\dot\alpha}_{\dot\beta}
    \end{pmatrix}\indices{_{{\mkern-6mu}A}^{B}}.
\end{equation}
Hence, $\gamma^5$ may be used to form projection operators $P_{\pm} = (\1 \pm \gamma^5)/2$ onto the two chiralities.

\subsubsection{A real basis}
In eleven-dimensional supergravity, we will use Majorana spinors.
Thus, for the compactification to four dimensions, it is convenient with a Majorana basis, in which the $\gamma$-matrices are real.
Such a basis can easily be constructed for instance by letting
\begin{equation}
    \gamma^0 = \i\sigma^2 \otimes \sigma^1,
    \qquad
    \gamma^1 = \sigma^1 \otimes \sigma^1,
    \qquad
    \gamma^2 = \sigma^3 \otimes \sigma^1,
    \qquad
    \gamma^3 = \1 \otimes \sigma^3,
\end{equation}
where the Pauli matrices are numerically the same as in \cref{eq:conventions:4d-spinors:Pauli}.
These satisfy the Dirac algebra since the Pauli matrices anticommute and square to $\1$.
From \cref{eq:conventions:4d-spinors:gamma5}, it is clear that the chirality projectors $P_{\pm} = (\1 \pm \gamma^5)/2$ are not real in any basis in which $\gamma^a$ are real.
Hence, there are no Majorana--Weyl spinors in four dimensions with Lorentzian signature.

\section{Spinors in eleven dimensions} \label{app:conventions:11d-spinors}%
A spinor in eleven dimensions has 32 components.
In a basis in which the eleven-dimensional $\eta_{\hat A \hat B}$ splits block-diagonally to a four-dimensional $\eta_{ab}$ and a seven-dimensional $\delta_{AB}$, we may construct eleven-dimensional $\Gamma$-matrices $\hat\Gamma^{\hat A}$ as
\begin{equation}
    \hat\Gamma^{\hat A} = (\gamma^a \otimes \1,\ -\gamma^5 \otimes \Gamma^A)^{\hat A},
\end{equation}
where $\gamma^a$ are the four-dimensional $\gamma$-matrices and $\Gamma^A$ are the seven-dimensional $\Gamma$-matrices.%
\footnote{That $\gamma^5$ enters in $\hat\Gamma^A$ corresponds to the fact that the Clifford algebra of the eleven-dimensional space is the $\ZZ_2$-\emph{graded} tensor product of the four and seven-dimensional Clifford algebras \cite{ref:Lawson--Michelsohn}.}
If we use the Majorana basis from \cref{app:conventions:4d-spinors} for $\gamma^a$ and the basis for $\Gamma^A$ given in \cref{app:octonions:spin7}, we get a basis for $\hat\Gamma^{\hat A}$ in which they are real, that is, a Majorana basis.%
\footnote{Note that $\gamma^a$ are real while $\Gamma^a$ and $\gamma^5$ are imaginary in these bases.}

Note that
\begin{equation}
\label{eq:conventions:11d-spinors:Gamma_parity}
    \hat\Gamma^{\hat A_1 \hdots \hat A_{11}}
    = - \epsilon^{\hat A_1 \hdots \hat A_{11}} \1.
\end{equation}
This specifies which of the two inequivalent representations of the Clifford algebra $\hat\Gamma^{\hat A}$ generate.

Since there is precisely one irreducible Dirac spinor representation in eleven dimensions, there must be precisely one irreducible invertible tensor (up to a constant factor) $C^{\alpha \beta}$, where $\alpha$ and $\beta$ are spinor indices, by Shur's lemma.
Suppose that there exists $C_\pm$ such that $C_\pm \hat\Gamma^{\hat A} = \pm (\hat\Gamma^{\hat A})^\transpose C_\pm$.
Since $\pm (\hat\Gamma^{\hat A})^\transpose$ satisfy the Clifford algebra, $C_\pm$ are intertwiners between different representations of the Clifford algebra.
However, with $\hat\Gamma^{\transpose\, \hat A_1 \hdots \hat A_n} \coloneqq (\hat\Gamma^{[\hat A_1})^\transpose \hdots (\hat\Gamma^{\hat A_n]})^\transpose$,
\begin{equation}
\label{eq:conventions:11d-spinors:GammaT_parity}
    \hat\Gamma^{\transpose\, \hat A_1 \hdots \hat A_{11}}
    = (\hat\Gamma^{\hat A_{11} \hdots \hat A_1})^\transpose
    = + \epsilon^{\hat A_1 \hdots \hat A_{11}} \1,
\end{equation}
whence $\hat\Gamma^{\transpose\, \hat A}$ generate the other, inequivalent, irreducible representation of the Clifford algebra.
Thus, $C_+$ cannot exist.%
\footnote{Assuming that $C_+$ exists, we find $\hat\Gamma^{\transpose\, \hat A_1 \hdots \hat A_{11}} = C_+ \hat\Gamma^{[A_1} C_+^{-1} \hdots C_+ \hat\Gamma^{A_n]} C_+^{-1} = - \epsilon^{\hat A_1 \hdots \hat A_{11}} \1$ which contradicts \cref{eq:conventions:11d-spinors:GammaT_parity}.}
By a similar argument, $C_-$ must exist since there are only two inequivalent irreducible representations of the Clifford algebra and $\hat\Gamma^{\hat A}$ and $-\hat\Gamma^{\transpose\, \hat A}$ generate equivalent representations since the representations are distinguished by the sign difference between \cref{eq:conventions:11d-spinors:Gamma_parity,eq:conventions:11d-spinors:GammaT_parity}.
In the following, we write $C$ instead of $C_-$ since $C_+$ does not exist.

Since $C^{\alpha \beta}$ is the only nonvanishing invariant with that index structure, it must be either symmetric or antisymmetric.
Also, if $C^\transpose = \pm C$,%
\footnote{The formula analogous to \cref{eq:conventions:11d-spinors:Gamma_symmetry} for $C_+$ has a factor $(-1)^{n(n-1)/2}$ instead of $(-1)^{n(n+1)/2}$.}
\begin{equation}
\label{eq:conventions:11d-spinors:Gamma_symmetry}
    (C \hat\Gamma^{\hat A_1 \hdots \hat A_n})^\transpose
    = \pm (\Gamma^{\hat A_n})^\transpose \hdots (\Gamma^{\hat A_1})^\transpose C
    %= \pm (-1)^n C \hat \Gamma^{\hat A_n \hdots A_1}
    = \pm (-1)^{n(n+1)/2} C \hat\Gamma^{\hat A_1 \hdots \hat A_n}.
\end{equation}
Using that $(C \hat\Gamma^{\hat A_{(i)}})_{i=0}^5$, where $\hat A_{(i)} = (\hat A_1, \hdots \hat A_i)$ is a multi-index, is a basis for all linear maps from the space of spinors to itself (that is, $32\times32$ matrices once a basis has been chosen), we find that $C^{\alpha \beta} = - C^{\beta \alpha}$, see \cref{tab:conventions:11d-spinors:C_symmetry}.
\begin{table}[H]
    \centering
    \caption[Symmetric and antisymmetric matrices in the $D=11$ $\Gamma$-basis.]{Number of symmetric (S) and antisymmetric (A) matrices in the $\Gamma$-basis, found by \cref{eq:conventions:11d-spinors:Gamma_symmetry}, depending on the sign in $C^\transpose = \pm C$. Since there are $528$ ($496$) (anti)symmetric $32\times 32$ matrices, we conclude that $C^\transpose = - C$.}
\label{tab:conventions:11d-spinors:C_symmetry}
    \begin{tabular}{l rr rr}
        \toprule
        & \multicolumn{2}{c}{$C^\transpose = + C$}
        & \multicolumn{2}{c}{$C^\transpose = - C$} \\
        \cmidrule(rl){2-3} \cmidrule(rl){4-5}
        & S & A & S & A \\ \midrule
        $C$ & 1 &  &  & 1 \\
        $C \hat\Gamma^{\hat A_{(1)}}$ &  & 11 & 11 &  \\
        $C \hat\Gamma^{\hat A_{(2)}}$ &  & 55 & 55 &  \\
        $C \hat\Gamma^{\hat A_{(3)}}$ & 165 &  &  & 165 \\
        $C \hat\Gamma^{\hat A_{(4)}}$ & 330 &  &  & 330 \\
        $C \hat\Gamma^{\hat A_{(5)}}$ &  & 462 & 462 & \\
        \addlinespace[\aboverulesep] \bottomrule
    \end{tabular}
\end{table}

The above algebraic properties of $C$ is all that we will need in calculations.
However, they only define $C$ up to a nonzero constant factor.
In the above Majorana basis, a particular choice of $C$ coincides numerically with $\hat\Gamma^{\hat 0}$.
This is, however, merely a coincidence since $C$ and $\hat\Gamma^{\hat 0}$ transform differently under a change of basis.

We use $C^{\alpha \beta}$ and its inverse to raise and lower spinor indices.
Indices on spinors are raised and lowered by left-multiplication ($\psi^\alpha = C^{\alpha \beta} \psi_\beta$).
We define raising and lowering of the left (right) spinor index on a linear map $M\indices{_\alpha^\beta}$ by multiplication from the left (right), that is $M\indices{^\alpha^\beta} = C^{\alpha \gamma} M\indices{_\gamma^\beta}$ and $M\indices{_\alpha_\beta} = M\indices{_\alpha^\gamma} C_{\gamma \beta}$.
This ensures that contracted indices can be raised and lowered without picking up a sign as long they stand next to each other, for instance, $M\indices{_\alpha^\beta} \psi_\beta = M\indices{_\alpha_\beta} \psi^\beta$ and $M\indices{_\alpha_\beta} \tilde M\indices{^\beta^\alpha} = M\indices{^\alpha^\beta} \tilde M\indices{_\beta_\alpha}$.

\chapter{Octonions} \label{app:octonions}%
Here, we introduce the octonions and, in subsequent sections, relate them to $\Spin(7)$ and the exceptional Lie group $G_2$.
The interested reader can find more details in \cite{ref:Baez}.
The octonions $\OO$ are a real vector space spanned by one real unit, $1$, and seven imaginary units, $o_a$, with multiplication defined by
\begin{equation}
    o_a o_b = -\delta_{ab} + \tensor{a}{_a_b^c} o_c
\end{equation}
and $1$ acting as a multiplicative identity both from the left and right.
Here, the structure constants $\tensor{a}{_a_b_c}$ are totally antisymmetric with independent nonzero components%
\footnote{We use $\delta_{ab}$ and its inverse to raise and lower indices.}
\begin{equation}
\label{eq:octonions:a_def}
    a_{abc} = 1,\quad \text{for}\ abc = 123, 257, 536, 374, 761, 642, 415,
\end{equation}
and the multiplication is extended to all of $\OO$ as to be distributive over addition.
Using the index split $a=(\hati, 0, i)$, this may be written as
\begin{equation}
\label{eq:octonions:a_split}
    a_{\hati \hatj \hat k} = \epsilon_{ijk},\qquad
    a_{i j \hat k} = -\epsilon_{ijk},\qquad
    a_{0 i \hatj} = -\delta_{ij}.
\end{equation}
As presented, the construction might seem arbitrary but the octonions fit into the sequence $\RR$, $\CC$, $\HH$, $\OO$, $\mathbb{S}$, $\hdots$ where $\HH$ are the quaternions, $\mathbb{S}$ the sedenions and every entry is obtained from the previous one through the Cayley--Dickson construction \cite{ref:Baez}.
In each step in this sequence, some structure is lost.
For instance, the complex numbers cannot be ordered in a way compatible with multiplication and the quaternions do not commute.%
\footnote{Another, purely algebraic, structure that is lost is that not every complex number is real in the sense that $x^* = x$ is not generally true for $x\in\CC$.}
In the step to the octonions, associativity is lost.
This means that the associator
\begin{equation}
    [x,y,z] = (xy)z - x(yz)
\end{equation}
is, in general, nonzero for $x,y,z\in\OO$.
However, the associator is completely antisymmetric, whence
\begin{equation}
    x(xy) = (xx) y,
    \qquad
    (xy)y = x(yy),
\end{equation}
for arbitrary $x,y\in\OO$.
Thus, the octonions are said to be alternative.
This property is lost in the next step; the sedenions are nonalternative.%
\footnote{The sedenions do, however, satisfy the weaker property of power-associativity, $x(xx)=(xx)x$.}

Octonion conjugation is defined as
\begin{equation}
    1^\ast = 1,\qquad
    o_a^\ast = -o_a.
\end{equation}
This lets us define a scalar product as
\begin{equation}
    \langle x, y\rangle = \Re(x^\ast y),
\end{equation}
which coincides with the standard scalar product on $\RR^8$ in the basis we have taken and induces a norm
\begin{equation}
    \|x\|^2 = \langle x, x \rangle
    = x^\ast x = x x^\ast,
\end{equation}
which is then the standard norm on $\RR^8$.
This norm satisfies
\begin{equation}
\label{eq:octonions:normed}
    \|xy\| = \|x\| \|y\|,
\end{equation}
whence the octonions are said to be a normed division algebra \cite{ref:Baez}.
This implies that there are no zero-divisors; if $x$ and $y$ are nonzero, $xy$ is also nonzero.
That there are no zero-divisors can also be seen by noting that, due to alternativity,
\begin{equation}
    x^\ast (xy) = (x^\ast x) y = \|x\|^2 y,
\end{equation}
whence multiplication by a nonzero $x$ is inverted by multiplying by
\begin{equation}
    x^{-1} = \frac{x^\ast}{\|x\|^2}.
\end{equation}
Similarly, the proof of \cref{eq:octonions:normed} is
\begin{equation}
    \|x y\|^2
    = (xy)(y^\ast x^\ast)
    = x(y y^\ast) x^\ast
    = \|x\|^2 \|y\|^2.
\end{equation}
Here, we have used $(xy)^\ast = y^\ast x^\ast$ and the fact that $x$, $y$, $x^\ast$ and $y^\ast$ all belong to the associative subalgebra generated by $\Im x$ and $\Im y$ \cite{ref:Baez}.

An automorphism of the octonion algebra is, per definition, an $\RR$-linear invertible map $g\from\OO\to\OO$ preserving the octonion multiplication, that is,
\begin{equation}
    \forall\, x,y\in\OO\qquad g(xy) = g(x) g(y).
\end{equation}
The automorphisms naturally form a group, $\operatorname{Aut}\OO$, with composition as group multiplication.
This is one way to define the exceptional Lie group $G_2$, and the definition we choose in this thesis.
It is worth pointing out that $g(1)=1$ for any $g\in G_2$, which is immediate from the definition.
Also, automorphisms preserve the scalar product, $\delta_{ab}$ and $a_{abc}$.
The Lie algebra $\g_2 = \operatorname{Lie}(G_2)$ is the derivation algebra $\der(\OO)$ of $G_2 = \operatorname{Aut}(\OO)$, that is, the linear transformations $D\from\OO\to\OO$ satisfying
\begin{equation}
    \forall\, x,y\in\OO\qquad D(xy) = x D(y) + D(x) y,
\end{equation}
and is the compact real form of the exceptional Lie algebra with the same name.

\section{\texorpdfstring{$\Spin(7)$}{Spin(7)}, octonions and \texorpdfstring{$G_2$}{G2}}%
\label{app:octonions:spin7}
Denote the $\RR$-linear map from $\OO$ to itself defined by left-multiplication by $x\in\OO$ by $L_x$.
Two such maps can be composed but since the associator is nonvanishing
\begin{equation}
    L_x L_y (z)
    \coloneqq L_x(L_y(z))
    = x(yz)
    \neq (xy)z
    = L_{xy}(z),
\end{equation}
in general.
However, due to alternativity
\begin{align}
\nonumber
    \{L_x, L_y\} z
    &= x(yz) + y(xz)
    = (xy)z - [x,y,z] + (yx)z - [y,x,z]
    =\\
    &= (xy+yx)z
    = L_{\{x,y\}} z.
\end{align}
Thus, since $\{o_a, o_b\} = -2\delta_{ab}$,
\begin{equation}
    \{L_{o_a}, L_{o_b}\}
    = -2\delta_{ab}.
\end{equation}
This is \emph{almost} identical to the anticommutator of two gamma matrices.
To fix the sign, consider the complexified octonions $\CC\otimes \OO$ and define
\begin{equation}
    \Gamma_a \coloneqq -\i L_{o_a}.
\end{equation}
Since these satisfy the correct anticommutation relations, $\CC\otimes \OO$ can be identified with the Dirac spinor representation of $\operatorname{Spin}(7)$.
To derive the matrix representation of $\Gamma_a$ consider $\Gamma_a o_A$ where $A=(\hat 0, a)$ and $o_{\hat 0} = 1\in \OO$.
From the definition, $\Gamma_a 1 = -\i o_a$, whence
\begin{equation}
    \tensor{(\Gamma_a)}{^{\hat 0}_{\hat 0}} = 0,
    \qquad
    \tensor{(\Gamma_a)}{^b_{\hat 0}} = -\i \delta_a^b,
\end{equation}
and $\Gamma_a o_b = -\i o_a o_b = \i\delta_{ab} - \i \tensor{a}{_a_b^c} o_c$, whence
\begin{equation}
    \tensor{(\Gamma_a)}{^{\hat 0}_b} = \i\delta_{ab},
    \qquad
    \tensor{(\Gamma_a)}{^c_b} = -\i \tensor{a}{_a_b^c}.
\end{equation}
Thus, $(\Gamma_a)_{AB}$ is antisymmetric in its spinor indices and the independent nonvanishing components are%
\footnote{With $\Gamma_a \coloneqq \pm \i L_{o_a}$ one gets $(\Gamma_a)_{b {\hat 0}} = \pm \i \delta_{ab}$ and $(\Gamma_a)_{bc} = \mp \i a_{abc}$ where the latter sign changes to $\pm$ if one uses right-multiplication ($\Gamma_a \coloneqq \pm \i R_{o_a}$ where $R_x y = yx$) instead of left-multiplication.}
\begin{equation}
\label{eq:octonions:gamma_def:1}
    (\Gamma_a)_{b {\hat 0}} = -\i \delta_{ab},
    \qquad
    (\Gamma_a)_{bc} = +\i a_{abc}.
\end{equation}
Here, we have used the invariant $\delta_{AB}$ to lower indices.
Note that
\begin{equation}
\label{eq:octonions:C_and_A}
    \delta_{AB} \tensor{(\Gamma_a)}{^B_C}
    = - \tensor{(\Gamma_a)}{^B_A} \delta_{BC},
    \qquad
    \delta_{\bar A B} \tensor{(\Gamma_a)}{^B_C}
    = \tensor{(\bar\Gamma_a)}{^{\bar B}_{\bar A}} \delta_{\bar B C},
\end{equation}
where bars denote complex conjugation, compare to \cref{eq:spinors:AC}.
Thus, $C_{AB} = \delta_{AB}$ is symmetric.
$\Gamma_a$ and $\Gamma_{ab}$ are antisymmetric while $\Gamma_{abc}$ is symmetric and the ones with more indices are related to these using $\epsilon^{a_1\hdots a_7}$.

\Cref{eq:octonions:C_and_A} also implies that $\OO$ is identified with Majorana spinors.
Note, however, that if $\psi$ is Majorana, then $\Gamma_a \psi$ is not Majorana due to the different signs in \cref{eq:octonions:C_and_A}.

Since there is only one spinor representation of $\operatorname{Spin}(7)$ and $[\Gamma_{a_1\hdots a_7}, \Gamma_b]=0$, it follows that $\Gamma_{a_1\hdots a_7} \propto \epsilon_{a_1\hdots a_7} \1$ due to Shur's lemma.
Using \cref{eq:octonions:a_def} we find
\begin{equation}
    (\Gamma_{1\hdots 7})_{{\hat 0} {\hat 0}}
    = (-\i)^7 \Re\bigl[1(o_1(o_2(o_3(o_4(o_5(o_6(o_7 1)))))))\bigr]
    = \i,
\end{equation}
whence
\begin{equation}
\label{eq:octonions:gamma_1-7:1}
    \Gamma_{a_1\hdots a_7}
    = \i \epsilon_{a_1\hdots a_7} \1.
\end{equation}

Consider now the subgroup $H$ of $\operatorname{Spin}(7)$ leaving $o_{\hat 0} = 1 \in\OO$ invariant.
Since the gamma matrices are invariant under $\operatorname{Spin}(7)$, it follows that $H$ is a subgroup of $G_2$.
To find the generators of the corresponding Lie algebra we want to find linear combinations $k^{ab} \Gamma_{ab}$ of $\Gamma_{ab}$ such that $k^{ab} \Gamma_{ab} \eta = 0$ where $\eta = o_{\hat 0}$.
Note that $k^{ab} \Gamma_{ab} \eta$ is an arbitrary homogeneous quadratic polynomial in the imaginary units where in no term $o_c o_d$ is $c=d$, due to the antisymmetrisation.
The polynomial must vanish using the octonion multiplication but should be nontrivial in formal variables.
Using this, it is straightforward to construct the generators by inspecting \cref{eq:octonions:a_split}.
If we start from $o_{0} o_{i}$ we can cancel the result using
\begin{equation}
    2 o_{0} o_i + \tensor{\epsilon}{_i^j^k} o_{\hatj} o_{\hat k}
    = (-2+2) o_{\hati}
    = 0,
\end{equation}
whence
\begin{equation}
\label{eq:octonions:G2:0i}
    T_{0 i} = 2 \Gamma_{0 i} + \tensor{\epsilon}{_i^j^k} \Gamma_{\hatj \hat k}
\end{equation}
are three generators.%
\footnote{Note that $i,j,k,\hdots$ and $\hati, \hatj, \hat k,\hdots$ transform under the same $\SU(2)$ whence, for instance, $\hati$ can be contracted with either $i$ or $\hati$.}
Starting from $o_{[i} o_{j]} = -\epsilon_{ij}{}^k o_{\hat k}$ we can similarly cancel the result using $o_{[\hati} o_{\hatj]}$, resulting in generators
\begin{equation}
\label{eq:octonions:G2:ij}
    T_{i j} = \Gamma_{ij} + \Gamma_{\hati \hatj}.
\end{equation}
Since $o_{0} o_i$, $o_i o_j$ and $o_{\hati} o_{\hatj}$ are the only ways to produce $o_{\hat k}$, all vanishing linear combinations of them can be expressed as linear combinations of $T_{0 i}$ and $T_{ij}$.
Analogously, we find
\begin{equation}
\label{eq:octonions:G2:0hati}
    T_{0 \hati} = 2 \Gamma_{0 \hati} + \tensor{\epsilon}{_i^j^k} \Gamma_{j \hat k},
\end{equation}
which is the only way to produce cancelling $o_i$ terms.
Lastly, $o_0$ may be produced from $o_i o_{\hatj}$, $o_{\hati} o_j$ and $o_k o^{\hat k}$.
However, the two former also produce an $o_k$ term.
We can demand that this vanishes since the only other way to cancel it is by using $o_{0} o_{\hati}$ which can always be substituted by $o_{i} o_{\hatj}$ and $o_{\hati} o_j$ terms by adding an appropriate multiple of $T_{0 \hati}$.
Cancellation of the remaining $o_0$ terms determines the prefactor of $o_k o^{\hat k}$.
We find
\begin{equation}
\label{eq:octonions:G2:ihatj}
    T_{i \hatj} = 3 \Gamma_{i \hatj} - 3 \Gamma_{\hati j} - 2 \delta_{ij} \tensor{\Gamma}{_k^{\hat k}}.
\end{equation}
$T_{0 i}$, $T_{0 \hati}$, $T_{ij}$ and $T_{i \hatj}$ clearly span $\operatorname{Lie}(H)$.
Since $T_{0 i}$ and $T_{0 \hati}$ are the only ones containing $\Gamma_{0 i}$ and $\Gamma_{0 \hati}$, respectively, these are linearly independent.
$T_{ij}$ is antisymmetric and independent of the previous ones while $T_{i \hatj}$ is symmetric, traceless and independent of the other generators.
In total, we get $3+3+3+5 = 14$ generators for the subgroup $H$.
Recalling that $H$ is a subgroup not only of $\operatorname{Spin}(7)$ but also of $G_2$, the corresponding Lie algebra is a subalgebra of $\g_2$.
Thus, since $\dim\g_2 = 14$, the Lie algebras are actually the same.
This holds at the group level as well: $G_2$ is the subgroup of $\operatorname{Spin}(7)$ leaving $1\in\OO$ invariant.

\section{Structure constant identities}%
\label{app:octonions:structure_constant_identities}
In this section we present some useful relations for the octonion structure constants.
Recall \cref{eq:octonions:gamma_def:1,eq:octonions:gamma_1-7:1}, which we repeat here for convenience%
\begin{subequations}
\begin{gather}
\label{eq:octonions:gamma_def:2}
    (\Gamma_a)_{b {\hat 0}} = -\i \delta_{ab},
    \qquad
    (\Gamma_a)_{bc} = +\i a_{abc},\\
\label{eq:octonions:gamma_1-7:2}
    \Gamma_{a_1\hdots a_7}
    = \i \epsilon_{a_1\hdots a_7} \1.
\end{gather}
\end{subequations}
Let $\eta = o_{\hat 0} = 1\in\OO$ be the $G_2$-invariant spinor.
We then have
\begin{equation}
\label{eq:octonions:gamma_a_abc}
    a_{abc} = \i \bar\eta \Gamma_{abc} \eta,
\end{equation}
since
\begin{equation}
    \Gamma_{abc} \eta
    = (-\i)^3 o_a (o_b o_c)
    = -\i \delta_{bc} o_a - \i a_{abc} + \i a_{bcd} a_{ade} o_e
\end{equation}
and $\bar\eta$ picks out the $o_{\hat 0}$ term.

Now define the dual of the structure constants
\begin{equation}
    c_{abcd} \coloneqq (\hodge a)_{abcd}
    = \frac{1}{3!} \epsilon_{efg abcd} a^{efg}
    = \frac{1}{6} \epsilon_{abcd efg} a^{efg}.
\end{equation}
Clearly, $c_{abcd}$ is completely antisymmetric and the independent components are
\begin{equation}
    c_{abcd} = 1,\quad \text{for}\ abc = 4567, 1274, 2354, 3164, 1265, 1375, 2376,
\end{equation}
or, with the index split $a=(\hati, 0, i)$,
\begin{equation}
    c_{0ijk} = \epsilon_{ijk},
    \qquad
    c_{0 \hati \hatj k} = -\epsilon_{ijk},
    \qquad
    \tensor{c}{_{\hati \hatj}^{kl}} = -2\delta_{ij}^{kl}.
\end{equation}
From \cref{eq:octonions:gamma_1-7:2} it follows that
\begin{equation}
    \Gamma_{abcd} = -\frac{\i}{6} \epsilon_{abcd efg} \Gamma^{efg},
\end{equation}
whence
\begin{equation}
\label{eq:octonions:gamma_c_abcd}
    c_{abcd} = -\bar\eta \Gamma_{abcd} \eta.
\end{equation}

The outer product $\eta \bar\eta$ can be expanded in terms of gamma matrices.
Lowering the index on $\bar\eta$, this product is $\eta_A \eta_B$ which is symmetric in $A B$ since we consider commuting spinors.%
\footnote{In the compactification of $D=11$ supergravity, $\eta$ is a Grassmann-even spinor on the internal 7-dimensional manifold while spinors on the 4-dimensional spacetime are Grassmann-odd.}
Hence, only terms containing $\1$ and $\Gamma_{abc}$ can enter in the expansion.
Thus we write
\begin{equation}
    \eta \bar\eta = x \1 + x^{abc} \Gamma_{abc}.
\end{equation}
Contracting the spinor indices with $\1$ and $\Gamma_{abc}$, respectively, gives
\begin{subequations}
\begin{alignat}{2}
    &\bar\eta \eta = 8 x
    &&\implies\quad x = \frac{1}{8} \bar\eta \eta,\\
    &\bar\eta \Gamma^{abc} \eta
    = x^{def} \trace(\Gamma^{abc}\Gamma_{def})
    = - 48 x^{abc}
    \quad&&\implies\quad x^{abc} = - \frac{1}{48} \bar\eta \Gamma^{abc} \eta.
\end{alignat}
\end{subequations}
Using the normalisation $\bar\eta \eta = 1$ we get
\begin{equation}
    \eta \bar\eta = \frac{1}{8} - \frac{1}{48} \Gamma^{abc} (\bar\eta \Gamma_{abc} \eta).
\end{equation}
This implies, using $\Gamma_a \Gamma_{bcd} \Gamma^a = -(4-3) \Gamma_{bcd}$,
\begin{equation}
    \Gamma_a \eta \bar\eta \Gamma^a = \frac{7}{8} + \frac{1}{48} \Gamma^{abc} (\bar\eta \Gamma_{abc} \eta).
\end{equation}
Adding these yields the Fierz identity
\begin{equation}
\label{eq:octonions:Fierz}
    \Gamma_a \eta \bar\eta \Gamma^a = 1 - \eta \bar\eta.
\end{equation}
Since, $\Gamma_a$ and $\Gamma_{ab}$ are antisymmetric, $\bar\eta \Gamma_a \eta = 0 = \bar\eta \Gamma_{ab} \eta$.
Using the Fierz identity, we find
\begin{equation}
    \bar\eta \Gamma_{abc} \eta \bar\eta \Gamma^{cde} \eta
    = \bar\eta \Gamma_{ab} (1 - \eta \bar\eta) \Gamma^{de} \eta
    = -2 \delta_{ab}^{de}
    + \bar\eta \tensor{\Gamma}{_a_b^d^e} \eta.
\end{equation}
Thus, using \cref{eq:octonions:gamma_a_abc,eq:octonions:gamma_c_abcd},
\begin{subequations}
\begin{alignat}{2}
    & a_{abc} a^{cde}
    &&= 2\delta_{ab}^{de} + \tensor{c}{_{ab}^{de}},\\
    & a_{abc} a^{bcd}
    &&= 6 \delta_a^d,\\
    & a_{abc} a^{abc}
    &&= 42,
\end{alignat}
\end{subequations}
where the latter two follows from contracting the former.
Through analogous calculations, we find
\begin{subequations}
\begin{alignat}{2}
    & c_{abcd} c^{defg}
    &&= a_{abc} a^{efg} - 9 \tensor{c}{_{[ab}^{[ef}} \delta_{c]}^{g]} - 6 \delta_{abc}^{efg},\\
    & c_{abcd} c^{cdef}
    &&= 8 \delta_{ab}^{ef} + 2 \tensor{c}{_{ab}^{ef}},\\
    & c_{abcd} c^{bcde}
    &&= -24 \delta_a^e,\\
    & c_{abcd} c^{abcd}
    &&= 168,
\end{alignat}
\end{subequations}
and
\begin{subequations}
\begin{alignat}{2}
    & c_{abcd} a^{def}
    &&= 6 \tensor{a}{_{[ab}^{[e}} \delta_{c]}^{f]},\\
    & c_{abcd} a^{cde}
    &&= 4 \tensor{a}{_{ab}^e},\\
    & c_{abcd} a^{bcd}
    &&= 0.
\end{alignat}
\end{subequations}
Using the above, it is easy to show that
\begin{equation}
    \delta_{a_1 a_2}
    = \frac{1}{4!} \tensor{a}{_{a_1}^{b_1 b_2}} \tensor{a}{_{a_2}^{b_3 b_4}} \tensor{c}{_{b_1 b_2 b_3 b_4}}.
\end{equation}

Lastly, there is a useful identity
\begin{equation}
\label{eq:octonions:structure_constant_identity_asym_aa}
    \tensor{a}{_{[a_1 a_2}^{[b_1}} \tensor{a}{_{a_3]}^{b_2 b_3]}}
    = \frac{1}{3} \tensor{a}{_{a_1 a_2 a_3}} \tensor{a}{^{b_1 b_2 b_3}}
    - 2 \tensor{c}{_{[a_1 a_2}^{[b_1 b_2}} \tensor*{\delta}{_{a_3]}^{b_3]}},
\end{equation}
which can be proven straightforwardly by checking it for all index combinations.

\chapter{Differential forms}\label{app:conventions:diff_forms}%
Differential forms are essentially antisymmetric tensors.
Here, we give a brief introduction to some relevant concepts, state our conventions and introduce an index-free formalism.
Consider a connected oriented pseudo-Riemannian $d$-dimensional manifold $\M$ without boundary and let $\Omega^p(\M)$ denote the space of (sufficiently smooth) $p$-forms on $\M$.
On a coordinate chart, we may write $\alpha \in \Omega^p(\M)$ as
\begin{equation}
    \alpha = \frac{1}{p!} \d x^{m_1} \wedge \hdots \wedge \d x^{m_p} \alpha_{m_1\hdots m_p}(x),
\end{equation}
where $\alpha_{m_1\hdots m_p}$ is completely antisymmetric.%
\footnote{Note that we use different conventions for superdifferential forms in superspace, see \cref{sec:superspace:superforms}.}
Since the manifold is oriented, there is a canonical volume form
\begin{equation}
    \vol = \frac{\sqrt{|g|}}{d!} \dperm_{m_1\hdots m_d} \d x^{m_1} \wedge \hdots \wedge \d x^{m_d},
\end{equation}
where $\dperm_{m_1\hdots m_d}$ is the covariant Levi-Civita symbol, the covariant tensor density of weight $+1$ with $\dperm_{1\hdots d} = +1$, and $g$ is the determinant of the metric of tensor-density-weight $-2$.
Equivalently, this may be written as
\begin{equation}
    \sqrt{|g|}\, \d x^{m_1} \wedge \hdots \wedge \d x^{m_d}
    = \uperm^{m_1 \hdots m_d} \vol,
\end{equation}
where $\uperm^{m_1 \hdots m_d}$ is the contravariant Levi-Civita symbol, that is, the contravariant tensor density of weight $-1$ with $\uperm^{1\hdots d} = +1$.
Note that, since $\sqrt{|g|}$ is a pseudo-tensor density of weight $-1$, the volume form is a pseudo-tensor density of weight $0$.
Also, we use different symbols $\dperm$ and $\uperm$ to distinguish between the covariant and contravariant permutation symbols since
\begin{equation}
    g_{m_1 n_1} \hdots g_{m_d n_d} \uperm^{n_1 \hdots n_d}
    = \uperm_{m_1\hdots m_d}
    = g\, \dperm_{m_1\hdots m_d}.
\end{equation}

Using the metric, we may define a pointwise inner product of $p$-forms $\alpha$ and $\beta$ as%
\footnote{For this to be positive definite, we need to restrict to Euclidean signature.}
\begin{equation}
\label{eq:conventions:diff_forms:form_pointwise_inner_product}
    (\alpha, \beta)
    = \frac{1}{p!} \alpha_{m_1\hdots m_p} \beta_{n_1\hdots n_p} g^{m_1 n_1}\hdots g^{m_p n_p}.
\end{equation}
The normalisation here is chosen such that, in the flat Euclidean case, $g_{mn} = \delta_{mn}$, the pointwise norm $\|\alpha\|^2 = (\alpha, \alpha)$ is $1$ for $\alpha = \d x^1 \wedge \hdots \wedge \d x^p$ where $p \leq d$.
To see this, note that the tensor representation of $\d x^{1} \wedge \hdots \wedge \d x^{p}$ has a $\pm 1$ on every position which is a permutation of $1,\hdots,p$, whence the contraction with the metric in \cref{eq:conventions:diff_forms:form_pointwise_inner_product} gives $p!$ which is then cancelled by the prefactor $1/p!$.

\section{The Hodge dual}
Using the volume form and the pointwise inner product, we can define the Hodge star operator $\hodge\from \Omega^{p}(\M) \to \Omega^{d-p}(\M)$ by
\begin{equation}
    \forall \alpha,\beta \in \Omega^{p}(\M)\colon\qquad
    \alpha \wedge (\hodge\beta) = (\alpha, \beta) \vol.
\end{equation}
Here, $\hodge\beta$ is referred to as the Hodge dual of $\beta$.
The Hodge star operator is linear and well defined \cite{ref:Madsen--Tornehave}.
Note that $\hodge 1 = \vol$ since, by the definition,
\begin{equation}
    \hodge 1 = 1 \wedge \hodge 1 = (1,1)\, \vol = \vol,
\end{equation}
where $1$ is interpreted as the constant function on $\M$ with value $1$.
Explicitly,
\begin{equation}
    \hodge \alpha
    = \frac{\sqrt{|g|}}{p!(d-p)!}
    \dperm_{m_1\hdots m_d} g^{m_1 n_1} \hdots g^{m_p n_p} \alpha_{n_1 \hdots n_p} \d x^{m_{p+1}} \wedge\hdots\wedge \d x^{m_d}.
\end{equation}
This is seen from
\begin{align}
\nonumber
    \alpha \wedge \hodge\beta
    &= \frac{1}{p!} \alpha_{m_1\hdots m_p} \d x^{m_1} \wedge\hdots \d x^{m_p}
    \wedge\\ \nonumber
    &\quad\wedge
    \frac{\sqrt{|g|}}{p!(d-p)!} \dperm_{n_1\hdots n_d} g^{n_1 q_1} \hdots g^{n_p q_p} \beta_{q_1\hdots q_p} \d x^{n_{p+1}} \wedge \hdots \wedge \d x^{n_d}
    =\\ \nonumber
    &= \frac{1}{p!^2 (d-p)!} \alpha_{m_1\hdots m_p} \beta^{n_1\hdots n_p} \dperm_{n_1\hdots n_d} \uperm^{m_1\hdots m_p n_{p+1} \hdots n_d} \vol
    =\\
    &= \frac{1}{p!} \alpha_{m_1\hdots m_p} \beta^{m_1\hdots m_p} \vol
    = (\alpha, \beta) \vol,
\end{align}
where, in the second to last step, we have used
\begin{equation}
    \dperm_{n_1\hdots n_d} \uperm^{m_1\hdots m_p n_{p+1} \hdots n_d}
    = p! (d-p)!\, \delta_{n_1}^{m_1}{}_{\hdots}^{\hdots}{}_{n_p}^{m_p}.
\end{equation}
Since $\hodge\alpha$ is a $(d-p)$-form, its components are read off as
\begin{equation}
\label{eq:convetions:diff_forms:hodge_components}
    (\hodge\alpha)_{m_{p+1} \hdots m_d}
    = \frac{\sqrt{|g|}}{p!} \dperm_{m_1\hdots m_d} \alpha^{m_1 \hdots m_p}.
\end{equation}
From this, it follows that
\begin{align}
\nonumber
    \hodge^2 \alpha
    &= \frac{\sqrt{|g|}}{p!(d-p)!} \dperm_{m_1\hdots m_d} (\hodge\alpha)^{m_1\hdots m_{d-p}}
    \d x^{m_{d-p+1}} \wedge \hdots \wedge \d x^{m_d}
    =\\ \nonumber
    &= \frac{|g|}{p!^2(d-p)!} \dperm_{m_1\hdots m_d} \dperm^{n_1\hdots n_p m_1\hdots m_{d-p}}
    \alpha_{n_1\hdots n_p} \d x^{m_{d-p+1}} \wedge \hdots \wedge \d x^{m_d}
    =\\ \nonumber
    &= (-1)^{p(d-p)} \frac{\sign{g}}{p!^2(d-p)!} \dperm_{m_{d-p+1}\hdots m_d m_1 \hdots m_{d-p}} \uperm^{n_1\hdots n_p m_1\hdots m_{d-p}}
    \alpha_{n_1\hdots n_p}
    \cdot\\ \nonumber
    &\quad \cdot \d x^{m_{d-p+1}} \wedge \hdots \wedge \d x^{m_d}
    =\\ \nonumber
    &= (-1)^{p(d-p)} \sign{g}\, \frac{1}{p!}
    \alpha_{n_1\hdots n_p} \d x^{n_1} \wedge\hdots\wedge \d x^{n_p}
    = (-1)^{p(d-p)} \sign{g}\: \alpha,
    \\[2pt]
    \therefore\quad
    \hodge^2 &= (-1)^{p(d-p)} \sign{g}.
\end{align}
Thus, the Hodge star operator gives a natural isomorphism $\Omega^p(\M) \simeq \Omega^{d-p}(\M)$.
Note that $\sign g$ only depends on the signature of the metric and is $+1$ ($-1$) for Euclidean (Lorentzian) signature.

\section{The exterior derivative and de Rham cohomology}
Differential forms can be differentiated in a coordinate-independent manner without the use of a covariant derivative.
The exterior derivative, $\d$, is the unique linear operator $\d\from \Omega^p(\M) \to \Omega^{p+1}(\M)$ satisfying
\begin{subequations}
\begin{alignat}{2}
    &f\in \Omega^0(U),\quad
    &&\d f = \d x^m \partial_m f,
    \\
    & &&\d^2 = 0,
    \\
    &\alpha\in \Omega^p(\M),\quad
    &&\d (\alpha \wedge \beta)
    = \d \alpha \wedge \beta + (-1)^p \alpha \wedge \d \beta,
\end{alignat}
\end{subequations}
where $U \subseteq \M$ is any coordinate patch \cite{ref:Madsen--Tornehave}.
In local coordinates, for a $p$-form $\alpha$,
\begin{subequations}
\label{eq:conventions:diff_forms:exterior_derivative_components}
\begin{align}
    &\d \alpha
    = \frac{1}{p!} \partial_n \alpha_{m_1 \hdots m_p} \d x^n \wedge \d x^{m_1} \wedge \hdots \wedge \d x^{m_p},
    \\
    & (\d \alpha)_{n m_1 \hdots m_p} = (p+1) \partial_{[n} \alpha_{m_1 \hdots m_p]}.
\end{align}
\end{subequations}
Note that we may replace the partial derivative, $\partial_n$, by the Levi-Civita connection, that is, the unique metric-compatible torsion-free affine connection, $\nabla_n$.
This follows from the Christoffel symbols being symmetric in their lower indices.

A $p$-form $\alpha$ is said to be closed if $\d \alpha = 0$ and exact if $\alpha = \d \beta$ for some $(p-1)$-form $\beta$.
Note that all exact $p$-forms are closed since $\d^2 = 0$.
Hence, the image of $\d\from \Omega^{p-1}(\M) \to \Omega^p(\M)$ is a linear subspace of the kernel of $\d\from \Omega^{p}(\M) \to \Omega^{p+1} (\M)$ and we may define
\begin{equation}
    H^p(\M) \coloneqq \frac{\ker\bigl(\d\from \Omega^{p}(\M) \to \Omega^{p+1} (\M)\bigr)}{\im\bigl(\d\from \Omega^{p-1}(\M) \to \Omega^p(\M)\bigr)}.
\end{equation}
$H^p(\M)$ is a (quotient) vector space known as the $p$'th de Rham cohomology group of $\M$ \cite{ref:Madsen--Tornehave}.
Each element of $H^p(\M)$ is an equivalence class, known as a cohomology class, $[\alpha]=\set{\alpha + \d \beta\colon \beta \in \Omega^{p-1}(\M)}$ where $\d \alpha = 0$, that is, it is a closed $p$-form modulo exact $p$-forms.
Based on intuition from $\RR^n$, one may think that all closed $p$-forms, for $p>0$, are exact, which would render $H^p(\M)$ trivial.
Indeed, this is true for sufficiently nice open subsets of $\RR^n$ (star-shaped) by Poincaré's lemma \cite{ref:Madsen--Tornehave}.
However, it is not true in general, the most obvious counterexample being $\d \theta$ on a circle, where $\theta$ is the usual angular coordinate.%
\footnote{Note that $\theta$ is not a global coordinate on $S^1$. Clearly, $\d\theta$ is exact on the coordinate chart. However, when writing $\d\theta$, we refer to the unique smooth global extension of this local form.}

Due to properties of the exterior derivative and since we do not need a metric on $\M$ to define the cohomology groups, $b_p = \dim H^p(\M)$ are topological invariants of the manifold \cite{ref:Madsen--Tornehave}.
For compact manifolds, $b_p < \infty$ is known as the $p$'th Betti number.
Since there are no $-1$-forms and the only closed $0$-forms are locally constant, $\dim H^0(\M)$ is the number of connected components of $\M$ (1 for connected $\M$).

Another useful result, which we will use below, is the (generalised) Stokes' theorem
\begin{equation}
    \int\limits_{\M} \d \alpha
    =
    \int\limits_{\partial \M} \alpha,
\end{equation}
where $\partial \M$ is the boundary of $\M$ and $\alpha$ is a $(d-1)$-form.%
\footnote{Note that any top form $\alpha$, that is, a $d$-form, can be written as $\alpha(x) = f(x) \vol$.}
In particular, if $\M$ has no boundary, the integral of an exact form vanishes.
This explains why $\d \theta$ on the circle cannot be exact.

\section{The codifferential}
Define the codifferential on $p$-forms by \cite{ref:Joyce}
\begin{equation}
    \dd
    \coloneqq (-1)^p \hodge^{-1} \d \hodge
    = \sign g\, (-1)^{d(p+1)+1} \hodge \d \hodge.
\end{equation}
Note that, while $\d$ raises the form-degree by one unit, $\dd$ lowers it by one unit.
Clearly, $\dd^2 = 0$ by the analogous property of $\d$.
Using the (indefinite) pointwise inner product on $\Omega^p(\M)$, we may define%
\footnote{When we do not write out the integration domain, it should be understood that the integral is over all of $\M$.}
\begin{equation}
\label{eq:conventions:diff_forms:L2_inner_product}
    \langle \alpha, \beta \rangle
    = \int \vol\, (\alpha, \beta).
\end{equation}
The codifferential is the formal adjoint of $\d$ since, by Stokes' theorem and the definition of $\hodge$,
\begin{align}
\nonumber
    0
    &= \int \d (\alpha \wedge \hodge \beta)
    = \int \d \alpha \wedge \hodge \beta
    - (-1)^p \int \alpha \wedge \d (\hodge \beta)
    =\\
    &= \langle \d\alpha, \beta \rangle
    - \langle \alpha, \dd\beta \rangle.
\end{align}
where $\alpha$ is a $(p-1)$-form and $\beta$ a $p$-form.
If $\delta \alpha = 0$, $\alpha$ is said to be coclosed and, if $\alpha = \delta \beta$, it is said to be coexact.

To find an index-expression for $\delta \alpha$, we first define the pseudo-tensor
\begin{equation}
    \epsilon^{m_1 \hdots m_d} = \frac{1}{\sqrt{|g|}} \uperm^{m_1 \hdots m_d}.
\end{equation}
It easy to see that $\epsilon^{m_1 \hdots m_d}$ is covariantly constant with respect to any metric-compatible connection since $\epsilon_{m_1 \hdots m_d} \epsilon^{m_1 \hdots m_d}$ is a constant.
From \cref{eq:conventions:diff_forms:exterior_derivative_components,eq:convetions:diff_forms:hodge_components}, we see that
\begin{align}
\nonumber
    (\hodge \d\alpha)_{m_{p+2}\hdots m_d}
    &= \frac{\sqrt{|g|}}{(p+1)!} \dperm_{m_1 \hdots m_d} (p+1) \nabla^{[m_1} \alpha^{\hdots m_{p+1}]}
    =\\
    &= \frac{\sign g}{p!} \epsilon_{m_1 \hdots m_d} \nabla^{[m_1} \alpha^{\hdots m_{p+1}]}.
\end{align}
By using this to compute $\hodge \d \hodge \alpha$, keeping track of all signs, one finds
\begin{equation}
    (\dd \alpha)_{m_1 \hdots m_{p-1}}
    = - \nabla^n \alpha_{n m_1 \dots m_{p-1}}.
\end{equation}

\section{The Hodge--de Rham operator and harmonic forms} \label{app:Hodge_de_Rham_harmonic_forms}%
Here, we restrict to the case of compact manifolds with Euclidean signature, so that \cref{eq:conventions:diff_forms:L2_inner_product} is the positive definite inner product of the Hilbert space of $L^2$-integrable $p$-forms \cite{ref:Joyce}.
Using the differential, $\d$, and the codifferential, $\dd$, we can define the Hodge--de Rham operator, or Hodge Laplacian,
\begin{equation}
    \Delta_p \coloneqq \dd \d + \d \dd,
\end{equation}
which is a second-order differential operator from $\Omega^{p}(\M)$ to itself.
Note that $\Delta_p$ is self-adjoint.
Also, $\Delta_p$ is nonnegative in the sense that $\langle \alpha, \Delta_p \alpha \rangle \geq 0$ since $\langle \alpha, \dd \d \alpha \rangle = \langle \d \alpha, \d \alpha \rangle \geq 0$ and, similarly, $\langle \alpha, \d \dd \alpha \rangle = \langle \dd \alpha, \dd \alpha \rangle \geq 0$.

If a $p$-form $\alpha$ satisfies $\Delta_p \alpha = 0$, we say that it is harmonic.%
\footnote{This should not be confused with the harmonics of \cref{coset:analysis}.}
Clearly, a closed, coclosed $p$-form $\alpha$ is harmonic.
The converse is also true since, if $\Delta_p \alpha = 0$,
\begin{equation}
    0
    = \langle \alpha, \Delta_p \alpha \rangle
    = \langle \alpha, \dd \d \alpha \rangle
    + \langle \alpha, \d \dd \alpha \rangle
    = \langle \d \alpha, \d \alpha \rangle
    + \langle \dd \alpha, \dd \alpha \rangle,
\end{equation}
which implies $\d \alpha = 0 = \dd \alpha$, since the $L^2$ inner product is positive definite \cite{ref:Joyce}.

There is an orthogonal decomposition $\Omega^{p}(\M) = \ker \d_p \oplus \im \dd_{p+1}$, where the subscripts denote the form-degrees of the differential forms $\d$ and $\dd$ are acting on.
This is seen by noting that $\alpha \in \ker \d$, that is, $\d_p \alpha = 0$, is equivalent to $0 = \langle \beta, \d_p \alpha \rangle = \langle \dd_{p+1} \beta, \alpha \rangle$ for all $\beta$, which per definition means that $\alpha \in (\im \dd_{p+1})^\perp$.
Hence, $\ker \d_p = (\im \dd_{p+1})^\perp$, the orthogonal complement of $\im \dd_{p+1}$.
By a completely analogous argument, $\ker \dd_p = (\im \d_{p-1})^\perp$.
Since all exact forms are closed, $\im \d_{p-1}$ is a linear subspace of $\ker \d_p$.
Thus, we can make the decomposition $\ker \d_p = \im \d_{p-1} \oplus \mathcal{H}^p$, where $\mathcal{H}^p$ is the orthogonal complement of $\im \d_{p-1}$ in $\ker \d_p$, that is, $\mathcal{H}^p = \ker \d_p \cap (\im \d_{p-1})^\perp$.%
\footnote{Similarly, $\ker \dd_p = \im \dd_{p+1} \oplus \mathcal{H}^p$.}
Putting this together, we have found the orthogonal decomposition
\begin{equation}
    \Omega^p(\M) = \mathcal{H}^p \oplus \im \d_{p-1} \oplus \im \dd_{p+1},
    \qquad
    \mathcal{H}^p = \ker \d_p \cap \ker \dd_p,
\end{equation}
known as the Hodge decomposition \cite{ref:Joyce}.
Since $\mathcal{H}^p$ contains all closed, coclosed $p$-forms, it is the space of harmonic $p$-forms.
Also, $\ker \d_p = \im \d_{p-1} \oplus \mathcal{H}^p$ implies that every cohomology class $[\alpha]$ contains precisely one harmonic form and
\begin{equation}
    H^p(\M) = \frac{\ker \d_p}{\im \d_{p-1}} \simeq \mathcal{H}^p.
\end{equation}
Hence, the Betti number $b_p$ is the dimension of the space of harmonic forms or, equivalently, the dimension of the $0$-eigenspace of $\Delta_p$.
Since connected manifolds have $b_0 = 1$, the only harmonic functions on $\M$ are constants.
This depends crucially on $\M$ being compact so that the integrals converge; the space of harmonic functions on $\RR^n$ is infinite-dimensional.

\chapter{Bundles, gauge theory and gravity} \label{app:bundles_gauge_theory_gravity}%
In this appendix, we give a brief introduction to the concept of bundles and Einstein--Cartan gravity, also known as Cartan's formulation of general relativity.
To set the stage for Einstein--Cartan gravity, we present a brief review of some aspects of gauge theory after the introduction to fibre bundles.
Although we give some mathematical details, we do not attempt at a complete or mathematically rigorous presentation but rather to give some intuition for the concepts.

\section{Fibre bundles} \label{app:bundles}%
A fibre bundle over a manifold is a space that locally looks like the product of the manifold and a fibre but may have a different structure globally.
Formally, it consists of a total space $E$, a base space $\M$, a typical fibre $F$ and a projection map $\pi\from E \to \M$ such that for each $x \in \M$ there is an open neighbourhood $U_x$ of $x$ and a diffeomorphism%
\footnote{In the setting of topological spaces this is instead a homeomorphism.}
$\varphi\from \pi^{-1}(U_x) \to U_x \times F$ satisfying $\pi_1 \circ \varphi = \pi$, where $\pi_1\from U_x \times F \to U_x,$ is the natural projection onto the first factor, $(y, f) \mapsto y$, \cite{ref:Madsen--Tornehave,ref:Kobayashi--Nomizu}.
This means that $U_x \times F$ and the preimage $\pi^{-1}(U_x)$, that is, the subset of $E$ that maps onto $U_x$ under the projection, are indistinguishable spaces.
Such a diffeomorphism $\psi$ is called a local trivialisation and is the bundle analogue of a coordinate chart of a manifold.
We think of the total space $E$ as glued-together fibres with one fibre $F_x \simeq F$ for each $x\in \M$.

A simple example of a fibre bundle is a cylinder, $S^1 \times [0, 1]$.
Here, the (typical) fibre is $[0,1]$ and the base space $S^1$.
Since this bundle is globally, and not only locally, a product, it is said to be a trivial bundle.
An example of a nontrivial bundle with the same base space, $S^1$, and the same fibre, $[0, 1]$, is provided by the Möbius loop.
This only looks like a product $S^1 \times [0,1]$ locally.

There are two types of bundles we are especially interested in, namely, vector bundles and principal bundles.
A vector bundle $E$ is a fibre bundle whose fibres, $V_x = \pi^{-1}(x)$, and typical fibre, $V=F$, are vector spaces.
Further, it is required that there is a trivialising cover, that is, an open cover of $\M$ consisting of local trivialisations, such that the maps $v\mapsto \varphi^{-1}(x, v)$ are linear maps between the vector spaces $V$ and $\pi^{-1}(x)$ \cite{ref:Madsen--Tornehave}.
An example of a vector bundle is the tangent bundle $T\M$, consisting of all tangent spaces of the manifold.
Given two local trivialisations $\varphi_{i,j}$ on a pair $U_{i,j}$ of intersecting open sets in $\M$, we may consider $\varphi_{ij} = \varphi_i \circ \varphi_j^{-1}$ which is a map from $U\times F$ to $U \times F$.
By the properties of $\varphi$, we see that $\varphi_{ij}(x, v) = (x, t_{ij}(x) v)$.
Here, $t_{ij}$ is known as a transition function \cite{ref:Kobayashi--Nomizu}.
For the tangent bundle, the transition functions are $\GL(d, \RR)$-valued, where $d = \dim \M$, and correspond to local changes of bases on the tangent spaces.

A vector field is a special case of what is known as a section of a bundle.
Technically, it is a map $X\from \M \to T\M$ such that $\pi \circ X = \id_\M$.
Whereas an arbitrary function from $\M$ to $T\M$ can assign a tangent vector in $T_y \M$ to a point $x\in \M$, the last requirement ensures that $X$ assigns a tangent vector in $T_x \M$ to $x$.
Differential forms are sections of exterior powers of the cotangent bundle.

The other type of bundles we are interested in is principal bundles.
For these, the fibre is a Lie group $F=G$, known as the structure group, which acts transitively, freely and smoothly from the right on the total space $P$ \cite{ref:Kobayashi--Nomizu}.
The group action is required to be compatible with the bundle structure in the sense that the fibres are preserved, that is, $\pi(p\cdot g) = \pi(p)$, and that, for the local trivialisations $\varphi_i\from \pi^{-1}(U_i) \to U_i \times G $, if $\varphi_i(p)=(x, g_1)$ then $\varphi_i(p\cdot g_2) = (x, g_1 g_2)$ or, equivalently, $\varphi_i^{-1}(x, g_1 g_2) = \varphi_i^{-1}(x, g_1) \cdot g_2$.
This means that the fibres $G_x = \pi^{-1}(x)$ are $G$-torsors, that is, they are diffeomorphic to $G$ but lack a preferred choice of identity element.
The transition functions of a principal $G$-bundle are $G$-valued.

A local trivialisation of a principal bundle $\varphi\from P \to U \times G$ determines an embedding $\phi$ of $U$ in $P$ by $\phi(x) = \varphi^{-1}(x, e)$, where $e$ is the identity element of $G$.%
\footnote{This is true for any fibre bundle and one may choose any element of the fibre. For principal bundles, there is a canonical choice provided by $e$. What follows does, however, not hold for arbitrary fibre bundles.}
This goes the other way too, given an embedding $\phi\from U \to P$ there is a (unique) local trivialisation $\varphi$ defined by $\varphi^{-1}(x, g) = \phi(x) \cdot g$.
Hence, a principal bundle is trivial if, and only if, it admits a global smooth section.

An example of a principal $\GL(d, \RR)$-bundle is the frame bundle, consisting of all bases of all tangent spaces of $\M$ \cite{ref:Joyce}.
If we have a metric $g$ on $\M$, we may consider the bundle of orthonormal frames.
The transition functions are then restricted to $\O(d, \RR)$.
The orthonormal frame bundle is a subbundle of the frame bundle and we say that we have a reduction of the structure group from $\GL(d, \RR)$ to $\O(d, \RR)$.
Such a reduction is always possible since all manifolds admit a Riemannian metric.
In general, there may, however, be obstructions to structure group reductions.
For instance, consider the further reduction from $\O(d, \RR)$ to $\SO(d, \RR)$ of the structure group of $T\M$.
This is only possible if there is a bundle of frames such that for every $x\in \M$, all frames in the fibre over $x$ have the same orientation, which is equivalent to the manifold being orientable.
The Möbius loop, now considered as the base manifold, is not orientable.
Another example is provided by considering a reduction of the tangent group to $\set{e}$, the trivial group.
If such a reduction exists, the tangent bundle is trivial and we say that the manifold is parallelisable.

Similar to the above, one can construct a frame bundle associated with any vector bundle, not only the tangent bundle.
It is also possible to go in the opposite direction and construct an associated vector bundle from a principal bundle.
To this end, suppose that we have a principal $G$-bundle with total space $P$.
To construct the associated vector bundle, we additionally need a vector space $V$ and a left-representation $\rho\from G\to \GL(V)$.
Define the right-action $(p, v) \cdot g = (p \cdot g, \rho(g^{-1}) v)$ of $G$ on $P \times V$.
This gives an equivalence relation $(p, v) \sim (p, v) \cdot g$ and we denote the set of equivalence classes $[p, v]$ by $E = P \times_\rho V$.
$E$ can be given a differentiable structure \cite{ref:Kobayashi--Nomizu} and is a vector bundle over $\M$ with fibre $V$ \cite{ref:Joyce}.
The projection $\pi_E\from E \to \M$ is given by $\pi_E([p, v]) = \pi(p)$.
This is well-defined since, choosing another representative, $\pi_E([p\cdot g, \rho(g^{-1}) v]) = \pi(p \cdot g) = \pi(p)$.
To give a vector space structure to the fibres of $E$, note that any two points in the fibre over $x$ can uniquely be written as $[p, v_1]$ and $[p, v_2]$, for any $p \in \pi^{-1}(x)$.
This follows immediately from the requirements on the group action of $G$ on $P$.
Now define $[p, v_1] + c [p, v_2] = [p, v_1 + c v_2]$.
That this addition and scalar multiplication are well-defined is easily seen by taking other representatives of $[p, v_{1,2}]$ and using that $\rho$ is a linear representation.

A local trivialisation $\varphi\from \pi^{-1}(U)\to U\times G$ of $P$ induces a local trivialisation of $E$, $\varphi_E\from \pi_E^{-1}(U) \to U\times V$, by
\begin{equation}
\label{conventions:bundles:associated_trivialisation}
    \varphi_E([p, v]) = \bigl(\pi_1\circ \varphi(p),\ \rho(\pi_2\circ \varphi(p)) v \bigr),
\end{equation}
where $\pi_{1,2}$ are the natural projections onto the first and second factors of $U \times G$, respectively.
By the properties of $\varphi$ and $\rho$, this is well-defined.
Also, by restricting to a single fibre, which amounts to fixing $p\in \pi^{-1}(x)$ by the above remark, we get a linear map $\pi_2\circ\varphi_E\from \pi_E^{-1}(x) \to V$ (the inverse is $v\mapsto \varphi_E^{-1}(x, v)$), which shows that the fibres are isomorphic to $V$.
Lastly, the transition functions of $E$ are $\rho(t_{ij})$, where $t_{ij}$ are the transition functions of $P$, and take values in $\rho(G) \subseteq \GL(V)$.

Sections of an associated bundle $E = P \times_\rho V$ are in one-to-one correspondence with $G$-equivariant functions $P\to V$.
A function $f\from P\to V$ is said to be $G$-equivariant if $f(p\cdot g) = \rho(g^{-1}) f(p)$.
To see the correspondence, suppose that we have such a function $f$.
To construct a section of $E$, simply let $x\mapsto[p, f(p)]$ for any $p \in P_x$.
This is well-defined since any other point in the same fibre $P_x$ is of the form $p\cdot g$ and $[p\cdot g, f(p\cdot g)] = [p, \rho(g) f(p\cdot g)] = [p, f(p)]$ by the equivariance of $f$.
Since $\pi_E([p, f(p)]) = \pi(p) = x$, this is indeed a section.
To go in the other direction, suppose that we have a section $s$ of $E$.
Note that $s\circ\pi(p)$, which is an element of $E_{\pi(p)}$, has a unique representative of the form $(p, v)$ for some $v\in V$.
Thus, we can define $f(p) = v$.
Now, $[p, v] = [p\cdot g, \rho(g^{-1}) v]$ whence $f(p\cdot g) = \rho(g^{-1}) v$, that is, $f$ is equivariant.
These two constructions are clearly inverse.

As explained above, a local trivialisation $\varphi\from \pi^{-1}(U) \to U \times G$ of $P$ induces a local trivialisation $\varphi_E\from \pi_E^{-1}(U) \to U \times V$.
For $x\in U$ we then have $\varphi_E \circ s(x) = (x, v(x))$, whence we may, locally, think of the section $s$ of $E$ as a $V$-valued function $v(x)$ on $U$.
Using that $s(x) = [p, f(p)]$ for any $p\in P_x$ and \cref{conventions:bundles:associated_trivialisation}, we find $v(x) = \rho(\pi_2\circ \varphi(p)) f(p)$.
By choosing $p = \varphi^{-1}(x, e) \eqqcolon \phi(x)$, where $\phi$ is the canonical embedding of $U$ in $P$ as above, we get
\begin{equation}
\label{conventions:bundles:local_section_associated_function}
    v(x) = f(\phi(x)).
\end{equation}
\clearpage\section{Gauge theory} \label{app:gauge_theory}%
In a gauge theory with structure group $G$, the fundamental geometrical object is a principal $G$-bundle over $\M$.%
\footnote{In physics, $G$ is often referred to as the gauge group (for instance the $\U(1)\times\SU(2)\times\SU(3)$ of the Standard Model). We reserve the term gauge group for the gauged structure group, that is, the group of gauge transformations.}
Fields that are charged under the gauge group correspond to sections of associated vector bundles.
Thus, to be able to, for instance, write down a kinetic Lagrangian, we need a way of differentiating such sections.

The tangent space $T_p P$ of the principal bundle $P$ at a point $p$ contains a subspace that is tangent to the fibre of $P$ through $p$, called the vertical subspace and denoted $V_p$.
A principal connection on $P$ is defined by assigning horizontal subspaces $H_p$ to each $p$ such that $V_p \oplus H_p = T_p P$, $H_p$ depends smoothly on $p$ and is equivariant in the sense that $H_{p\cdot g}$ agrees with the pushforward of $H_p$ along $p \mapsto p\cdot g$ \cite{ref:Kobayashi--Nomizu}.
There is a natural way of identifying $V_p$ with $\g = \operatorname{Lie}(G)$ and one can, by using this, define a $\g$-valued 1-form on $P$ as the projection onto $V_p$ in the decomposition $T_p P = V_p \oplus H_p$ \cite{ref:Kobayashi--Nomizu}.
Given a local trivialisation over $U \subseteq \M$, this 1-form can be pulled back to a local $\g$-valued 1-form $A$ on $U$.
In what follows, we use this local description since it is more well-suited for the calculations we are concerned with in the main text, even though this somewhat obscures the geometrical nature of the subject.

Using the local connection form $A$, we define a covariant exterior differential,
\begin{equation}
    \D = \d + A,
\end{equation}
acting on the tensor product of the bundle of differential $p$-forms and a bundle associated with the principal bundle.
Locally, a section of such a bundle can be written as $V^i = \d x^{m_1} \wedge \hdots \wedge \d x^{m_p} V_{m_1 \hdots m_p}{}^i/p!$, where $i$ is an index of some representation of $G$.%
\footnote{We always put the form-indices closest to the symbol on components of $p$-forms with additional indices.}
The covariant exterior differential acting on $V^i$ reads
\begin{equation}
    \D V^i
    = \d V^i + A\indices{^i_j} \wedge V^j.
\end{equation}
Note that the special case of a 0-form is a section of an associated bundle.
A change of local trivialisation of the principal bundle induces a change of trivialisation of the associated vector bundles corresponding to $V^i \mapsto V'^i = g\indices{^i_j}V^j$, where $g\in G$ depends on the spacetime point.%
\footnote{A change of local trivialisation is a passive gauge transformation.}
Demanding that the covariant exterior derivative is covariant, that is, $\D V \mapsto \D' V' = g \D V$, we get, dropping indices,
\begin{equation}
    \D' V' = \d (g V) + A' \wedge gV
    = g\, \d V + \d g \wedge V + A' \wedge gV
    = g(\d V + A \wedge V)
\end{equation}
whence
\begin{equation}
\label{eq:gauge_theory:connection_transformation}
    A' = g A g^{-1} + g\, \d g^{-1},
\end{equation}
where we have used that $0=\d(g g^{-1}) = (\d g) g^{-1} + g\, \d g^{-1}$.
This transformation law can also be derived from the properties of the global connection form on the principal bundle \cite{ref:Kobayashi--Nomizu}.

Given a principal connection, we define its field strength, or curvature 2-form, as
\begin{equation}
    F = \d A + A \wedge A.
\end{equation}
Define the operator $\mathcal{F}$ by $\mathcal{F} V = F \wedge V$.
Then,
\begin{equation}
    \D^2 V
    = \D(\d V + A \wedge V)
    = (\d A \wedge V - A \wedge \d V) + (A \wedge \d V + A \wedge A \wedge V)
    = F \wedge V,
\end{equation}
that is, $\D^2 = \mathcal{F}$, which is known as the Bianchi identity of the first type.
Since $\D \mapsto \D' = g \D g^{-1}$, this implies that $F \mapsto F' = g F g^{-1}$ under a change of local trivialisation.
Thus, $F$ is a tensorial 2-form transforming under the adjoint representation.
This can also be seen by a straightforward calculation,
\begin{equation}
    F'
    = \d(g A g^{-1} + g\, \d g^{-1})
    + (g A g^{-1} + g\, \d g^{-1}) \wedge (g A g^{-1} + g\, \d g^{-1})
    = g F g^{-1}.
\end{equation}
Clearly, $[\D, \mathcal{F}] = 0$, which for the 2-form $F$ is equivalent to $\D F = 0$.
This is the Bianchi identity of the second type and can also be shown as
\begin{equation}
    \D F
    = \D (\d A + A \wedge A)
    = (\d A \wedge A - A \wedge \d A) + [A \wedge F]
    = 0,
\end{equation}
where $[A \wedge F] = A \wedge F - F \wedge A$ appears since $F$ transforms under the adjoint representation.

\section{Einstein--Cartan gravity} \label{app:Cartan_gravity}%
In Cartan's formulation of general relativity, one uses the language of principal bundles and gauge theory to formulate Einstein's theory of gravity.
On a spacetime of dimension $d$, the structure group is $\Spin(d-1, 1)$.
The principal connection is referred to as the spin connection and is denoted by $\omega$.
The curvature 2-form is denoted by $R$.
Given a metric of signature $(d-1, 1)$, the structure group $\GL(d)$ of $T\M$ may always be reduced to $\SO(d-1, 1)$, provided $\M$ is orientable.
Essentially, this amounts to restricting the frame bundle to the subbundle of positively oriented orthonormal frames.
Spinor fields are, however, sections of a vector bundle associated not to $\SO(d-1, 1)$ but $\Spin(d-1, 1)$.
Thus, we need a lift of the structure group $\SO(d-1, 1)$ to the double cover $\Spin(d-1, 1)$, that is, a spin structure, to be able to define spinors.
Apart from the obstructions to orientability \cite{ref:Lawson--Michelsohn,ref:ONeil} and the existence of a Lorentzian metric \cite{ref:ONeil}, there can be further obstructions to such a lift.
Specifically, there is such a lift to $\Spin(d-1, 1)$ if and only if the second Stiefel--Whitney class of $\M$ vanishes \cite{ref:Lawson--Michelsohn}.

Assuming that we have a spin structure, there is a bundle of spin frames and we can take a local section $e_a = e_a{}^m \partial_m$.
There is a dual frame of the cotangent bundle $e^a = \d x^m e_m{}^a$ such that $e_a{}^m e_m{}^b = \delta_a^b$ and $e_m{}^a e_a{}^n = \delta_m^n$.
These are known as vielbeins or, in the case $d=4$, vierbeins.%
\footnote{``Viel'' is German for many while ``vier'' means four.}
The Lorentz indices $a, b, c, \hdots$ are referred to as flat and coordinate indices $m, n, p, \hdots$ as curved.
The vielbeins are used to convert between the two types of indices.
However, the Lorentz covariant derivate $\D$ acts only on flat indices.
This means that we distinguish between, for instance, $T \M$ and the associated vector bundle of $\Spin(d-1, 1)$ carrying the vector representation and view $e_a{}^m$ as an isomorphism between them.
In Einstein--Cartan gravity, we therefore only use curved indices as form-indices.

Since we have reduced $\GL(d)$ to $\SO(d-1, 1)$ by restricting to orthonormal frames, the vielbeins are orthonormal and
\begin{equation}
    g_{mn} = \tensor{e}{_m^a} \tensor{e}{_n^b} \eta_{ab}.
\end{equation}
The word orthonormal is perhaps only appropriate for $\eta_{ab} = \diag(-1, +1, \hdots, +1)_{ab}$.
However, any metric tensor with flat indices and the correct signature can be used.

The torsion 2-form of the spin connection is defined by
\begin{equation}
    T^a \coloneqq \D e^a,
    \quad\implies\quad
    \tensor{T}{_m_n^a}
    = 2 \partial_{[m} \tensor{e}{_{n]}^a}
    + 2 \tensor{\omega}{_{[m|}^a_b} \tensor{e}{_{|n]}^b},
\end{equation}
which implies that $T_{abc} = 2 e_{[b}{}^n \partial_{a]} e_{nc} - 2 \omega_{[ab]c}$.
Since the spin connection is antisymmetric in its last two indices, which follows from it being a principal $\so(d-1,1)$ connection, we get
\begin{equation}
    \omega_{abc}
    = \mathring\omega_{abc} + \kappa_{abc},
    \qquad
    \left\{
    \begin{aligned} % abc - acb - bca
        &\mathring\omega_{abc}
        = e_{[b}{}^n \partial_{a]} e_{nc} - e_{[c}{}^n \partial_{a]} e_{nb} - e_{[c}{}^n \partial_{b]} e_{na},
        \\
        &\kappa_{abc}
        = - \frac{1}{2} \bigl(T_{abc} - T_{acb} - T_{bca}\bigr),
    \end{aligned}
    \right.
\end{equation}
where $\mathring\omega$ is the unique torsion-free spin connection and $\kappa_{abc}$ is referred to as the contorsion tensor.
Note that $T_{abc} = -2 \kappa_{[ab]c}$.
As we will see, this formalism reduces to the ordinary formalism of general relativity if the torsion is constrained to $0$ so that $\omega = \mathring\omega$.

Next, we define the affine connection%
\footnote{We refer to a linear connection on the tangent bundle as an affine connection. The term affine connection could perhaps more appropriately be used for what \cite{ref:Kobayashi--Nomizu} refers to as a generalised affine connection, namely a principal connection on the bundle of affine frames with an $\mathfrak{aff}(d)$-valued connection form.}
$\nabla$ by
\begin{equation}
    \nabla_m V^n = \tensor{e}{_a^n} \D_m V^a.
\end{equation}
We can express $\nabla_m V^n$ as
\begin{align}
\nonumber
    \nabla_m V^n
    &= \tensor{e}{_a^n} (\partial_m V^a + \tensor{\omega}{_m^a_b} V^b)
    = \tensor{e}{_a^n} \partial_m (V^p \tensor{e}{_p^a})
    + \tensor{e}{_a^n} \tensor{\omega}{_m^a_b} \tensor{e}{_p^b} V^p
    =\\ \nonumber
    &= \partial_m V^n
    + (\tensor{e}{_a^n} \tensor{\omega}{_m^a_b} \tensor{e}{_p^b} + \tensor{e}{_a^n} \partial_m  \tensor{e}{_p^a}) V^p
    =\\
    &= \partial_m V^n + \tensor{\affine}{_m^n_p} V^n,
\end{align}
where $\affine$ is the $\gl(d)$-valued connection form of the affine connection.%
\footnote{We use $\affine$ rather than $\Gamma$ to distinguish the Christoffel symbols from $\Gamma$-matrices with three indices.}
Thus, the Christoffel symbols of the affine connection, that is, components of the connection form, are%
\footnote{The placement of indices on $\affine$ is not conventional. We follow the previously stated convention that form-indices are placed first.}
\begin{equation}
\label{eq:Cartan_grav:affine}
    \tensor{\affine}{_m^n_p} = \tensor{e}{_a^n} \tensor{\omega}{_m^a_b} \tensor{e}{_p^b} + \tensor{e}{_a^n} \partial_m  \tensor{e}{_p^a}.
\end{equation}
We write $\nabla$ also for the total covariant derivative, $\nabla_m = \partial_m + \affine_m + \omega_m$, acting on both flat and curved indices.
Since the covariant derivate obeys Leibniz's rule and the definition of the affine connection reads, in terms of the total covariant derivative, $\nabla_m V^n = \tensor{e}{_a^n} \nabla_m V^a$, this definition implies that
\begin{equation}
    \nabla_m \tensor{e}{_a^n} = 0,
    \qquad\quad
    \nabla_m \tensor{e}{_n^a} = 0.
\end{equation}
This is sometimes referred to as the vielbein postulate.
Since $g_{mn} = \eta_{ab} e_m{}^a e_n{}^b$, it immediately follows that $\nabla_m g_{np} = 0$, that is, the affine connection is metric compatible.
Note that the Lorentz covariant derivate $\D$ can always be replaced by the total covariant derivate $\nabla$ but the opposite direction is only possible if all indices that $\nabla$ acts on are $\Spin(d-1, 1)$-indices, that is, flat.
In particular, $T^a = \nabla e^a$ and, by the vielbein postulate,
\begin{equation}
    T^m \coloneqq e_a{}^m T^a
    = \nabla \d x^m
    = \tensor{\affine}{^m_n} \wedge \d x^n.
\end{equation}
Hence, the torsion vanishes precisely when $\affine_{[m}{}^p{}_{n]} = 0$.
This is just the statement that the unique metric-compatible torsion-free affine connection, that is, the Levi-Civita connection, is symmetric in its lower indices.

The curvature 2-form $R_a{}^b = \d \omega_a{}^b + \omega_a{}^c \wedge \omega_c{}^b$ has components given by
\begin{equation}
    \tensor{R}{_m_n_a^b}
    = 2 \partial_{[m} \tensor{\omega}{_{n]}_a^b}
    + 2 \tensor{\omega}{_{[m|a}^c} \tensor{\omega}{_{|n]}_c^b}.
\end{equation}
This expression is structurally identical to the expression for the Riemann tensor in terms of the Christoffel symbols of the Levi-Civita connection.
As remarked in \cite{ref:Jackiw--et_al}, the relation \cref{eq:Cartan_grav:affine} between the affine connection and the spin connection has the structure of the gauge transformation in \cref{eq:gauge_theory:connection_transformation}.
The calculation that the curvature 2-form transforms tensorially goes through even though $e^n{}_a$ is not a local Lorentz transformation.
Thus, the Riemann tensor is obtained by converting the flat indices on the curvature 2-form of the torsion-free spin connection to curved ones using the vielbeins, as expected.
Hence, the framework of general relativity is obtained by demanding that the torsion vanishes.%

Lastly, consider the action
\begin{equation}
    S = \frac{1}{2\kappa} \int \d^D x\, e R,
\end{equation}
where $e$ is the determinant of $e_m{}^a$, $e = \sqrt{|g|}$, and $R = R_a{}^a = R_{ab}{}^{ab}$ is the curvature scalar.
This is known as the Palatini action.
Clearly, it reduces to the Einstein--Hilbert action under the constraint of vanishing torsion since, then, all quantities can be expressed in terms of the metric.
However, we may view the vielbein and spin connection as a priori independent, thereby obtaining a first-order formulation of gravity.
To see that this is classically equivalent to the Einstein--Hilbert action, we need to show that the equation of motion for the spin connection forces it to be torsion-free.
To this end, let $\mathring\omega$ be the torsion-free spin connection and write $\omega = \mathring\omega + \kappa$, where $\kappa$ is the contorsion.
We can make a change of variables and view the contorsion and the vielbein as the fundamental quantities.
Thus, we wish to show that $\kappa = 0$ on-shell.
By the definition of the curvature,
\begin{equation}
\label{eq:gauge_Cartan:arbitrary_R}
    R
    = \d(\mathring\omega + \kappa) + (\mathring\omega + \kappa) \wedge (\mathring\omega + \kappa)
    = \mathring R + \mathring\D \kappa + \kappa \wedge \kappa
    = \mathring R + \D \kappa - \kappa \wedge \kappa,
\end{equation}
where $\mathring R$ is the curvature 2-form of $\mathring\omega$.%
\footnote{Since $\kappa$ is a 1-form in the adjoint representation, $\D \kappa = \d \kappa + \omega \wedge_{\ad} \kappa = \d \kappa + \omega \wedge \kappa + \kappa \wedge \omega$.}
Thus, the action can be written as
\begin{equation}
\label{eq:gauge_Cartan:Palatini:2}
    S = \frac{1}{2 \kappa} \int \d^D x\, e \tensor{e}{_a^m} \tensor{e}{_b^n} \bigl(
        \mathring R\indices{_m_n^a^b}
        + 2 \mathring \D_{[m} \tensor{\kappa}{_{n]}^{ab}}
        + 2 \tensor{\kappa}{_{[m}^{ac}} \tensor{\kappa}{_{n] c}^b}
    \bigr).
\end{equation}
Since the contorsion is antisymmetric in its last two indices, the middle term can be written as $2 \mathring\nabla_m \kappa_n{}^{mn}$, which is a total derivative not contributing to the equation of motion.
Thus, the equation of motion for $\kappa$ is given by the variation of the third term in \cref{eq:gauge_Cartan:Palatini:2}.
A straightforward calculation shows that $\kappa_{mab} = 0$ on-shell.

When matter is added to the Palatini action, there can be extra terms in the equation of motion for the contorsion.
As long as there are no new terms containing derivatives of $\kappa$, we can still solve for the contorsion in terms of the other fields through its equation of motion.
This means that there are no independent propagating degrees of freedom in the contorsion and that we can eliminate it from the theory by substitution in the Lagrangian.
The interaction terms arising in this way would not be present if we instead constrain the torsion to 0.
However, keeping $\kappa$ dynamical, we could just add the negative of these terms to the original Lagrangian to eliminate these interactions ad hoc.
This is essentially the method of Lagrange multipliers.

\chapter{Grassmann numbers} \label{app:Grassmann}%
Grassmann numbers are graded-commutative objects crucial to, for instance, the path integral formulation of theories with fermions and superspace formulations of supersymmetric theories and supergravity.
The Grassmann numbers form a graded algebra over the real or complex numbers.
Here, we introduce some conventions related to Grassmann numbers, in particular concerning differentiation and integration with respect to Grassmann variables.

Let $\theta^\alpha$, $\alpha=1,\hdots,n$, be Grassmann variables with a multiplication satisfying
\begin{equation}
    \theta^\alpha \theta^\beta = - \theta^\beta \theta^\alpha,
\end{equation}
generating the Grassmann algebra.
The grading of the algebra simply counts the numbers of Grassmann variables appearing multiplicatively in an expression.
Elements that are even in the grading, for instance $1$ and $\theta^\alpha \theta^\beta$, are called Grassmann-even while elements that are odd in the grading, for instance $\theta^\alpha$ and $\theta^\alpha \theta^\beta \theta^\gamma$, are Grassmann-odd.
For instance in $D=11$ supergravity, we use real Grassmann variables but here we focus on the complex case since the real case is easily inferred.
We define complex conjugation such that
\begin{equation}
\label{eq:Grassmann_conjugation}
    (\theta^\alpha \theta^\beta)^* = \bar{\theta}^{\dot\beta} \bar{\theta}^{\dot\alpha}.
\end{equation}

We define differentiation with respect to $\theta^\alpha$ and $\bar{\theta}^{\dot\alpha}$ by
\begin{equation}
\begin{alignedat}{3}
    &\partial_\alpha \coloneqq \frac{\partial}{\partial \theta^\alpha},
    \qquad\quad
    &&\frac{\partial}{\partial \theta^\alpha} \theta^\beta \coloneqq \delta_\alpha^\beta,
    \qquad\quad
    &&\frac{\partial}{\partial \theta^\alpha} \bar{\theta}^{\dot\beta} \coloneqq 0,
    \\
    &\bar{\partial}_{\dot\alpha} \coloneqq \frac{\partial}{\partial \bar{\theta}^{\dot\alpha}},
    &&\frac{\partial}{\partial \bar{\theta}^{\dot\alpha}} \bar{\theta}^{\dot\beta} \coloneqq \delta_{\dot\alpha}^{\dot\beta},
    &&\frac{\partial}{\partial \bar{\theta}^{\dot\alpha}} \theta^\beta \coloneqq 0,
\end{alignedat}
\end{equation}
linearity and the graded Leibniz rule.
Thus,
\begin{equation}
    \partial_\alpha (\theta^{\beta_1} \hdots \theta^{\beta_n})
    = \frac{\partial}{\partial \theta^\alpha} (\theta^{\beta_1} \hdots \theta^{\beta_n})
    = n \delta_\alpha^{[\beta_1} \theta^{\beta_2} \hdots \theta^{\beta_n]}.
\end{equation}
If we insist on \cref{eq:Grassmann_conjugation} being valid for Grassmann operators and functions as well, this is consistent with
\begin{equation}
    (\partial_\alpha)^* = -\bar{\partial}_{\dot\alpha}
\end{equation}
since then
\begin{align}
    \nonumber
    \bigl( \partial_\alpha (\theta^{\beta_1} \hdots \theta^{\beta_n}) \bigr)^*
    &= \bigl( n \delta_\alpha^{[\beta_1} \theta^{\beta_2} \hdots \theta^{\beta_n]} \bigr)^*
    = n \delta_{\dot\alpha}^{[\dot\beta_1} \bar{\theta}^{\dot\beta_n} \hdots \bar{\theta}^{\dot\beta_2]}
    \\
    &= -(-1)^n \bar{\partial}_{\dot\alpha} (\bar{\theta}^{\dot\beta_n} \hdots \bar{\theta}^{\dot\beta_1})
    = (-1)^{n+1} n \delta_{\dot\alpha}^{[\dot\beta_n} \bar{\theta}^{\dot\beta_{n-1}} \hdots \bar{\theta}^{\dot\beta_1]},
\end{align}
and an $n$-cycle has parity $(-1)^{n+1}$.

In $d=4$, we can raise and lower the indices using the antisymmetric $\epsilon_{\alpha\beta}$.
Contrary to how we raise and lower indices on other quantities, the indices on $\partial_\alpha$ and $\bar{\partial}_{\dot\alpha}$ are raised and lowered from the right.
This ensures that
\begin{equation}
    \partial^\alpha \theta_\beta = \delta^\alpha_\beta
    \qquad\qquad
    \bar{\partial}^{\dot\alpha} \bar{\theta}_{\dot\beta} = \delta^{\dot\alpha}_{\dot\beta}.
\end{equation}

We also want to be able to integrate over Grassmann variables.
For this, we use the Berezin integral
\begin{equation}
    \int \d\theta_\alpha\, \theta^\beta \coloneqq \delta_\alpha^\beta,
    \qquad\qquad
    \int \d\bar{\theta}_{\dot\alpha}\, \bar{\theta}^{\dot\beta} \coloneqq \delta_{\dot\alpha}^{\dot\beta}.
\end{equation}
Note that $\d\theta_\alpha$ should be interpreted as the integration measure for $\theta^\alpha$, even though the index position is different.
Also, the dimension of the measure is opposite to that of the Grassmann variable to make the integral dimensionless.
The integration measure $\d\theta_\alpha$ should not be confused with the superdifferential form $\d \theta^\alpha$; it should always be clear from the context which of the two is being referred to.
If we make a change of variables $\theta^\alpha \mapsto \tensor{S}{^\alpha_\beta} \theta^\beta$ we see that the measure has to transform like $\d\theta_\alpha \mapsto \d\theta_\beta \tensor{{S^{-1}}}{^\beta_\alpha}$.
This also motivates
\begin{equation}
    \d \theta^\alpha = \d \theta_\beta \epsilon^{\beta\alpha},
\end{equation}
similar to how the index on $\partial_\alpha$ was raised.

By requiring graded linearity we get
\begin{equation}
    \int\d\theta_\alpha\, \theta^{\beta_1}\hdots \theta^{\beta_n}
    = n \delta_\alpha^{[\beta_1} \theta^{\beta_2} \hdots \theta^{\beta_n]}.
\end{equation}
Note that this means that the integration operator and differential operator can be identified
\begin{equation}
    \int \d\theta_\alpha = \partial_\alpha.
\end{equation}
These considerations go through completely analogously for dotted indices.
In particular, this means that
\begin{equation}
    \bigl( \int\d\theta_\alpha \bigr)^* = -\int\d\bar{\theta}_{\dot\alpha}.
\end{equation}

In $d=4$ dimensions we also define
\begin{equation}
    \int\d^2\theta\, \theta^2 \coloneqq 1,
    \qquad\qquad
    \int\d^2\bar{\theta}\, \bar{\theta}^2 \coloneqq 1.
\end{equation}
This is consistent with
\begin{equation}
    \int\d^2 \theta = \frac{1}{4} \epsilon^{\alpha\beta} \int\d\theta_\alpha \int\d\theta_\beta
\end{equation}
since
\begin{equation}
    \epsilon^{\alpha\beta} \int\d\theta_\alpha \int\d\theta_\beta \epsilon_{\gamma\delta} \theta^\gamma \theta^\delta
    = 2 \epsilon^{\alpha\beta} \epsilon_{\gamma\delta} \int\d\theta_\alpha\, \delta_{\beta}^{[\gamma} \theta^{\delta]}
    = 2 \epsilon^{\alpha\beta} \epsilon_{\gamma\delta} \delta_{\alpha\beta}^{\delta\gamma}
    = 4.
\end{equation}
From the above, it is also clear that
\begin{equation}
    \Bigl( \int\d^2\theta \Bigl)^\ast = \int\d^2\bar{\theta}.
\end{equation}
Hence, we can identify
\begin{equation}
    \int \d^2\theta = \frac{1}{4} \partial^\alpha \partial_\alpha = \frac{1}{4} \partial^2,
    \qquad\quad
    \int \d^2\bar{\theta} = \frac{1}{4} \bar{\partial}_{\dot\alpha} \bar{\partial}^{\dot\alpha} = \frac{1}{4} \bar{\partial}^2.
\end{equation}

\chapter{Solving the supergravity Bianchi identities} \label{app:sugra_calc}%
In this appendix, we solve the Bianchi identities
\begin{equation}
\label{eq:sugra_calc:tensor_BI}
    \D_{[A} \tensor{T}{_{BC)}^D}
    + \tensor{T}{_{[AB}^E} \tensor{T}{_{|E|C)}^D}
    = \tensor{R}{_{[ABC)}^D},
    \qquad
    \D_{[A} H_{BCDE)}
    + 2 \tensor{T}{_{[AB}^F} H_{|F|CDE)}
    = 0,
\end{equation}
see \cref{eq:sugra:tensor_BI}, of eleven-dimensional supergravity subject to the constraints that the only nonzero components of $H_{ABCD}$ and $\tensor{T}{_A_B^C}$ are
\begin{subequations}
\begin{gather}
    H_{abcd},
    \qquad\quad
    H_{ab \gamma\delta} = 2\i (\Gamma_{ab})\indices{_\gamma_\delta},
    \\
\label{eq:sugra_calc:T_constraints}
    \tensor{T}{_a_b^\gamma},
    \qquad
    \tensor{T}{_\alpha_\beta^c}
    = 2\i (\Gamma^c)\indices{_\alpha_\beta},
    \qquad
    \tensor{T}{_a_\beta^\gamma}
    = H_{bcde} \bigl(
        k_1 \delta_a^{[b} (\Gamma^{cde]})\indices{_\beta^\gamma}
        + k_2 (\tensor{\Gamma}{_a^{bcde}})\indices{_\beta^\gamma}
    \bigr),
\end{gather}
\end{subequations}
as in \cref{eq:sugra:H_nonzero,eq:sugra:T_nonzero}.
We begin by writing the Bianchi identities for all combinations of bosonic and fermionic indices.
Using the constraints, these read
\begin{subequations}
\label{eq:sugra_calc:BI_T}
\begin{alignat}{7}
    \nonumber
    &(ABC, D)\colon\qquad\quad
    &&\D_{[A} \tensor{T}{_{BC)}^D} \qquad
    &&+ \quad
    &&\tensor{T}{_{[AB}^E} \tensor{T}{_{|E|C)}^D} \quad
    &&= \quad
    &&\tensor{R}{_{[ABC)}^D}, \qquad\qquad
    \\
\label{eq:sugra_calc:BI_T:a}
    &  (\alpha\beta\gamma, d)\colon
    && 0
    && +
    && 0
    && =
    && 0,
    \\
\label{eq:sugra_calc:BI_T:b}
    &  (\alpha\beta\gamma, \delta)\colon
    && 0
    && +
    && 2\i \Gamma^e_{(\alpha\beta} \tensor{T}{_{|e|\gamma)}^\delta}
    && =
    && \tensor{R}{_{(\alpha\beta\gamma)}^\delta},
    \\
\label{eq:sugra_calc:BI_T:c}
    &  (a\beta\gamma, d)\colon
    && 0
    && +
    && 4\i \tensor{T}{_{a (\beta}^\varepsilon} \Gamma^d_{|\varepsilon|\gamma)}
    && =
    && \tensor{R}{_{\beta\gamma a}^d},
    \\
\label{eq:sugra_calc:BI_T:d}
    &  (a\beta\gamma, \delta)\colon
    && 2\D_{(\beta} \tensor{T}{_{\gamma)}_a^\delta}
    && +
    && 2\i \Gamma^e_{\beta\gamma} \tensor{T}{_e_a^\delta}
    && =
    && 2\tensor{R}{_a_{(\beta\gamma)}^\delta},
    \\
\label{eq:sugra_calc:BI_T:e}
    & (ab\gamma, d)\colon
    && 0
    && +
    && 2\i \tensor{T}{_a_b^\varepsilon} \Gamma_{\varepsilon \gamma}^d
    && =
    && 2 \tensor{R}{_{\gamma [a b]}^d},
    \\
\label{eq:sugra_calc:BI_T:f}
    &  (ab\gamma, \delta)\colon
    && 2 \D_{[a} \tensor{T}{_{b]}_\gamma^\delta} + \D_\gamma \tensor{T}{_{ab}^\delta} \quad
    && +
    && 2 \tensor{T}{_{\gamma [a}^\varepsilon} \tensor{T}{_{|\varepsilon|b]}^\delta}
    && =
    && \tensor{R}{_{ab}_\gamma^\delta},
    \\
\label{eq:sugra_calc:BI_T:g}
    & (abc, d)\colon
    && 0
    && +
    && 0
    && =
    && \tensor{R}{_{[abc]}^d},
    \\
\label{eq:sugra_calc:BI_T:h}
    & (abc, \delta)\colon
    && \D_{[a} \tensor{T}{_{bc]}^\delta}
    && +
    && \tensor{T}{_{[ab}^\varepsilon} \tensor{T}{_{|\varepsilon|c]}^\delta}
    && =
    && 0,
\end{alignat}
\end{subequations}
and
\begin{subequations}
\label{eq:sugra_calc:BI_H}
\begin{alignat}{7}
    \nonumber
    &  (ABCDE)\colon\qquad\quad
    && \D_{[A} H_{BCDE]} \quad
    && + \quad
    && 2 \tensor{T}{_{[AB}^F} H_{|F|CDE]} \quad
    && = \quad
    && 0, \qquad\qquad
    \\
\label{eq:sugra_calc:BI_H:a}
    &(\alpha\beta\gamma\delta\varepsilon)\colon
    && 0
    && +
    && 0
    && =
    && 0,
    \\
\label{eq:sugra_calc:BI_H:b}
    & (a\beta\gamma\delta\varepsilon)\colon
    && 0
    && -
    && 8 \Gamma^f_{(\beta\gamma} \Gamma_{|fa|\delta\varepsilon)}
    && =
    && 0,
    \\
\label{eq:sugra_calc:BI_H:c}
    &  (ab\gamma\delta\varepsilon)\colon
    && 0
    && +
    && 0
    && =
    && 0,
    \\
\label{eq:sugra_calc:BI_H:d}
    &  (abc\delta\varepsilon)\colon
    && 0
    && +
    && 4\i (\Gamma^f_{\delta\varepsilon} H_{fabc} - 6 \tensor{T}{_{(\varepsilon|[a}^\zeta} \Gamma_{bc]|\delta)\zeta}) \quad
    && =
    && 0,
    \\
\label{eq:sugra_calc:BI_H:e}
    &  (abcd\varepsilon)\colon
    && \D_\varepsilon H_{abcd}
    && +
    && 12\i \tensor{T}{_{[ab}^\zeta} \Gamma_{cd]\zeta\varepsilon}
    && =
    && 0,
    \\
\label{eq:sugra_calc:BI_H:f}
    &  (abcde)\colon
    && \D_{[a} H_{bcde]}
    && +
    && 0
    && =
    && 0,
\end{alignat}
\end{subequations}
where we have dropped the parentheses around $\Gamma$ to save space.
The result of the following analysis is presented in \cref{sec:sugra_calc:results}.

\subsubsection{\Cref{eq:sugra_calc:BI_H:b}}
We start with \cref{eq:sugra_calc:BI_H:b}, which does not even contain any dynamical field.
Contracting with all symmetric matrices $\Gamma^{c_I \gamma\delta}$, where $c_I$ is a multi-index,
\begin{align}
    \nonumber
    \Gamma^{c_I \gamma\delta} \Gamma^b_{(\beta\gamma} \Gamma_{|ba|\delta\varepsilon)}
    &=
    \frac{1}{6} \Gamma^b_{\beta\gamma} \Gamma^{c_I \gamma\delta} \Gamma_{ba\delta\varepsilon}
    + \frac{1}{6} \Gamma^b_{\beta\delta} \Gamma^{c_I \delta\gamma} \Gamma_{ba\gamma\varepsilon}
    + \frac{1}{6} \Gamma^b_{\beta\varepsilon} \Gamma^{c_I \gamma\delta} \Gamma_{ba\delta\gamma}
    \\ \nonumber &\quad
    + \frac{1}{6} \Gamma^b_{\varepsilon\gamma} \Gamma^{c_I \gamma\delta} \Gamma_{ba\delta\beta}
    + \frac{1}{6} \Gamma^b_{\varepsilon\delta} \Gamma^{c_I \delta\gamma} \Gamma_{ba\gamma\beta}
    + \frac{1}{6} \Gamma^b_{\gamma \delta} \Gamma^{c_I \delta\gamma} \Gamma_{ba\beta\varepsilon}
    =\\
    &= \frac{1}{6} \bigl[
    4 (\Gamma^b \Gamma^{c_I} \Gamma_{ba})_{(\beta\varepsilon)}
    + \trace{(\Gamma^{c_I} \Gamma_{ba})} \Gamma^b_{\beta\varepsilon}
    + \trace{(\Gamma^b \Gamma^{c_I})} \Gamma_{ba \beta\varepsilon}
    \bigr],
\end{align}
and computing the terms
\begingroup
\allowdisplaybreaks
\begin{subequations}
\begin{align}
    &\trace{(\Gamma^b \Gamma^{c_1})} \Gamma_{ba\, \beta\varepsilon}
    = 32\tensor{\Gamma}{^{c_1}_{a\,}_\beta_\varepsilon},
    \\
    &\trace{(\Gamma^b \Gamma^{c_1 c_2})} \Gamma_{ba\, \beta\varepsilon}
    = 0,
    \\
    &\trace{(\Gamma^b \Gamma^{c_1 \hdots c_5})} \Gamma_{ba\, \beta\varepsilon}
    = 0,
    \\[4pt]
    &\trace{(\Gamma^{c_1} \Gamma_{ba})} \Gamma^b_{\beta\varepsilon}
    = 0,
    \\
    &\trace{(\Gamma^{c_1 c_2} \Gamma_{ba})} \Gamma^b_{\beta\varepsilon}
    = 64 \delta_a^{[c_1} \Gamma^{c_2]}_{\beta\varepsilon},
    \\
    &\trace{(\Gamma^{c_1 \hdots c_5} \Gamma_{ba})} \Gamma^b_{\beta\varepsilon}
    = 0,
    \\[4pt]
    &(\Gamma^b \Gamma^{c_1} \Gamma_{ba})_{(\beta\varepsilon)}
    = - 8 \tensor{\Gamma}{^{c_1}_{a\,}_\beta_\varepsilon},
    \\
    &(\Gamma^b \Gamma^{c_1 c_2} \Gamma_{ba})_{(\beta\varepsilon)}
    = - 16 \delta_a^{[c_1} \Gamma^{c_2]}_{\beta\varepsilon},
    \\
    &(\Gamma^b \Gamma^{c_1 \hdots c_5} \Gamma_{ba})_{(\beta\varepsilon)}
    = -(5-5) \tensor{\Gamma}{^{c_1\hdots c_5}_{a\,\beta\varepsilon}}
    = 0,
\end{align}
\end{subequations}
\endgroup
we see that \cref{eq:sugra_calc:BI_H:b} is indeed an identity.

\subsubsection{\Cref{eq:sugra_calc:BI_H:d}}
Next, we use \cref{eq:sugra_calc:BI_H:d} to solve for $k_1$ and $k_2$ in \cref{eq:sugra_calc:T_constraints}.
Expanding $\tensor{T}{_a_\varepsilon^\zeta}$, using \cref{eq:sugra_calc:T_constraints}, and contracting with all symmetric $\Gamma^{d_I \varepsilon \delta}$
\begin{align}
\label{eq:sugra_calc:BI_H:d:solving}
    0
    &=
    \Gamma^{d_I \varepsilon \delta} \Gamma^a_{\delta \varepsilon} H_{a b_1 b_2 b_3}
    + 6 H_{c_1\hdots c_4} \Gamma^{d_I \delta \varepsilon} (k_1 \delta_{[b_1|}^{c_1} \tensor{\Gamma}{^{\hdots c_4}_{\varepsilon}^{\zeta}}
    + k_2 \tensor{\Gamma}{_{[b_1|}^{c_1\hdots c_4}_\varepsilon^\zeta}
    ) \tensor{\Gamma}{_{|b_2 b_3]}_{\zeta\delta}}
    =\\ \nonumber
    &= \trace{(\Gamma^{d_I} \Gamma^a)} H_{a b_1 b_2 b_3}
    + 6 H_{c_1\hdots c_4} \bigl(
    k_1 \delta_{[b_1}^{[c_1} \trace{(\Gamma^{|d_I|} \Gamma^{c_2 c_3 c_4]} \Gamma_{b_2 b_3]})}
    +
    k_2 \trace{(\Gamma^{d_I} \tensor{\Gamma}{_{[b_1}^{c_1 \hdots c_4}} \Gamma_{b_2 b_3]})}
    \bigr).
\end{align}
Calculating the terms
\begingroup
\allowdisplaybreaks
\begin{subequations}
\begin{align}
    &\begin{aligned}
    \trace{(\Gamma^{d_1} \Gamma^a)} H_{a b_1 b_2 b_3}
    = 32 \tensor{H}{^{d_1}_{b_1 b_2 b_3}},
    \end{aligned}
    \\
    &\begin{aligned}
    \trace{(\Gamma^{d_1 d_2} \Gamma^a)} H_{a b_1 b_2 b_3}
    = 0,
    \end{aligned}
    \\
    &\begin{aligned}
    \trace{(\Gamma^{d_1 \hdots d_5} \Gamma^a)} H_{a b_1 b_2 b_3}
    = 0,
    \end{aligned}
    \\[4pt]
    &\begin{aligned}
    \delta_{[b_1}^{[c_1} \trace{(\Gamma^{|d_1|} \Gamma^{c_2 c_3 c_4]} \Gamma_{b_2 b_3]})}
    = -32\cdot 6 \delta_{[b_1}^{[c_1} \eta^{|d_1| c_2} \delta^{c_3 c_4]}_{b_2 b_3]}
    = 192 \eta^{d_1 [c_1} \delta^{c_2 c_3 c_4]}_{b_1 b_2 b_3},
    \end{aligned}
    \\
    &\begin{aligned}
    \delta_{[b_1}^{[c_1} \trace{(\Gamma^{|d_1 d_2|} \Gamma^{c_2 c_3 c_4]} \Gamma_{b_2 b_3]})}
    = 0,
    \end{aligned}
    \\
    &\begin{aligned}
    \eta_{[b_1 |[c_1} \trace{(\Gamma^{|d_1 \hdots d_5|} \Gamma_{c_2 c_3 c_4]|} \Gamma_{b_2 b_3]})}
    =
    -32\cdot 5! \eta_{[b_1 |[c_1} \delta^{d_1 d_2 d_3 d_4 d_5}_{c_2 c_3 c_4]| b_2 b_3]},
    \end{aligned}
    \\[4pt]
    &\begin{aligned}
    \trace{(\Gamma^{d_1} \tensor{\Gamma}{_{[b_1}^{c_1 \hdots c_4}} \Gamma_{b_2 b_3]})}
    = 0,
    \end{aligned}
    \\
    &\begin{aligned}
    \trace{(\Gamma^{d_1 d_2} \tensor{\Gamma}{_{[b_1}^{c_1 \hdots c_4}} \Gamma_{b_2 b_3]})}
    = 0,
    \end{aligned}
    \\
    &\begin{aligned}[b]
    \trace{(\Gamma^{d_1 \hdots d_5} \tensor{\Gamma}{_{[b_1}_{|c_1 \hdots c_4|}} \Gamma_{b_2 b_3]})}
    &= -8 \eta_{[b_1 |[ c_1} \trace{(\Gamma^{d_1 \hdots d_5} \Gamma_{c_2 c_3 c_4]| b_2 b_3]})}
    =\\
    &= 32\cdot 5!\cdot 8  \eta_{[b_1 |[c_1} \delta^{d_1 d_2 d_3 d_4 d_5}_{c_2 c_3 c_4]| b_2 b_3]}.
    \end{aligned}
\end{align}
\end{subequations}
\endgroup
Thus, \cref{eq:sugra_calc:BI_H:d:solving} becomes
\begin{subequations}
\label{eq:sugra_calc:k1_k2}
\begin{alignat}{3}
\label{eq:sugra_calc:k1}
    &0 = 32 \tensor{H}{^{d_1}_{b_1 b_2 b_3}} (1 + 36 k_1)
    \quad &&\implies\quad
    &&k_1 = -\frac{1}{36},
    \\
    &0 = 0, &&
    \\
\label{eq:sugra_calc:k1_k2_relation}
    &0 = -192\cdot 5!
    \delta^{[d_1 d_2}_{[b_1 b_2} \tensor{H}{_{b_3]}^{\hdots d_5]}}
    (k_1 - 8 k_2)
    \quad &&\implies\quad
    &&k_2 = \frac{k_1}{8} = - \frac{1}{288}.
\end{alignat}
\end{subequations}
This solves \cref{eq:sugra_calc:BI_H:d} completely.

\subsubsection{\Cref{eq:sugra_calc:BI_H:e}}
From \cref{eq:sugra_calc:BI_H:e}, we immediately find
\begin{equation}
\label{eq:sugra_calc:BI_H:e:solved}
    \D_\varepsilon H_{abcd}
    = -12\i \tensor{T}{_{[ab}^\zeta}  \Gamma_{cd] \zeta \varepsilon}.
\end{equation}
Here, one could act with another covariant derivative and use the Bianchi identity of the first type.
However, $\D H = 0$ implies $\D^2 H = 0$ whence no information not already contained in \cref{eq:sugra_calc:BI_T,eq:sugra_calc:BI_H} can be extracted in this way.
Similar remarks apply to \cref{eq:sugra_calc:BI_H:f} whence we now turn to \cref{eq:sugra_calc:BI_T}.

\subsubsection{\Cref{eq:sugra_calc:BI_T:c}}
This equation gives $\tensor{R}{_{\beta\gamma}_a^d}$ in terms of $H_{abcd}$ as
\begin{align}
    \nonumber
    \tensor{R}{_{\beta\gamma}_a_d}
    &=
    -\frac{\i}{72} H_{b_1 \hdots b_4}
    (8 \delta_a^{b_1}\tensor{\Gamma}{^{b_2 b_3 b_4}_{(\beta|}^\varepsilon}
    + \tensor{\Gamma}{_a^{b_1 \hdots b_4}_{(\beta|}^{\varepsilon}})
    \Gamma_{d\,\varepsilon|\gamma)}
    =\\
\label{eq:sugra_calc:BI_T:c:solved}
    &=
    -\frac{\i}{72} H_{b_1 \hdots b_4}
    (
    24 \tensor*{\delta}{*^{b_1}_a^{b_2}_d} \Gamma^{b_3 b_4}
    + \tensor{\Gamma}{_{ad}^{b_1 \hdots b_4}}
    )_{\beta\gamma}.
\end{align}
Note that the right-hand side is antisymmetric in $a\ d$, which means that \cref{eq:sugra_calc:BI_T:c} puts no constraint on $H$ and that we have solved it completely.
Had we not put in $H$ in the theory by hand, \cref{eq:sugra_calc:BI_T:c} would have constrained some irreducible components of $\tensor{T}{_a_\alpha^\beta}$.

\subsubsection{\Cref{eq:sugra_calc:BI_T:b}}
Since $R$ is Lie algebra-valued in its two last indices, $\tensor{R}{_{\beta\gamma}_\alpha^\delta} = \frac{1}{4} R_{\beta\gamma ad} \tensor{\Gamma}{^{ad}_\alpha^\delta}$.
Hence, using \cref{eq:sugra_calc:BI_T:c:solved}, \cref{eq:sugra_calc:BI_T:b} becomes
\begin{align}
\nonumber
    &2 H_{a_1\hdots a_4} \Gamma^b_{(\alpha \beta|} (
    k_1 \delta_b^{a_1} \tensor{\Gamma}{^{\hdots a_4}_{|\gamma)}^\delta}
    + k_2 \tensor{\Gamma}{_b^{a_1\hdots a_4}_{|\gamma)}^\delta}
    )
    =\\
    &\qquad =
    H_{a_1\hdots a_4} (
    3 k_1 \delta^{a_1 a_2}_{bc} \tensor{\Gamma}{^{a_3 a_4}_{(\alpha\beta|}}
    + k_2 \tensor{\Gamma}{_{bc}^{a_1 \hdots a_4}_{(\alpha\beta|}}
    )
    \tensor{\Gamma}{^{bc}_{|\gamma)}^\delta}.
\end{align}
Contracting with all symmetric $\Gamma^{d_I \beta \gamma}$, using \cref{eq:sugra_calc:k1_k2_relation} and
\begin{equation}
    \Gamma^{d_I \beta \gamma} \Gamma^{b_J}_{(\alpha \beta} \tensor{\Gamma}{^{a_K}_{\gamma)}^\delta}
    =
    \frac{1}{3}
    \tensor{
    (
    2 \Gamma^{b_J} \Gamma^{d_I} \Gamma^{a_K}
    + \trace{(\Gamma^{b_J} \Gamma^{d_I})} \Gamma^{a_K}
    )}{_\alpha^\delta},
\end{equation}
we get
\begin{alignat}{2}
    \nonumber
    0 = H_{a_1 \hdots a_4}
    &\bigl(\
    32 \Gamma^{a_1} \Gamma_{d_I} \Gamma^{\hdots a_4}
    &&+ 16 \trace{(\Gamma^{a_1} \Gamma_{d_I})} \Gamma^{\hdots a_4}
    +\\ \nonumber
    &+ 4 \Gamma^b \Gamma_{d_I} \tensor{\Gamma}{_b^{a_1\hdots a_4}}
    &&+ 2 \trace{(\Gamma^b \Gamma_{d_I})} \tensor{\Gamma}{_b^{a_1\hdots a_4}}
    +\\ \nonumber
    &- 48 \Gamma^{a_1 a_2} \Gamma_{d_I} \Gamma^{a_3 a_4}
    &&- 24 \trace{(\Gamma^{a_1 a_2} \Gamma_{d_I})} \Gamma^{a_3 a_4}
    +\\
\label{eq:sugra_calc:BI_T:b:solving}
    &-2 \tensor{\Gamma}{_{bc}^{a_1\hdots a_4}}  \Gamma_{d_I} \Gamma^{bc}
    &&- \trace{(\tensor{\Gamma}{_{bc}^{a_1\hdots a_4}}  \Gamma_{d_I})} \Gamma^{bc}
    \bigr).
\end{alignat}
There are eight terms to compute for each number of $d$-indices.
The first one is
\begingroup
\allowdisplaybreaks
\begin{subequations}
\label{eq:sugra_calc:BI_T:b:solving:term_1}
\begin{align}
    &\begin{aligned}[b]
        \Gamma^{[a_1} \Gamma_{d_1} \Gamma^{\hdots a_4]}
        &=
        - \tensor{\Gamma}{_{d_1}^{a_1\hdots a_4}}
        + \delta_{d_1}^{[a_1} \Gamma^{\hdots a_4]}
        - 3 \delta_{d_1}^{[a_1} \Gamma^{\hdots a_4]}
        =\\
        &=
        - \tensor{\Gamma}{_{d_1}^{a_1\hdots a_4}}
        - 2 \delta_{d_1}^{[a_1} \Gamma^{\hdots a_4]},
    \end{aligned}
    \\[2pt]
    &\begin{aligned}[b]
        \Gamma^{[a_1} \Gamma_{d_1 d_2} \Gamma^{\hdots a_4]}
        &=
        \tensor{\Gamma}{_{d_1 d_2}^{a_1 \hdots a_4}}
        + 2 \delta_{[d_1}^{[a_1} \tensor{\Gamma}{_{d_2]}^{\hdots a_4]}}
        + 6 \delta_{[d_1}^{[a_1} \tensor{\Gamma}{^{a_2}_{d_2]}^{\hdots a_4]}}
        +\\
        &\quad
        + 6 \delta_{d_1 d_2}^{[a_1 a_2} \Gamma^{\hdots a_4]}
        - 6 \delta_{d_1 d_2}^{[a_1 a_2} \Gamma^{\hdots a_4]}
        =\\
        &=
        \tensor{\Gamma}{_{d_1 d_2}^{a_1 \hdots a_4}}
        - 4 \delta_{[d_1}^{[a_1} \tensor{\Gamma}{_{d_2]}^{\hdots a_4]}},
    \end{aligned}
    \\[2pt]
    &\begin{aligned}[b]
        \Gamma^{[a_1} \Gamma_{d_1 \hdots d_5} \Gamma^{\hdots a_4]}
        &=
        - \tensor{\Gamma}{_{d_1 \hdots d_5}^{a_1 \hdots a_4}}
        + 5 \delta_{[d_1}^{[a_1} \tensor{\Gamma}{_{\hdots d_5]}^{\hdots a_4]}}
        - 15 \delta_{[d_1}^{[a_1} \tensor{\Gamma}{^{a_2}_{\hdots d_5]}^{\hdots a_4]}}
        +\\
        &\quad
        -60 \delta_{[d_1 d_2}^{[a_1 a_2} \tensor{\Gamma}{_{\hdots d_5]}^{\hdots a_4]}}
        -60 \delta_{[d_1 d_2}^{[a_1 a_2} \tensor{\Gamma}{^{a_3}_{\hdots d_5]}^{a_4]}}
        +\\
        &\quad
        - 180 \delta_{[d_1 d_2 d_3}^{[a_1 a_2 a_3} \tensor{\Gamma}{_{\hdots d_5]}^{a_4]}}
        + 60 \delta_{[d_1 d_2 d_3}^{[a_1 a_2 a_3} \tensor{\Gamma}{^{a_4]}_{\hdots d_5]}}
        + 120 \delta_{[d_1 d_2 d_3 d_4}^{a_1 a_2 a_3 a_3} \Gamma_{d_5]}
        =\\
        &=
        - \tensor{\Gamma}{_{d_1 \hdots d_5}^{a_1 \hdots a_4}}
        - 10 \delta_{[d_1}^{[a_1} \tensor{\Gamma}{_{\hdots d_5]}^{\hdots a_4]}}
        - 120 \delta_{[d_1 d_2 d_3}^{[a_1 a_2 a_3} \tensor{\Gamma}{_{\hdots d_5]}^{a_4]}}
        +\\
        &\quad
        + 120 \delta_{[d_1 d_2 d_3 d_4}^{a_1 a_2 a_3 a_4} \Gamma_{d_5]}.
    \end{aligned}
\end{align}
\end{subequations}
\endgroup
The second one is
\begingroup
\allowdisplaybreaks
\begin{subequations}
\label{eq:sugra_calc:BI_T:b:solving:term_2}
\begin{align}
    &\trace{(\Gamma^{[a_1} \Gamma_{d_1})} \Gamma^{\hdots a_4]}
    = 32 \delta_{d_1}^{[a_1} \Gamma^{\hdots a_4]},
    \\
    &\trace{(\Gamma^{[a_1} \Gamma_{d_1 d_2})} \Gamma^{\hdots a_4]}
    = 0,
    \\
    &\trace{(\Gamma^{[a_1} \Gamma_{d_1 \hdots d_5})} \Gamma^{\hdots a_4]}
    = 0.
\end{align}
\end{subequations}
\endgroup
The third one is
\begingroup
\allowdisplaybreaks
\begin{subequations}
\label{eq:sugra_calc:BI_T:b:solving:term_3}
\begin{align}
    &\begin{aligned}[b]
        \Gamma^b \Gamma_{d_1} \tensor{\Gamma}{_b^{a_1\hdots a_4}}
        &=
        -(6-1) \tensor{\Gamma}{_{d_1}^{a_1 \hdots a_4}}
        - 4\cdot 7 \delta_{d_1}^{[a_1} \Gamma^{\hdots a_4]}
        =\\
        &=
        -5 \tensor{\Gamma}{_{d_1}^{a_1 \hdots a_4}}
        - 28 \delta_{d_1}^{[a_1} \Gamma^{\hdots a_4]},
    \end{aligned}
    \\[2pt]
    &\begin{aligned}[b]
        \Gamma^b \Gamma_{d_1 d_2} \tensor{\Gamma}{_b^{a_1\hdots a_4}}
        &=
        (5-2) \tensor{\Gamma}{_{d_1 d_2}^{a_1 \hdots a_4}}
        - 8 (6-1) \delta_{[d_1}^{[a_1} \tensor{\Gamma}{_{d_2]}^{\hdots a_4]}}
        - 12 \cdot 7 \delta_{d_1 d_2}^{[a_1 a_2} \Gamma^{\hdots a_4]}
        =\\
        &=
        3 \tensor{\Gamma}{_{d_1 d_2}^{a_1 \hdots a_4}}
        - 40 \delta_{[d_1}^{[a_1} \tensor{\Gamma}{_{d_2]}^{\hdots a_4]}}
        - 84 \delta_{d_1 d_2}^{[a_1 a_2} \Gamma^{\hdots a_4]},
    \end{aligned}
    \\[2pt]
    &\begin{aligned}[b]
        \Gamma^b \Gamma_{d_1 \hdots d_5} \tensor{\Gamma}{_b^{a_1\hdots a_4}}
        &=
        - (2-5) \tensor{\Gamma}{_{d_1 \hdots d_5}^{a_1 \hdots a_4}}
        - 20(3-4) \delta_{[d_1}^{[a_1} \tensor{\Gamma}{_{\hdots d_5]}^{\hdots a_4]}}
        +\\
        &\quad
        + 120(4-3) \delta_{[d_1 d_2}^{[a_1 a_2} \tensor{\Gamma}{_{\hdots d_5]}^{\hdots a_4]}}
        + 240(5-2) \delta_{[d_1 d_2 d_3}^{[a_1 a_3 a_3} \tensor{\Gamma}{_{\hdots d_5]}^{a_4]}}
        +\\
        &\quad
        - 120(6-1) \delta_{[d_1 d_2 d_3 d_4}^{a_1 a_2 a_3 a_4} \Gamma_{d_5]}
        =\\
        &=
        3 \tensor{\Gamma}{_{d_1 \hdots d_5}^{a_1 \hdots a_4}}
        + 20 \delta_{[d_1}^{[a_1} \tensor{\Gamma}{_{\hdots d_5]}^{\hdots a_4]}}
        + 120 \delta_{[d_1 d_2}^{[a_1 a_2} \tensor{\Gamma}{_{\hdots d_5]}^{\hdots a_4]}}
        +\\
        &\quad
        + 720 \delta_{[d_1 d_2 d_3}^{[a_1 a_3 a_3} \tensor{\Gamma}{_{\hdots d_5]}^{a_4]}}
        - 600 \delta_{[d_1 d_2 d_3 d_4}^{a_1 a_2 a_3 a_4} \Gamma_{d_5]}.
    \end{aligned}
\end{align}
\end{subequations}
\endgroup
The fourth one is
\begingroup
\allowdisplaybreaks
\begin{subequations}
\label{eq:sugra_calc:BI_T:b:solving:term_4}
\begin{align}
    & \trace{(\Gamma^b \Gamma_{d_1})} \tensor{\Gamma}{_b^{a_1\hdots a_4}}
    = 32 \tensor{\Gamma}{_{d_1}^{a_1 \hdots a_4}},
    \\
    & \trace{(\Gamma^b \Gamma_{d_1 d_2})} \tensor{\Gamma}{_b^{a_1\hdots a_4}}
    = 0,
    \\
    & \trace{(\Gamma^b \Gamma_{d_1 \hdots d_5})} \tensor{\Gamma}{_b^{a_1\hdots a_4}}
    = 0.
\end{align}
\end{subequations}
\endgroup
The fifth term is
\begingroup
\allowdisplaybreaks
\begin{subequations}
\label{eq:sugra_calc:BI_T:b:solving:term_5}
\begin{align}
    &\begin{aligned}[b]
         \Gamma^{[a_1 a_2} \Gamma_{d_1} \Gamma^{a_3 a_4]}
         &=
         \tensor{\Gamma}{_{d_1}^{a_1 \hdots a_4}}
         -2 \delta_{d_1}^{[a_1} \tensor{\Gamma}{^{\hdots a_4]}}
         +2 \delta_{d_1}^{[a_1} \tensor{\Gamma}{^{\hdots a_4]}}
         =\\
         &=
         \tensor{\Gamma}{_{d_1}^{a_1 \hdots a_4}},
    \end{aligned}
    \\[2pt]
    &\begin{aligned}[b]
         \Gamma^{[a_1 a_2} \Gamma_{d_1 d_2} \Gamma^{a_3 a_4]}
         &=
         \tensor{\Gamma}{_{d_1 d_2}^{a_1 \hdots a_4}}
         - 4 \delta_{[d_1}^{[a_1} \tensor{\Gamma}{^{a_2}_{d_2]}^{\hdots a_4]}}
         - 4 \delta_{[d_1}^{[a_1} \tensor{\Gamma}{^{a_2 a_3}_{d_2]}^{a_4]}}
         +\\
         &\quad
         - 2 \delta_{d_1 d_2}^{[a_1 a_2} \Gamma^{\hdots a_4]}
         - 2 \delta_{d_1 d_2}^{[a_1 a_2} \Gamma^{\hdots a_4]}
         + 8 \delta_{d_1 d_2}^{[a_1 a_2} \Gamma^{\hdots a_4]}
         =\\
         &=
         \tensor{\Gamma}{_{d_1 d_2}^{a_1 \hdots a_4}}
         + 4 \delta_{d_1 d_2}^{[a_1 a_2} \Gamma^{\hdots a_4]}
    \end{aligned}
    \\[2pt]
    &\begin{aligned}[b]
         \Gamma^{[a_1 a_2} \Gamma_{d_1 \hdots d_5} \Gamma^{a_3 a_4]}
         &=
         \tensor{\Gamma}{_{d_1\hdots d_5}^{a_1 \hdots a_4}}
         - 5 \delta_{[d_1}^{[a_1} \tensor{\Gamma}{^{a_2}_{\hdots d_5]}^{\hdots a_4]}}
         + 5 \delta_{[d_1}^{[a_1} \tensor{\Gamma}{^{a_2 a_3}_{\hdots d_5]}^{a_4]}}
         +\\
         &\quad
         - 20 \delta_{[d_1 d_2}^{[a_1 a_2} \tensor{\Gamma}{_{\hdots d_5]}^{\hdots a_4]}}
         - 20 \delta_{[d_1 d_2}^{[a_1 a_2} \tensor{\Gamma}{^{\hdots a_4]}_{\hdots d_5]}}
         - 80 \delta_{[d_1 d_2}^{[a_1 a_2} \tensor{\Gamma}{^{a_3}_{\hdots d_5]}^{a_4]}}
         +\\
         &\quad
         - 120 \delta_{[d_1 d_2 d_3}^{[a_1 a_2 a_3} \tensor{\Gamma}{_{\hdots d_5]}^{a_4]}}
         + 120 \delta_{[d_1 d_2 d_3}^{[a_1 a_2 a_3} \tensor{\Gamma}{^{a_4]}_{\hdots d_5]}}
         + 120 \delta_{[d_1 d_2 d_3 d_4}^{a_1 a_2 a_3 a_4} \tensor{\Gamma}{_{d_5]}}
         =\\
         &=
         \tensor{\Gamma}{_{d_1\hdots d_5}^{a_1 \hdots a_4}}
         + 40 \delta_{[d_1 d_2}^{[a_1 a_2} \tensor{\Gamma}{_{\hdots d_5]}^{\hdots a_4]}}
         + 120 \delta_{[d_1 d_2 d_3 d_4}^{a_1 a_2 a_3 a_4} \tensor{\Gamma}{_{d_5]}}
    \end{aligned}
\end{align}
\end{subequations}
\endgroup
The sixth one is
\begingroup
\allowdisplaybreaks
\begin{subequations}
\label{eq:sugra_calc:BI_T:b:solving:term_6}
\begin{align}
    &\trace{(\Gamma^{[a_1 a_2} \Gamma_{d_1})} \Gamma^{a_3 a_4]}
    = 0,
    \\
    &\trace{(\Gamma^{[a_1 a_2} \Gamma_{d_1 d_2})} \Gamma^{a_3 a_4]}
    = - 64 \delta_{d_1 d_2}^{[a_1 a_2} \Gamma^{\hdots a_4]},
    \\
    &\trace{(\Gamma^{[a_1 a_2} \Gamma_{d_1 \hdots d_5})} \Gamma^{a_3 a_4]}
    = 0.
\end{align}
\end{subequations}
\endgroup
The seventh term is
\begingroup
\allowdisplaybreaks
\begin{subequations}
\label{eq:sugra_calc:BI_T:b:solving:term_7}
\begin{align}
    &\begin{aligned}[b]
        \tensor{\Gamma}{_{bc}^{a_1\hdots a_4}}  \Gamma_{d_1} \Gamma^{bc}
        &=
        -(6\cdot 5 - 2\cdot 6) \tensor{\Gamma}{^{a_1 \hdots a_4}_{d_1}}
        + 4(7\cdot 6) \delta_{d_1}^{[a_1} \Gamma^{\hdots a_4]}
        =\\
        &=
        - 18 \tensor{\Gamma}{_{d_1}^{a_1 \hdots a_4}}
        + 168 \delta_{d_1}^{[a_1} \Gamma^{\hdots a_4]},
    \end{aligned}
    \\[2pt]
    &\begin{aligned}[b]
        \tensor{\Gamma}{_{bc}^{a_1\hdots a_4}}  \Gamma_{d_1 d_1} \Gamma^{bc}
        &=
        -(5\cdot 4 - 4\cdot 5 + 2) \tensor{\Gamma}{^{a_1 \hdots a_4}_{d_1 d_2}}
        + 8 (6\cdot 5 - 2\cdot 6) \delta_{[d_1}^{[a_1} \tensor{\Gamma}{^{\hdots a_4]}_{d_2]}}
        +\\
        &\quad
        + 12 (7\cdot 6) \delta_{d_1 d_2}^{[a_1 a_2} \Gamma^{\hdots a_4]}
        =\\
        &=
        - 2 \tensor{\Gamma}{_{d_1 d_2}^{a_1 \hdots a_4}}
        - 144 \delta_{[d_1}^{[a_1} \tensor{\Gamma}{_{d_2]}^{\hdots a_4]}}
        + 504 \delta_{d_1 d_2}^{[a_1 a_2} \Gamma^{\hdots a_4]},
    \end{aligned}
    \\[2pt]
    &\begin{aligned}[b]
        \tensor{\Gamma}{_{bc}^{a_1\hdots a_4}}  \Gamma_{d_1 \hdots d_5} \Gamma^{bc}
        &=
        - (2\cdot 1 - 10 \cdot 2 + 5\cdot 5) \tensor{\Gamma}{_{d_1 \hdots d_5}^{a_1 \hdots a_4}}
        +\\
        &\quad
        + 20 (3\cdot 2 - 8\cdot 3 + 4\cdot 3) \delta_{[d_1}^{[a_1} \tensor{\Gamma}{^{\hdots a_4]}_{\hdots d_5]}}
        +\\
        &\quad
        + 120 (4\cdot 3 - 6\cdot 4 + 3\cdot 2) \delta_{[d_1 d_2}^{[a_1 a_2} \tensor{\Gamma}{^{\hdots a_4]}_{\hdots d_5]}}
        +\\
        &\quad
        - 240 (5\cdot 4 - 4\cdot 5 + 2\cdot 1) \delta_{[d_1 d_2 d_3}^{[a_1 a_2 a_3} \tensor{\Gamma}{^{a_4]}_{\hdots d_5]}}
        +\\
        &\quad
        - 120 (6\cdot 5 - 2\cdot 6 + 1\cdot 0) \delta_{[d_1 d_2 d_3 d_4}^{a_1 a_2 a_3 a_4} \Gamma_{d_5]}
        =\\
        &=
        - 2 \tensor{\Gamma}{_{d_1\hdots d_5}^{a_1 \hdots a_5}}
        - 120 \delta_{[d_1}^{[a_1} \tensor{\Gamma}{_{\hdots d_5]}^{\hdots a_4]}}
        - 720 \delta_{[d_1 d_2}^{[a_1 a_2} \tensor{\Gamma}{_{\hdots d_5]}^{\hdots a_4]}}
        +\\
        &\quad
        - 480 \delta_{[d_1 d_2 d_3}^{[a_1 a_2 a_3} \tensor{\Gamma}{_{\hdots d_5]}^{a_4]}}
        - 2160 \delta_{[d_1 d_2 d_3 d_4}^{a_1 a_2 a_3 a_4} \Gamma_{d_5]}.
    \end{aligned}
\end{align}
\end{subequations}
\endgroup
Finally, the eighth term is
\begingroup
\allowdisplaybreaks
\begin{subequations}
\label{eq:sugra_calc:BI_T:b:solving:term_8}
\begin{align}
    &\begin{aligned}[b]
        \trace{(\tensor{\Gamma}{_{bc}^{a_1\hdots a_4}}  \Gamma_{d_1})} \Gamma^{bc}
        = 0,
    \end{aligned}
    \\
    &\begin{aligned}[b]
        \trace{(\tensor{\Gamma}{_{bc}^{a_1\hdots a_4}}  \Gamma_{d_1 d_2})} \Gamma^{bc}
        = 0,
    \end{aligned}
    \\
    &\begin{aligned}[b]
        \trace{(\tensor{\Gamma}{_{bc}^{a_1\hdots a_4}}  \Gamma_{d_1 \hdots d_5})} \Gamma^{bc}
        &=
        - 32 \tensor{\epsilon}{_{bc}^{a_1\hdots a_4}_{d_1\hdots d_5}} \Gamma^{bc}
        =\\
        &=
        - 64 \tensor{\Gamma}{_{d_1\hdots d_5}^{a_1\hdots a_4}},
    \end{aligned}
\end{align}
\end{subequations}
\endgroup
where we have used
\begin{equation}
    n!\, \Gamma^{a_1 \hdots a_{11-n}}
    = - \Gamma^{b_1 \hdots b_{n}} \tensor{\epsilon}{_{b_{n} \hdots b_1}^{a_1 \hdots a_{11-n}}},
\end{equation}
which follows immediately from \cref{eq:conventions:11d-spinors:Gamma_parity}.
Inserting \cref{eq:sugra_calc:BI_T:b:solving:term_1,eq:sugra_calc:BI_T:b:solving:term_2,eq:sugra_calc:BI_T:b:solving:term_3,eq:sugra_calc:BI_T:b:solving:term_4,eq:sugra_calc:BI_T:b:solving:term_5,eq:sugra_calc:BI_T:b:solving:term_6,eq:sugra_calc:BI_T:b:solving:term_7,eq:sugra_calc:BI_T:b:solving:term_8} in \cref{eq:sugra_calc:BI_T:b:solving} the parenthesis vanishes in all three cases.
Hence, \cref{eq:sugra_calc:BI_T:b} follows from what we already knew and does not constrain $H_{abcd}$.

\subsubsection{\Cref{eq:sugra_calc:BI_T:e}}
Since $R_{\gamma abd}$ is antisymmetric in its last two indices, $R_{\gamma abd} = R_{\gamma [ab] d} - R_{\gamma [ad] b} - R_{\gamma [bd] a}$.
Thus, \cref{eq:sugra_calc:BI_T:e} can equivalently be written as
\begin{equation}
\label{eq:sugra_calc:BI_T:e:solved}
    R_{\gamma abd}
    =
    \i \tensor{T}{_{ab}^\varepsilon} \Gamma_{d \varepsilon \gamma}
    - \i \tensor{T}{_{ad}^\varepsilon} \Gamma_{b \varepsilon \gamma}
    - \i \tensor{T}{_{bd}^\varepsilon} \Gamma_{a \varepsilon \gamma}.
\end{equation}
Since the right-hand side is antisymmetric in $b\, d$, this solves \cref{eq:sugra_calc:BI_T:e} completely.

\subsubsection{\Cref{eq:sugra_calc:BI_T:d}}
This equation can be expressed only in terms of $\tensor{T}{_a_b^\gamma}$ since, from \cref{eq:sugra_calc:BI_T:e:solved,eq:sugra_calc:BI_H:e:solved}
\begin{subequations}
\label{eq:sugra_calc:BI_T:d:pre-solving}
\begin{align}
    &\begin{aligned}[b]
        \tensor{R}{_a_{\beta\gamma}^\delta}
        &= -\frac{1}{4} \tensor{R}{_{\beta a c d}} \tensor{\Gamma}{^{cd}_\gamma^\delta}
        =\\
        &=\frac{\i}{4} \bigl(
        \tensor{T}{_{cd}^\varepsilon} \Gamma_{a \varepsilon \beta}
        - \tensor{T}{_{ac}^\varepsilon} \Gamma_{d \varepsilon \beta}
        + \tensor{T}{_{ad}^\varepsilon} \Gamma_{c \varepsilon \beta}
        \bigr)
        \tensor{\Gamma}{^{cd}_\gamma^\delta},
    \end{aligned}
    \\[2pt]
    &\begin{aligned}[b]
        \D_\beta \tensor{T}{_\gamma_a^\delta}
        &= -\D_\beta H_{c_1\hdots c_4}
        (k_1 \delta_a^{c_1} \tensor{\Gamma}{^{c_2 c_3 c_4}_\gamma^\delta} + k_2 \tensor{\Gamma}{_a^{c_1 c_2 c_3 c_4}_\gamma^\delta})
        =\\
        &=
        12\i \tensor{T}{_{[c_1 c_2}^\zeta}  \Gamma_{c_3 c_4] \zeta \beta}
        (k_1 \delta_a^{c_1} \tensor{\Gamma}{^{c_2 c_3 c_4}_\gamma^\delta} + k_2 \tensor{\Gamma}{_a^{c_1 c_2 c_3 c_4}_\gamma^\delta}).
    \end{aligned}
\end{align}
\end{subequations}
Inserting \cref{eq:sugra_calc:BI_T:d:pre-solving} in \cref{eq:sugra_calc:BI_T:d}, contracting with all symmetric $\tensor{\Gamma}{_{d_I}^{\beta \gamma}}$ and suppressing the spinor indices, we get
\begin{align}
    \nonumber
    &48 T_{[c_1 c_2} \Gamma_{c_3 c_4]} \Gamma_{d_I} (k_1 \delta_b^{[c_1} \Gamma^{\hdots c_4]} + k_2 \tensor{\Gamma}{_b^{c_1 \hdots c_4}})
    + 4 \trace{(\Gamma^a \Gamma_{d_I})} T_{ab}
    =\\
\label{eq:sugra_calc:BI_T:d:solving}
    &=
    T_{ca} \Gamma_b \Gamma_{d_I} \Gamma^{ca}
    - 2 T_{ab} \Gamma_c \Gamma_{d_I} \Gamma^{ca}.
\end{align}
Splitting the first term as
\begin{equation}
    48 \delta_b^{[c_1} T_{[c_1 c_2} \Gamma_{c_3 c_4]} \Gamma_{d_I}  \Gamma^{\hdots c_4]}
    =
    24 T_{b c_2} \Gamma_{c_3 c_4} \Gamma_{d_I} \Gamma^{c_2 c_3 c_4}
    + 24 T_{c_3 c_4} \Gamma_{b c_2} \Gamma_{d_I} \Gamma^{c_2 c_3 c_4},
\end{equation}
and using \cref{eq:sugra_calc:k1,eq:sugra_calc:k1_k2_relation}, \cref{eq:sugra_calc:BI_T:d:solving} becomes
\begin{align}
    \nonumber
    0 &=
    - 4 T_{b c_2} \Gamma_{c_3 c_4} \Gamma_{d_I} \Gamma^{c_2 c_3 c_4}
    - 4 T_{c_3 c_4} \Gamma_{b c_2} \Gamma_{d_I} \Gamma^{c_2 c_3 c_4}
    - T_{c_1 c_2} \Gamma_{c_3 c_4} \Gamma_{d_I} \tensor{\Gamma}{_b^{c_1 \hdots c_4}}
    +\\
\label{eq:sugra_calc:BI_T:d:solving:2}
    &\quad
    + 24 \trace{(\Gamma^a \Gamma_{d_I})} T_{ab}
    - 6 T_{c_1 c_2} \Gamma_b \Gamma_{d_I} \Gamma^{c_1 c_2}
    + 12 T_{c_2 b} \Gamma_{c_1} \Gamma_{d_I} \Gamma^{c_1 c_2}.
\end{align}
We have six terms to compute for each number of $d$-indices.
First, we decompose $\tensor{T}{_a_b^\gamma}$ into its irreducible components
\begin{equation}
    \tensor{T}{_a_b^\gamma}
    = \tensor{\tilde{T}}{_a_b^\gamma}
    + 2 \tensor{\tilde{T}}{_{[a}^\delta} \tensor{\Gamma}{_{b]}_\delta^\gamma}
    + \tilde{T}^\delta \tensor{\Gamma}{_{ab}_\delta^\gamma},
\end{equation}
where
\begin{equation}
    \tensor{\tilde{T}}{_a_b^\gamma} \Gamma^b_{\gamma \alpha}
    =
    0,
    \qquad\quad
    \tensor{\tilde{T}}{_a^\beta} \Gamma^a_{\beta \alpha}
    = 0.
\end{equation}
When computing the six terms above, we will need to contract one or both bosonic indices on $T$ with $\Gamma$-matrices with various numbers of indices.
To not have to redo the calculation, we compute the general contractions here.
First,
\begin{align}
    T_{ab} \Gamma^b
    =
    (\tilde{T}_{ab} + \tilde{T}_a \Gamma_b - \tilde{T}_b \Gamma_a + \tilde{T} \Gamma_{ab}) \Gamma^b
    =
    9 \tilde{T}_a + 10 \tilde{T} \Gamma_a.
\end{align}
Now,
\begin{align}
    \nonumber
    T_{ab} \Gamma^{b c_1 \hdots c_n}
    &=
    T_{ab} (\Gamma^b \Gamma^{c_1 \hdots c_n} - n \eta^{b[c_1} \Gamma^{\hdots c_n]})
    =\\ \nonumber
    &= (9 \tilde{T}_a + 10 \tilde{T} \Gamma_a) \Gamma^{c_1 \hdots c_n}
    - n \tensor{T}{_a^{[c_1}} \Gamma^{\hdots c_n]}
    =\\ \nonumber
    &= 9 \tilde{T}_a \Gamma^{c_1 \hdots c_n}
    + 10 \tilde{T} \tensor{\Gamma}{_a^{c_1\hdots c_n}}
    + 10 n \tilde{T} \delta_a^{[c_1} \Gamma^{\hdots c_n]}
    +\\ \nonumber
    &\quad
    -n \tensor{\tilde{T}}{_a^{[c_1}} \Gamma^{\hdots c_n]}
    -n \tensor{\tilde{T}}{_a} \Gamma^{[c_1} \Gamma^{\hdots c_n]}
    +n \tensor{\tilde{T}}{^{[c_1}} \Gamma_{a} \Gamma^{\hdots c_n]}
    -n \tilde{T} \tensor{\Gamma}{_a^{[c_1}} \Gamma^{\hdots c_n]}
    =\\ \nonumber
    &=
    -n \tensor{\tilde{T}}{_a^{[c_1}} \Gamma^{\hdots c_n]}
    +\\ \nonumber
    &\quad
    + (9-n) \tilde{T}_a \Gamma^{c_1\hdots c_n}
    + n \tilde{T}^{[c_1} \tensor{\Gamma}{_a^{\hdots c_n]}}
    + n(n-1) \tilde{T}^{[c_1} \delta^{c_2}_a \Gamma^{\hdots c_n]}
    +\\
\label{eq:sugra_calc:T_single_contraction}
    &\quad
    + (10-n) \tilde{T} \tensor{\Gamma}{_a^{c_1 \hdots c_n}}
    + n(11-n) \tilde{T} \delta_a^{[c_1} \Gamma^{\hdots c_n]}.
\end{align}
When contracting both indices, we get
\begin{align}
\label{eq:sugra_calc:T_double_contraction}
    T_{ab} \tensor{\Gamma}{^b^a_{c_1\hdots c_n}}
    &=
    T_{ab} \bigl(
    \Gamma^{ba} \Gamma_{c_1\hdots c_n}
    - 2 n \delta_{[c_1}^{[a} \tensor{\Gamma}{^{b]}_{\hdots c_n]}}
    - n (n-1) \tensor*{\delta}{*_{[c_1}^a_{c_2}^b} \Gamma_{\hdots c_n]}
    \bigr)
    =\\ \nonumber
    &=
    n(n-1) \tilde{T}_{[c_1 c_2} \Gamma_{\hdots c_n]}
    - 2n (10-n) \tilde{T}_{[c_1} \Gamma_{\hdots c_n]}
    + (110-21n+n^2) \tilde{T} \Gamma_{c_1\hdots c_n},
\end{align}
where we have used
\begin{subequations}
\begin{align}
    &T_{ab} \Gamma^{ba}
    = T_{ab} \Gamma^b \Gamma^a
    = (9 \tilde{T}_a + 10 \tilde{T} \Gamma_a) \Gamma^a
    = 110 \tilde{T},
    \\
    &T_{ab} \delta_{[c_1}^{[a} \tensor{\Gamma}{^{b]}_{\hdots c_n]}}
    =
    - (n-1) \tilde{T}_{[c_1 c_2} \Gamma_{\hdots c_n]}
    + (11-2n) \tilde{T}_{[c_1} \Gamma_{\hdots c_n]}
    + (11-n) \tilde{T} \Gamma_{c_1 \hdots c_n},
    \\
    &T_{ab} \tensor*{\delta}{*_{[c_1}^{\; a}_{c_2}^b} \Gamma_{\hdots c_n]}
    =
    \tilde{T}_{[c_1 c_2} \Gamma_{\hdots c_n]}
    + 2 \tilde{T}_{[c_1} \Gamma_{\hdots c_n]}
    + \tilde{T} \Gamma_{c_1 \hdots c_n}.
\end{align}
\end{subequations}
Since we might get constraints on some of the irreducible components of $T$, we calculate all six terms with a single $d$-index first
\begingroup
\allowdisplaybreaks
\begin{subequations}
\begin{align}
    &\begin{aligned}[b]
        T_{b c_2} \Gamma_{c_3 c_4} \Gamma_{d_1} \Gamma^{c_2 c_3 c_4}
        &=
        T_{b c_2} (-90 \delta_{d_1}^{c_2} + 54 \tensor{\Gamma}{^{c_2}_{d_1}})
        =\\
        &=
        - 90 (\tilde{T}_{b{d_1}} + \tilde{T}_b \Gamma_{d_1} - \tilde{T}_{d_1} \Gamma_b + \tilde{T} \Gamma_{b{d_1}})
        +\\
        &\quad
        + 54 (-\tilde{T}_{b{d_1}} + 8 \tilde{T}_b \Gamma_{d_1} + \tilde{T}_{d_1} \Gamma_b + 9 \tilde{T} \Gamma_{b{d_1}} + 10 \tilde{T} \eta_{b{d_1}})
        =\\
        &=
        - 144 \tilde{T}_{b{d_1}}
        + 342 \tilde{T}_b \Gamma_{d_1}
        + 144 \tilde{T}_{d_1} \Gamma_b
        + 396 \tilde{T} \Gamma_{b{d_1}}
        + 540 \tilde{T} \eta_{b{d_1}},
    \end{aligned}
    \\[2pt]
    &\begin{aligned}[b]
        T_{c_3 c_4} \Gamma_{b c_2} \Gamma_{d_1} \Gamma^{c_2 c_3 c_4}
        &=
        T_{c_3 c_4} (
        - 6 \tensor{\Gamma}{_{b{d_1}}^{c_3 c_4}}
        - 8 \eta_{b{d_1}} \Gamma^{c_3 c_4}
        - 14 \delta_b^{[c_3} \tensor{\Gamma}{^{c_4]}_{d_1}}
        + 16 \delta_{d_1}^{[c_3} \tensor{\Gamma}{^{c_4]}_b}
        + 18 \delta_{b\ {d_1}}^{c_3 c_4}
        )
        =\\
        &=
        +6 (2 \tilde{T}_{b{d_1}} - 16 \tilde{T}_b \Gamma_{d_1} + 16 \tilde{T}_{d_1} \Gamma_{b} + 72 \tilde{T} \Gamma_{b{d_1}})
        +\\
        &\quad
        +8 (110 \tilde{T} \eta_{b d_1})
        +\\
        &\quad
        - 14 (-\tilde{T}_{b{d_1}} + 8 \tilde{T}_b \Gamma_{d_1} + \tilde{T}_{d_1} \Gamma_b + 9 \tilde{T} \Gamma_{b{d_1}} + 10 \tilde{T} \eta_{b{d_1}})
        +\\
        &\quad
        + 16 (\tilde{T}_{b{d_1}} + \tilde{T}_b \Gamma_{d_1} + 8 \tilde{T}_{d_1} \Gamma_b - 9 \tilde{T} \Gamma_{b{d_1}} + 10 \tilde{T} \eta_{b{d_1}})
        +\\
        &\quad
        + 18 (\tilde{T}_{b{d_1}} + \tilde{T}_b \Gamma_{d_1} - \tilde{T}_{d_1} \Gamma_b + \tilde{T} \Gamma_{b{d_1}})
        =\\
        &=
        60 \tilde{T}_{b{d_1}}
        - 174 \tilde{T}_b \Gamma_{d_1}
        + 192 \tilde{T}_{d_1} \Gamma_b
        + 180 \tilde{T} \Gamma_{b{d_1}}
        + 900 \tilde{T} \eta_{b{d_1}},
    \end{aligned}
    \\[2pt]
    &\begin{aligned}[b]
        T_{c_1 c_2} \Gamma_{c_3 c_4} \Gamma_{d_1} \tensor{\Gamma}{_b^{c_1 \hdots c_4}}
        &=
        T_{c_1 c_2} (
        28 \tensor{\Gamma}{^{c_1 c_2}_{b{d_1}}}
        - 112 \delta_{d_1}^{[c_1} \tensor{\Gamma}{^{c_2]}_b}
        - 56 \eta_{b{d_1}} \Gamma^{c_1 c_2}
        )
        =\\
        &=
        -28 (2 \tilde{T}_{b{d_1}} - 16 \tilde{T}_b \Gamma_{d_1} + 16 \tilde{T}_{d_1} \Gamma_{b} + 72 \tilde{T} \Gamma_{b{d_1}})
        +\\
        &\quad
        - 112 (\tilde{T}_{b{d_1}} + \tilde{T}_b \Gamma_{d_1} + 8 \tilde{T}_{d_1} \Gamma_b - 9 \tilde{T} \Gamma_{b{d_1}} + 10 \tilde{T} \eta_{b{d_1}})
        +\\
        &\quad
        + 56 (110 \tilde{T} \eta_{b d_1})
        =\\
        &=
        - 168 \tilde{T}_{b{d_1}}
        + 336 \tilde{T}_b \Gamma_{d_1}
        - 1344 \tilde{T}_{d_1} \Gamma_b
        - 1008 \tilde{T} \Gamma_{b{d_1}}
        + 5040 \tilde{T} \eta_{b{d_1}},
    \end{aligned}
    \\[2pt]
    &\begin{aligned}[b]
        \trace{(\Gamma^a \Gamma_{d_1})} T_{ab}
        &=
        -32 T_{b{d_1}}
        =\\
        &=
        - 32 \tilde{T}_{b{d_1}}
        - 32 \tilde{T}_b \Gamma_{d_1}
        + 32 \tilde{T}_{d_1} \Gamma_b
        - 32 \tilde{T} \Gamma_{b{d_1}},
    \end{aligned}
    \\[2pt]
    &\begin{aligned}[b]
        T_{c_1 c_2} \Gamma_b \Gamma_{d_1} \Gamma^{c_1 c_2}
        &=
        T_{c_1 c_2} (
        \tensor{\Gamma}{_{bd_1}^{c_1 c_2}}
        + \eta_{bd_1} \Gamma^{c_1 c_2}
        - 4 \delta_{[b}^{[c_1} \tensor{\Gamma}{_{d_1]}^{c_2]}}
        - 2 \delta_{b\ d_1}^{c_1 c_2}
        )
        =\\
        &=
        - 1 (2 \tilde{T}_{b{d_1}} - 16 \tilde{T}_b \Gamma_{d_1} + 16 \tilde{T}_{d_1} \Gamma_{b} + 72 \tilde{T} \Gamma_{b{d_1}})
        +\\
        &\quad
        - 1 (110 \tilde{T} \eta_{b d_1})
        +\\
        &\quad
        + 4 (-\tilde{T}_{b d_1} + 7 \tilde{T}_{[b} \Gamma_{d_1]})
        +\\
        &\quad
        - 2 (\tilde{T}_{b{d_1}} + \tilde{T}_b \Gamma_{d_1} - \tilde{T}_{d_1} \Gamma_b + \tilde{T} \Gamma_{b{d_1}})
        =\\
        &=
        - 8 \tilde{T}_{b{d_1}}
        + 28 \tilde{T}_b \Gamma_{d_1}
        - 28 \tilde{T}_{d_1} \Gamma_b
        - 38 \tilde{T} \Gamma_{b{d_1}}
        - 110 \tilde{T} \eta_{b d_1},
    \end{aligned}
    \\[2pt]
    &\begin{aligned}[b]
        T_{c_2 b} \Gamma_{c_1} \Gamma_{d_1} \Gamma^{c_1 c_2}
        &=
        T_{c_2 b} (
        -8 \tensor{\Gamma}{_{d_1}^{c_2}}
        - 10 \delta_{d_1}^{c_2}
        )
        =\\
        &=
        - 8 (-\tilde{T}_{b{d_1}} + 8 \tilde{T}_b \Gamma_{d_1} + \tilde{T}_{d_1} \Gamma_b + 9 \tilde{T} \Gamma_{b{d_1}} + 10 \tilde{T} \eta_{b{d_1}})
        +\\
        &\quad
        - 10 (\tilde{T}_{b{d_1}} + \tilde{T}_b \Gamma_{d_1} - \tilde{T}_{d_1} \Gamma_b + \tilde{T} \Gamma_{b{d_1}})
        =\\
        &=
        18 \tilde{T}_{b{d_1}}
        - 54 \tilde{T}_b \Gamma_{d_1}
        - 18 \tilde{T}_{d_1} \Gamma_b
        - 62 \tilde{T} \Gamma_{b{d_1}}
        - 80 \tilde{T} \eta_{b d_1}.
    \end{aligned}
\end{align}
\end{subequations}
\endgroup
Collecting the terms, \cref{eq:sugra_calc:BI_T:d:solving:2} becomes
\begin{equation}
\label{eq:sugra_calc:BI_T:d:solving:3}
    0
    =
    0 \tilde{T}_{b{d_1}}
    - 2592 \tilde{T}_b \Gamma_{d_1}
    + 720 \tilde{T}_{d_1} \Gamma_b
    - 2580 \tilde{T} \Gamma_{b{d_1}}
    - 11100 \tilde{T} \eta_{b d_1}.
\end{equation}
Contracting $b\ d_1$ immediately gives $\tilde T = 0$ whence $2592 \tilde{T}_b \Gamma_{d_1} = 720 \tilde{T}_{d_1} \Gamma_b$.
Contracting the latter with $\Gamma^b$ we find $\tilde T_a = 0$ since $2\cdot 2592 \neq 11 \cdot 720$.

Having found that $\tensor{T}{_a_b^\gamma} = \tensor{\tilde T}{_a_b^\gamma}$, we move on to the case with two $d$-indices.
When contracting $T$ with $\Gamma$-matrices, we now only get the first terms in \cref{eq:sugra_calc:T_single_contraction,eq:sugra_calc:T_double_contraction}.
Hence,
\begingroup
\allowdisplaybreaks
\begin{subequations}
\begin{align}
    &\begin{aligned}[b]
        T_{b c_2} \Gamma_{c_3 c_4} \Gamma_{d_1 d_2} \Gamma^{c_2 c_3 c_4}
        &=
        T_{b c_2} (
        - 26 \tensor{\Gamma}{_{d_1 d_2}^{c_2}}
        + 108 \delta_{[d_1}^{c_2} \Gamma_{d_2]}
        )
        =\\
        &=
        - 26 (-2 \tilde{T}_{b[d_1} \Gamma_{d_2]})
        + 108 (\tilde{T}_{b[d_1}) \Gamma_{d_2]}
        =\\
        &= 160 \tilde{T}_{b[d_1} \Gamma_{d_2]},
    \end{aligned}
    \\[2pt]
    &\begin{aligned}[b]
        T_{c_3 c_4} \Gamma_{b c_2} \Gamma_{d_1 d_2} \Gamma^{c_2 c_3 c_4}
        &=
        T_{c_3 c_4} \bigl(
        4 \tensor{\Gamma}{_{d_1 d_2 b}^{c_3 c_4}}
        +\\
        &\qquad\quad\
        + 12 \eta_{b[d_1} \tensor{\Gamma}{_{d_2]}^{c_3 c_4}}
        + 10 \delta_b^{[c_3} \tensor{\Gamma}{_{d_1 d_2}^{c_4]}}
        + 24 \delta_{[d_1}^{[c_3} \tensor{\Gamma}{_{d_2] b}^{c_4]}}
         +\\
        &\qquad\quad\
        + 32 \eta_{b[d_1} \delta_{d_2]}^{[c_2} \Gamma^{c_4]}
        - 28 \delta_{b\ [d_1}^{c_3 c_4} \Gamma_{d_2]}
        - 16 \delta_{d_1 d_2}^{c_3 c_4} \Gamma_b
        \bigr)
        =\\
        &=
        4 (-6 \tilde{T}_{[d_1 d_2} \Gamma_{b]})
        + 12 \cdot 0
        + 10 (-2 \tilde{T}_{b[d_1} \Gamma_{d_2]})
        +\\
        &\quad
        + 24 (- \tilde{T}_{d_1 d_2} \Gamma_b + \tilde{T}_{[d_1|b|} \Gamma_{d_2]})
        + 32 \cdot 0
        - 28 \tilde{T}_{b[d_1} \Gamma_{d_2]}
        +\\
        &\quad
        - 16 \tilde{T}_{d_1 d_2} \Gamma_b
        =\\
        &=
        - 48 \tilde{T}_{d_1 d_2} \Gamma_b
        - 88 \tilde{T}_{b [d_1} \Gamma_{d_2]},
    \end{aligned}
    \\[2pt]
    &\begin{aligned}[b]
        T_{c_1 c_2} \Gamma_{c_3 c_4} \Gamma_{d_1 d_2} \tensor{\Gamma}{_b^{c_1 \hdots c_4}}
        &=
        T_{c_1 c_2} \bigl(
        - 8 \tensor{\Gamma}{_{d_1 d_2 b}^{c_1 c_2}}
        + 56 \eta_{b [d_1} \tensor{\Gamma}{_{d_2]}^{c_1 c_2}}
        - 112 \delta_{[d_1}^{[c_1} \tensor{\Gamma}{_{d_2] b}^{c_2]}}
        +\\
        &\qquad\quad\
        + 224 \eta_{b[d_1} \delta_{d_2]}^{[c_1} \Gamma^{c_2]}
        + 112 \delta_{d_1 d_2}^{c_1 c_2} \Gamma_b
        \bigr)
        =\\
        &=
        - 8 (-6 \tilde{T}_{[d_1 d_2} \Gamma_{b]})
        + 56 \cdot 0
        - 112 (-\tilde{T}_{d_1 d_2} \Gamma_b + \tilde{T}_{[d_1|b|} \Gamma_{d_2]})
        +\\
        &\quad
        + 224 \cdot 0
        + 112 (\tilde{T}_{d_1 d_2} \Gamma_b)
        =\\
        &=
        240 \tilde{T}_{d_1 d_2} \Gamma_b
        + 144 \tilde{T}_{b [d_1} \Gamma_{d_2]},
    \end{aligned}
    \\[2pt]
    &\begin{aligned}[b]
        \trace{(\Gamma^a \Gamma_{d_1 d_2})} T_{ab}
        &= 0,
    \end{aligned}
    \\[2pt]
    &\begin{aligned}[b]
        T_{c_1 c_2} \Gamma_b \Gamma_{d_1 d_2} \Gamma^{c_1 c_2}
        &=
        T_{c_1 c_2} \bigl(
        \tensor{\Gamma}{_{b d_1 d_2}^{c_1 c_2}}
        + 2 \eta_{b [d_1} \tensor{\Gamma}{_{d_2]}^{c_1 c_2}}
        + 2 \delta_b^{[c_1} \tensor{\Gamma}{_{d_1 d_2}^{c_2]}}
        - 4 \delta_{[d_1}^{[c_1} \tensor{\Gamma}{_{d_2]}^{c_2]}_b}
        +\\
        &\qquad\quad\
        + 4 \eta_{b [d_1} \delta_{d_2]}^{[c_1} \Gamma^{c_2]}
        - 4 \delta_{b\ [d_1}^{c_1 c_2} \Gamma_{d_2]}
        - 2 \delta_{d_1 d_2}^{c_1 c_2} \Gamma_b
        \bigr)
        =\\
        &=
        1 (-6 \tilde{T}_{[d_1 d_2} \Gamma_{b]})
        + 2 \cdot 0
        + 2 (-2 \tilde{T}_{b[d_1} \Gamma_{d_2]})
        +\\
        &\quad
        - 4 (-\tilde{T}_{d_1 d_2} \Gamma_b + \tilde{T}_{[d_1|b|} \Gamma_{d_2]})
        + 4 \cdot 0
        - 4 (\tilde{T}_{b[d_1} \Gamma_{d_2]})
        - 2 (\tilde{T}_{d_1 d_2} \Gamma_b)
        =\\
        &=
        - 8 \tilde{T}_{d_1 d_2} \Gamma_b
        - 16 \tilde{T}_{b [d_1} \Gamma_{d_2]},
    \end{aligned}
    \\[2pt]
    &\begin{aligned}[b]
        T_{c_2 b} \Gamma_{c_1} \Gamma_{d_1 d_2} \Gamma^{c_1 c_2}
        &=
        T_{c_2 b} (
        6 \tensor{\Gamma}{^{c_2}_{d_1 d_2}}
        -16 \delta_{[d_1}^{c_2} \Gamma_{d_2]}
        )
        =\\
        &=
        6 (2 \tilde{T}_{b[d_1} \Gamma_{d_2]})
        - 16 (-\tilde{T}_{b[d_1} \Gamma_{d_2]})
        =\\
        &=
        28 \tilde{T}_{b[d_1} \Gamma_{d_2]}.
    \end{aligned}
\end{align}
\end{subequations}
\endgroup
The terms sum to $0$, so we get no new information.

Lastly, we do the calculation with five $d$-indices
\begingroup
\allowdisplaybreaks
\begin{subequations}
\begin{align}
    &\begin{aligned}[b]
        T_{b c_2} \Gamma_{c_3 c_4} \Gamma_{d_1 \hdots d_5} \Gamma^{c_2 c_3 c_4}
        &=
        T_{b c_2} (
        10 \tensor{\Gamma}{_{d_1 \hdots d_5}^{c_2}}
        + 30 \delta_{[d_1}^{c_2} \Gamma_{\hdots d_5]}
        )
        =\\
        &=
        10 (5 \tilde{T}_{b [d_1} \Gamma_{\hdots d_5]})
        + 30 (\tilde{T}_{b [d_1} \Gamma_{\hdots d_5]})
        =\\
        &=
        80 \tilde{T}_{b [d_1} \Gamma_{\hdots d_5]},
    \end{aligned}
    \\[2pt]
    &\begin{aligned}[b]
        T_{c_3 c_4} \Gamma_{b c_2} \Gamma_{d_1 \hdots d_5} \Gamma^{c_2 c_3 c_4}
        &=
        T_{c_3 c_4} \bigl(
        - 2 \tensor{\Gamma}{_{d_1 \hdots d_5 b}^{c_3 c_4}}
        + 0 \eta_{b [d_1} \tensor{\Gamma}{_{\hdots d_5]}^{c_3 c_4}}
        - 2 \delta_b^{[c_3} \tensor{\Gamma}{_{d_1\hdots d_5}^{c_4]}}
        +\\
        &\qquad\quad\
        + 0 \delta_{[d_1}^{[c_3} \tensor{\Gamma}{_{\hdots d_5] b}^{c_4]}}
        + 10 \delta_{b\ [d_1}^{c_3 c_4} \Gamma_{\hdots d_5]}
        + 80 \eta_{b[d_1} \delta_{d_2}^{[c_3} \tensor{\Gamma}{_{\hdots d_5]}^{c_4]}}
        +\\
        &\qquad\quad\
        - 40 \delta_{[d_1 d_2}^{c_3 c_4} \Gamma_{\hdots d_5] b}
        + 240 \eta_{b [d_1} \delta_{d_2 d_3}^{c_3 c_4} \Gamma_{\hdots d_5]}
        \bigr)
        =\\
        &=
        - 2 (-30 \tilde{T}_{[d_1 d_2} \Gamma_{\hdots d_5] b})
        - 2 (5 \tilde{T}_{b[d_1} \Gamma_{\hdots d_5]})
        + 10 (\tilde{T}_{b[d_1} \Gamma_{\hdots d_5]})
        \\
        &\quad
        + 80 (3 \eta_{b[d_1} \tilde{T}_{d_2 d_3} \Gamma_{\hdots d_5]})
        - 40 (\tilde{T}_{[d_1 d_2} \Gamma_{\hdots d_5] b})
        \\
        &\quad
        -240 (\eta_{b [d_1} \tilde{T}_{d_2 d_3} \Gamma_{\hdots d_5]})
        =\\
        &=
        - 20 \tilde{T}_{b [d_1} \Gamma_{\hdots d_5]}
        + 0 \tilde{T}_{[d_1 d_2} \Gamma_{\hdots d_5] b}
        + 480 \eta_{b[d_1} \tilde{T}_{d_2 d_3} \Gamma_{\hdots d_5]},
    \end{aligned}
    \\[2pt]
    &\begin{aligned}[b]
        T_{c_1 c_2} \Gamma_{c_3 c_4} \Gamma_{d_1 \hdots d_5} \tensor{\Gamma}{_b^{c_1 \hdots c_4}}
        &=
        T_{c_1 c_2} \bigl(
        4 \tensor{\Gamma}{_{d_1 \hdots d_5 b}^{c_1 c_2}}
        + 40 \eta_{b [d_1} \tensor{\Gamma}{_{\hdots d_5]}^{c_1 c_2}}
        - 80 \delta_{[d_1}^{[c_1} \tensor{\Gamma}{_{\hdots d_5] b}^{c_2]}}
        +\\
        &\qquad\quad\
        - 160 \eta_{b[d_1} \delta_{d_2}^{[c_1} \tensor{\Gamma}{_{\hdots d_5]}^{c_2]}}
        - 80 \delta_{[d_1 d_2}^{c_1 c_2} \tensor{\Gamma}{_{\hdots d_5] b}}
        +\\
        &\qquad\quad\
        + 480 \eta_{b[d_1} \delta_{d_2 d_3}^{c_1 c_2} \Gamma_{\hdots d_5]}
        \bigr)
        =\\
        &=
        4 (-30 \tilde{T}_{[d_1 d_2} \Gamma_{\hdots d_5 b]})
        + 40 (-12 \eta_{b[d_1} \tilde{T}_{d_2 d_3} \Gamma_{\hdots d_5]})
        \\
        &\quad
        - 80 (5 \tilde{T}_{[d_1 d_2} \Gamma_{\hdots d_5 b]})
        - 160 (3 \eta_{b[d_1} \tilde{T}_{d_2 d_3} \Gamma_{\hdots d_5]})
        \\
        &\quad
        - 80 (\tilde{T}_{[d_1 d_2} \Gamma_{\hdots d_5] b})
        + 480 (\eta_{b[d_1} \tilde{T}_{d_2 d_3} \Gamma_{\hdots d_5]})
        =\\
        &=
        120 \tilde{T}_{b [d_1} \Gamma_{\hdots d_5]}
        - 480 \tilde{T}_{[d_1 d_2} \Gamma_{\hdots d_5] b}
        - 480 \eta_{b [d_1} \tilde{T}_{d_2 d_3} \Gamma_{\hdots d_5]},
    \end{aligned}
    \\[2pt]
    &\begin{aligned}[b]
        \trace{(\Gamma^a \Gamma_{d_1 \hdots d_5})} T_{ab}
        &= 0,
    \end{aligned}
    \\[2pt]
    &\begin{aligned}[b]
        T_{c_1 c_2} \Gamma_b \Gamma_{d_1 \hdots d_5} \Gamma^{c_1 c_2}
        &=
        T_{c_1 c_2} \bigl(
        \tensor{\Gamma}{_{b d_1 \hdots d_5}^{c_1 c_2}}
        - 2 \delta_b^{[c_1} \tensor{\Gamma}{_{d_1 \hdots d_5}^{c_2]}}
        + 10 \delta_{[d_1}^{[c_1} \tensor{\Gamma}{_{\hdots d_5] b}^{c_2]}}
        +\\
        &\qquad\quad\
        + 5 \eta_{b [d_1} \tensor{\Gamma}{_{\hdots d_5]}^{c_1 c_2}}
        - 10 \delta_{b\ [d_1}^{c_1 c_2} \Gamma_{\hdots d_5]}
        + 20 \delta_{[d_1 d_2}^{c_1 c_2} \Gamma_{\hdots d_5] b}
        +\\
        &\qquad\quad\
        - 40 \eta_{b [d_1} \delta_{d_2}^{[c_1} \tensor{\Gamma}{_{\hdots d_5]}^{c_2]}}
        - 60 \eta_{b[d_1} \delta_{d_2 d_3}^{c_1 c_2} \Gamma_{\hdots d_5]}
        \bigr)
        =\\
        &=
        (-30 \tilde{T}_{[b d_1} \Gamma_{\hdots d_5]})
        - 2 (5 \tilde{T}_{b [d_1} \Gamma_{\hdots d_5]})
        + 10 (5 \tilde{T}_{[d_1 d_2} \Gamma_{\hdots d_5 b]})
        \\
        &\quad
        + 5 (-60 \eta_{b [d_1} \tilde{T}_{d_2 d_3} \Gamma_{\hdots d_5]})
        - 10 (\tilde{T}_{b[d_1} \Gamma_{\hdots d_5]})
        + 20 (\tilde{T}_{[d_1 d_2} \Gamma_{\hdots d_5] b})
        \\
        &\quad
        - 40 (3 \eta_{b[d_1} \tilde{T}_{d_2 d_3} \Gamma_{\hdots d_5]})
        - 60 (\eta_{b[d_1} \tilde{T}_{d_2 d_3} \Gamma_{\hdots d_5]})
        =\\
        &=
        - 40 \tilde{T}_{b [d_1} \Gamma_{\hdots d_5]}
        + 80 \tilde{T}_{[d_1 d_2} \Gamma_{\hdots d_5] b}
        - 240 \eta_{b [d_1} \tilde{T}_{d_2 d_3} \Gamma_{\hdots d_5]},
    \end{aligned}
    \\[2pt]
    &\begin{aligned}[b]
        T_{c_2 b} \Gamma_{c_1} \Gamma_{d_1 \hdots d_5} \Gamma^{c_1 c_2}
        &=
        T_{c_2 b} (
        0 \tensor{\Gamma}{_{d_1 \hdots d_5}^{c_2}}
        - 10 \delta_{[d_1}^{c_2} \Gamma_{\hdots d_5]}
        )
        =\\
        &= 10 \tilde{T}_{b[d_1} \Gamma_{\hdots d_5]}.
    \end{aligned}
\end{align}
\end{subequations}
\endgroup
Again, the terms sum to $0$.
To conclude, we have solved \cref{eq:sugra_calc:BI_T:d} completely and found that $\tensor{T}{_a_b^\gamma}$ consists of the single irreducible part $\tensor{\tilde T}{_a_b^\gamma}$.
Equivalently, this can be expressed as
\begin{equation}
\label{eq:sugra_calc:T_eom}
    \tensor{T}{_a_b^\gamma} \tensor{\Gamma}{^a^b^c_\gamma^\delta}
    = 0,
\end{equation}
as can be seen from \cref{eq:sugra_calc:T_single_contraction,eq:sugra_calc:T_double_contraction}.
This is the equation of motion for $\tensor{T}{_a_b^\gamma}$.

\subsubsection{\Cref{eq:sugra_calc:BI_T:f}}
The equation reads
\begin{equation}
\label{eq:sugra_calc:BI_T:f:solving}
    \D_\gamma \tensor{T}{_a_b^\delta}
    = \tensor{R}{_a_b_\gamma^\delta}
    + 2 \tensor{T}{_\gamma_{[a}^\varepsilon} \tensor{T}{_{b]}_\varepsilon^\delta}
    - 2 \D_{[a} \tensor{T}{_{b]}_\gamma^\delta},
\end{equation}
where the right-hand side can be expressed in terms of $R_{abcd}$ and $H_{abcd}$ by using \cref{eq:sugra_calc:T_constraints} and that $R_{abcd}$ is Lie algebra-valued.%
\todo[disable]{}

Contracting \cref{eq:sugra_calc:BI_T:f:solving} with $\delta_\delta^\gamma$, we find
\begin{equation}
\label{eq:sugra_calc:BI_T:f:solving_G0_contraction}
    \D_\gamma \tensor{T}{_a_b^\gamma}
    + 2 \tensor{T}{_\gamma_{[a}^\varepsilon} \tensor{T}{_{|\varepsilon|b]}^\gamma}
    = 0.
\end{equation}
Contracting \cref{eq:sugra_calc:BI_T:f:solving} with $\tensor{\Gamma}{^b_d_\delta^\gamma}$, using $T_{ab} \Gamma^{bc} = - \tensor{T}{_a^c}$, \cref{eq:sugra_calc:BI_T:f:solving_G0_contraction} and $\tensor{R}{_a_b} \coloneqq \tensor{R}{_d_a_b^d}$,%
\footnote{We contract the outermost indices of the curvature tensor to define the Ricci tensor due to the right-action convention. Thus, $R$ gets its usual sign: for instance, $R < 0$ for AdS.}
\begin{equation}
\label{eq:sugra_calc:BI_T:f:solving_G2_contraction}
    -16 R_{ad}
    = 2 \tensor{T}{_\gamma_{[a}^\varepsilon} \tensor{T}{_{|\varepsilon|d]}^\gamma}
    + 2 \tensor{T}{_\gamma_{[a}^\varepsilon} \tensor{T}{_{|\varepsilon|b]}^\delta} \tensor{\Gamma}{^b_d_\delta^\gamma}
    + 2 \D_{[a} \tensor{T}{_{b]}_\gamma^\delta} \tensor{\Gamma}{^b_d_\delta^\gamma}.
\end{equation}
Using the previous results, only $\Gamma$-algebra remain to get the equation of motion for $R$.
First,
\begin{equation}
    \D_{[a} \tensor{T}{_{b]}_\gamma^\delta} \tensor{\Gamma}{^b_d_\delta^\gamma} = 0,
\end{equation}
since $\tensor{T}{_b_\gamma^\delta}$ only contains $\Gamma^{(3)}$ and $\Gamma^{(5)}$, that is, $\Gamma$-matrices with three and five indices.
Next,
\begin{align}
\nonumber
    \tensor{T}{_\gamma_{[a}^\varepsilon} \tensor{T}{_{|\varepsilon|d]}^\gamma}
    &= H_{b_1\hdots b_4} H^{c_1\hdots c_4} \bigl[
    k_1^2 \trace{(\delta_{[a}^{b_1} \Gamma^{\hdots b_4} \eta_{d] c_1} \Gamma_{\hdots c_4} )}
    + k_2^2 \trace{( \tensor{\Gamma}{_{[a}^{b_1 \hdots b_4}} \Gamma_{d]c_1\hdots c_4} )}
    \bigr]
    =\\ \nonumber
    &=
    H_{b_1\hdots b_4} H^{c_1\hdots c_4} \bigl[
    -32\cdot 6 k_1^2 \delta_{[a}^{b_1} \eta_{d] c_1} \delta_{c_2}^{b_2}{}_{c_3}^{b_3}{}_{c_4}^{b_4}
    + 32 \frac{4\cdot 5!}{5} k_2^2 \eta_{c_1 [a} \delta_{d]}^{b_1}{}_{c_2}^{b_2}{}_{c_3}^{b_3}{}_{c_4}^{b_4}
    \bigr]
    =\\ \nonumber
    &=
    \frac{192}{36^2} H_{c_2 c_3 c_3 [a} \tensor{H}{_{d]}^{c_2 c_3 c_4}}
    - \frac{3072}{288^2} H_{c_2 c_3 c_3 [a} \tensor{H}{_{d]}^{c_2 c_3 c_4}}
    =\\
    &= 0.
\end{align}
The second term in the right-hand side of \cref{eq:sugra_calc:BI_T:f:solving_G2_contraction} is
\begin{equation}
    2 \tensor{T}{_\gamma_{[a}^\varepsilon} \tensor{T}{_{|\varepsilon|b]}^\delta} \tensor{\Gamma}{^b_d_\delta^\gamma}
    =
    \trace{(T_a T_b \tensor{\Gamma}{^b_d} - T_b T_a \tensor{\Gamma}{^b_d})}
    =
    \trace{\bigl[T_a (T_b \tensor{\Gamma}{^b_d} - \tensor{\Gamma}{^b_d} T_b) \bigr]},
\end{equation}
where we have suppressed spinor indices and treated $\tensor{T}{_a_\beta^\gamma}$ as a matrix $\tensor{(T_a)}{_\beta^\gamma}$.
Since $T_a$ only contains $\Gamma^{(3)}$ and $\Gamma^{(5)}$, we need only keep $\Gamma^{(3,5,6,8)}$-terms in the parenthesis.
Dropping other terms that do not contribute to the trace, indicated by $\simeq$ below, we compute
\begingroup
\allowdisplaybreaks
\begin{subequations}
\begin{align}
    &\begin{aligned}[b]
        T_b \tensor{\Gamma}{^b_d}
        &= H_{a_1\hdots a_4} \bigl(
        k_1 \delta_b^{a_1} \Gamma^{\hdots a_4} \tensor{\Gamma}{^b_d}
        + k_2 \tensor{\Gamma}{_b^{a_1\hdots a_4}} \tensor{\Gamma}{^b_d}
        \bigr)
        \simeq\\
        &\simeq
        H_{a_1\hdots a_4} \bigr[
        (3 k_1 - 28 k_2) \delta_d^{a_1} \Gamma^{\hdots a_4}
        + (-k_1 + 6 k_2) \tensor{\Gamma}{_d^{a_1\hdots a_4}}
        \bigl],
    \end{aligned}
    \\[2pt]
    &\begin{aligned}[b]
        \tensor{\Gamma}{^b_d} T_b
        &= H_{a_1\hdots a_4} \bigl(
        k_1 \delta_b^{a_1} \tensor{\Gamma}{^b_d} \Gamma^{\hdots a_4}
        + k_2 \tensor{\Gamma}{^b_d} \tensor{\Gamma}{_b^{a_1\hdots a_4}}
        \bigr)
        \simeq\\
        &\simeq
        H_{a_1\hdots a_4} \bigr[
        (- 3 k_1 - 28 k_2) \delta_d^{a_1} \Gamma^{\hdots a_4}
        + (-k_1 - 6 k_2) \tensor{\Gamma}{_d^{a_1\hdots a_4}}
        \bigl],
    \end{aligned}
\end{align}
\end{subequations}
\endgroup
whence
\begin{align}
    \nonumber
    &\negphantom{\trace}\trace{\bigl[T_a (T_b \tensor{\Gamma}{^b_d} - \tensor{\Gamma}{^b_d} T_b) \bigr]}
    = \trace{\bigl[T_a H_{c_1\hdots c_4} (6k_1 \delta_d^{c_1} \Gamma^{\hdots c_4} + 12 k_2 \tensor{\Gamma}{_d^{c_1\hdots c_4}}) \bigr]}
    =\\ \nonumber
    &=
    H_{e_1\hdots e_4} H^{c_1\hdots c_4} \bigl[
    6 k_1^2 \delta_a^{e_1} \eta_{d c_1} \trace{(\Gamma^{e_2 e_3 e_4} \Gamma_{c_2 c_3 c_4})}
    + 12 k_2^2 \trace{(\tensor{\Gamma}{_a^{e_1\hdots e_4}}\tensor{\Gamma}{_{d c_1\hdots  c_4}})}
    \bigr]
    =\\ \nonumber
    &=
    -3!\cdot 32\cdot 6 k_1^2 H_{a c_2 c_3 c_4} H_d{}^{c_2 c_3 c_4}
    +\\ \nonumber
    &\quad
    + 4!\cdot 32\cdot 12 k_2^2 \eta_{ab} H^2
    - 4!\cdot 32 \cdot 4 \cdot 12 k_2^2 H_{d c_2 c_3 c_4} H_a{}^{c_2 c_3 c_4}
    =\\
    &=
    -\frac{4}{3} H_{a c_2 c_3 c_4} H_d{}^{c_2 c_3 c_4}
    +\frac{1}{9} \eta_{ad} H^2,
\end{align}
where $H^2 \coloneqq H_{a_1\hdots a_4} H^{a_1\hdots a_4}$ and we used, in the second to last step,
\begin{equation}
    \trace{(\tensor{\Gamma}{_a^{e_1\hdots e_4}}\tensor{\Gamma}{_{d c_1\hdots  c_4}})}
    = \eta_{ad} \trace{(\Gamma^{e_1\hdots e_4}\Gamma_{c_1\hdots c_4})}
    - 4 \delta_d^{[e_1} \trace{(\tensor{\Gamma}{_a^{\hdots e_4]}} \Gamma_{c_1\hdots c_4})}.
\end{equation}
Inserting the above in \cref{eq:sugra_calc:BI_T:f:solving_G2_contraction}, we find
\begin{equation}
    -16 R_{ad}
    = \frac{1}{9} \eta_{ad} H^2
    - \frac{4}{3} H_{a c_2 c_3 c_4} H_d{}^{c_2 c_3 c_4}.
\end{equation}
To write this with the Einstein tensor in the left-hand side, we contract this and find $-16 R = - H^2/9$.
Thus,
\begin{equation}
\label{eq:sugra_calc:R_eom}
    R_{ab} - \frac{1}{2} \eta_{ab} R
    = \frac{1}{12} H_{a cde} H_b{}^{cde}
    - \frac{1}{96} \eta_{ab} H^2.
\end{equation}

We now turn to the equation of motion for $H$.
The strategy is similar to the $R$-equation but we contract \cref{eq:sugra_calc:BI_T:f:solving} with other combinations of $\Gamma$-matrices.
First contract with $\tensor{\Gamma}{_c_\delta^\gamma}$ and then antisymmetrise $a\, b\, c$.
Since the curvature tensor is Lie algebra-valued in its last pair of indices, $\tensor{R}{_{[ab|\gamma|}^\delta} \tensor{\Gamma}{_{c]}_\delta^\gamma} = 0$.
Similarly, $2 \D_{[a} \tensor{T}{_{b|\gamma|}^\delta} \tensor{\Gamma}{_{c]}_\delta^\gamma} = 0$ since $\tensor{T}{_b_\gamma^\delta}$ only contains $\Gamma^{(3,5)}$.
Lastly,
\begin{align}
\nonumber
    2 \tensor{T}{_\gamma_{[a}^\varepsilon} \tensor{T}{_{|\varepsilon|b}^\delta} \tensor{\Gamma}{_{c]}_\delta^\gamma}
    &= 2 k_2^2 H_{d_1\hdots d_4} H_{e_1\hdots e_4} \trace{(\tensor{\Gamma}{_{[a}^{d_1\hdots d_4}} \tensor{\Gamma}{_d^{e_1\hdots e_4}} \Gamma_{c]})}
    =\\ \nonumber
    &=
    -64 k_2^2 H_{d_1\hdots d_4} H_{e_1\hdots e_4} \epsilon_a{}^{d_1\hdots d_4}{}_b{}^{e_1\hdots e_4}{}_c
    =\\
    &=
    -\frac{1}{1296} \epsilon_{abc}{}^{d_1\hdots d_4 e_1\hdots e_4} H_{d_1\hdots d_4} H_{e_1\hdots e_4},
\end{align}
whence the contracted and antisymmetrised \cref{eq:sugra_calc:BI_T:f:solving} becomes
\begin{equation}
\label{eq:sugra_calc:BI_T:f:solving_G1_contraction}
    \D_\gamma T_{[ab}{}^\delta \Gamma_{c]\delta}{}^\gamma
    = \frac{1}{1296} \epsilon_{abc}{}^{d_1\hdots d_4 e_1\hdots e_4} H_{d_1\hdots d_4} H_{e_1\hdots e_4}.
\end{equation}
Now we contract \cref{eq:sugra_calc:BI_T:f:solving} with $\tensor{\Gamma}{^b_{cd\delta}^\gamma}$ and then antisymmetrise $a\, c\, d$.
The term containing $R$ is again zero since $R_{ab}$ is Lie algebra-valued.
Using \cref{eq:sugra_calc:BI_T:f:solving_G1_contraction}, the term in the left-hand side of \cref{eq:sugra_calc:BI_T:f:solving} gives
\begin{equation}
    \D_\gamma \tensor{T}{_{[a|b|}^\delta} \tensor{\Gamma}{^b_{cd]\delta}^\gamma}
    = -2 \D_\gamma \tensor{T}{_{[ac}^\delta} \tensor{\Gamma}{_{d]}_\delta^\gamma}
    = - \frac{1}{648} \epsilon_{acd}{}^{e_1\hdots e_4 f_1\hdots f_4} H_{e_1\hdots e_4} H_{f_1\hdots f_4}.
\end{equation}
The last term in the right-hand side of \cref{eq:sugra_calc:BI_T:f:solving} splits into
\begin{align}
\nonumber
    \D_b \tensor{T}{_{[a|\gamma|}^\delta} \tensor{\Gamma}{^b_{cd]\delta}^\gamma}
    &= k_1 \D^b H_{[a|e_2 e_3 e_4} \trace{(\Gamma^{e_2 e_3 e_4} \Gamma_{b|cd]})}
    = -32\cdot 6 k_1 \D^b H_{abcd}
    =\\
    &= -\frac{16}{3} \D^b H_{bacd},
\end{align}
and
\begin{equation}
    -\D_{[a|} \tensor{T}{_{b\gamma}^\delta} \tensor{\Gamma}{^b_{|cd]\delta}^\gamma}
    = -k_1 \D_{[a|} H_{b e_2 e_3 e_4} \trace{(\Gamma^{e_2 e_3 e_4}\Gamma_{b|cd]})}
    = 0.
\end{equation}
The second term in the right-hand side of \cref{eq:sugra_calc:BI_T:f:solving} splits into $\tensor{T}{_\gamma_{[a|}^\varepsilon} \tensor{T}{_b_\varepsilon^\delta} \tensor{\Gamma}{^b_{|cd]\delta}^\gamma}$ and $\tensor{T}{_\gamma_{b}^\varepsilon} \tensor{T}{_\varepsilon_{[a}^\delta} \tensor{\Gamma}{^b_{cd]\delta}^\gamma}$.
Since the components of $T$ only contain $\Gamma^{(3)}$ and $\Gamma^{(5)}$ the only potentially nonvanishing contributions come from products $\Gamma^{(3)} \Gamma^{(5)} \Gamma^{(3)}$ and $\Gamma^{(5)} \Gamma^{(3)} \Gamma^{(3)}$.
However,
\begin{equation}
    H_{e_1\hdots e_4} H_{f_1\hdots f_4} \trace{( \delta_{[a}^{e_1}\Gamma^{\hdots e_4} \tensor{\Gamma}{_{|b|}^{f_1\hdots f_4}} \tensor{\Gamma}{^b_{cd]}} )}
    = 0,
\end{equation}
since this is really only 9 $\Gamma$-matrices due to the contracted $b$'s.
On the other hand,
\begin{equation}
    H_{e_1\hdots e_4} H_{f_1\hdots f_4} \trace{( \tensor{\Gamma}{_{[a}^{e_1\hdots e_4}} \delta_{|b|}^{f_1} \Gamma^{\hdots f_4}  \tensor{\Gamma}{^b_{cd]}} )}
    = 32 \tensor{\epsilon}{_{acd}^{e_1\hdots e_4 f_1\hdots f_4}} H_{e_1\hdots e_4} H_{f_1\hdots f_4},
\end{equation}
whence
\begin{equation}
    \tensor{T}{_\gamma_{[a|}^\varepsilon} \tensor{T}{_b_\varepsilon^\delta} \tensor{\Gamma}{^b_{|cd]\delta}^\gamma}
    = -\frac{1}{324} \tensor{\epsilon}{_{acd}^{e_1\hdots e_4 f_1\hdots f_4}} H_{e_1\hdots e_4} H_{f_1\hdots f_4}.
\end{equation}
$\tensor{T}{_\gamma_{b}^\varepsilon} \tensor{T}{_\varepsilon_{[a}^\delta} \tensor{\Gamma}{^b_{cd]\delta}^\gamma}$ is similar.
As above, only the term where the $b$-index is on the $\delta$ is nonzero.
Hence
\begin{align}
    \tensor{T}{_\gamma_{b}^\varepsilon} \tensor{T}{_\varepsilon_{[a}^\delta} \tensor{\Gamma}{^b_{cd]\delta}^\gamma}
    &=
    k_1 k_2 H_{e_1\hdots e_4} H_{f_1\hdots f_4} \trace{( \delta_b^{e_1}\tensor{\Gamma}{^{\hdots e_4}} \tensor{\Gamma}{_{[a}^{f_1\hdots f_4}} \tensor{\Gamma}{^b_{cd]}} )}
    =\\
    &= -\frac{1}{324} \tensor{\epsilon}{_{acd}^{e_1\hdots e_4 f_1\hdots f_4}} H_{e_1\hdots e_4} H_{f_1\hdots f_4}.
\end{align}
Inserting the above terms in \cref{eq:sugra_calc:BI_T:f:solving}, we get
\begin{equation}
\label{eq:sugra_calc:H_eom}
    \D^d H_{dabc} = - \frac{1}{1152} \tensor{\epsilon}{_a_b_c^{e_1 \hdots e_4}^{f_1 \hdots f_4}} H_{e_1 \hdots e_4} H_{f_1 \hdots f_4}.
\end{equation}

\subsubsection{The Bianchi identity of the second type}
Although the Bianchi identity of the second type, $\D \tensor{R}{_B^A} = 0$, is not independent of the above, it can be used to extract $\D_\varepsilon R_{abcd}$.
Alternatively, this could be obtained from \cref{eq:sugra_calc:BI_T:f:solving} by applying another covariant derivative and using the Bianchi identity of the first type.
From \cref{eq:sugra:tensor_BI:R},
\begin{equation}
\label{eq:sugra_calc:D_alpha_R}
    \D_\varepsilon R_{abcd}
    + 2 \D_{[b} R_{|\varepsilon| a] c d}
    + \tensor{T}{_a_b^\zeta} R_{\zeta \varepsilon c d}
    + 2 \tensor{T}{_\varepsilon_{[a}^\zeta} R_{|\zeta|b] cd}
    = 0.
\end{equation}
Using \cref{eq:sugra_calc:BI_T:c:solved,eq:sugra_calc:BI_T:e:solved,eq:sugra_calc:T_constraints}, we see that $\D_\varepsilon R_{abcd}$ can be expressed in terms of $H_{abcd}$, $\tensor{T}{_a_b^\gamma}$ and their $\D_a$-derivatives.

\clearpage
\section{Solution to the supergravity Bianchi identities} \label{sec:sugra_calc:results}%
To conclude, we have found (\cref{eq:sugra_calc:T_constraints,eq:sugra_calc:k1_k2})
\begin{equation}
    \tensor{T}{_a_\beta^\gamma}
    = - \frac{1}{288} H_{bcde} \bigl(
        8 \delta_a^{[b} \Gamma\indices{^{cde]}_\beta^\gamma}
        +  \Gamma\indices{_a^{bcde}_\beta^\gamma}
    \bigr),
\end{equation}
the Bianchi identities for the field strengths $\tensor{T}{_a_b^\gamma}$, $\tensor{R}{_a_b_c^d}$ and $\tensor{H}{_a_b_c_d}$ (\cref{eq:sugra_calc:BI_T:g,eq:sugra_calc:BI_T:h,eq:sugra_calc:BI_H:f})
\begin{subequations}
\begin{align}
    & \D_{[a} \tensor{T}{_{bc]}^\delta}
    + \tensor{T}{_{[ab}^\varepsilon} \tensor{T}{_{|\varepsilon|c]}^\delta}
    = 0,
    \\
    & \tensor{R}{_{[abc]}^d} = 0,
    \\
    & \D_{[a} H_{bcde]} = 0,
\end{align}
\end{subequations}
the equations of motion (\cref{eq:sugra_calc:T_eom,eq:sugra_calc:R_eom,eq:sugra_calc:H_eom})
\begin{subequations}
\begin{align}
    & \tensor{T}{_a_b^\gamma} \tensor{\Gamma}{^a^b^c_\gamma^\delta}
    = 0,
    \\
    & R_{ab} - \frac{1}{2} \eta_{ab} R
    = \frac{1}{12} H_{a cde} H_b{}^{cde}
    - \frac{1}{96} \eta_{ab} H^2,
    \\
    & \D^d H_{dabc}
    = - \frac{1}{1152} \tensor{\epsilon}{_a_b_c^{e_1 \hdots e_4}^{f_1 \hdots f_4}} H_{e_1 \hdots e_4} H_{f_1 \hdots f_4},
\end{align}
\end{subequations}
as well as equations relating the $\D_\alpha$-derivatives of the field strengths to their values and $\D_a$-derivatives (\cref{eq:sugra_calc:BI_H:e:solved,eq:sugra_calc:BI_T:f:solving,eq:sugra_calc:D_alpha_R}).
This implies that only the $\theta = 0$ components of the field strengths are independent component fields.

\if\thesisStatus d
    \include{appendices/additional_material/additional_material}
\fi

% REFERENCES / BIBLIOGRAPHY
% Use a separate file to be able to exclude with \includeonly
\printbibliography[title={\thesisBibName}, heading=bibintoc]

% LAST PAGE
\if\thesisArXiv y
\else
\if\thesisPrint y
\else
\if\thesisStatus f
    \include{template/pages/lastpage}
\fi
\fi
\fi

\end{document}